\documentclass[prx, twocolumn, superscriptaddress, longbibliography]{revtex4-1}
\usepackage{bm, amsmath, amsfonts, amssymb, braket}
\usepackage{multirow}
\usepackage[dvipdfmx]{graphicx}
\usepackage{float}
\usepackage{color}
\usepackage{comment}

\newcommand{\ii}{\text{i}}
\newcommand{\Z}{$\mathbb{Z}$}
\newcommand{\Zt}{$\mathbb{Z}_{2}$}
\newcommand{\Tp}{{\cal T}_{+}}
\newcommand{\Tm}{{\cal T}_{-}}
\newcommand{\Cp}{{\cal C}_{+}}
\newcommand{\Cm}{{\cal C}_{-}}
\newcommand{\CS}{\Gamma}
\newcommand{\SLS}{{\cal S}}
\newcommand{\Lr}{~${\rm L}_{\rm r}$}
\newcommand{\Li}{~${\rm L}_{\rm i}$}

\newcommand{\Ca}{${\cal C}_{0}$ & \Z & $0$ & \Z & $0$ & \Z & $0$ & \Z & $0$}
\newcommand{\Caa}{${\cal C}_{0} \times {\cal C}_{0}$ & \Z\,$\oplus$\,\Z & $0$ & \Z\,$\oplus$\,\Z & $0$ & \Z\,$\oplus$\,\Z & $0$ & \Z\,$\oplus$\,\Z & $0$}
\newcommand{\Cb}{${\cal C}_{1}$ & $0$ & \Z & $0$ & \Z & $0$ & \Z & $0$ & \Z}
\newcommand{\Cbb}{${\cal C}_{1} \times {\cal C}_{1}$ & $0$ & \Z\,$\oplus$\,\Z & $0$ & \Z\,$\oplus$\,\Z & $0$ & \Z\,$\oplus$\,\Z & $0$ & \Z\,$\oplus$\,\Z}
\newcommand{\Ra}{${\cal R}_{0}$ & \Z & $0$ & $0$ & $0$ & $2$\Z & $0$ & \Zt & \Zt}
\newcommand{\Raa}{${\cal R}_{0} \times {\cal R}_{0}$ & \Z\,$\oplus$\,\Z & $0$ & $0$ & $0$ & $2$\Z\,$\oplus$\,$2$\Z & $0$ & \Zt\,$\oplus$\,\Zt & \Zt\,$\oplus$\,\Zt}
\newcommand{\Rb}{${\cal R}_{1}$ & \Zt & \Z & $0$ & $0$ & $0$ & $2$\Z & $0$ & \Zt}
\newcommand{\Rbb}{${\cal R}_{1} \times {\cal R}_{1}$ & \Zt\,$\oplus$\,\Zt & \Z\,$\oplus$\,\Z & $0$ & $0$ & $0$ & $2$\Z\,$\oplus$\,$2$\Z & $0$ & \Zt\,$\oplus$\,\Zt}
\newcommand{\Rc}{${\cal R}_{2}$ & \Zt & \Zt & \Z & $0$ & $0$ & $0$ & $2$\Z & $0$}
\newcommand{\Rcc}{${\cal R}_{2} \times {\cal R}_{2}$ & \Zt\,$\oplus$\,\Zt & \Zt\,$\oplus$\,\Zt & \Z\,$\oplus$\,\Z & $0$ & $0$ & $0$ & $2$\Z\,$\oplus$\,$2$\Z & $0$}
\newcommand{\Rd}{${\cal R}_{3}$ & $0$ & \Zt & \Zt & \Z & $0$ & $0$ & $0$ & $2$\Z}
\newcommand{\Rdd}{${\cal R}_{3} \times {\cal R}_{3}$ & $0$ & \Zt\,$\oplus$\,\Zt & \Zt\,$\oplus$\,\Zt & \Z\,$\oplus$\,\Z & $0$ & $0$ & $0$ & $2$\Z\,$\oplus$\,$2$\Z}
\newcommand{\Ree}{${\cal R}_{4}$ & $2$\Z & $0$ & \Zt & \Zt & \Z & $0$ & $0$ & $0$}
\newcommand{\Reee}{${\cal R}_{4} \times {\cal R}_{4}$ & $2$\Z\,$\oplus$\,$2$\Z & $0$ & \Zt\,$\oplus$\,\Zt & \Zt\,$\oplus$\,\Zt & \Z\,$\oplus$\,\Z & $0$ & $0$ & $0$}
\newcommand{\Rf}{${\cal R}_{5}$ & $0$ & $2$\Z & $0$ & \Zt & \Zt & \Z & $0$ & $0$}
\newcommand{\Rff}{${\cal R}_{5} \times {\cal R}_{5}$ & $0$ & $2$\Z\,$\oplus$\,$2$\Z & $0$ & \Zt\,$\oplus$\,\Zt & \Zt\,$\oplus$\,\Zt & \Z\,$\oplus$\,\Z & $0$ & $0$}
\newcommand{\Rg}{${\cal R}_{6}$ & $0$ & $0$ & $2$\Z & $0$ & \Zt & \Zt & \Z & $0$}
\newcommand{\Rgg}{${\cal R}_{6} \times {\cal R}_{6}$ & $0$ & $0$ & $2$\Z\,$\oplus$\,$2$\Z & $0$ & \Zt\,$\oplus$\,\Zt & \Zt\,$\oplus$\,\Zt & \Z\,$\oplus$\,\Z & $0$}
\newcommand{\Rh}{${\cal R}_{7}$ & $0$ & $0$ & $0$ & $2$\Z & $0$ & \Zt & \Zt & \Z}
\newcommand{\Rhh}{${\cal R}_{7} \times {\cal R}_{7}$ & $0$ & $0$ & $0$ & $2$\Z\,$\oplus$\,$2$\Z & $0$ & \Zt\,$\oplus$\,\Zt & \Zt\,$\oplus$\,\Zt & \Z\,$\oplus$\,\Z}

\begin{document}

\title{Symmetry and Topology in Non-Hermitian Physics}

\author{Kohei Kawabata}
\email{kawabata@cat.phys.s.u-tokyo.ac.jp}
\affiliation{Department of Physics, University of Tokyo, 7-3-1 Hongo, Bunkyo-ku, Tokyo 113-0033, Japan}

\author{Ken Shiozaki}
\email{ken.shiozaki@yukawa.kyoto-u.ac.jp}
\affiliation{Yukawa Institute for Theoretical Physics, Kyoto University, Kyoto 606-8502, Japan}

\author{Masahito Ueda}
\email{ueda@phys.s.u-tokyo.ac.jp}
\affiliation{Department of Physics, University of Tokyo, 7-3-1 Hongo, Bunkyo-ku, Tokyo
113-0033, Japan}
\affiliation{RIKEN Center for Emergent Matter Science (CEMS), Wako, Saitama 351-0198, Japan}

\author{Masatoshi Sato}
\email{msato@yukawa.kyoto-u.ac.jp}
\affiliation{Yukawa Institute for Theoretical Physics, Kyoto University, Kyoto 606-8502, Japan}

\date{\today}

\begin{abstract}
Non-Hermiticity enriches topological phases beyond the existing Hermitian framework. Whereas their unusual features with no Hermitian counterparts were extensively explored, a full understanding about the role of symmetry in non-Hermitian physics has still been elusive and there has remained an urgent need to establish their topological classification in view of rapid theoretical and experimental progress. Here, we develop a complete theory of symmetry and topology in non-Hermitian physics. We demonstrate that non-Hermiticity ramifies the celebrated Altland-Zirnbauer symmetry classification for insulators and superconductors. In particular, charge conjugation is defined in terms of transposition rather than complex conjugation due to the lack of Hermiticity, and hence chiral symmetry becomes distinct from sublattice symmetry. It is also shown that non-Hermiticity enables a Hermitian-conjugate counterpart of the Altland-Zirnbauer symmetry. Taking into account sublattice symmetry or pseudo-Hermiticity as an additional symmetry, the total number of symmetry classes is 38 instead of 10, which describe intrinsic non-Hermitian topological phases as well as non-Hermitian random matrices. Furthermore, due to the complex nature of energy spectra, non-Hermitian systems feature two different types of complex-energy gaps, point-like and line-like vacant regions. On the basis of these concepts and \textit{K}-theory, we complete classification of non-Hermitian topological phases in arbitrary dimensions and symmetry classes. Remarkably, non-Hermitian topology depends on the type of complex-energy gaps and multiple topological structures appear for each symmetry class and each spatial dimension, which are also illustrated in detail with concrete examples. Moreover, the bulk-boundary correspondence in non-Hermitian systems is elucidated within our framework, and symmetries preventing the non-Hermitian skin effect are identified. Our classification not only categorizes recently observed lasing and transport topological phenomena, but also predicts a new type of symmetry-protected topological lasers with lasing helical edge states and dissipative topological superconductors with nonorthogonal Majorana edge states. Furthermore, our theory provides topological classification of Hermitian and non-Hermitian free bosons. Our work establishes a theoretical framework for the fundamental and comprehensive understanding of non-Hermitian topological phases and paves the way toward uncovering unique phenomena and functionalities that emerge from the interplay of non-Hermiticity and topology.
\end{abstract}

\maketitle

\section{Introduction}

While Hermiticity is a common assumption that underlies physics of isolated systems, non-Hermitian Hamiltonians~\cite{Bender-98, *Bender-02, *Bender-review} have recently attracted growing attention. In fact, non-Hermiticity is ubiquitous in nature: it appears in nonequilibrium open systems with gain and/or loss~\cite{Konotop-review, Feng-review, Christodoulides-review, Alu-review, Ozdemir-review} and correlated electron systems as a result of finite-lifetime quasiparticles~\cite{Kozii-17, *Papaj-18, *Shen-18-QO, *Kozii-18, Zyuzin-18, *Moors-19, Yoshida-18, *Yoshida-19, Philip-18, Zhai-18, Bergholtz-19}. Moreover, effective non-Hermitian matrices are significant, for instance, in superconductors that undergo the depinning transition accompanying the localization transition~\cite{Hatano-96, *Hatano-97, *Hatano-98, Efetov-97, Brouwer-97, Goldsheid-98, Mudry-98, Shnerb-98, Nelson-98, *Amir-16, Fukui-98, Feinberg-99, LeClair-00, Hamazaki-18}, noninteracting bosonic systems that can exhibit dynamical instability~\cite{Katsura-10, *Onose-10, Shindou-13, Barnett-13, *Galilo-15, Engelhardt-15, *Engelhardt-16, Peano-16-nc, *Peano-16-x, Lieu-18, Clerk-18, Kondo-18, Kawaguchi-review}, and mechanical metamaterials~\cite{Kane-Lubensky-14, Roychowdhury-18-L, *Roychowdhury-18-B}. Non-Hermitian matrices exhibit unconventional characteristics compared with Hermitian ones~\cite{Moiseyev-11}: eigenstates are, in general, nonorthogonal~\cite{Brody-14} and a complex-energy spectrum can possess exceptional points~\cite{Berry-04, Heiss-12}. It has been shown that these mathematical properties lead to a number of unique phenomena and functionalities with no counterparts in Hermitian systems in both theory~\cite{Bender-07, Musslimani-08, *Makris-08, Klaiman-08, Graefe-08, Gunther-08, Mostafazadeh-09, Longhi-09, Longhi-10, Chong-11, Lin-11, Brody-12, Wiersig-14, Jing-14, Zhu-14, Lee-14X, Lee-14L, Dana-15, *Kirikchi-18, Longhi-15-SR, *Longhi-15-B, Kominis-15, *Kominis-16, Liu-16, Ge-17, Ashida-17, KK-QI-17, Ghatak-18, Qi-18, Konotop-18, Quijandria-18, Lourenco-18, Lau-18, Nakagawa-18, Dora-19, Shibata-19-Kitaev, *Shibata-19-AT} and experiments~\cite{Guo-09, Ruter-10, Schindler-11, Regensburger-12, Feng-13, Bender-13, Peng-14-NP, Peng-14-S, Feng-14, Hodaei-14, Fleury-15, Gao-15, Peng-16-PNAS, Miao-16, Peng-16, Doppler-16, Xu-16, Zhang-16, Assawaworrarit-17, Hodaei-17, Chen-17, Martins-18, Rivet-18, Mullers-18, Yoon-18, Joglekar-Luo-19, Li-19, Wu-19, Xiao-18}. Examples include power oscillations~\cite{Musslimani-08, *Makris-08, Regensburger-12}, unidirectional invisibility~\cite{Lin-11, Regensburger-12, Feng-13, Peng-14-NP}, high-performance lasers~\cite{Longhi-10, Chong-11, Jing-14, Ge-17, Peng-14-S, Feng-14, Hodaei-14, Miao-16}, exceptional-point encirclement~\cite{Gao-15, Doppler-16, Xu-16, Yoon-18}, and enhanced sensitivity~\cite{Wiersig-14,  Liu-16, Lau-18, Hodaei-17, Chen-17}.

Much research in recent years has focused on the topological characterization of non-Hermitian systems~\cite{Rudner-09, *Rudner-10, Hu-11, Esaki-11, *Sato-12, Pikulin-12, *Pikulin-13, Liang-13, Schomerus-13, Malzard-15, SanJose-16, Lee-16, Gonzalez-16, *Gonzalez-17, *Molina-18, Harter-16, Leykam-17, Xu-17, Menke-17, Lieu-18-SSH, Cerjan-18, MartinezAlvarez-18, Shen-18, Yin-18, Kunst-18, Kawabata-18-Kitaev, Yao-18-SSH, *Yao-18-Chern, Gong-18, *Bandres-Segev-18, Carlstrom-18, Kawabata-18-Chern, Takata-18, Kawabata-18, Okugawa-19, Budich-19, Yang-19, Jin-19, Zhou-19, Wang-19, Liu-19, Edvardsson-19, Carlstrom-19, Lee-19, Kunst-19, Longhi-19, Lee-Li-19, Rudner-16, Qiu-18, Zeng-19, Yokomizo-19, Okuma-19, Poli-15, Zeuner-15, Zhen-15, Weimann-17, Obuse-17, *Xiao-17, St-Jean-17, Bahari-17, Zhou-18-exp, Zhao-18, Parto-18, Harari-18, *Bandres-18, Wang-18, Cerjan-18-exp} beyond the existing Hermitian framework for condensed matter such as insulators~\cite{SSH-79, TKNN-82, Kohmoto-85, Haldane-88, Kane-Mele-05-Z2, *Kane-Mele-05-QSH, Fu-Kane-06, BHZ-06, Fu-Kane-Mele-07, Moore-Balents-07, Fu-Kane-07, Qi-08, Roy-09, Konig-07, Hsieh-08, Kane-review, Zhang-review} and superconductors~\cite{Read-00, Kitaev-01, Ivanov-01, Fu-Kane-08, Sato-09, *Sato-10, Sau-10, *Lutchyn-10, Oreg-10, Alicea-11, Alicea-review, Sato-review}, as well as photonic systems~\cite{Haldane-08, *Raghu-08, Wang-08, *Wang-09, Hafezi-11, Fang-12, Khanikaev-13, Rechtsman-13, Hafezi-13, Bliokh-15, Lu-review, Ozawa-review} and ultracold atoms~\cite{Aidelsburger-11, Aidelsburger-13, Atala-13, Jotzu-14, Mancini-15, Stuhl-15, Aidelsburger-15, Nakajima-16, Lohse-16, Goldman-review, Cooper-review}. Remarkably, certain topological phases survive even in the presence of non-Hermiticity~\cite{Rudner-09, Esaki-11, *Sato-12}, including non-Hermitian extensions of the Su-Schrieffer-Heeger model (i.e., one-dimensional system with chiral or sublattice symmetry~\cite{SSH-79})~\cite{Lieu-18, Esaki-11, *Sato-12, Schomerus-13, Lee-16, Lieu-18-SSH, MartinezAlvarez-18, Yin-18, Kunst-18, Yao-18-SSH, *Yao-18-Chern, St-Jean-17, Weimann-17, Parto-18}, the Chern insulator (i.e., two-dimensional system without any symmetry~\cite{TKNN-82, Kohmoto-85, Haldane-88})~\cite{Philip-18, Zhai-18, Shen-18, Kunst-18, Yao-18-SSH, *Yao-18-Chern, Kawabata-18-Chern, Bahari-17, Harari-18, *Bandres-18}, and the quantum spin Hall insulator (i.e., two-dimensional system with time-reversal symmetry~\cite{Kane-Mele-05-Z2, *Kane-Mele-05-QSH})~\cite{Kawabata-18}. In spite of their persistence, non-Hermiticity drastically changes the properties of topological boundary states. For instance, non-Hermiticity amplifies the edge states, which enables a novel laser topologically protected against disorder and defects~\cite{St-Jean-17, Zhao-18, Parto-18, Bahari-17, Harari-18, *Bandres-18}. It also makes the Majorana edge states nonorthogonal, which leads to nonlocal particle transport~\cite{Kawabata-18-Kitaev}. Moreover, the conventional bulk-boundary correspondence breaks down in certain non-Hermitian lattice models~\cite{Lee-16, MartinezAlvarez-18, Kawabata-18-Chern, Liu-19}. Nevertheless, recent researches~\cite{Kunst-18, Yao-18-SSH, *Yao-18-Chern} establish the modified bulk-boundary correspondence that works even in the presence of non-Hermiticity.

Symmetry plays a pivotal role in topological phases. The most fundamental symmetry is internal (nonspatial) symmetry, which does not rely on any specific spatial structure. In Hermitian systems, key internal symmetries culminate in the Altland-Zirnbauer (AZ) symmetry~\cite{AZ-97}: time-reversal symmetry, particle-hole symmetry, and chiral symmetry. Whereas symmetry protects non-Hermitian topological phases as well, non-Hermiticity can alter the nature of symmetry in a fundamental manner. In particular, Ref.~\cite{Kawabata-18} shows that the two antiunitary symmetries that are disparate in Hermitian systems can be equivalent to each other for non-Hermitian systems. This symmetry unification results in emergent non-Hermitian topological phases that are absent in Hermitian systems. However, it has remained unclear whether the AZ symmetry fully describes all internal symmetries in non-Hermitian physics.

Here, an essential distinction between Hermitian and non-Hermitian systems is the degrees of freedom that we have access to; nonunitary operations forbidden in Hermitian systems can be performed in non-Hermitian systems. In other words, a change in the spectrum from real to complex increases the number of parameters that describe the system. Since topology crucially depends on the underlying manifold, non-Hermiticity is expected to alter the topological classification of insulators and superconductors~\cite{Schnyder-08, Kitaev-09, Ryu-10, Teo-10, Slager-13, Chiu-13, Morimoto-13, Shiozaki-Sato-14, Shiozaki-16, *Shiozaki-17, *Shiozaki-18, Po-17, *Watanabe-18, *Ono-18, *Tang-19, *OYW-18, Bradlyn-17, *Vergniory-19, Kruthoff-17, Zhang-19, Schnyder-Ryu-review}. In fact, the emergent non-Hermitian topological phases~\cite{Kawabata-18} do imply such a change in the topological classification. A recent work~\cite{Gong-18, *Bandres-Segev-18} proposed classification of non-Hermitian topological systems on the basis of two antiunitary symmetries. Under this classification, however, topological phases are absent in two dimensions due to its strict definition of the complex-energy gap, which seems to conflict with the recent theoretical~\cite{Philip-18, Zhai-18, Esaki-11, *Sato-12, Shen-18, Kunst-18, Yao-18-SSH, *Yao-18-Chern, Kawabata-18-Chern, Kawabata-18} and experimental~\cite{Bahari-17, Harari-18, *Bandres-18} works in two dimensions. Moreover, Ref.~\cite{Gong-18, *Bandres-Segev-18} does not take into account the so-called pseudo-Hermiticity~\cite{Mostafazadeh-02-1, *Mostafazadeh-02-2, *Mostafazadeh-02-3,  Mostafazadeh-03, Brody-16}, which is a generalization of Hermiticity and parity-time symmetry~\cite{Bender-98, *Bender-02, *Bender-review}. As pointed out in Ref.~\cite{Esaki-11, *Sato-12}, pseudo-Hermiticity is a possible constraint unique to non-Hermitian systems and may provide a novel topological feature. Therefore, it has still been elusive how non-Hermiticity alters topology of insulators and superconductors. In view of the rapid theoretical and experimental advances in non-Hermitian physics, there has been a great interest and an urgent need for comprehensive topological classification that provides a reference point for experiments and predicts novel non-Hermitian topological phases.

\subsection{Summary of the results}

This work provides a complete theory of symmetry and topology in non-Hermitian physics. Non-Hermiticity dramatically changes fundamental concepts such as symmetry and energy gaps compared with the conventional ones in Hermitian physics. We first organize the internal symmetries in Sec.~\ref{sec: symmetry}. It is shown that symmetry ramifies due to the distinction between complex conjugation and transposition for non-Hermitian Hamiltonians, which culminates in the 38-fold symmetry in contrast to the 10-fold AZ symmetry in Hermitian systems. In particular, we demonstrate that particle-hole symmetry in non-Hermitian Bogoliubov-de Gennes (BdG) Hamiltonians should be defined with transposition as Eq.~(\ref{eq: PHS}), rather than complex conjugation. Similarly, chiral symmetry and sublattice symmetry become distinct from each other in non-Hermitian physics, although they are equivalent in the presence of Hermiticity. Moreover, the 38-fold symmetry naturally includes pseudo-Hermiticity~\cite{Mostafazadeh-02-1, *Mostafazadeh-02-2, *Mostafazadeh-02-3,  Mostafazadeh-03, Brody-16}, which provides a novel topological structure unique to non-Hermitian systems~\cite{Esaki-11, *Sato-12}. We note that the Bernard-LeClair symmetry classification~\cite{Bernard-LeClair-02}, which was previously considered to describe non-Hermitian random matrices~\cite{Bernard-LeClair-02, Magnea-08} and non-Hermitian topological phases~\cite{Lieu-18, Esaki-11, *Sato-12, Budich-19}, only partially reproduces our 38-fold symmetry classification. In fact, the previous symmetry classification overcounted some and overlooked others of our non-Hermitian symmetry classes, as discussed in detail in Sec.~\ref{sec: BL symmetry}. Our 38-fold symmetry classification thus serves as a non-Hermitian generalization of the renowned AZ symmetry classification for Hermitian Hamiltonians. We next show in Sec.~\ref{sec: complex gap} that an extension of the energy gap for non-Hermitian Hamiltonians is not unique due to the complex nature of the energy spectrum. It can be either point-like (zero-dimensional) or line-like (one-dimensional) in the complex-energy plane (Fig.~\ref{fig: complex gap}). Importantly, the definition that should be adopted depends on individual physical situations, and the two definitions are independent of and complementary to each other; non-Hermitian topology relies on the type of complex-energy gaps.

On the basis of these symmetry and complex-energy gaps in addition to \textit{K}-theory~\cite{Karoubi}, we provide in Sec.~\ref{sec: topological classification} complete topological classification of non-Hermitian insulators and superconductors for all the 38 symmetry classes and two types of the complex-energy gap. The results are summarized as periodic tables~\ref{tab: complex AZ}-\ref{tab: real AZ + pH}. The crucial idea behind this topological classification is that the complex-spectral-flattening procedures differ according to the type of the complex-energy gap: a non-Hermitian Hamiltonian can be flattened to a unitary matrix in the presence of a point gap, whereas it can be flattened to a Hermitian or an anti-Hermitian matrix in the presence of a line gap (Fig.~\ref{fig: flattening}). The corresponding topological invariants are systematically obtained in Sec.~\ref{sec: topological invariants}. We also elucidate the non-Hermitian bulk-boundary correspondence in terms of our classification in Sec.~\ref{sec: BEC}. Remarkably, whereas the conventional bulk-boundary correspondence can break down in generic non-Hermitian systems, we demonstrate that it is restored by certain classes of symmetry including parity-time symmetry and pseudo-Hermiticity. As a unique non-Hermitian feature, there appear multiple topological structures in each symmetry class and each spatial dimension, which is illustrated with an example in Sec.~\ref{sec: simple examples}. As discussed in Sec.~\ref{sec: experiment}, our classification describes the non-Hermitian topological phases observed in recent experiments~\cite{Poli-15, Zeuner-15, Weimann-17, Obuse-17, *Xiao-17, St-Jean-17, Bahari-17, Zhao-18, Parto-18, Harari-18, *Bandres-18, Wang-18, Cerjan-18-exp}, which are not fitted into the previous classification scheme~\cite{Gong-18, *Bandres-Segev-18}. Furthermore, our classification systematically predicts a new type of symmetry-protected topological lasers that support lasing helical edge states and dissipative topological superconductors that support nonorthogonal Majorana edge states. As a crucial byproduct, our non-Hermitian theory also provides the topological classification of Hermitian and non-Hermitian free bosons as shown in Sec.~\ref{sec: free boson}. We conclude this work in Sec.~\ref{sec: conclusions}.

\subsection{Distinction from the previous work}

The general theory in the present work supersedes and encompasses the results in the previous work~\cite{Gong-18}. In particular, this work provides the following fundamental insights into symmetry and topology in non-Hermitian physics:

\begin{itemize}
\item {\it Symmetry ramification.\,---} We discover that non-Hermiticity ramifies symmetry due to the distinction between complex conjugation and transposition, which are equivalent for Hermitian Hamiltonians, as described in Sec.~\ref{sec: symmetry}. Consequently,  we have a lot of new non-Hermitian symmetries, culminating in the 38-fold symmetry beyond the celebrated AZ symmetry in Hermitian physics. In particular, whereas a number of recent works including Ref.~\cite{Gong-18} focused on symmetry in terms of complex conjugation (such as parity-time symmetry), the crucial significance of the transposition symmetry has not been appreciated, and we have found its special role, for instance, in symmetry-protected topological lasers and dissipative superconductors.
\item {\it Complex-energy gaps and non-Hermitian topology.\,---} The definition of an energy gap is nontrivial in non-Hermitian systems and essential for the nature of non-Hermitian topological phases. We clarify this fundamental issue and find that the complex nature of the energy spectrum leads to the two types of complex-energy gaps, a point gap and a line gap (Fig.~\ref{fig: complex gap}), as described in Sec.~\ref{sec: complex gap}. Whereas a point gap and the corresponding topological classification were considered in Ref.~\cite{Gong-18}, a line gap was not considered there, and the present work has developed the unified understanding of complex-energy gaps. Importantly, the two types of complex-energy gaps enrich non-Hermitian topological phases in a fundamental manner that has no analogs to the Hermitian ones; non-Hermitian topology strongly depends on the type of complex-energy gaps.
\end{itemize}

Our complete classification of non-Hermitian topological phases relies on these fundamental insights in non-Hermitian physics. Crucially, although the previous classification~\cite{Gong-18} cannot correctly describe the recent experiments on non-Hermitian topological systems, the present classification encompasses them because of the above fundamental insights into symmetry and energy gaps, as described in Sec.~\ref{sec: recent experiment}. Moreover, our work systematically predicts novel non-Hermitian topological phases that enable richer phenomena and functionalities due to the interplay of non-Hermiticity and topology. For example, our theory predicts novel symmetry-protected topological lasers and dissipative topological superconductors, as described in Secs.~\ref{sec: SPT laser} and \ref{sec: NH TSC}.

\section{Symmetry}
	\label{sec: symmetry}

For Hermitian Hamiltonians, internal (nonspatial) symmetries fall into the AZ symmetry class~\cite{AZ-97}: time-reversal symmetry (TRS), particle-hole symmetry (PHS), and chiral symmetry (CS), where TRS and PHS are antiunitary, whereas CS is unitary. These symmetries lead to the 10-fold classification of Hermitian topological insulators and superconductors~\cite{Schnyder-08, Kitaev-09, Ryu-10}. On the other hand, it is nontrivial whether the AZ symmetry fully describes all the internal symmetries even in the presence of non-Hermiticity. In fact, PHS is defined with transposition as Eq.~(\ref{eq: PHS}) and cannot be described in terms of complex conjugation any longer for non-Hermitian BdG Hamiltonians due to the distinction between complex conjugation and transposition. Correspondingly, CS does not coincide with sublattice symmetry (SLS), although they are equivalent in the presence of Hermiticity. As a consequence, the total number of symmetry classes is $38$ as shown below, each of which describes intrinsic non-Hermitian topological phases as well as non-Hermitian random matrices.

\subsection{Symmetry ramification and unification}

Before describing our 38-fold symmetry in detail, we summarize the changes in the nature of symmetry in non-Hermitian physics. In fact, non-Hermiticity ramifies and unifies symmetry in a fundamental manner. First, to see the symmetry ramification, let us consider PHS as an example. For Hermitian systems, PHS is defined by
\begin{equation}
{\cal C}\,H^{*}\,{\cal C}^{-1} = - H,
	\label{eq: PHS cc}
\end{equation}
where ${\cal C}$ is a unitary matrix. Thus, PHS can be generalized by Eq.~(\ref{eq: PHS cc}) for non-Hermitian systems. However, we can generalize PHS in another way. The key property is that complex conjugation coincides with transposition for Hermitian systems by definition: $H^{*} = H^{T}$. As a result, for Hermitian systems, Eq.~(\ref{eq: PHS cc}) is equivalent to the following equation defined with transposition:
\begin{equation}
{\cal C}\,H^{T}\,{\cal C}^{-1} = - H.
	\label{eq: PHS t}
\end{equation}
Importantly, Eqs.~(\ref{eq: PHS cc}) and (\ref{eq: PHS t}) are not equivalent for non-Hermitian systems; thus, non-Hermiticity ramifies PHS. Since non-Hermitian BdG Hamiltonians for superconductors and superfluids satisfy Eq.~(\ref{eq: PHS t}) as shown below, we denote the symmetry in Eq.~(\ref{eq: PHS t}) [Eq.~(\ref{eq: PHS cc})] as PHS ($\mathrm{PHS}^{\dag}$) for non-Hermitian systems.

Such symmetry ramification occurs also for all the other symmetries. Another crucial example is CS (SLS), which is defined for Hermitian systems by 
\begin{equation}
\Gamma\,H\,\Gamma^{-1} = - H,
	\label{eq: CS no}
\end{equation}
where $\Gamma$ is a unitary matrix. Equation~(\ref{eq: CS no}) can be directly generalized to non-Hermitian systems, but again, CS can be generalized in a different manner. In fact, for Hermitian systems, Eq.~(\ref{eq: CS no}) is equivalent to 
\begin{equation}
\Gamma\,H^{\dag}\,\Gamma^{-1} = - H,
	\label{eq: CS dag}
\end{equation}
due to $H = H^{\dag}$. Importantly, Eqs.~(\ref{eq: CS no}) and (\ref{eq: CS dag}) are not equivalent to each other for non-Hermitian systems, although they are equivalent in the presence of Hermiticity. Since the physical CS, which is a combined symmetry of TRS and PHS, is described by Eq.~(\ref{eq: CS dag}) as shown below, we denote the symmetry in Eq.~(\ref{eq: CS dag}) [Eq.~(\ref{eq: CS no})] as CS ($\mathrm{CS}^{\dag}$) for non-Hermitian systems. We also denote $\mathrm{CS}^{\dag}$ as SLS because bipartite lattice systems often realize this symmetry even in the presence of non-Hermiticity.

Non-Hermiticity not only ramifies but also unifies symmetry~\cite{Kawabata-18}. To see this symmetry unification, we consider the following antiunitary symmetries:
\begin{equation}
{\cal T}_{+}\,H^{*}\,{\cal T}_{+}^{-1} = H,\quad
{\cal T}_{-}\,H^{*}\,{\cal T}_{-}^{-1} = - H,
\end{equation}
where ${\cal T}_{\pm}$ are unitary matrices. Here ${\cal T}_{+}$ denotes TRS, while ${\cal T}_{-}$ denotes ${\rm PHS}^{\dag}$, which are clearly disparate from each other for Hermitian systems. However, when a non-Hermitian system $H$ respects TRS, another non-Hermitian system $\ii H$ respects ${\rm PHS}^{\dag}$. Thus, a set of all the non-Hermitian systems having TRS coincides with another set of all the non-Hermitian systems having ${\rm PHS}^{\dag}$; non-Hermiticity unifies TRS and ${\rm PHS}^{\dag}$. As a consequence of the symmetry ramification and unification, the 10-fold AZ symmetry class for Hermitian systems is replaced by our 38-fold symmetry class for non-Hermitian systems, as demonstrated in the following.

\subsection{AZ symmetry}

\begin{table*}[t]
	\centering
	\caption{AZ and $\text{AZ}^{\dag}$ symmetry classes for non-Hermitian Hamiltonians. Time-reversal symmetry (TRS) and particle-hole symmetry (PHS) are defined by $\Tp H^{*} \left( {\bm k} \right) \Tp^{-1} = H \left( - {\bm k} \right)$ with $\Tp \Tp^{*} = \pm 1$ and $\Cm H^{T} \left( {\bm k} \right) \Cm^{-1} = - H \left( - {\bm k} \right)$ with $\Cm \Cm^{*} = \pm 1$, respectively. Chiral symmetry (CS) is a combined symmetry of TRS and PHS defined by $\CS H^{\dag} \left( {\bm k} \right) \CS^{-1} = - H \left( {\bm k} \right)$ with $\CS^{2} = 1$. The 10-fold AZ symmetry class is divided into two complex classes that only involve CS and eight real classes where TRS and PHS are relevant. Moreover, $\text{TRS}^{\dag}$ and $\text{PHS}^{\dag}$ are respectively defined by $\Cp H^{T} \left( {\bm k} \right) \Cp^{-1} = H \left( - {\bm k} \right)$ with $\Cp \Cp^{*} = \pm 1$ and $\Tm H^{*} \left( {\bm k} \right) \Tm^{-1} = - H \left( - {\bm k} \right)$ with $\Tm \Tm^{*} = \pm 1$, which constitute the $\text{AZ}^{\dag}$ symmetry classes. Class AI (AII) in the real AZ symmetry class and class $\text{D}^{\dag}$ ($\text{C}^{\dag}$) in the real $\text{AZ}^{\dag}$ symmetry class are equivalent to each other. \\}
		\label{tab: AZ}
     \begin{tabular}{ccccccc} \hline \hline
    \multicolumn{2}{c}{~Symmetry class~} & ~TRS ($\Tp$)~ & ~PHS ($\Cm$)~ & ~$\text{TRS}^{\dag}$ ($\Cp$)~ & ~$\text{PHS}^{\dag}$ ($\Tm$)~ & ~CS ($\CS$)~ \\ \hline
    \multirow{2}{*}{~Complex AZ~} 
    & A & $0$ & $0$ & $0$ & $0$ & $0$ \\
    & AIII & $0$ & $0$ & $0$ & $0$ & $1$ \\ \hline
    \multirow{9}{*}{Real AZ} 
    & AI & $+1$ & $0$ & $0$ & $0$ & $0$ \\
    & BDI & $+1$ & $+1$ & $0$ & $0$ & $1$ \\
    & D & $0$ & $+1$ & $0$ & $0$ & $0$ \\
    & DIII & $-1$ & $+1$ & $0$ & $0$ & $1$ \\
    & AII & $-1$ & $0$ & $0$ & $0$ & $0$ \\
    & CII & $-1$ & $-1$ & $0$ & $0$ & $1$ \\
    & C & $0$ & $-1$ & $0$ & $0$ & $0$ \\
    & CI & $+1$ & $-1$ & $0$ & $0$ & $1$ \\ \hline
    \multirow{9}{*}{Real $\text{AZ}^{\dag}$} 
    & $\text{AI}^{\dag}$ & $0$ & $0$ & $+1$ & $0$ & $0$ \\
    & $\text{BDI}^{\dag}$ & $0$ & $0$ & $+1$ & $+1$ & $1$ \\
    & $\text{D}^{\dag}$ & $0$ & $0$ & $0$ & $+1$ & $0$ \\
    & $\text{DIII}^{\dag}$ & $0$ & $0$ & $-1$ & $+1$ & $1$ \\
    & $\text{AII}^{\dag}$ & $0$ & $0$ & $-1$ & $0$ & $0$ \\
    & $\text{CII}^{\dag}$ & $0$ & $0$ & $-1$ & $-1$ & $1$ \\
    & $\text{C}^{\dag}$ & $0$ & $0$ & $0$ & $-1$ & $0$ \\
    & $\text{CI}^{\dag}$ & $0$ & $0$ & $+1$ & $-1$ & $1$ \\ \hline \hline
    \end{tabular}
\end{table*}

We consider a generic noninteracting fermionic system described by the following second-quantized non-Hermitian Hamiltonian 
\begin{equation}
\hat{H} = \sum_{ij}
\hat{\psi}_{i}^{\dag} H_{ij} \hat{\psi}_{j},
\end{equation}
where the matrix $H$ is a first-quantized (single-particle) non-Hermitian Hamiltonian. In addition, $( \hat{\psi}_{i} )_{i=1,2,\cdots}$ is a set of fermion annihilation operators for a normal system or Nambu spinors for a superconductor. Time-reversal operation is described by an antiunitary operator $\hat{\cal T}$ that acts on the fermion operators as
\begin{equation}
\hat{\cal T} \hat{\psi}_{i} \hat{\cal T}^{-1}
= \sum_{j} \left( \Tp \right)_{ij} \hat{\psi}_{j},~~
\hat{\cal T}\,\ii\,\hat{\cal T}^{-1}
= - \ii,
\end{equation}
where $\Tp$ is a unitary matrix ($\Tp \Tp^{\dag} = \Tp^{\dag} \Tp = 1$). This operation serves as time reversal even in non-Hermitian systems. In fact, if an operator $\hat{O}$ is invariant under $\hat{\cal T}$ (i.e., $\hat{\cal T} \hat{O} \hat{\cal T}^{-1} = \hat{O}$), $\hat{\cal T} \hat{H} \hat{\cal T}^{-1} = \hat{H}$ implies
\begin{eqnarray}
\hat{\cal T} \hat{O} \left( t \right) \hat{\cal T}^{-1} 
&=& \hat{\cal T}\,( e^{\ii \hat{H}^{\dag} t} \hat{O} e^{-\ii \hat{H} t} )\,\hat{\cal T}^{-1} \nonumber \\
&=& e^{-\ii \hat{H}^{\dag} t} \hat{O} e^{\ii \hat{H} t}
= \hat{O} \left( -t \right),
\end{eqnarray}
where $\hat{O} \left( t \right) := e^{\ii \hat{H}^{\dag} t} \hat{O} e^{-\ii \hat{H} t}$ is the time-evolved operator under the non-Hermitian Hamiltonian $\hat{H}$~\cite{Brody-12, KK-QI-17}. Then time-reversal invariance of the second-quantized Hamiltonian (i.e., $\hat{\cal T} \hat{H} \hat{\cal T}^{-1} = \hat{H}$) leads to
\begin{equation}
\Tp^{-1} H^{*} \Tp = H,~~\Tp \Tp^{*} = \pm 1
\end{equation}
in real space, and 
\begin{equation}
\Tp H^{*} \left( {\bm k} \right) \Tp^{-1}
= H \left( -{\bm k} \right),~~\Tp \Tp^{*} = \pm 1
	\label{eq: TRS}
\end{equation}
in momentum space, where $H \left( {\bm k} \right)$ is a Bloch-BdG Hamiltonian. This action on a single-particle non-Hermitian Hamiltonian by TRS is the same as that on a Hermitian one~\cite{AZ-97}. Our discussion can also be applied to the generalized non-Bloch wave functions~\cite{Yao-18-SSH, *Yao-18-Chern, Kunst-19, Yokomizo-19}, as long as the corresponding symmetry is respected with complex wavenumbers (see Sec.~\ref{sec: BEC} for more details).

PHS is associated with charge conjugation that mixes fermion creation and annihilation operators and generally appears in superconductors and superfluids. It is described by a unitary operator $\hat{\cal C}$ that acts on the fermion operators as
\begin{equation}
\hat{\cal C}\,\hat{\psi}_{i}\,\hat{\cal C}^{-1}
= \sum_{j} \left( \Cm^{*} \right)_{ij} \hat{\psi}_{j}^{\dag},
\end{equation}
where $\Cm$ is a unitary matrix ($\Cm \Cm^{\dag} = \Cm^{\dag} \Cm = 1$). Then the presence of PHS for the second-quantized Hamiltonian $\hat{\cal C} \hat{H} \hat{\cal C}^{-1} = \hat{H}$ leads to
\begin{equation}
\Cm^{-1} H^{T} \Cm
= - H,~~\Cm \Cm^{*} = \pm 1
\end{equation}
in real space, and 
\begin{equation}
\Cm H^{T} \left( {\bm k} \right) \Cm^{-1}
= - H \left( -{\bm k} \right),~~\Cm \Cm^{*} = \pm 1
	\label{eq: PHS}
\end{equation}
in momentum space. Remarkably, in the presence of Hermiticity ($\hat{H}^{\dag} = \hat{H}$), this PHS condition is equivalent to $\Cm H^{*} \left( {\bm k} \right) \Cm^{-1} = - H \left( -{\bm k} \right)$~\cite{AZ-97}. For non-Hermitian Hamiltonians, however, complex conjugation and transposition do not coincide with each other and thus PHS is defined in terms of not complex conjugation but transposition.

As a combination of TRS and PHS, CS is defined by an antiunitary operator $\hat{\Gamma} := \hat{\cal T} \hat{\cal C}$. The invariance of the Hamiltonian $\hat{H}$ under $\hat{\Gamma}$ imposes the following condition on a single-particle Hamiltonian:
\begin{equation}
\CS^{-1} H^{\dag} \CS
= - H,~~\CS^{2} = 1
\end{equation}
in real space, and 
\begin{equation}
\CS H^{\dag} \left( {\bm k} \right) \CS^{-1}
= - H \left( {\bm k} \right),~~\CS^{2} = 1
	\label{eq: CS}
\end{equation}
in momentum space. This CS condition is equivalent to $\CS H \left( {\bm k} \right) \CS^{-1} = - H \left( {\bm k} \right)$ in the presence of Hermiticity ($\hat{H}^{\dag} = \hat{H}$)~\cite{AZ-97}, but it is not for non-Hermitian Hamiltonians. For instance, the graphene~\cite{Esaki-11, *Sato-12} and the Su-Schrieffer-Heeger model~\cite{Schomerus-13, Lieu-18-SSH, Weimann-17, Parto-18} with balanced gain and loss respect CS.

The three symmetries $\Tp$, $\Cm$, and $\CS$ constitute a natural and physical extension of the AZ symmetry class for non-Hermitian Hamiltonians (Table~\ref{tab: AZ}), which respectively act on a Bloch-BdG non-Hermitian Hamiltonian as Eqs.~(\ref{eq: TRS}), (\ref{eq: PHS}), and (\ref{eq: CS}). The 10-fold AZ symmetry class is divided into two complex classes that only involve CS and eight real classes where TRS and PHS are relevant. We again emphasize that PHS is defined in terms of not complex conjugation but transposition for non-Hermitian Hamiltonians, and that the definition of CS also changes correspondingly.

\subsection{$\text{AZ}^{\dag}$ symmetry}
	\label{sec: AZ-dag}

In contrast to the Hermitian case, internal symmetries arise other than the AZ symmetry. In fact, as a result of the distinction between complex conjugation and transposition for non-Hermitian Hamiltonians ($H^{*} \neq H^{T}$), a variant of TRS can be defined with transposition by
\begin{eqnarray}
\Cp H^{T} \left( {\bm k} \right) \Cp^{-1}
= H \left( -{\bm k} \right),~~\Cp \Cp^{*} = \pm 1,
	\label{eq: TRS-dag}
\end{eqnarray}
where $\Cp$ is a unitary matrix ($\Cp \Cp^{\dag} = \Cp^{\dag} \Cp = 1$). Similarly, a variant of PHS can be defined with complex conjugation by 
\begin{equation}
\Tm H^{*} \left( {\bm k} \right) \Tm^{-1}
= - H \left( -{\bm k} \right),~~\Tm \Tm^{*} = \pm 1,
	\label{eq: PHS-dag}
\end{equation}
where $\Tm$ is a unitary matrix ($\Tm \Tm^{\dag} = \Tm^{\dag} \Tm = 1$). In the following, we denote the symmetry described by Eq.~(\ref{eq: TRS-dag}) as $\text{TRS}^{\dag}$ and the symmetry described by Eq.~(\ref{eq: PHS-dag}) as $\text{PHS}^{\dag}$, since $\text{TRS}^{\dag}$ ($\text{PHS}^{\dag}$) is defined by Hermitian conjugation of TRS (PHS). For Hermitian Hamiltonians ($H = H^{\dag}$), TRS and PHS respectively coincide with $\text{TRS}^{\dag}$ and $\text{PHS}^{\dag}$; however, this is not the case in the presence of non-Hermiticity. This Hermitian-conjugate counterpart of the AZ symmetry also appears naturally in non-Hermitian systems. For instance, onsite dissipation often breaks Hermiticity and TRS at the same time, but their combination can be retained as $\text{TRS}^{\dag}$. Furthermore, effective non-Hermitian Hamiltonians in the scattering theory support these symmetries~\cite{HS-pc}. We note that $\text{PHS}^{\dag}$ is equivalent to ``non-Hermitian particle-hole symmetry" in Refs.~\cite{Kawabata-18-Kitaev, Kawabata-18, Pikulin-12, *Pikulin-13, Malzard-15, Ge-17, Qi-18, Gong-18, *Bandres-Segev-18}, which plays an important role in a single-mode laser~\cite{Ge-17} and a flatband~\cite{Qi-18} in photonics.

An important consequence of $\text{TRS}^{\dag}$ with $\mathcal{C}_{+} \mathcal{C}_{+}^{*} = -1$ is the twofold degeneracy of the complex spectrum as a non-Hermitian generalization of the Kramers theorem~\cite{Esaki-11, *Sato-12}. To see this non-Hermitian Kramers degeneracy, we consider a non-Hermitian Hamiltonian $H$ that satisfies $\mathcal{C}_{+} H^{T} \mathcal{C}_{+}^{-1} = H$ with $\mathcal{C}_{+} \mathcal{C}_{+}^{*} = -1$ and denote its complex eigenenergy and the corresponding right (left) eigenstate as $E_{n} \in \mathbb{C}$ and $\ket{u_{n}}$ ($| u_{n} \rangle\!\rangle$), respectively. We then have
\begin{equation}
H \left( \mathcal{C}_{+} | u_{n}^{*} \rangle\!\rangle \right)
= E \left( \mathcal{C}_{+} | u_{n}^{*} \rangle\!\rangle \right),
\end{equation}
which implies that $\mathcal{C}_{+} | u_{n}^{*} \rangle\!\rangle$ is also an eigenstate that belongs to the same eigenenergy $E$. Because of $\mathcal{C}_{+}^{T} = - \mathcal{C}_{+}$, we also have
\begin{equation}
\langle\!\langle u_{n} |\,\mathcal{C}_{+} | u_{n}^{*} \rangle\!\rangle
= \langle\!\langle u_{n} |\,\mathcal{C}_{+}^{T} | u_{n}^{*} \rangle\!\rangle
= - \langle\!\langle u_{n} |\,\mathcal{C}_{+} | u_{n}^{*} \rangle\!\rangle,
\end{equation}
which leads to $\langle\!\langle u_{n} |\,\mathcal{C}_{+} | u_{n}^{*} \rangle\!\rangle = 0$. This fact in turn indicates that $\ket{u_{n}}$ and $\mathcal{C}_{+} | u_{n}^{*} \rangle\!\rangle$ are biorthogonal to each other~\cite{Brody-14} and linearly independent of each other; all the eigenstates are thus at least twofold degenerate. This non-Hermitian Kramers degeneracy holds even for complex eigenenergies. In the presence of TRS with $\mathcal{T}_{+} \mathcal{T}_{+}^{*} = -1$, by contrast, eigenenergies are twofold degenerate if and only if they are real~\cite{Kawabata-18}.

$\text{TRS}^{\dag}$ and $\text{PHS}^{\dag}$ in addition to CS also constitute the 10-fold symmetry class, which we call the $\text{AZ}^{\dag}$ symmetry class (Table~\ref{tab: AZ}). This $\text{AZ}^{\dag}$ symmetry class is again divided into two complex classes that only involve CS and eight real classes where $\text{TRS}^{\dag}$ and $\text{PHS}^{\dag}$ are relevant. Here, each complex $\text{AZ}^{\dag}$ class coincides with the corresponding complex AZ class. Moreover, class AI in the real AZ class and class $\text{D}^{\dag}$ in the real $\text{AZ}^{\dag}$ class are equivalent due to the topological unification of TRS and $\text{PHS}^{\dag}$~\cite{Kawabata-18}: when a non-Hermitian Hamiltonian $H$ respects TRS, another non-Hermitian Hamiltonian $\ii H$ respects $\text{PHS}^{\dag}$. Similarly, class AII in the real AZ class and class $\text{C}^{\dag}$ in the real $\text{AZ}^{\dag}$ class are equivalent.

\subsection{Sublattice symmetry}

Another important internal symmetry is SLS, which is defined for a Bloch-BdG Hamiltonian by 
\begin{equation}
\SLS H \left( {\bm k} \right) \SLS^{-1}
= - H \left( {\bm k} \right),~~\SLS^{2} = 1,
	\label{eq: SLS}
\end{equation}
where $\SLS$ is a unitary matrix ($\SLS \SLS^{\dag} = \SLS^{\dag} \SLS = 1$). For instance, SLS appears in a bipartite lattice where particle hopping only connects sites on different sublattices, such as the Su-Schrieffer-Heeger model~\cite{SSH-79} with asymmetric hopping~\cite{Lee-16, Lieu-18-SSH, MartinezAlvarez-18, Yin-18, Kunst-18, Yao-18-SSH, *Yao-18-Chern}. Remarkably, SLS coincides with CS defined by Eq.~(\ref{eq: CS}) in the presence of Hermiticity ($H = H^{\dag}$)~\cite{AZ-97}, but this is not the case for non-Hermitian Hamiltonians. 

SLS can be considered as an additional symmetry to the AZ symmetry~\cite{Shiozaki-Sato-14} (see Tables~\ref{tab: symmetry - complex AZ + SLS} and \ref{tab: symmetry - real AZ + SLS} in Appendix~\ref{appendix: SLS} for details). There are 3 symmetry classes for the complex AZ class with SLS (Table~\ref{tab: symmetry - complex AZ + SLS}) and 19 symmetry classes for the real AZ class with SLS (Table~\ref{tab: symmetry - real AZ + SLS}). Here classes AI, BDI, and CII with SLS that anticommutes with TRS are respectively equivalent to classes AII, DIII, and CI with SLS that obeys the same algebra. Moreover, each real AZ class with SLS is equivalent to the corresponding real ${\rm AZ}^{\dag}$ class with SLS (see Table~\ref{tab: SLS - AZ & AZ-dag} in Appendix~\ref{appendix: SLS} for details).

\subsection{Pseudo-Hermiticity}

In non-Hermitian physics, pseudo-Hermiticity serves as another key internal symmetry~\cite{Mostafazadeh-02-1, *Mostafazadeh-02-2, *Mostafazadeh-02-3,  Mostafazadeh-03, Brody-16}, which is defined by
\begin{equation}
\eta H^{\dag} \left( {\bm k} \right) \eta^{-1}
= H \left( {\bm k} \right),~~\eta^{2} = 1,
	\label{eq: def pseudo-Hermiticity}
\end{equation}
with a unitary and Hermitian matrix $\eta$ ($\eta \eta^{\dag} = \eta^{\dag} \eta = 1$ and $\eta^{\dag} = \eta$). Here, pseudo-Hermiticity is a generalization of Hermiticity, in that it is trivially satisfied with $\eta = 1$ in the presence of Hermiticity. In addition, it has a similar role to parity-time symmetry~\cite{Bender-98, *Bender-02, *Bender-review} because positivity of $\eta$ is equivalent to the real spectrum of a non-Hermitian Hamiltonian~\cite{Mostafazadeh-02-1, *Mostafazadeh-02-2, *Mostafazadeh-02-3}. Pseudo-Hermiticity can also be considered as an additional symmetry to the AZ or ${\rm AZ}^{\dag}$ symmetry class. Moreover, the AZ or ${\rm AZ}^{\dag}$ class with pseudo-Hermiticity is equivalent to the AZ or ${\rm AZ}^{\dag}$ class with SLS (see Table~\ref{tab: pH & SLS} in Appendix~\ref{appendix: pseudo-Hermiticity} for details).

\subsection{38-fold classification}
	\label{sec: BL symmetry}

The symmetries discussed above constitute all the internal symmetries in non-Hermitian physics, which generalize and extend the AZ symmetry classification~\cite{AZ-97} for Hermitian Hamiltonians to that for non-Hermitian ones. This symmetry classification is 38-fold: the 10 AZ symmetry classes with the additional 6 ${\rm AZ}^{\dag}$ symmetry classes, as well as the 22 AZ symmetry classes with SLS. Notably, the 4 (i.e., A, AIII, $\mathrm{D}^{\dag}$, and $\mathrm{C}^{\dag}$) in the ${\rm AZ}^{\dag}$ symmetry class also appear in the AZ symmetry class, and each AZ symmetry class with SLS is equivalent to the corresponding ${\rm AZ}^{\dag}$ symmetry class with SLS (see Appendix~\ref{appendix: SLS} for details) or the AZ symmetry class with pseudo-Hermiticity (see Appendix~\ref{appendix: pseudo-Hermiticity} for details). Our 38-fold symmetry classification is applicable to a number of non-Hermitian systems, as discussed in detail below.

We can confirm that the 38-fold symmetry exhausts all the internal symmetries for non-Hermitian systems. We first note that it is sufficient to consider a single symmetry for each type without loss of generality. For instance, if a Hamiltonian respects two TRSs, $\mathcal{T}_{+}$ and $\tilde{\cal T}_{+}$, the combined symmetry $\mathcal{T}_{+} \tilde{\cal T}_{+}^{*}$ gives a unitary symmetry that commutes with the Hamiltonian. In the block-diagonal form of $\mathcal{T}_{+} \tilde{\cal T}_{+}^{*}$, $\mathcal{T}_{+}$ and $\tilde{\cal T}_{+}$ are trivially related to each other in each block and it suffices to consider a single TRS. The 38 symmetry classes are divided according to the number $N$ of their generators $\mathrm{TRS}$, $\mathrm{PHS}$, $\mathrm{TRS}^{\dag}$, and $\mathrm{PHS}^{\dag}$ as follows:
\begin{itemize}
\item $N=0$.\,--- There are $5$ classes: no symmetry, CS, SLS, and both CS and SLS that commute or anticommute with each other. 
\item $N=1$.\,--- Each symmetry has two different signs of the square (e.g., $\mathcal{T}_{+} \mathcal{T}_{+}^{*} = +1$ or $\mathcal{T}_{+} \mathcal{T}_{+}^{*} = -1$). Moreover, $\mathrm{TRS}$ and $\mathrm{PHS}^{\dag}$ are equivalent to each other~\cite{Kawabata-18}. As a result, we have $2 \times \left( 4-1 \right) = 6$ classes.
\item $N=2$.\,--- Whereas there are six combinations to choose two from the four symmetries (i.e., TRS, PHS, $\mathrm{TRS}^{\dag}$, and $\mathrm{PHS}^{\dag}$), the combinations $\mathrm{TRS}\,\&\,\mathrm{PHS}$ and $\mathrm{TRS}\,\&\,\mathrm{TRS}^{\dag}$ are equivalent to $\mathrm{PHS}\,\&\,\mathrm{PHS}^{\dag}$ and $\mathrm{TRS}^{\dag}\,\&\,\mathrm{PHS}^{\dag}$, respectively, due to the topological unification of $\mathrm{TRS}$ and $\mathrm{PHS}^{\dag}$~\cite{Kawabata-18}. Moreover, the combination $\mathrm{TRS}$ ($\mathcal{T}_{+} \mathcal{T}_{+}^{*} = +1$) \& $\mathrm{PHS}^{\dag}$ ($\mathcal{T}_{-} \mathcal{T}_{-}^{*} = -1$) is equivalent to $\mathrm{TRS}$ ($\mathcal{T}_{+} \mathcal{T}_{+}^{*} = -1$) \& $\mathrm{PHS}^{\dag}$ ($\mathcal{T}_{-} \mathcal{T}_{-}^{*} = +1$). We thus have $2^{2} \times \left( 6-2 \right) - 1 = 15$ classes.
\item $N=3$.\,--- Without loss of generality, we can consider the generators to be $\mathrm{TRS}$, $\mathrm{PHS}$, and $\mathrm{PHS}^{\dag}$, and assume that $\mathrm{TRS}$ and $\mathrm{PHS}$ commute with each other. There are three combinations of $\mathrm{TRS}$ and $\mathrm{PHS}^{\dag}$, in a similar manner to the case with $N=2$. Furthermore, $\mathrm{PHS}$ has two different signs of the square and can commute or anticommute with $\mathrm{PHS}^{\dag}$. Consequently, we have $3 \times 2 \times 2 = 12$ classes.
\item $N=4$.\,--- If we have independent $\mathrm{TRS}$, $\mathrm{PHS}$, $\mathrm{TRS}^{\dag}$, and $\mathrm{PHS}^{\dag}$, the combination of all the symmetries gives a unitary symmetry that commutes with the Hamiltonian. In the block-diagonal form of the unitary operator, this case reduces to $N \leq 3$.
\end{itemize}
We have, in total, $5 + 6 + 15 + 12 + 0 =38$ classes.

\begin{table}[t]
	\centering
	\caption{Relationship between the Bernard-LeClair (BL) symmetry and our 38-fold symmetry discussed in the present work. Here, TRS, PHS, CS, and SLS respectively stand for time-reversal symmetry, particle-hole symmetry, chiral symmetry, and sublattice symmetry; ${\rm TRS}^\dag$ (${\rm PHS}^\dag$) denotes the symmetry defined by Hermitian conjugation of TRS (PHS). \\}
	\label{tab: BL symmetry}
\begin{tabular}{cc} 
\hline \hline
~BL symmetry~ & ~The 38-fold symmetry~ \\ 
\hline
~C sym.~ & PHS, TRS$^\dagger$\\
~P sym.~ & SLS \\
~Q sym.~ & ~CS, pseudo-Hermiticity~ \\
~K sym.~ & TRS, PHS$^{\dagger}$\\
\hline \hline
  \end{tabular}
\end{table}

Our 38-fold classification is basically equivalent to the Bernard-LeClair symmetry classification that describes non-Hermitian random matrices~\cite{Bernard-LeClair-02, Magnea-08, Esaki-11, *Sato-12, Lieu-18, Budich-19}:
\begin{align}
&\mbox{C sym.}:\quad c H^{T} c^{-1} = \epsilon_{c} H ,~c^{T} c^{-1}=\pm 1, \label{eq: BL - C} \\
&\mbox{P sym.}:\quad p H p^{-1} = - H,~p^{2} = 1,\\
&\mbox{Q sym.}:\quad q H^{\dag} q^{-1} = H,~q^{\dag} q^{-1} =1, \label{eq: BL - Q} \\
&\mbox{K sym.}:\quad k H^{*} k^{-1} = H,~k k^{*} = \pm 1,
\end{align}
with $\epsilon_{c} = \pm 1$ and unitary operators $c$, $p$, $q$, and $k$. Table~\ref{tab: BL symmetry} summarizes the relationship between the Bernard-LeClair symmetry and ours. Whereas our classification is 38-fold, the one of Bernard and LeClair is 43-fold. This disagreement originates from overcounting and overlooking non-Hermitian symmetry classes in their classification. In particular, they distinguished the pseudo-Hermiticity [Q symmetry defined by Eq.~(\ref{eq: BL - Q})] with positivity from generic pseudo-Hermiticity without positivity. However, it is known that the pseudo-Hermiticity with positivity is equivalent to Hermiticity~\cite{Mostafazadeh-02-1, *Mostafazadeh-02-2, *Mostafazadeh-02-3,  Mostafazadeh-03, Brody-16}. Thus, the former pseudo-Hermiticity just gives the Hermitian symmetry classes. Here the following 5 pairs of symmetry classes distinguished in the Bernard-LeClair classification are considered to be the same in our classification:
\begin{equation} \begin{split}
&\left( q = 1 \right)~\&~\left( q = \sigma_{z} \right);~\\
&\left( q = 1,\,c = 1 \right)_{\epsilon_{c} = \pm 1}~\&~\left( q = \sigma_{z},\,c = 1 \right)_{\epsilon_{c} = \pm 1};~\\
&\left( q = 1,\,c=\ii \sigma_{y} \right)_{\epsilon_{c} = \pm 1}~\&~\left( q = \sigma_{z} \otimes 1,\,c = 1 \otimes \ii \sigma_{y} \right)_{\epsilon_{c} = \pm 1}.~
\end{split} \end{equation}
We recall that the Hermitian symmetry class is the 10-fold AZ symmetry class. Subtracting these Hermitian 10 classes from their 43 classes, we only have 33 classes as intrinsic non-Hermitian symmetry classes. However, they overlooked the following 5 symmetry classes, which should be added when the aforementioned distinction is made:
\begin{equation} \begin{split}
&\left( p = \sigma_{z} \otimes 1,\,q = 1 \otimes \sigma_{x} \right);\\
&\left( p = \sigma_{z} \otimes 1,\,q = 1 \otimes \sigma_{y},\,c = 1 \otimes 1 \right);~\\
&\left( p = \sigma_{z} \otimes 1,\,q = 1 \otimes \sigma_{y},\,c = 1 \otimes \ii \sigma_{y} \right);~\\
&\left( p = \sigma_{z} \otimes 1,\, q = 1 \otimes \sigma_{x},\,c = \sigma_{x} \otimes 1 \right)_{\epsilon_{c} = \pm 1}.~
	\label{eq: BL - not included}
\end{split} \end{equation}
Adding these 5 classes to 33 classes reproduces our 38-fold symmetry class.

We complete the non-Hermitian 38-fold symmetry class, in which the 5 classes in Eq.~(\ref{eq: BL - not included}) were overlooked by Bernard and LeClair. Remarkably, our 38 classes present different classifying spaces and give different topological phases, as shown in Sec.~\ref{sec: topological classification}. Importantly, although their symmetries are mathematically the same as ours after correctly including the 5 classes in Eq.~(\ref{eq: BL - not included}), the physical insight into these symmetry classes has remained elusive until the present work. Therefore, our symmetries give a more fundamental framework in the study of non-Hermitian physics.

\section{Complex-energy gaps}
	\label{sec: complex gap}

\begin{figure}[t]
\centering
\includegraphics[width=86mm]{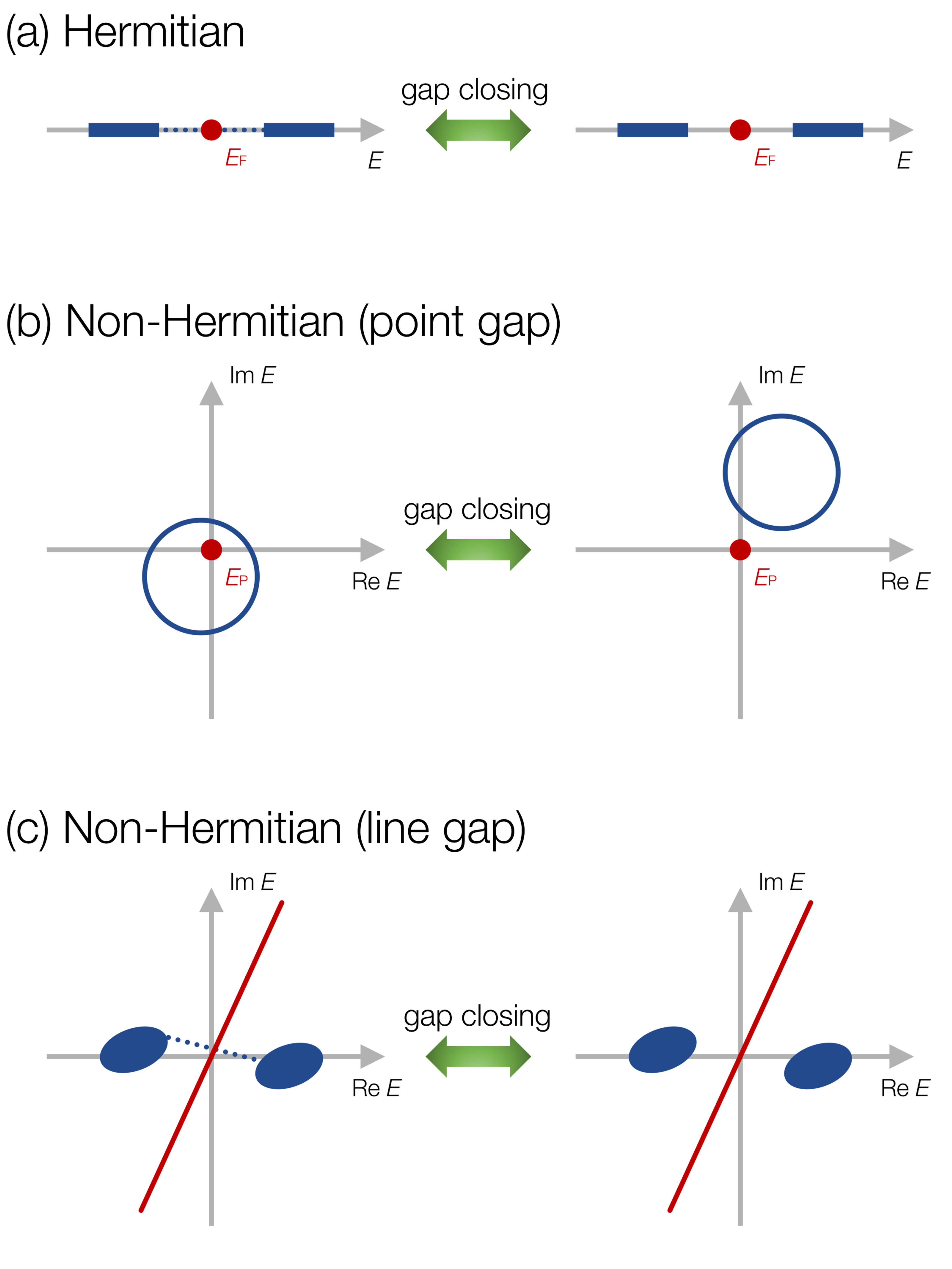} 
\caption{Definition of the energy gap for Hermitian and non-Hermitian Hamiltonians. (a)~Energy gap for a Hermitian Hamiltonian. A Hermitian Hamiltonian is defined to be gapped if and only if its energy bands do not cross the Fermi energy $E_{\rm F}$ (red dot), and gap closing associated with a topological phase transition occurs between the trivial and topological phases. (b)~Point gap for a non-Hermitian Hamiltonian. A non-Hermitian Hamiltonian is defined to have a point gap if and only if its complex-energy bands do not cross a reference point $E = E_{\rm P}$ in the complex-energy plane (red dot). (c)~Line gap for a non-Hermitian Hamiltonian. A non-Hermitian Hamiltonian is defined to have a line gap if and only if its complex-energy bands do not cross a reference line in the complex-energy plane (red line).}
	\label{fig: complex gap}
\end{figure}

In the topological classification of Hermitian insulators and superconductors, two Hermitian Hamiltonians are defined to be topologically equivalent if and only if they are continuously deformed into each other while retaining symmetry and an energy gap. In the non-Hermitian case, on the other hand, it is nontrivial how the energy gap is defined since the spectrum is complex for a generic non-Hermitian Hamiltonian.

Here, we recall that an energy gap means the energy region where no states are present. In the Hermitian case, such a vacant region in the spectrum should be contractible to a zero-dimensional point $E = E_{\rm F}$ called the Fermi energy since the spectrum is entirely real and one-dimensional. Thus it is naturally and uniquely defined to have an energy gap if and only if its energy bands do not cross the Fermi energy $E = E_{\rm F}$ [Fig.~\ref{fig: complex gap}\,(a)]. In the non-Hermitian case, by contrast, the forbidden energy range where no states exist is not necessarily contractible to a zero-dimensional point since the complex spectrum of a generic non-Hermitian Hamiltonian is two-dimensional. As a result, such a forbidden energy region can be either a zero-dimensional point or a one-dimensional line, and accordingly the definition of the complex-energy gap in a non-Hermitian Hamiltonian is not unique. It can be defined to have a zero-dimensional point gap if and only if its complex-energy bands do not cross a reference point $E = E_{\rm P}$ in the complex-energy plane [Fig.~\ref{fig: complex gap}\,(b)], and independently, it can also be defined to have a one-dimensional line gap if and only if its complex-energy bands do not cross a reference line in the complex-energy plane [Fig.~\ref{fig: complex gap}\,(c)]. The precise definitions of these complex-energy gaps are provided later in this section.

Importantly, two definitions are independent of each other, and which one should be adopted depends on the individual physical situations that we are interested in. For instance, the localization (Anderson) transition in a one-dimensional non-Hermitian system can be captured by topology in terms of a point gap~\cite{Hatano-96, *Hatano-97, *Hatano-98, Shnerb-98,  Gong-18, *Bandres-Segev-18, Longhi-19}. On the other hand, the topologically protected edge states experimentally observed in non-Hermitian optical and photonic systems~\cite{Poli-15, Zeuner-15, Weimann-17, Obuse-17, *Xiao-17, St-Jean-17, Bahari-17, Zhao-18, Parto-18, Harari-18, *Bandres-18} can be understood by a line gap. The two definitions of the complex-energy gaps are thus complementary to each other. Moreover, the topological classification drastically changes according to the definition of the complex-energy gap, as discussed in detail in the next section. In the absence of symmetry, for example, a topological phase characterized by a point gap is present only in odd spatial dimensions, whereas a topological phase characterized with a line gap is present only in even spatial dimensions (see class A of Table~\ref{tab: complex AZ} in Sec.~\ref{sec: topological classification}). We note that Refs.~\cite{Esaki-11, *Sato-12, Shen-18, Kawabata-18} explicitly adopt a line gap, whereas Ref.~\cite{Gong-18, *Bandres-Segev-18} adopts a point gap.

\subsection{Point gap}
Although a complex-energy point $E = E_{\rm P}$ that serves as an obstacle in the complex-energy plane is arbitrary in the absence of symmetry, it is subject to restrictions in the presence of symmetry. For instance, it should be taken as ${\rm Im}\,E_{\rm P} = 0$ in the presence of TRS since eigenenergies come in $\left( E, E^{*} \right)$ pairs; it should be taken as $E_{\rm P} = 0$ in the presence of SLS since eigenenergies come in $\left( E, -E \right)$ pairs. Thus it is convenient to choose $E_{\rm P}$ to be zero energy, which leads to the precise definition of the point gap as follows:

\bigskip
{\it Definition 1 (point gap) ---}
A non-Hermitian Hamiltonian $H \left( {\bm k} \right)$ is defined to have a point gap if and only if it is invertible (i.e., $\forall\,{\bm k}~\det H \left( {\bm k} \right) \neq 0$) and all the eigenenergies are nonzero (i.e., $\forall\,{\bm k}~~E \left( {\bm k} \right) \neq 0$).

\bigskip
Under this definition, a gapless system possesses a zero-energy state for some ${\bm k}$. A point gap helps understand the localization-delocalization transition in non-Hermitian systems in one dimension~\cite{Hatano-96, *Hatano-97, *Hatano-98, Shnerb-98,  Gong-18, *Bandres-Segev-18, Longhi-19} that occurs due to the competition between disorder and non-Hermiticity. Since one-dimensional Hermitian systems always show the Anderson localization, the delocalization is unique to non-Hermitian systems. Here, a topological invariant [i.e., the winding number in Eq.~(\ref{eq: topological invariant - winding (1D)})] can be assigned to a generic non-Hermitian system in one dimension. In the Hatano-Nelson model (i.e., a non-Hermitian extension of the one-dimensional Anderson model with asymmetric hopping)~\cite{Hatano-96, *Hatano-97, *Hatano-98}, wave functions are delocalized (localized) and the system is metallic (insulating) if the winding number is nonzero (zero)~\cite{Gong-18}. Moreover, it has recently been reported that localization (delocalization) of wave functions corresponds to the nontrivial (trivial) topology in a non-Hermitian quasicrystal (Aubry-Andr\'e-Harper model)~\cite{Longhi-19}.

\subsection{Line gap}
A complex-energy line that serves as an obstacle in the complex-energy plane can also be subject to restrictions in the presence of symmetry, whereas such a line is arbitrary in the absence of symmetry. In particular, it should be either the imaginary axis (${\rm Re}\,E = 0$) or the real axis (${\rm Im}\,E = 0$) when symmetry imposes a real structure on the complex spectrum. For instance, the real axis should be considered when pairs of eigenenergies $\left( E, E^{*} \right)$ appear with TRS; the imaginary axis should be considered when pairs of eigenenergies $\left( E, - E^{*} \right)$ appear with CS. In contrast to the point gap, there are no restrictions in the presence of SLS since SLS does not give the complex spectrum real structures [eigenenergies just come in $\left( E, -E \right)$ pairs]. Thus, it is convenient to choose the line that determines the complex gap as the imaginary axis (real gap) or the real axis (imaginary gap), which leads to the precise definition of the line gap in the following:

\bigskip
{\it Definition 2 (line gap) ---}
A non-Hermitian Hamiltonian $H \left( {\bm k} \right)$ is defined to have a line gap in the real (imaginary) part of its complex spectrum [real (imaginary) gap] if and only if it is invertible (i.e., $\forall\,{\bm k}~\det H \left( {\bm k} \right) \neq 0$) and the real (imaginary) part of all the eigenenergies is nonzero [i.e., $\forall\,{\bm k}~~{\rm Re}\,E \left( {\bm k} \right) \neq 0$ (${\rm Im}\,E \left( {\bm k} \right) \neq 0$)].

\bigskip
Under this definition of a real (imaginary) gap, a gapless system includes an eigenenergy with ${\rm Re}\,E \left( {\bm k} \right) = 0$ (${\rm Im}\,E \left( {\bm k} \right) = 0$) for some ${\bm k}$. Line gaps are employed explicitly in Refs.~\cite{Esaki-11, *Sato-12, Shen-18, Kawabata-18} and implicitly in many other pieces of work, and characterize topologically protected boundary states, which were also observed in experiments~\cite{Poli-15, Zeuner-15, Weimann-17, Obuse-17, *Xiao-17, St-Jean-17, Bahari-17, Zhao-18, Parto-18, Harari-18, *Bandres-18}. Remarkably, topologically protected boundary states in Hermitian systems are immune to non-Hermiticity as long as a real gap is open and relevant symmetry is respected, which is generally ensured by the nontrivial non-Hermitian topology in terms of line gaps. Furthermore, the presence of an imaginary gap has a significant influence on the nonequilibrium wave dynamics~\cite{Kawabata-18}, although it has no counterparts in the Hermitian band theory.

\section{Topological classification}
	\label{sec: topological classification}

\begin{figure*}[t]
\centering
\includegraphics[width=\linewidth]{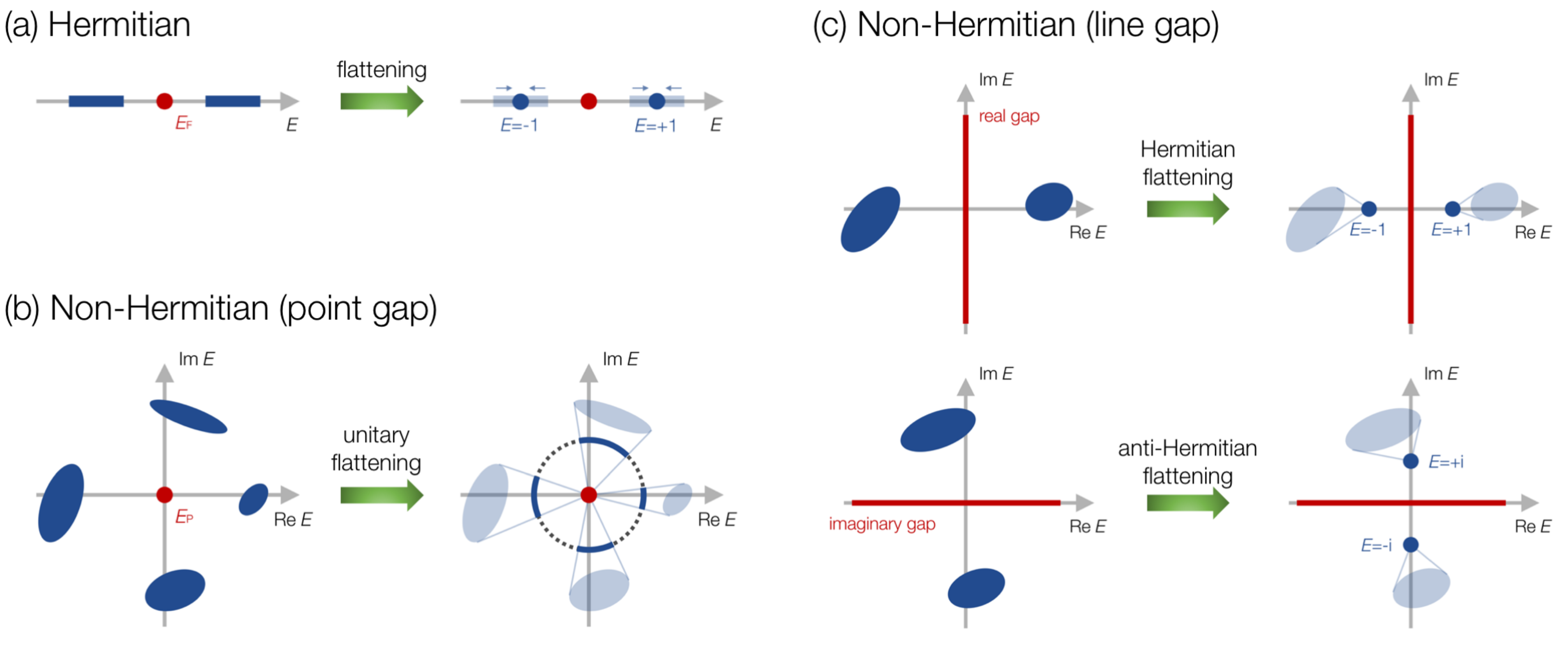} 
\caption{Flattening procedures of Hermitian and non-Hermitian Hamiltonians. (a)~Flattening of a Hermitian Hamiltonian with an energy gap. A Hermitian Hamiltonian can be flattened to another Hermitian Hamiltonian with $H^{2} = 1$ without closing the energy gap. (b)~Unitary flattening of a non-Hermitian Hamiltonian with a point gap. A non-Hermitian Hamiltonian can be flattened to a unitary Hamiltonian with $H^{\dag} H = 1$ without closing the point gap. (c)~Hermitian flattening of a non-Hermitian Hamiltonian with a line gap. A non-Hermitian Hamiltonian can be flattened to a Hermitian (an anti-Hermitian) Hamiltonian with $H^{2} = +1$ ($H^{2} = -1$) in the presence of a real (an imaginary) gap.}
	\label{fig: flattening}
\end{figure*}

We provide topological classification of non-Hermitian insulators and superconductors according to all the 38 symmetry classes discussed in Sec.~\ref{sec: symmetry} and the two types of the complex-energy gaps discussed in Sec.~\ref{sec: complex gap}. Here, non-Hermitian Hamiltonians $H_{0} \left( {\bm k} \right)$ and $H_{1} \left( {\bm k} \right)$ are defined to be topologically equivalent if and only if there exists a family of non-Hermitian Hamiltonians $H_{\lambda} \left( {\bm k} \right)$ ($0 \leq \lambda \leq 1$) that interpolates between them, i.e.,  
\begin{equation}
H_{\lambda = 0} \left( {\bm k} \right) = H_{0} \left( {\bm k} \right),\quad
 H_{\lambda = 1} \left( {\bm k} \right) = H_{1} \left( {\bm k} \right)
\end{equation}
with certain symmetries and a complex-energy gap for all $\lambda \in \left[ 0, 1 \right]$. Our strategy is to reduce this non-Hermitian problem to the established topological classification of Hermitian Hamiltonians in the AZ symmetry class without~\cite{Schnyder-08, Kitaev-09, Ryu-10} and with~\cite{Shiozaki-Sato-14} additional symmetries. In particular, we demonstrate that a non-Hermitian Hamiltonian can be continuously deformed into a unitary matrix and hence a larger Hermitian matrix in the presence of a point gap [Fig.~\ref{fig: flattening}\,(b); see also Theorem 1 below and its proof in Appendix~\ref{appendix: unitary flattening} for details] and a Hermitian or an anti-Hermitian matrix in the presence of a line gap [Fig.~\ref{fig: flattening}\,(c); see also Theorem 2 below and its proof in Appendix~\ref{appendix: Hermitian flattening} for details]. The \textit{K}-theory classification for point gaps is also discussed in Appendix~\ref{appendix: K-theory}.

Our results are listed in the periodic tables for the complex AZ symmetry class (Table~\ref{tab: complex AZ}), the real AZ symmetry class (Table~\ref{tab: real AZ}), the real ${\rm AZ}^{\dag}$ symmetry class (Table~\ref{tab: real AZ*}), the complex AZ symmetry class with SLS (Table~\ref{tab: complex AZ + SLS}), and the real AZ symmetry class with SLS (Table~\ref{tab: real AZ + SLS}). In addition to this 38-fold topological classification, we provide the periodic tables for the AZ symmetry class with pseudo-Hermiticity (Tables~\ref{tab: complex AZ + pH} and \ref{tab: real AZ + pH}). The 7-fold periodic table based on two antiunitary symmetries ($\Tp$ and $\Tm$) and unitary symmetry ($\SLS$) is also shown in Table~\ref{tab: two antiunitary symmetries} in Appendix~\ref{appendix: two antiunitary symmetries}.

\subsection{Unitary flattening for point gaps}
	\label{sec: classification (unitary flattening)}

In the presence of a point gap, a non-Hermitian Hamiltonian can be flattened into a unitary matrix without point-gap closing. This property is guaranteed by the following theorem (see Appendix~\ref{appendix: unitary flattening} for a proof):

\bigskip
{\it Theorem 1 (unitary flattening for point gaps) ---}
If a non-Hermitian Hamiltonian $H \left( {\bm k} \right)$ has a point gap, it can be continuously deformed into a unitary matrix $U \left( {\bm k} \right)$ while keeping the point gap and its symmetry [Fig.~\ref{fig: flattening}\,(b)].

\bigskip
Theorem 1 reduces the topological classification of a non-Hermitian Hamiltonian to that of a unitary matrix. Furthermore, with the flattened unitary matrix $U \left( {\bm k} \right)$, we have a flattened Hermitian matrix
\begin{equation}
\tilde{H} \left( {\bm k} \right)
:= \left( \begin{array}{@{\,}cc@{\,}} 
	0 & U \left( {\bm k} \right) \\
	U^{\dag} \left( {\bm k} \right) & 0 \\ 
	\end{array} \right),~~\tilde{H}^{2} \left( {\bm k} \right) = 1.
	\label{eq: def - tilde H}
\end{equation}
Here the presence of symmetry for the original non-Hermitian Hamiltonian $H \left( {\bm k} \right)$ discussed in Sec.~\ref{sec: symmetry} imposes the following constraints on the extended Hermitian Hamiltonian $\tilde{H} \left( {\bm k} \right)$:
\begin{eqnarray}
\tilde{\cal T}_{\pm} \tilde{H}^{*} \left( {\bm k} \right) \tilde{\cal T}_{\pm}^{-1}
&=& \pm \tilde{H} \left( -{\bm k} \right),~
\tilde{\cal T}_{\pm} := \left( \begin{array}{@{\,}cc@{\,}} 
	{\cal T}_{\pm} & 0 \\
	0 & {\cal T}_{\pm} \\ 
	\end{array} \right); 
	\label{eq: T - tilde}
\end{eqnarray}
\begin{eqnarray}
\tilde{\cal C}_{\pm} \tilde{H}^{*} \left( {\bm k} \right) \tilde{\cal C}_{\pm}^{-1}
&=& \pm \tilde{H} \left( -{\bm k} \right),~
\tilde{\cal C}_{\pm} := \left( \begin{array}{@{\,}cc@{\,}} 
	0 & {\cal C}_{\pm} \\
	{\cal C}_{\pm} & 0 \\ 
	\end{array} \right); 
	\label{eq: C - tilde}
\end{eqnarray}
\begin{eqnarray}
\tilde{\CS} \tilde{H} \left( {\bm k} \right) \tilde{\CS}^{-1}
&=& - \tilde{H} \left( {\bm k} \right),~
\tilde{\CS} := \left( \begin{array}{@{\,}cc@{\,}} 
	0 & {\CS} \\
	{\CS} & 0 \\ 
	\end{array} \right); 
	\label{eq: CS - tilde}
\end{eqnarray}
\begin{eqnarray}
\tilde{\SLS} \tilde{H} \left( {\bm k} \right) \tilde{\SLS}^{-1}
&=& - \tilde{H} \left( {\bm k} \right),~
\tilde{\SLS} := \left( \begin{array}{@{\,}cc@{\,}} 
	\SLS & 0 \\
	0 & \SLS \\ 
	\end{array} \right); 
	\label{eq: SLS - tilde}
\end{eqnarray}
\begin{eqnarray}
\tilde{\eta} \tilde{H} \left( {\bm k} \right) \tilde{\eta}^{-1}
&=& \tilde{H} \left( {\bm k} \right),~
\tilde{\eta} := \left( \begin{array}{@{\,}cc@{\,}} 
	0 & {\eta} \\
	{\eta} & 0 \\ 
	\end{array} \right).
	\label{eq: pH - tilde} 
\end{eqnarray}
Moreover, $\tilde{H} \left( {\bm k} \right)$ respects additional CS (SLS):
\begin{equation}
\Sigma \tilde{H} \left( {\bm k} \right) \Sigma^{-1}
= -\tilde{H} \left( {\bm k} \right),~~
\Sigma := \left( \begin{array}{@{\,}cc@{\,}} 
	1 & 0 \\
	0 & -1 \\ 
	\end{array} \right).
	\label{eq: sigma - tilde}
\end{equation}
Importantly, there exists a one-to-one correspondence between a unitary matrix $U \left( {\bm k} \right)$ and an extended Hermitian matrix $\tilde{H} \left( {\bm k} \right)$ that satisfies Eq.~(\ref{eq: sigma - tilde})~\cite{Gong-18, *Bandres-Segev-18, Roy-17}, and hence topology of $H \left( {\bm k} \right)$ can also be captured by the extended Hermitian Hamiltonian $\tilde{H} \left( {\bm k} \right)$. Therefore, the topological classification of a non-Hermitian Hamiltonian $H \left( {\bm k} \right)$ with a point gap and symmetry reduces to that of a Hermitian Hamiltonian that respects symmetry given by Eqs.~(\ref{eq: T - tilde})-(\ref{eq: sigma - tilde}), which was already obtained in Refs.~\cite{Schnyder-08, Kitaev-09, Ryu-10, Shiozaki-Sato-14}. In this manner, the periodic tables under point gaps are obtained as Tables~\ref{tab: complex AZ}-\ref{tab: real AZ + pH}. Notably, a similar theorem was proved in Ref.~\cite{Gong-18}. However, it is not applicable in the presence of ${\cal C}_{\pm}$, and Theorem 1 in the present work is a nontrivial generalization of the theorem in Ref.~\cite{Gong-18}.

Let us consider class DIII as an example (Table~\ref{tab: real AZ}). The original non-Hermitian Hamiltonian $H \left( {\bm k} \right)$ respects both TRS and PHS:
\begin{eqnarray}
\Tp H^{*} \left( {\bm k} \right) \Tp^{-1}
= H \left( -{\bm k} \right),~~\Tp \Tp^{*} = -1; \label{eq: DIII - TRS} \\
\Cm H^{T} \left( {\bm k} \right) \Cm^{-1}
= - H \left( -{\bm k} \right),~~\Cm \Cm^{*} = +1. \label{eq: DIII - PHS}
\end{eqnarray}
As a result, the extended and flattened Hermitian Hamiltonian $\tilde{H} \left( {\bm k} \right)$ respects TRS described by Eq.~(\ref{eq: T - tilde}) with $\tilde{\cal T}_{+} \tilde{\cal T}^{*}_{+} = -1$ and PHS described by Eq.~(\ref{eq: C - tilde}) with $\tilde{\cal C}_{-} \tilde{\cal C}^{*}_{-} = +1$, as well as additional CS (SLS) described by Eq.~(\ref{eq: sigma - tilde}). Therefore, the topological classification of the original non-Hermitian Hamiltonian reduces to that of the Hermitian Hamiltonian in class DIII with additional CS that commutes with TRS and anticommutes with PHS. The topology of such Hermitian Hamiltonians is characterized by the classifying space ${\cal R}_{4}$~\cite{Shiozaki-Sato-14}.

\subsection{Hermitian flattening for line gaps}
	\label{sec: classification (Hermitian flattening)}

In contrast to the unitary flattening for point gaps, the flattening procedure changes for line gaps. In fact, a non-Hermitian Hamiltonian can be flattened into a Hermitian matrix in the presence of a real gap and an anti-Hermitian matrix in the presence of an imaginary gap. This property is guaranteed by the following theorem (see Appendix~\ref{appendix: Hermitian flattening} for a proof):

\bigskip
{\it Theorem 2 (Hermitian flattening for line gaps) ---}
If a non-Hermitian Hamiltonian $H \left( {\bm k} \right)$ has a line gap in the real (imaginary) part of its complex spectrum [real (imaginary) gap], it can be continuously deformed into a Hermitian (an anti-Hermitian) matrix while keeping the line gap and its symmetry [Fig.~\ref{fig: flattening}\,(c)]. 

\bigskip
Theorem 2 also reduces the topological classification of a non-Hermitian Hamiltonian to that of a Hermitian matrix~\cite{Schnyder-08, Kitaev-09, Ryu-10, Shiozaki-Sato-14}. Here, we note that topology of an anti-Hermitian Hamiltonian $H \left( {\bm k} \right)$ [i.e, $H^{\dag} \left( {\bm k} \right) = - H \left( {\bm k} \right)$] under an imaginary gap is equivalent to that of a Hermitian Hamiltonian $\ii H \left( {\bm k} \right)$ under a real gap~\cite{Kawabata-18}. The periodic tables under line gaps are also obtained as Tables~\ref{tab: complex AZ}-\ref{tab: real AZ + pH}.

Let us again consider class DIII as an example (Table~\ref{tab: real AZ}). The original non-Hermitian Hamiltonian $H \left( {\bm k}\right)$ respects both TRS and PHS as Eqs.~(\ref{eq: DIII - TRS}) and (\ref{eq: DIII - PHS}), respectively. In the presence of a real gap, $H \left( {\bm k}\right)$ can be flattened to a Hermitian Hamiltonian $\bar{H} \left( {\bm k}\right)$ that belongs to class DIII, which is characterized by the classifying space ${\cal R}_{3}$~\cite{Schnyder-08, Kitaev-09, Ryu-10}. In the presence of an imaginary gap, on the other hand, $H \left( {\bm k}\right)$ can be flattened to an anti-Hermitian Hamiltonian $\bar{H} \left( {\bm k}\right)$ that respects Eqs.~(\ref{eq: DIII - TRS}) and (\ref{eq: DIII - PHS}). Importantly, the topology of $\bar{H} \left( {\bm k}\right)$ is equivalent to that of $\bar{\bar{H}} \left( {\bm k}\right) := \ii \bar{H} \left( {\bm k}\right)$, which respects Hermiticity and  
\begin{eqnarray}
\Tp \bar{\bar{H}}^{*} \left( {\bm k} \right) \Tp^{-1}
= - \bar{\bar{H}} \left( -{\bm k} \right),~~\Tp \Tp^{*} = -1; \label{eq: DIII - TRS - bb} \\
\Cm \bar{\bar{H}}^{T} \left( {\bm k} \right) \Cm^{-1}
= - \bar{\bar{H}} \left( -{\bm k} \right),~~\Cm \Cm^{*} = +1. \label{eq: DIII - PHS - bb}
\end{eqnarray}
Here complex conjugation and transposition coincide with each other due to the presence of Hermiticity, and Eq.~(\ref{eq: DIII - PHS - bb}) reduces to the antiunitary constraint given by
\begin{eqnarray}
\Cm \bar{\bar{H}}^{*} \left( {\bm k} \right) \Cm^{-1}
= - \bar{\bar{H}} \left( -{\bm k} \right),~~\Cm \Cm^{*} = +1. \label{eq: DIII - PHS - bbb}
\end{eqnarray}
Thus, the non-Hermitian Hamiltonian $H \left( {\bm k} \right)$ under an imaginary gap reduces to the Hermitian Hamiltonian $\bar{\bar{H}} \left( {\bm k}\right)$ that respects the two antiunitary symmetries as Eqs.~(\ref{eq: DIII - TRS - bb}) and (\ref{eq: DIII - PHS - bbb}). The topology of such Hermitian Hamiltonians is characterized by the classifying space ${\cal C}_{0}$~\cite{Shiozaki-Sato-14}.

\onecolumngrid \clearpage
\begin{table*}[h]
	\centering
	\caption{Topological classification table for non-Hermitian systems in the complex AZ symmetry class. Non-Hermitian topological phases are classified according to the AZ symmetry class, the spatial dimension $d$, and the definition of complex-energy point (P) or line (L) gaps. The subscript of L specifies the line gap for the real or imaginary part of the complex spectrum. \\}
	\label{tab: complex AZ}
     \begin{tabular}{ccccccccccc} \hline \hline
    ~AZ class~ & ~Gap~ & ~Classifying space~ & ~$d=0$~ & ~$d=1$~ & ~$d=2$~ & ~$d=3$~ & ~$d=4$~ & ~$d=5$~ & ~$d=6$~ & ~$d=7$~ \\ \hline
    \multirow{2}{*}{A}
    & P & \Cb \\
    & L & \Ca \\ \hline
    \multirow{3}{*}{AIII}
    & P & \Ca \\    
    & \Lr & \Cb \\ 
    & \Li & \Caa \\ \hline \hline
  \end{tabular}
\end{table*}

\bigskip \bigskip \bigskip
\begin{table*}[h]
	\centering
	\caption{Topological classification table for non-Hermitian systems in the real AZ symmetry class. Non-Hermitian topological phases are classified according to the AZ symmetry class, the spatial dimension $d$, and the definition of complex-energy point (P) or line (L) gaps. The subscript of L specifies the line gap for the real or imaginary part of the complex spectrum. \\}
		\label{tab: real AZ}
     \begin{tabular}{ccccccccccc} \hline \hline
    ~AZ class~ & ~Gap~ & ~Classifying space~ & ~$d=0$~ & ~$d=1$~ & ~$d=2$~ & ~$d=3$~ & ~$d=4$~ & ~$d=5$~ & ~$d=6$~ & ~$d=7$~ \\ \hline
    \multirow{3}{*}{AI}
    & P & \Rb \\
    & \Lr & \Ra \\
    & \Li & \Rc \\ \hline
    \multirow{3}{*}{BDI}
    & P & \Rc \\
    & \Lr & \Rb \\
    & \Li & \Rcc \\ \hline
    \multirow{2}{*}{D}
    & P & \Rd \\
    & L & \Rc \\ \hline
    \multirow{3}{*}{DIII}
    & P & \Ree \\
    & \Lr & \Rd \\
    & \Li & \Ca \\ \hline
    \multirow{3}{*}{AII}
    & P & \Rf \\
    & \Lr & \Ree \\
    & \Li & \Rg \\ \hline
    \multirow{3}{*}{CII}
    & P & \Rg \\
    & \Lr & \Rf \\
    & \Li & \Rgg \\ \hline
    \multirow{2}{*}{C}
    & P & \Rh \\
    & L & \Rg \\ \hline
    \multirow{3}{*}{CI}
    & P & \Ra \\
    & \Lr & \Rh \\
    & \Li & \Ca \\ \hline \hline
  \end{tabular}
\end{table*}

\begin{table*}[h]
	\centering
	\caption{Topological classification table for non-Hermitian systems in the real $\text{AZ}^{\dag}$ symmetry class. Non-Hermitian topological phases are classified according to the $\text{AZ}^{\dag}$ symmetry class, the spatial dimension $d$, and the definition of complex-energy point (P) or line (L) gaps. The subscript of L specifies the line gap for the real or imaginary part of the complex spectrum. \\}
		\label{tab: real AZ*}
     \begin{tabular}{ccccccccccc} \hline \hline
    ~$\text{AZ}^{\dag}$ class~ & ~Gap~ & ~Classifying space~ & ~$d=0$~ & ~$d=1$~ & ~$d=2$~ & ~$d=3$~ & ~$d=4$~ & ~$d=5$~ & ~$d=6$~ & ~$d=7$~ \\ \hline
    \multirow{2}{*}{$\text{AI}^{\dag}$}
    & P & \Rh \\
    & L & \Ra \\ \hline
    \multirow{3}{*}{$\text{BDI}^{\dag}$}
    & P & \Ra \\
    & \Lr & \Rb \\
    & \Li & \Raa \\ \hline
    \multirow{3}{*}{$\text{D}^{\dag}$}
    & P & \Rb \\
    & \Lr & \Rc \\
    & \Li & \Ra \\ \hline
    \multirow{3}{*}{$\text{DIII}^{\dag}$}
    & P & \Rc \\
    & \Lr & \Rd \\
    & \Li & \Ca \\ \hline
    \multirow{2}{*}{$\text{AII}^{\dag}$}
    & P & \Rd \\
    & L & \Ree \\ \hline
    \multirow{3}{*}{$\text{CII}^{\dag}$}
    & P & \Ree \\
    & \Lr & \Rf \\
    & \Li & \Reee \\ \hline
    \multirow{3}{*}{$\text{C}^{\dag}$}
    & P & \Rf \\
    & \Lr & \Rg \\
    & \Li & \Ree \\ \hline
    \multirow{3}{*}{$\text{CI}^{\dag}$}
    & P & \Rg \\
    & \Lr & \Rh \\
    & \Li & \Ca \\ \hline \hline
  \end{tabular}
\end{table*}

\begin{table*}[h]
	\centering
	\caption{Topological classification table for non-Hermitian systems in the complex AZ symmetry class with sublattice symmetry (SLS). Non-Hermitian topological phases are classified according to the AZ symmetry class with additional SLS, the spatial dimension $d$, and the definition of complex-energy point (P) or line (L) gaps. The subscript of L specifies the line gap for the real or imaginary part of the complex spectrum. The subscript of $\SLS_{\pm}$ specifies the commutation ($+$) or anticommutation ($-$) relation to chiral symmetry: $\CS \SLS_{\pm} = \pm \SLS_{\pm} \CS$. \\}
		\label{tab: complex AZ + SLS}
     \begin{tabular}{cccccccccccc} \hline \hline
    ~SLS~ & ~AZ class~ & ~Gap~ & ~Classifying space~ & ~$d=0$~ & ~$d=1$~ & ~$d=2$~ & ~$d=3$~ & ~$d=4$~ & ~$d=5$~ & ~$d=6$~ & ~$d=7$~ \\ \hline
    \multirow{3}{*}{$\SLS_{+}$} & \multirow{3}{*}{AIII}
    & P & \Cb \\
    & & \Lr & \Cbb \\
    & & \Li & \Cbb \\ \hline \hline
    \multirow{2}{*}{$\SLS$} & \multirow{2}{*}{A}
    & P & \Cbb \\
    & & L & \Cb \\ \hline
    \multirow{3}{*}{$\SLS_{-}$} & \multirow{3}{*}{AIII}
    & P & \Caa \\
    & & \Lr & \Ca \\
    & & \Li & \Ca \\ \hline \hline
  \end{tabular}
\end{table*}

\begin{table*}[h]
	\centering
	\caption{Topological classification table for non-Hermitian systems in the real AZ symmetry class with sublattice symmetry (SLS). Non-Hermitian topological phases are classified according to the AZ symmetry class with additional SLS, the spatial dimension $d$, and the definition of complex-energy point (P) or line (L) gaps. The subscript of L specifies the line gap for the real or imaginary part of the complex spectrum. The subscript of $\SLS_{\pm}$ specifies the commutation ($+$) or anticommutation ($-$) relation to time-reversal symmetry (TRS) and/or particle-hole symmetry (PHS). For the symmetry classes that involve both TRS and PHS (BDI, DIII, CII, and CI), the first subscript specifies the relation to TRS and the second one to PHS. \\}
	\label{tab: real AZ + SLS}
     \begin{tabular}{cccccccccccc} \hline \hline
    ~SLS~ & ~AZ class~ & ~Gap~ & ~Classifying space~ & ~$d=0$~ & ~$d=1$~ & ~$d=2$~ & ~$d=3$~ & ~$d=4$~ & ~$d=5$~ & ~$d=6$~ & ~$d=7$~ \\ \hline
    \multirow{3}{*}{$\SLS_{++}$} & \multirow{3}{*}{BDI}
    & P & \Rb \\
    & & \Lr & \Rbb \\
    & & \Li & \Rbb \\ \hline
    \multirow{3}{*}{$\SLS_{--}$} & \multirow{3}{*}{DIII}
    & P & \Rd \\
    & & \Lr & \Rdd \\
    & & \Li & \Cb \\ \hline
    \multirow{3}{*}{$\SLS_{++}$} & \multirow{3}{*}{CII}
    & P & \Rf \\
    & & \Lr & \Rff \\
    & & \Li & \Rff \\ \hline
    \multirow{3}{*}{$\SLS_{--}$} & \multirow{3}{*}{CI}
    & P & \Rh \\
    & & \Lr & \Rhh \\
    & & \Li & \Cb \\ \hline \hline
    \multirow{3}{*}{$\SLS_{-}$} & \multirow{3}{*}{AI}
    & P & \Cb \\
    & & \Lr & \Rh \\
    & & \Li & \Rd \\ \hline
    \multirow{3}{*}{$\SLS_{-+}$} & \multirow{3}{*}{BDI}
    & P & \Ca \\
    & & \Lr & \Ra \\
    & & \Li & \Rc \\ \hline
    \multirow{2}{*}{$\SLS_{+}$} & \multirow{2}{*}{D}
    & P & \Cb \\
    & & L & \Rb \\ \hline
    \multirow{3}{*}{$\SLS_{-+}$} & \multirow{3}{*}{DIII}
    & P & \Ca \\
    & & \Lr & \Rc \\
    & & \Li & \Ra \\ \hline
    \multirow{3}{*}{$\SLS_{-}$} & \multirow{3}{*}{AII}
    & P & \Cb \\
    & & \Lr & \Rd \\
    & & \Li & \Rh \\ \hline
    \multirow{3}{*}{$\SLS_{-+}$} & \multirow{3}{*}{CII}
    & P & \Ca \\
    & & \Lr & \Ree \\
    & & \Li & \Rg \\ \hline
    \multirow{2}{*}{$\SLS_{+}$} & \multirow{2}{*}{C}
    & P & \Cb \\
    & & L & \Rf \\ \hline
    \multirow{3}{*}{$\SLS_{-+}$} & \multirow{3}{*}{CI}
    & P & \Ca \\
    & & \Lr & \Rg \\
    & & \Li & \Ree \\ \hline \hline
    \multicolumn{12}{c}{continued on next page}
  \end{tabular}
\end{table*}

\begin{table*}[h]
	\centering
     \begin{tabular}{cccccccccccc}
    \multicolumn{12}{c}{TABLE \ref{tab: real AZ + SLS} --- continued} \\ \hline \hline
    \multirow{3}{*}{~$\SLS_{--}$~} & \multirow{3}{*}{~~BDI~~}
    & ~~P~~ & \Rd \\
    & & \Lr & \Cb \\
    & & \Li & \Rdd \\ \hline
    \multirow{3}{*}{$\SLS_{++}$} & \multirow{3}{*}{DIII}
    & P & \Rf \\
    & & \Lr & \Cb \\
    & & \Li & \Cb \\ \hline
    \multirow{3}{*}{$\SLS_{--}$} & \multirow{3}{*}{CII}
    & P & \Rh \\
    & & \Lr & \Cb \\
    & & \Li & \Rhh \\ \hline
    \multirow{3}{*}{$\SLS_{++}$} & \multirow{3}{*}{CI}
    & P & \Rb \\
    & & \Lr & \Cb \\
    & & \Li & \Cb \\ \hline \hline
    \multirow{3}{*}{$\SLS_{+}$} & \multirow{3}{*}{AI}
    & ~P~ & ~\Rbb~ \\
    & & \Lr & \Rb \\
    & & \Li & \Rb \\ \hline
    \multirow{3}{*}{$\SLS_{+-}$} & \multirow{3}{*}{~BDI~}
    & P & \Rcc \\
    & & \Lr & \Rc \\
    & & \Li & \Rc \\ \hline
    \multirow{2}{*}{$\SLS_{-}$} & \multirow{2}{*}{D}
    & P & \Rdd \\
    & & L & \Rd \\ \hline
    \multirow{3}{*}{$\SLS_{+-}$} & \multirow{3}{*}{DIII}
    & P & \Reee \\
    & & \Lr & \Ree \\
    & & \Li & \Ree \\ \hline
    \multirow{3}{*}{$\SLS_{+}$} & \multirow{3}{*}{AII}
    & P & \Rff \\
    & & \Lr & \Rf \\
    & & \Li & \Rf \\ \hline
    \multirow{3}{*}{$\SLS_{+-}$} & \multirow{3}{*}{CII}
    & P & \Rgg \\
    & & \Lr & \Rg \\
    & & \Li & \Rg \\ \hline
    \multirow{2}{*}{$\SLS_{-}$} & \multirow{2}{*}{C}
    & P & \Rhh \\
    & & L & \Rh \\ \hline
    \multirow{3}{*}{$\SLS_{+-}$} & \multirow{3}{*}{CI}
    & P & \Raa \\
    & & \Lr & \Ra \\
    & & \Li & \Ra \\ \hline \hline
  \end{tabular}
\end{table*}

\begin{table*}[h]
	\centering
	\caption{Topological classification table for non-Hermitian systems in the complex AZ symmetry class with pseudo-Hermiticity (pH). Non-Hermitian topological phases are classified according to the AZ symmetry class with additional pH, the spatial dimension $d$, and the definition of complex-energy point (P) or (L) gaps. The subscript of L specifies the line gap for the real or imaginary part of the complex spectrum. The subscript of $\eta_{\pm}$ specifies the commutation ($+$) or anticommutation ($-$) relation to chiral symmetry: $\Gamma \eta_{\pm} = \pm \eta_{\pm} \Gamma$. \\}
		\label{tab: complex AZ + pH}
     \begin{tabular}{cccccccccccc} \hline \hline
    ~pH~ & ~AZ class~ & ~Gap~ & ~Classifying space~ & ~$d=0$~ & ~$d=1$~ & ~$d=2$~ & ~$d=3$~ & ~$d=4$~ & ~$d=5$~ & ~$d=6$~ & ~$d=7$~ \\ \hline
    \multirow{3}{*}{$\eta$} & \multirow{3}{*}{A}
    & P & \Ca \\
    & & \Lr & \Caa \\
    & & \Li & \Cb \\ \hline
    \multirow{3}{*}{$\eta_{+}$} & \multirow{3}{*}{AIII}
    & P & \Cb \\
    & & \Lr & \Cbb \\
    & & \Li & \Cbb \\ \hline \hline
    \multirow{3}{*}{$\eta_{-}$} & \multirow{3}{*}{AIII}
    & P & \Caa \\
    & & \Lr & \Ca \\
    & & \Li & \Ca \\ \hline \hline
  \end{tabular}
\end{table*}

\begin{table*}[h]
	\centering
	\caption{Topological classification table for non-Hermitian systems in the real AZ symmetry class with pseudo-Hermiticity (pH). Non-Hermitian topological phases are classified according to the AZ symmetry class with additional pH, the spatial dimension $d$, and the definition of complex-energy point (P) or line (L) gaps. The subscript of L specifies the line gap for the real or imaginary part of the complex spectrum. The subscript of $\eta_{\pm}$ specifies the commutation ($+$) or anticommutation ($-$) relation to time-reversal symmetry (TRS) and/or particle-hole symmetry (PHS). For the symmetry classes that involve both TRS and PHS (BDI, DIII, CII, and CI), the first subscript specifies the relation to TRS and the second one to PHS. \\}
		\label{tab: real AZ + pH}
     \begin{tabular}{cccccccccccc} \hline \hline
    ~pH~ & ~AZ class~ & ~Gap~ & ~Classifying space~ & ~$d=0$~ & ~$d=1$~ & ~$d=2$~ & ~$d=3$~ & ~$d=4$~ & ~$d=5$~ & ~$d=6$~ & ~$d=7$~ \\ \hline
    \multirow{3}{*}{$\eta_{+}$} & \multirow{3}{*}{AI}
    & P & \Ra \\
    & & \Lr & \Raa \\
    & & \Li & \Rb \\ \hline
    \multirow{3}{*}{$\eta_{++}$} & \multirow{3}{*}{BDI}
    & P & \Rb \\
    & & \Lr & \Rbb \\
    & & \Li & \Rbb \\ \hline
    \multirow{3}{*}{$\eta_{+}$} & \multirow{3}{*}{D}
    & P & \Rc \\
    & & \Lr & \Rcc \\
    & & \Li & \Rb \\ \hline
    \multirow{3}{*}{$\eta_{++}$} & \multirow{3}{*}{DIII}
    & P & \Rd \\
    & & \Lr & \Rdd \\
    & & \Li & \Cb \\ \hline
    \multirow{3}{*}{$\eta_{+}$} & \multirow{3}{*}{AII}
    & P & \Ree \\
    & & \Lr & \Reee \\
    & & \Li & \Rf \\ \hline
    \multirow{3}{*}{$\eta_{++}$} & \multirow{3}{*}{CII}
    & P & \Rf \\
    & & \Lr & \Rff \\
    & & \Li & \Rff \\ \hline
    \multirow{3}{*}{$\eta_{+}$} & \multirow{3}{*}{C}
    & P & \Rg \\
    & & \Lr & \Rgg \\
    & & \Li & \Rf \\ \hline
    \multirow{3}{*}{$\eta_{++}$} & \multirow{3}{*}{CI}
    & P & \Rh \\
    & & \Lr & \Rhh \\
    & & \Li & \Cb \\ \hline \hline
    \multicolumn{12}{c}{continued on next page}
  \end{tabular}
\end{table*}

\begin{table*}[h]
	\centering
     \begin{tabular}{cccccccccccc}
    \multicolumn{12}{c}{TABLE \ref{tab: real AZ + pH} --- continued} \\ \hline \hline
    \multirow{3}{*}{$\eta_{+-}$} & \multirow{3}{*}{BDI}
    & P & \Ca \\
    & & \Lr & \Ra \\
    & & \Li & \Rc \\ \hline
    \multirow{3}{*}{$\eta_{-+}$} & \multirow{3}{*}{DIII}
    & P & \Ca \\
    & & \Lr & \Rc \\
    & & \Li & \Ra \\ \hline
    \multirow{3}{*}{$\eta_{+-}$} & \multirow{3}{*}{CII}
    & P & \Ca \\
    & & \Lr & \Ree \\
    & & \Li & \Rg \\ \hline
    \multirow{3}{*}{$\eta_{-+}$} & \multirow{3}{*}{CI}
    & P & \Ca \\
    & & \Lr & \Rg \\
    & & \Li & \Ree \\ \hline \hline
    \multirow{3}{*}{$\eta_{-}$} & \multirow{3}{*}{~~AI~~}
    & ~~P~~ & \Rc \\
    & & \Lr & \Ca \\
    & & \Li & \Rd \\ \hline
    \multirow{3}{*}{$\eta_{--}$} & \multirow{3}{*}{BDI}
    & P & \Rd \\
    & & \Lr & \Cb \\
    & & \Li & \Rdd \\ \hline
    \multirow{3}{*}{$\eta_{-}$} & \multirow{3}{*}{D}
    & P & \Ree \\
    & & \Lr & \Ca \\
    & & \Li & \Rd \\ \hline
    \multirow{3}{*}{$\eta_{--}$} & \multirow{3}{*}{DIII}
    & P & \Rf \\
    & & \Lr & \Cb \\
    & & \Li & \Cb \\ \hline
    \multirow{3}{*}{$\eta_{-}$} & \multirow{3}{*}{AII}
    & P & \Rg \\
    & & \Lr & \Ca \\
    & & \Li & \Rh \\ \hline
    \multirow{3}{*}{$\eta_{--}$} & \multirow{3}{*}{CII}
    & P & \Rh \\
    & & \Lr & \Cb \\
    & & \Li & \Rhh \\ \hline
    \multirow{3}{*}{$\eta_{-}$} & \multirow{3}{*}{C}
    & P & \Ra \\
    & & \Lr & \Ca \\
    & & \Li & \Rh \\ \hline
    \multirow{3}{*}{$\eta_{--}$} & \multirow{3}{*}{CI}
    & P & \Rb \\
    & & \Lr & \Cb \\
    & & \Li & \Cb \\ \hline \hline
    \multirow{3}{*}{~$\eta_{-+}$~} & \multirow{3}{*}{~BDI~}
    & ~P~ & \Rcc \\
    & & \Lr & \Rc \\
    & & \Li & \Rc \\ \hline
    \multirow{3}{*}{$\eta_{+-}$} & \multirow{3}{*}{DIII}
    & P & \Reee \\
    & & \Lr & \Ree \\
    & & \Li & \Ree \\ \hline
    \multirow{3}{*}{$\eta_{-+}$} & \multirow{3}{*}{CII}
    & P & \Rgg \\
    & & \Lr & \Rg \\
    & & \Li & \Rg \\ \hline
    \multirow{3}{*}{$\eta_{+-}$} & \multirow{3}{*}{CI}
    & P & \Raa \\
    & & \Lr & \Ra \\
    & & \Li & \Ra \\ \hline \hline
  \end{tabular}
\end{table*}

\clearpage
\twocolumngrid
\subsection{Topological invariants}
	\label{sec: topological invariants}
	
Based on our flattening procedures, the topological invariants for Tables~\ref{tab: complex AZ}-\ref{tab: real AZ + pH} are obtained in a systematic manner. For point gaps, the extended Hermitian Hamiltonian $\tilde{H} \left( {\bm k} \right)$ defined as Eq.~(\ref{eq: def - tilde H}) from a non-Hermitian Hamiltonian $H \left( {\bm k} \right)$ is relevant. The topological invariant of $H \left( {\bm k} \right)$ reduces to that of $\tilde{H} \left( {\bm k} \right)$, the latter of which is already obtained in the literature~\cite{Schnyder-Ryu-review}. For instance, let us consider a non-Hermitian system $H \left( {\bm k} \right)$ with a point gap and no symmetry, which can have topological phases in odd dimensions as shown in Table~\ref{tab: complex AZ}. Since the extended Hermitian system $\tilde{H} \left( {\bm k} \right)$ respects CS, the topological invariant that characterizes $H \left( {\bm k} \right)$ is given as the winding number
\begin{equation}
W_{2n+1} := \frac{n!}{\left( 2\pi \ii \right)^{n+1} \left( 2n+1 \right)!} \int_{\mathrm{BZ}^{d}} \mathrm{tr} \left( H^{-1} dH \right)^{2n+1}.
	\label{eq: topological invariant - winding}
\end{equation}
In one dimension, this formula reduces to~\cite{Gong-18, *Bandres-Segev-18}
\begin{equation} \begin{split}
W_{1} &= \oint_{\mathrm{BZ}} \frac{dk}{2\pi \ii}~\mathrm{tr} \left( H^{-1} \frac{dH}{dk} \right)
= \oint_{\mathrm{BZ}} \frac{dk}{2\pi \ii} \left( \frac{d}{dk} \log \det H \right).
	\label{eq: topological invariant - winding (1D)}
\end{split} \end{equation}
Similarly, the topological invariants for the other symmetry classes are readily determined (see Appendix~\ref{appendix: topological invariants - point} for more examples).

For line gaps, a non-Hermitian Hamiltonian $H \left( {\bm k} \right)$ can always be continuously deformed into a Hermitian (or an anti-Hermitian) Hamiltonian. As a result, the topological invariant of a non-Hermitian Hamiltonian $H \left( {\bm k} \right)$ with a line gap is obtained in a similar manner as in the Hermitian case. Remarkably, right and left eigenstates are different from each other in sharp contrast to the Hermitian case. Nevertheless, topological invariants are given by the combination of both eigenstates~\cite{Esaki-11, *Sato-12, Shen-18}. For instance, in even spatial dimensions $d=2n$, the $n$-th Chern number $C_{n}$ can be defined as
\begin{eqnarray}
C_{n} &:=& \frac{1}{n!} \left( \frac{\ii}{2\pi} \right)^{n} \int_{\mathrm{BZ}^{d}} \mathrm{tr}\,\mathcal{F}^{n} \nonumber \\
&=& - \frac{1}{2^{2n+1} n!} \left( \frac{\ii}{2\pi} \right)^{n} \int_{\mathrm{BZ}^{d}} \mathrm{tr} \left[ Q_{R} \left( dQ_{R} \right)^{2n} \right] \nonumber \\
&=& \frac{n!}{\left( 2\pi \ii \right)^{n+1} \left( 2n+1\right)!} \int_{\mathbb{R}_{\omega} \times \mathrm{BZ}^{d}} \mathrm{tr} \left( G dG^{-1} \right)^{2n+1},\qquad
	\label{eq: topological invariant - Chern - line}
\end{eqnarray}
as long as a line gap is open. Here, $\mathcal{F}$, $Q_{\rm R}$, and $G$ are non-Hermitian extensions of the Berry curvature, the $Q$ matrix, and the Green function, respectively (see Appendix~\ref{appendix: line - Chern} for details). In addition, in odd spatial dimensions $d=2n+1$, the winding number $W_{2n+1}$ can be defined in the presence of CS defined by Eq.~(\ref{eq: CS}) with a unitary operator $\Gamma$ as
\begin{equation}
W_{2n+1} := - \frac{n!}{2 \left( 2\pi \ii \right)^{n+1} \left( 2n+1 \right)!} \int_{\mathrm{BZ}^{d}} \mathrm{tr} \left[ \Gamma \left( Q^{-1} dQ \right)^{2n+1} \right],
	\label{eq: topological invariant - winding - line}
\end{equation}
as long as a real gap is open. Here, $Q$ is another non-Hermitian extension of the $Q$ matrix that respects CS. Whereas $Q_{\rm R}$ is non-Hermitian but $Q_{\rm R}^{2} = 1$, $Q$ is Hermitian but $Q^{2} \neq 1$ (see Appendix~\ref{appendix: CS - line - winding number} for details). In this manner, the topological invariants for the other symmetry classes are readily obtained (see Appendix~\ref{appendix: topological invariants - line} for more examples). Remarkably, it is shown that the single Chern number is defined in two-dimensional non-Hermitian systems although we have four different Berry connections and curvatures due to the distinction between right and left eigenstates~\cite{Shen-18}. Our classification indeed corroborates this fact, as well as generally demonstrates that the conventional topological invariants in Hermitian systems are uniquely generalized to non-Hermitian systems in the presence of a line gap.

We note in passing that the classification is based on homotopy over the sphere in momentum space and is thus modified if the full Brillouin-zone torus of lattice systems is considered. This modification is the same as that for the Hermitian case because the topological classification of non-Hermitian systems reduces to that of unitary, Hermitian, and anti-Hermitian systems, all of which in turn result in the topological classification of Hermitian systems. In fact, the classification becomes finer, and we have weak topological invariants as well as the strong topological invariants shown in Tables~\ref{tab: complex AZ}-\ref{tab: real AZ + pH}. In three dimensions, for instance, we may have three, three, and one weak topological invariants corresponding to the classification table for two, one, and zero dimensions, respectively, in addition to one strong topological invariant corresponding to the classification table for three dimensions.

\subsection{Dirac Hamiltonian}
	\label{sec: Dirac matrix}

Hermitian topological insulators and superconductors can be universally understood with continuum models that have the massive Dirac Hamiltonian representation~\cite{Schnyder-Ryu-review}: 
\begin{equation}
H \left( {\bm k} \right)
= \sum_{i=1}^{d} k_{i} \Gamma_{i} + m \Gamma_{0},
\end{equation}
where ${\bm k} = ( k_{1}, \cdots, k_{d} )$ is the momentum deviation from a relevant momentum reference point, and $\Gamma_{1}, \cdots, \Gamma_{d}$ are Dirac matrices that satisfy the Clifford algebra (i.e., $\{ \Gamma_{i}, \Gamma_{j} \} = 2 \delta_{ij}$). The mass term $m\Gamma_{0}$ anticommutes with all the Dirac matrices $\Gamma_{1}, \cdots, \Gamma_{d}$ in the kinetic term and determines the topology of the classifying space.

Our classification suggests that non-Hermitian topological systems can also be described by a non-Hermitian generalization of the Dirac Hamiltonians. However, the complex-spectral-flattening procedures distinct from the Hermitian case imply that non-Hermiticity can modify the proper representation of Dirac matrices. In fact, in the presence of a point gap, non-Hermitian Dirac matrices $\Gamma_{i}^{\rm P}$ ($i=1, \cdots, d$) are defined so that their Hermitianized matrices 
\begin{equation}
\tilde{\Gamma}_{i}^{\rm P} := \left( \begin{array}{@{\,}cc@{\,}} 
	0 & \Gamma_{i}^{\rm P} \\
	(\Gamma_{i}^{\rm P})^{\dag} & 0 \\ 
	\end{array} \right)
\end{equation}
obey the Clifford algebra (i.e., $\{ \tilde{\Gamma}_{i}^{\rm P}, \tilde{\Gamma}_{j}^{\rm P} \} = 2 \delta_{ij}$), which in turn leads to the relations for $\Gamma_{i}^{\rm P}$
\begin{equation}
\Gamma_{i}^{\rm P} (\Gamma_{j}^{\rm P})^{\dag} + \Gamma_{j}^{\rm P} (\Gamma_{i}^{\rm P})^{\dag} = 2 \delta_{ij}.
\end{equation}
This set of relations determines the proper non-Hermitian Dirac matrices in the presence of a point gap. Table~\ref{tab: Dirac matrix} shows an example of the representations of $\Gamma_{i}^{\rm P}$'s ($i=1, \cdots, n$) for small $n$, which is clearly distinct from the conventional Dirac matrices. In the presence of a line gap, on the other hand, Dirac matrices take the same representation as the Hermitian case since a non-Hermitian Hamiltonian can be flattened into a Hermitian one or an anti-Hermitian one.

\begin{table}[t]
	\centering
	\caption{Non-Hermitian Dirac matrices. A set of non-Hermitian Dirac matrices $\Gamma_{1}, \cdots, \Gamma_{n}$ is shown for both cases of point gaps and line gaps for $n = 1, \cdots, 5$ with Pauli matrices $\sigma_{i}$ and $\tau_{i}$ ($i=x, y, z$). \\}
	\label{tab: Dirac matrix}
\begin{tabular}{cll} 
\hline \hline
~$n$~ & ~Point gap~ & ~Line gap~ \\ 
\hline
~$1$~ & ~$1$ & ~$1$ \\
~$2$~ & ~$1, \ii$ & ~$\sigma_{x}, \sigma_{y}$ \\
~$3$~ & ~$\ii, \sigma_{x}, \sigma_{y}$ & ~$\sigma_{x}, \sigma_{y}, \sigma_{z}$ \\
~$4$~ & ~$\ii, \sigma_{x}, \sigma_{y}, \sigma_{z}$ & ~$\sigma_{x}, \sigma_{y}, \sigma_{z} \tau_{x}, \sigma_{z} \tau_{y}$ \\
~$5$~ & ~$\ii, \sigma_{x}, \sigma_{y}, \sigma_{z}\tau_{x}, \sigma_{z}\tau_{y}$ & ~$\sigma_{x}, \sigma_{y}, \sigma_{z} \tau_{x}, \sigma_{z} \tau_{y}, \sigma_{z} \tau_{z}$ \\
\hline \hline
  \end{tabular}
\end{table}

The non-Hermitian Dirac Hamiltonian provides a systematic way to have a model for the classification tables. In 1D class A in Table~\ref{tab: complex AZ}, for instance, a non-Hermitian Dirac Hamiltonian can be expressed as
\begin{equation}
H \left( k \right) = k + \ii m
\quad \left( m \in \mathbb{R} \right)
\end{equation}
in the presence of a point gap. With this continuum model, the $\mathbb{Z}$ topological invariant (winding number) in Eq.~(\ref{eq: topological invariant - winding (1D)}) can be readily obtained as
\begin{eqnarray}
W &=& \int_{-\infty}^{\infty} \frac{dk}{2\pi\ii} \frac{d}{dk} \log \left( k+\ii m \right)
= \frac{{\rm sgn} \left( m \right)}{2}.
\end{eqnarray}
Here the fractional topological invariant $W$ is common to continuum models for both Hermitian and non-Hermitian cases and should be complemented by the structure of wave functions away from the relevant momentum point. It becomes an integer $W={\rm sgn} \left( m \right)$ when we regularize the mass $m$ as $m-k^2$.

\section{Bulk-boundary correspondence}
	\label{sec: BEC}
	
\subsection{Sensitivity to the boundary conditions}

The hallmark of topological phases is the bulk-boundary correspondence: topologically protected boundary states emerge that correspond to the nontrivial topology of the bulk states. Importantly, non-Hermiticity alters the nature of the bulk-boundary correspondence~\cite{Lee-16, MartinezAlvarez-18, Kunst-18, Yao-18-SSH, *Yao-18-Chern, Kawabata-18-Chern, Jin-19, Wang-19, Liu-19, Edvardsson-19, Lee-19, Kunst-19, Lee-Li-19, Yokomizo-19, Okuma-19}. The essential distinction from the Hermitian case is that the bulk spectra of non-Hermitian systems can dramatically change according to the boundary conditions. Nevertheless, the modified bulk-boundary correspondence has recently been established~\cite{Kunst-18, Yao-18-SSH, *Yao-18-Chern} despite the sensitivity of the spectrum to the boundary conditions.

Remarkably, our general classification depends not on the boundary conditions but solely on symmetry and spatial dimension, and indeed predicts the emergence of topologically protected boundary states. As described above, the bulk spectrum of a non-Hermitian Hamiltonian $H_{\rm PBC}$ with periodic boundaries can be different from that of the corresponding Hamiltonian $H_{\rm OBC}$ with open boundaries. Whereas the topologically protected edge states may not be described by the topological invariant determined by $H_{\rm PBC}$, they are accompanied by the nontrivial topology of $H_{\rm OBC}$. In particular, since a non-Hermitian Hamiltonian with a real gap (line gap for the real part of the spectrum) can always be continuously deformed into a Hermitian Hamiltonian as described in Sec.~\ref{sec: classification (Hermitian flattening)}, the topologically protected edge states in a certain Hermitian system survive as long as the real gap is open. It is also noteworthy that the wavenumber for $H_{\rm OBC}$ can be complex due to $H_{\rm PBC} \neq H_{\rm OBC}$, which signals the localization of bulk states, i.e., the non-Hermitian skin effect~\cite{Yao-18-SSH, *Yao-18-Chern}. Despite the complex wavenumber, our classification works because the bulk Hamiltonian is still given as a map from the $d$-dimensional torus as a generalized Brillouin zone to a non-Hermitian system with certain symmetry.

\begin{figure}[t]
\centering
\includegraphics[width=86mm]{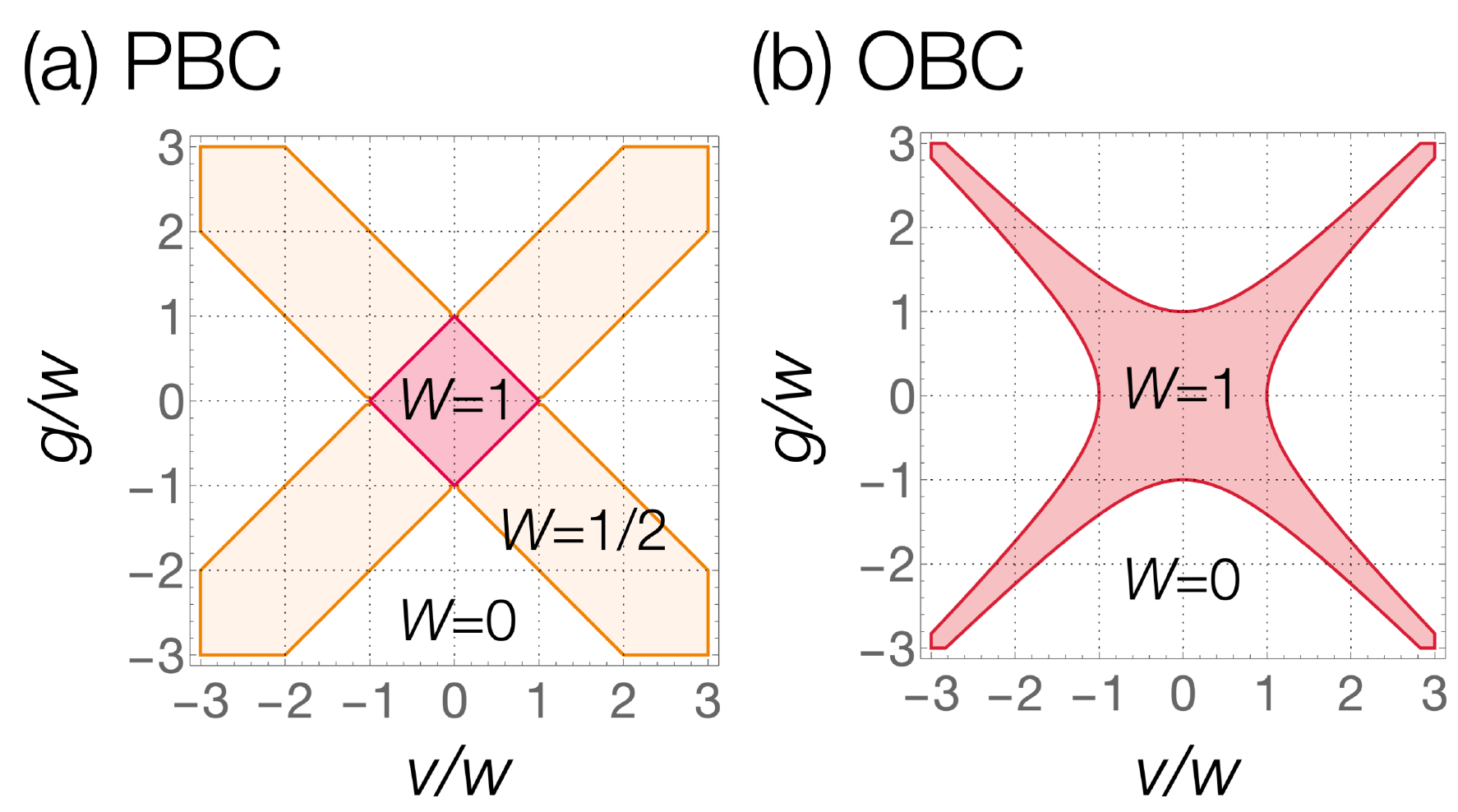} 
\caption{Phase diagrams of the non-Hermitian Su-Schrieffer-Heeger model with asymmetric hopping under (a)~the periodic boundary condition and (b)~the open boundary condition. The one-dimensional model is described by the intracell hopping $v$, the intercell hopping $w$, and the asymmetry $g$ of the intracell hopping $v$ as a degree of non-Hermiticity. Whereas the bulk Hamiltonian dramatically depends on the boundary conditions, the one-dimensional model respects sublattice symmetry and the winding number $W$ is well defined regardless of the boundary conditions. The emergence of the topologically protected edge states is predicted by the phase diagram (b) under the open boundary condition.}
	\label{fig: SSH}
\end{figure}

To illustrate the above ideas, we consider a non-Hermitian extension of the Su-Schrieffer-Heeger model (see Appendix~\ref{appendix: SSH} for details)~\cite{Lee-16, Lieu-18-SSH, MartinezAlvarez-18, Yin-18, Kunst-18, Yao-18-SSH, *Yao-18-Chern}:
\begin{eqnarray}
\hat{H}
&=& \sum_{i} \left[ \left( v-g \right) \hat{b}_{i}^{\dag} \hat{a}_{i} + \left( v+g \right) \hat{a}_{i}^{\dag} \hat{b}_{i} \right. \nonumber \\
&&\quad\quad\quad\quad\quad\quad\quad\left. + w\,( \hat{b}_{i-1}^{\dag} \hat{a}_{i} + \hat{a}_{i}^{\dag} \hat{b}_{i-1} ) \right],
	\label{eq: SSH - asymmetric hopping}
\end{eqnarray}
where $\hat{a}_{i}$ ($\hat{a}_{i}^{\dag}$) and $\hat{b}_{i}$ ($\hat{b}_{i}^{\dag}$) annihilate (create) a particle on site $i$ in sublattices, respectively, $v, w \in \mathbb{R}$ denote the hopping amplitudes, and $g \in \mathbb{R}$ denotes the asymmetry of the intracell hopping $v$ as a degree of non-Hermiticity. Regardless of the boundary conditions, the system respects SLS defined in Eq.~(\ref{eq: SLS}) with $\mathcal{S} := \sigma_{z}$:
\begin{equation}
\sigma_{z} H \left( k \right) \sigma_{z}^{-1} = - H \left( k \right),
\end{equation} 
where the bulk Hamiltonian $H \left( k \right)$ is obtained for each boundary condition. As a result of SLS, the $\mathbb{Z}$ topological invariant is well defined in the presence of a line gap (see Table~\ref{tab: complex AZ + SLS}), which is given as the winding number
\begin{equation}
W := \oint \frac{dk}{4\pi \ii}\,\mathrm{tr} \left[ \mathcal{S} H^{-1} \left( k\right) \frac{dH \left( k\right)}{dk} \right].
	\label{eq: SSH - winding}
\end{equation}
As discussed above, the bulk spectrum $E \left( k \right)$ crucially depends on the boundary conditions. In fact, under the periodic boundary condition, the bulk Hamiltonian is given as
\begin{equation}
H_{\rm PBC} \left( k \right) = \left( \begin{array}{@{\,}cc@{\,}} 
	0 & v + g + w\,e^{-\ii k} \\
	v - g + w\,e^{\ii k} & 0 \\ 
	\end{array} \right)
\end{equation}
with a real wavenumber $k \in \left[ 0, 2\pi \right)$. Under the open boundary condition, on the other hand, the wavenumber is complex and 
\begin{equation}
\left| e^{\ii k} \right|
= e^{- \mathrm{Im}\,k}
= \sqrt{\left| \frac{v-g}{v+g} \right|}.
\end{equation}
Correspondingly, the bulk Hamiltonian with open boundaries is given as
\begin{equation}
H_{\rm OBC} \left( q \right) = \left( \begin{array}{@{\,}cc@{\,}} 
	0 & \sqrt{v^{2}-g^{2}} + w\,e^{-\ii q} \\
	\sqrt{v^{2}-g^{2}} + w\,e^{\ii q} & 0 \\ 
	\end{array} \right)
	\label{eq: SSH - H - OBC}
\end{equation}
with a real wavenumber $q \in \left[ 0, 2\pi \right)$. Whereas the topologically protected edge states are not described by the winding number defined by $H_{\rm PBC}$, the bulk-edge correspondence holds for $H_{\rm OBC}$; there emerges a pair of edge states that are topologically protected to have zero energy for $W = 1$ determined by $H_{\rm OBC}$ (Fig.~\ref{fig: SSH}). In a similar manner, the emergence of topologically protected boundary states can be predicted on the basis of our classification. See also Sec.~\ref{sec: experiment} for the bulk-edge correspondence in other examples. 

\subsection{Symmetry restoration}
	\label{sec: BEC - symmetry}

Whereas generic non-Hermitian Hamiltonians are sensitive to the boundary conditions, certain non-Hermitian Hamiltonians may not depend on them. For example, the bulk wave functions are delocalized, and the non-Hermitian skin effect does not occur in a different non-Hermitian extension of the Su-Schrieffer-Heeger model with balanced gain and loss~\cite{Esaki-11, *Sato-12, Schomerus-13, Lieu-18-SSH, Weimann-17, Parto-18} (see also Sec.~\ref{sec: recent experiment} for details). Furthermore, a couple of recent works~\cite{Kunst-19, Yokomizo-19} show that bulk states are always delocalized in Hermitian systems, indicating the absence of the skin effect in the presence of Hermiticity. 

Here we show that symmetry plays a significant role in the non-Hermitian bulk-boundary correspondence. In particular, we find that the bulk Hamiltonian is generally insensitive to the boundary conditions in the presence of pseudo-Hermiticity in Eq.~(\ref{eq: def pseudo-Hermiticity}) or parity-time symmetry if the bulk spectrum is real. In one dimension, we also obtain the same property in the presence of $\mathrm{TRS}^{\dag}$ in Eq.~(\ref{eq: TRS-dag}) with $\mathcal{C}_{+} \mathcal{C}_{+}^{*} = +1$ or parity (inversion) symmetry, where the bulk spectrum can be complex. Parity symmetry is defined by
\begin{equation}
\mathcal{P} H \left( {\bm k} \right) \mathcal{P}^{-1} = H \left( - {\bm k} \right),\quad
\mathcal{P}^{2} = 1,
	\label{eq: def-parity}
\end{equation}
where $\mathcal{P}$ is a unitary matrix, and parity-time symmetry is defined by
\begin{equation}
(\mathcal{P}\mathcal{T}_{+})\,H^{*} \left( {\bm k} \right) (\mathcal{P}\mathcal{T}_{+})^{-1} 
= H \left( {\bm k} \right),~
(\mathcal{P}\mathcal{T}_{+}) (\mathcal{P}\mathcal{T}_{+})^{*} = \pm 1,
	\label{eq: def-PT}
\end{equation}
where $\mathcal{P}\mathcal{T}_{+}$ is a unitary matrix.

To demonstrate the insensitivity of the bulk Hamiltonian to the boundary conditions, we begin with the characteristic equation that determines the generalized Brillouin zone. For clarity, we consider a generic one-dimensional system with its length $L$. Suppose that the bulk Hamiltonian is given as 
\begin{equation}
H \left( k \right) = \sum_{n=-l}^{l} H_{n} e^{\ii kn},\quad
k \in \mathbb{C},
\end{equation}
where $l$ is the range of the hopping and $H_{n}$ is an $N \times N$ matrix that describes the internal degrees of freedom. If the non-Hermitian system is periodic and respects translational symmetry, the wavenumber $k$ is real due to Bloch's theorem. On the other hand, $k$ can be complex under the open boundary condition. The Schr\"odinger equation leads to
\begin{equation}
\det \left( \sum_{n=-l}^{l} H_{n} e^{\ii kn} - E \right) = 0,
	\label{eq: characteristic}
\end{equation}
which is a $2lN$-th order algebraic equation for $e^{\ii k}$. Its solutions are denoted as $e^{\ii k_{1}}, \cdots, e^{\ii k_{2lN}}$ ($| e^{\ii k_{1}} | \leq \cdots \leq | e^{\ii k_{2lN}} |$). It is shown that the condition
\begin{equation}
| e^{\ii k_{lN}} | = | e^{\ii k_{lN+1}} |
	\label{eq: YW-YM}
\end{equation}
is necessary so that the bulk spectrum is continuum for $L \to \infty$ and the generalized Brillouin zone is determined as the trajectory of $k_{lN}$ and $k_{lN+1}$~\cite{Yao-18-SSH, *Yao-18-Chern, Yokomizo-19}.

We demonstrate that the wavenumbers $k_{lN}$ and $k_{lN+1}$ should be real regardless of the boundary conditions if pseudo-Hermiticity is present (i.e., $\eta H^{\dag} \left( k \right) \eta^{-1} = H \left( k \right)$) and the eigenenergy $E$ is real. Under the periodic boundary condition, $k$ is real and the presence of pseudo-Hermiticity leads to
\begin{equation}
\eta H_{-n}^{\dag} \eta^{-1} = H_{n}.
\end{equation}
By contrast, the wavenumber $k$ can generally be complex under the open boundary condition. Nevertheless, from the characteristic equation~(\ref{eq: characteristic}), we have
\begin{equation}
\det \left( \sum_{n=-l}^{l} H_{n} e^{\ii k^{*} n} - E \right) = 0,
\end{equation}
where the reality of $E$ and pseudo-Hermiticity are used. This equation implies that the solution comes in $( e^{\ii k}, e^{\ii k^{*}} )$ pairs and we have $e^{\ii k_{2lN+1-i}} = e^{\ii k_{i}^{*}}$ due to $| e^{\ii k^{*}} | = 1/| e^{\ii k} |$ and $| e^{\ii k_{1}} | \leq \cdots \leq | e^{\ii k_{2lN}} |$. In particular, we have $e^{\ii k_{lN+1}} = e^{\ii k_{lN}^{*}}$ and Eq.~(\ref{eq: YW-YM}) reduces to
\begin{equation}
| e^{\ii k_{lN}} | = | e^{\ii k_{lN+1}} | = 1,
	\label{eq: real wavenumber}
\end{equation}
which indicates $k_{lN}, k_{lN+1} \in \mathbb{R}$, i.e., the reality of the wavenumbers even under the open boundary condition. Similarly, in the presence of parity-time symmetry, we have $(\mathcal{P}\mathcal{T}_{+})\,H_{-n}^{*}\,(\mathcal{P}\mathcal{T}_{+})^{-1} = H_{n}$ and the pair structure $( e^{\ii k}, e^{\ii k^{*}} )$ for $E \in \mathbb{R}$, leading to the reality of the wavenumbers regardless of the boundary conditions. The insensitivity of bulk systems with real spectra to the boundary conditions is consistent with recent experiments and relevant for symmetry-protected topological lasers (see Sec.~\ref{sec: experiment} for details). It is also notable that Ref.~\cite{Kunst-19} obtains a similar result for unbroken parity-time symmetry in a transfer-matrix perspective.

In one dimension, parity symmetry or $\mathrm{TRS}^{\dag}$ with $\mathcal{C}_{+} \mathcal{C}_{+}^{*} = +1$ leads to the real wavenumbers even for complex spectra $E \in \mathbb{C}$. In fact, we have
\begin{equation}
\mathcal{C}_{+}\,H_{-n}^{T}\,\mathcal{C}_{+}^{-1} = H_{n}
\end{equation}
in the presence of $\mathrm{TRS}^{\dag}$, which then leads to
\begin{equation}
\det \left( \sum_{n=-l}^{l} H_{n} e^{- \ii k n} - E \right) = 0
\end{equation}
from the characteristic equation~(\ref{eq: characteristic}). We stress that the reality of $E$ is not used to derive these equations. Consequently, the solution comes in $( e^{\ii k}, e^{-\ii k} )$ pairs and $e^{\ii k_{2lN+1-i}} = e^{-\ii k_{i}}$, resulting in Eq.~(\ref{eq: real wavenumber}). Similarly, in the presence of parity symmetry, we have $\mathcal{P}\,H_{-n}\,\mathcal{P}^{-1} = H_{n}$ and the real wavenumbers. Therefore, these types of symmetries restore the delocalization of bulk wave functions. We note that the above argument may not work for $\text{TRS}^{\dag}$ with $\mathcal{C}_{+} \mathcal{C}_{+}^{*} = -1$ since the $( e^{\ii k}, e^{-\ii k} )$ pairs form Kramers pairs and thus cannot mix with each other.

Before closing this section, a couple of comments are in order. First, in generic systems, we may have additional unitary symmetry that commutes with the Hamiltonian. In this case, the Hamiltonian is block-diagonal in the eigenbasis of the unitary symmetry, and we need one of the above symmetries in each eigensector to avoid the non-Hermitian skin effect. Second, similar symmetry protection of the conventional bulk-boundary correspondence occurs in the presence of other point group symmetry or magnetic point group symmetry. For instance, if reflection symmetry is respected, i.e.,
\begin{equation}
\mathcal{R}\,H\left( {\bm k_{\parallel}}, {\bm k_{\perp}} \right) \mathcal{R}^{-1}
= H \left( {\bm k_{\parallel}}, - {\bm k_{\perp}}\right),
\end{equation}
where $\mathcal{R}$ is a unitary matrix, the non-Hermitian skin effect does not occur for the boundaries parallel to the reflection plane.

\section{Simple examples}
	\label{sec: simple examples}

\subsection{Real and imaginary gaps}
	\label{sec: 1D AIII}

For each symmetry class and each spatial dimension, multiple topological structures appear in the classification tables. For instance, since CS acts as an anti-Hermitian conjugation for non-Hermitian Hamiltonians as Eq.~(\ref{eq: CS}), it distinguishes between real and imaginary gaps, both of which give different topological structures (Table~\ref{tab: complex AZ}). To understand this unique non-Hermitian feature in detail, we consider a $2 \times 2$ non-Hermitian Hamiltonian in one dimension,
\begin{align}
H \left( k \right) 
= h_{0} \left( k \right) \sigma_{0} + {\bm h} \left( k \right) \cdot {\bm \sigma},
\end{align}
where $\sigma_{0}$ is the $2 \times 2$ identity matrix and ${\bm \sigma} = \left( \sigma_{x}, \sigma_{y}, \sigma_{z} \right)$ is a set of Pauli matrices. Imposing CS with $\Gamma := \sigma_z$, we have the following constraints on $h_{i} \left( k \right)$ ($i=0, x, y, z$): 
\begin{align}
h_{0, z}^{*} \left( k \right) = - h_{0, z} \left( k \right), \quad
h_{x, y}^{*} \left( k \right) = h_{x, y} \left( k \right),
\end{align}
which imply that $h_{0} \left( k \right)$ and $h_{z} \left( k \right)$ [$h_{x} \left( k \right)$ and $h_{y} \left( k \right)$] are pure imaginary (real) for all $k$. Therefore, by redefining $h_{0} \left( k \right)$ and $h_{z} \left( k \right)$ as 
$h_{0} \left( k \right) \to \ii h_{0} \left( k \right)$ and $h_{z} \left( k \right) \to \ii h_{z} \left( k \right)$, we obtain the Hamiltonian with CS as 
\begin{align}
H \left( k \right) 
= \ii h_{0} \left( k \right) \sigma_{0} + h_{x} \left( k \right) \sigma_{x} + h_{y} \left( k \right) \sigma_{y} + \ii h_{z} \left( k \right)\sigma_{z}, 
\end{align} 
where $h_{i} \left( k \right)$'s ($i=0, x, y, z$) are real functions. The eigenenergies of $H \left( k \right)$ are given as
\begin{align}
E \left( k \right) = 
\ii h_{0} \left( k \right) \pm \sqrt{h_{x}^{2} \left( k \right) + h_{y}^{2} \left( k \right) - h_{z}^{2} \left( k \right)}, 
\end{align}
and thus the system supports a real (an imaginary) gap for $h^{2}_{x} \left( k \right) + h_{y}^{2} \left( k \right) > h_{z}^{2} \left( k \right)$ [$h^{2}_{x} \left( k \right) + h_{y}^{2} \left( k \right) < h_{z}^{2} \left( k \right)$]. 

\begin{figure}[t]
\centering
\includegraphics[width=86mm]{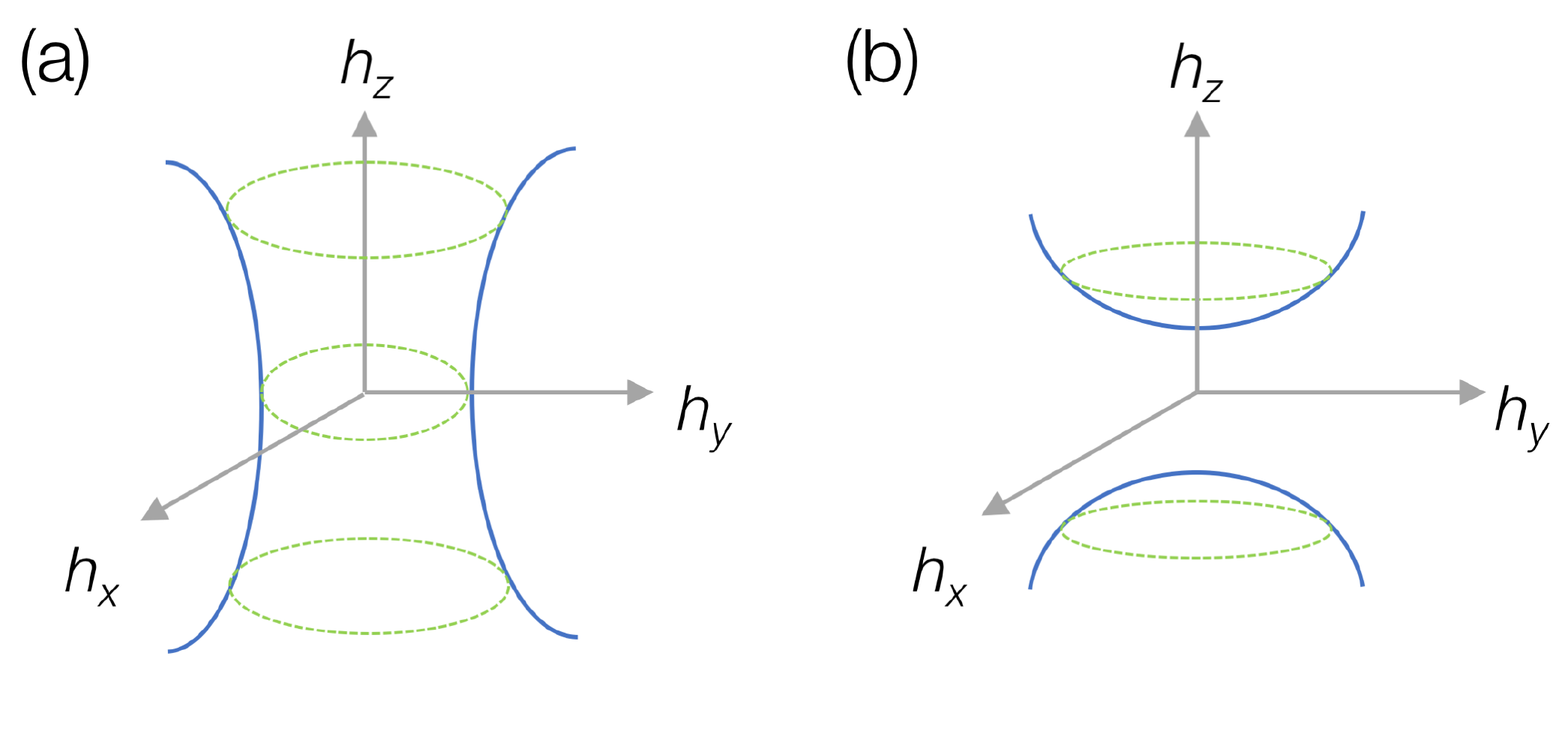} 
\caption{Topology of $2 \times 2$ non-Hermitian Hamiltonians in one dimension that respect chiral symmetry (1D class AIII). The blue curves delineate the surface determined by the complex-spectral flattening. (a)~Hermitian flattening with $E = \pm 1$ in the presence of a real gap. A winding number can be defined over the surface defined by $h_{x}^{2} + h_{y}^{2} - h_{z}^{2} = 1$. (b)~Anti-Hermitian flattening with $E = \pm \ii$ in the presence of an imaginary gap. Topologically stable loops are absent on the surface defined by $h_{x}^{2} + h_{y}^{2} - h_{z}^{2} = -1$, and hence no winding number is well defined.}
	\label{fig: classAIII}
\end{figure}

First we consider the case with a real gap. After the Hermitian flattening with $E \left( k \right) = \pm 1$, $h_{i} \left( k \right)$'s obey
\begin{align}
h_{0} \left( k \right) = 0,
\quad 
h_{x}^{2} \left( k \right) + h_{y}^{2} \left( k \right) - h_{z}^{2} \left( k \right) = 1.
\end{align}
These conditions define a surface in the parameter space $\left( h_{x}, h_{y}, h_{z} \right)$ of the Hamiltonian [Fig.~\ref{fig: classAIII}\,(a)]. The surface is open in the $h_{z}$-direction and circular in the other directions. Each Hamiltonian with a real gap gives an image from the one-dimensional Brillouin zone through ${\bm h} \left( k \right)$, which draws a circle on the surface. Obviously, a one-dimensional winding number can be defined just by counting how many times the circle winds the surface, which coincides with $W_{1}$ in Eq.~(\ref{eq: topological invariant - winding - line}).

By contrast, we cannot have such a winding number in the case with an imaginary gap. After the anti-Hermitian flattening with $E \left( k \right) = \pm \ii$, we have
\begin{align}
h_{0} \left( k \right) = 0, 
\quad 
h_{x}^{2} \left( k \right) + h_{y}^2 \left( k \right) - h_{z}^2 \left( k \right) = -1,
\end{align} 
which gives a surface in Fig.~\ref{fig: classAIII}\,(b). Since topologically stable loops are absent on this surface, no one-dimensional winding number is well defined. The above observation is fully consistent with the periodic table~\ref{tab: complex AZ}: for 1D class AIII, the topological invariant is characterized by an integer for a real gap [the winding number $W_{1}$ in Eq.~(\ref{eq: topological invariant - winding - line})], while it is trivial for an imaginary gap.

\subsection{Pseudo-Hermiticity}
	\label{sec : pH - topology}
	
Although there are topological phases characterized by the Chern number in a two-dimensional system without symmetry (Table~\ref{tab: complex AZ}), these topological phases vanish in the presence of TRS with $\Tp \Tp^{*} = +1$ under a real gap (Table~\ref{tab: real AZ}). However, in the presence of pseudo-Hermiticity $\eta_{-}$ that anticommutes with TRS ($\eta_{-} \Tp = - \Tp \eta_{-}^{*}$), a different type of topological phases emerges that is described by the time-reversal-invariant Chern number~\cite{Esaki-11, *Sato-12}, as shown in detail below. Therefore, pseudo-Hermiticity provides a novel topological structure as a unique non-Hermitian feature. 

To see this unique property of pseudo-Hermiticity, we start with the standard procedure for diagonalization of non-Hermitian Hamiltonians.
Let $|u_n\rangle$ ($|u_n\rangle\!\rangle$) be a right (left) eigenstate of $H$
\begin{align}
H\,|u_n\rangle
=E_{n}\,|u_n\rangle,\quad
H^{\dagger}\,|u_n\rangle\!\rangle
=E^{*}_{n}\,|u_n\rangle\!\rangle. 
\end{align}
The eigenstates form the biorthonormal basis~\cite{Brody-14}, which obey 
\begin{align}
\langle u_m|u_n\rangle\!\rangle=\langle\!\langle u_m|u_n\rangle=\delta_{mn}, 
\end{align}
with the completeness condition
\begin{align}
\sum_n|u_n\rangle\!\rangle \langle u_n|=\sum_n|u_n\rangle \langle\!\langle u_n|=1. 
\end{align}
We compactly express these biorthonormal relations as
\begin{align}
R^{\dagger}L=L^{\dagger}R=RL^\dagger =LR^{\dagger}=1
\end{align}
by introducing $2N\times 2N$ matrices
\begin{align}
R := (|u_1\rangle, |u_2\rangle, ...),\quad
L := (|u_1\rangle\!\rangle, |u_2\rangle\!\rangle, ...),
\end{align}
which diagonalize $H$ as
\begin{align}
H =R\left(
\begin{array}{ccc}
E_1 & & \\
& E_2 & \\
&& \ddots
\end{array}
\right) R^{-1}. 
\end{align} 

Now let us see how pseudo-Hermiticity imposes an additional constraint. From pseudo-Hermiticity defined by $\eta^{-1} H^{\dag} \eta = H$, we have
\begin{align}
H \left[ \eta\,|u_n\rangle\!\rangle \right]=E_n^* \left[ \eta\,|u_n\rangle\!\rangle \right],
\end{align}
which implies that $\eta\,|u_n\rangle\!\rangle$ is a right eigenstate of $H$ with eigenenergy $E_{n}^{*}$. Therefore, $E_n$, in general, has a complex-conjugate partner $E_n^*$ in the spectrum of $H$. This simple structure leads to significant consequences. In particular, an isolated real eigenenergy of $H$ remains real for any continuous deformation of $H$ unless it coalesces with other eigenenergies. In fact, the above constraint implies that eigenenergies should come in complex-conjugate pairs, and thus an isolated real eigenenergy cannot become complex by itself. Such reality of the spectrum is important to obtain stable states in non-Hermitian systems. For example, it is relevant for topological lasers as discussed in Sec.~\ref{sec: SPT laser} and for free bosons as discussed in Sec.~\ref{sec: free boson}.

Pseudo-Hermiticity also gives a nontrivial topological structure~\cite{Esaki-11, *Sato-12}. In the presence of a real gap at ${\rm Re}\,E=E_{\rm F}$, for instance, we can define an ``empty'' (``occupied'') state as a state with ${\rm Re}\,E_n>E_{\rm F}$ (${\rm Re}\,E_n<E_{\rm F}$).
If $|u_n\rangle$ is an occupied (empty) state, so is $\eta\,|u_n\rangle\!\rangle$ 
since $|u_n\rangle$ and $\eta\,|u_n\rangle\!\rangle$ have eigenenergies with the same real part.
From the completeness of the basis, we can relate them as
\begin{align}
\eta\,|u_n\rangle\!\rangle =\sum_m |u_m\rangle\,c_{mn},
	\label{eq: etau-u}
\end{align} 
with $c_{mn} := \langle\!\langle u_m|\eta|u_n\rangle\!\rangle$. Here, we have $c_{mn}= 0$ for $E_m\neq E_n^*$, and $c_{mn}$ is Hermitian with respect to the indices $m$ and $n$ (i.e., $c_{mn} = c_{nm}^{*}$)
since $\eta$ is Hermitian.
Thus, we can diagonalize $c_{mn}$ by a unitary matrix $G$
without mixing between occupied and empty states, 
\begin{align}
\sum_{lk}G^*_{lm}c_{lk}G_{kn}=\lambda_m\delta_{mn}, 
\end{align}
with real eigenvalues $\lambda_m$. Taking the new biorthogonal basis
\begin{align}
|\phi_n\rangle := \sum_m |u_m\rangle\, G_{mn} \sqrt{|\lambda_{n}|},~
|\phi_n\rangle\!\rangle := \sum_m |u_m\rangle\!\rangle\,\frac{G_{mn}}{\sqrt{|\lambda_{n}|}},
\end{align}
with $\langle \phi_m|\phi_n\rangle\!\rangle=\langle\!\langle \phi_m|\phi_n\rangle=\delta_{mn}$, 
we have 
\begin{align}
\eta\,|\phi_n\rangle = {\rm sgn} \left( \lambda_{n} \right) |\phi_n\rangle\!\rangle. 
\end{align}
Therefore, both occupied states and empty states in the new basis are divided into two subsectors, i.e., states with
\begin{align}
\eta\,|\phi_n\rangle = +|\phi_n\rangle\!\rangle, 
\end{align}
and states with
\begin{align}
\eta\,|\phi_n\rangle=-|\phi_n\rangle\!\rangle.
\end{align}
We denote these states as $|\phi_n^{\pm}\rangle$. This is a non-Hermitian generalization of the $\eta$-eigensector.
In fact, for Hermitian $H$, the right and left eigenstates coincide with each other, and thus the above equations reduce to the eigenequations of $\eta$: $\eta\,|\phi^{\pm}_n\rangle=\pm\,|\phi^{\pm}_n\rangle$.
Importantly, the new basis $\{ |\phi^{\pm}_n\rangle \}$ is no longer eigenstates of $H$ unless $E_n$ is real.
Indeed, $G_{mn}$ mixes the eigenstate with $E_n$ and that with $E_n^*$.
However, since $G_{mn}$ does not mix occupied and empty states, the new basis $\{ |\phi^{\pm}_n\rangle \}$ keeps the same topology as the original one $\{ |u_n\rangle \}$, while the subsector structure manifests itself only in the new basis. 

The presence of this subsector structure enables us to introduce a topological invariant for each subsector. For instance, in the case of class A with pseudo-Hermiticity in two dimensions, the Chern number $C_{1}^{\pm}$ in Eq.~(\ref{eq: topological invariant - Chern - line}) is defined for each subsector. As shown in Table \ref{tab: complex AZ + pH}, these two independent Chern numbers agree with two integer topological invariants (${\mathbb Z}\oplus{\mathbb Z}$) in 2D class A with pseudo-Hermiticity in the presence of a real gap (see Appendix~\ref{appendix: pH - invariant} for another formula). This subsector structure also makes it possible to define the nonzero Chern number even in time-reversal-symmetric systems. From time-reversal symmetry, the total Chern number vanishes (i.e., $C_{n}^{+}+C_{n}^{-}=0$), but their difference can be nonzero [i.e., $(C_n^{+}-C_n^{-})/2\neq 0\in {\mathbb Z}$], which is referred to as the time-reversal-invariant Chern number in Ref.~\cite{Esaki-11, *Sato-12}. As shown in Table~\ref{tab: real AZ + pH}, our classification correctly captures this integer invariant ($\mathbb Z$) in the presence of a real gap. It is also found that pseudo-Hermiticity is naturally imposed on free bosonic systems, which is discussed separately in Sec.~\ref{sec: free boson}.

\section{Experimental relevance}
	\label{sec: experiment}

\subsection{Recent experiments}
	\label{sec: recent experiment}

Our topological classification of non-Hermitian systems based on internal symmetry is directly relevant to various experiments in nonequilibrium open systems with gain and/or loss~\cite{Poli-15, Zeuner-15, Weimann-17, Obuse-17, *Xiao-17, St-Jean-17, Bahari-17, Zhao-18, Parto-18, Harari-18, *Bandres-18}. In fact, the observed topologically protected edge states are justified by the periodic tables~\ref{tab: complex AZ}-\ref{tab: real AZ + pH}. For instance, topologically protected bound states were observed in a passive dimerized photonic crystal in one dimension~\cite{Weimann-17}. Moreover, lasing topological edge states were observed in an active (pumped) array of microring resonators in one dimension~\cite{Parto-18}. Both systems are essentially described by the Su-Schrieffer-Heeger model~\cite{SSH-79} with balanced gain and loss~\cite{Esaki-11, *Sato-12, Schomerus-13, Lieu-18-SSH}:
\begin{eqnarray}
\hat{H}
&=& \sum_{i} \left[ v\,( \hat{b}_{i}^{\dag} \hat{a}_{i} + \hat{a}_{i}^{\dag} \hat{b}_{i} )
+ w\,( \hat{b}_{i-1}^{\dag} \hat{a}_{i} + \hat{a}_{i}^{\dag} \hat{b}_{i-1} ) \right. \nonumber \\
&&~~~~~~~~~~\left. + \ii \gamma\,( \hat{a}_{i}^{\dag} \hat{a}_{i} - \hat{b}_{i}^{\dag} \hat{b}_{i} ) \right],
\end{eqnarray}
where $\hat{a}_{i}$ ($\hat{a}_{i}^{\dag}$) and $\hat{b}_{i}$ ($\hat{b}_{i}^{\dag}$) annihilate (create) a photon on site $i$ in sublattices, respectively, $v, w \in \mathbb{R}$ denote the hopping amplitudes, and $\gamma \in \mathbb{R}$ denotes the balanced gain and loss as a degree of non-Hermiticity. In momentum space, the Bloch Hamiltonian is obtained as
\begin{equation}
H \left( k \right) = \left( \begin{array}{@{\,}cc@{\,}} 
	\ii \gamma & v + w\,e^{-\ii k} \\
	v + w\,e^{\ii k} & - \ii \gamma \\ 
	\end{array} \right).
\end{equation}
Although this system no longer respects SLS due to the presence of gain and loss, it remains to respect CS defined by Eq.~(\ref{eq: CS}) with $\CS := \sigma_{z}$:
\begin{equation}
\sigma_{z} H^{\dag} \left( k \right) \sigma_{z}^{-1} = - H \left( k \right).
\end{equation}
This system thus belongs to AZ symmetry class AIII, and our classification table~\ref{tab: complex AZ} predicts the topological phase characterized by integers under the definition of a real gap (line gap in the real part of the complex spectrum). This $\mathbb{Z}$ topological phase is characterized by the winding number $W_{1}$ in Eq.~(\ref{eq: topological invariant - winding - line}), where the Hermitian $Q$ matrix is given as~\cite{Esaki-11, *Sato-12}
\begin{equation}
Q \left( k \right) = \frac{1}{\sqrt{\left| v + we^{-\ii k}\right| - \gamma^{2}}} \begin{pmatrix}
0 & v+we^{-\ii k} \\
v+we^{\ii k} & 0 
\end{pmatrix}.
\end{equation} 
The nonzero winding number implies the emergence of topologically protected bound states with ${\rm Re}\,E = 0$, which are indeed observed in experiments~\cite{Weimann-17, Parto-18}. In stark contrast to Hermitian systems, these bound states can have eigenenergies with positive imaginary parts, which leads to their amplification (lasing) with time~\cite{Parto-18}. Notably, the bulk real spectrum does not depend on the boundary conditions, and the non-Hermitian skin effect does not occur in this system. This insensitivity to the boundary conditions originates from parity-time symmetry in Eq.~(\ref{eq: def-PT}) with $\mathcal{P}\mathcal{T}_{+} := \sigma_{x}$:
\begin{equation}
\sigma_{x} H^{*} \left( k \right) \sigma_{x}^{-1} = H \left( k \right).
\end{equation}
We also emphasize that the topological phase in this system cannot be captured by the classification in Ref.~\cite{Gong-18, *Bandres-Segev-18}, which considers neither line gap nor CS that is essential in this non-Hermitian topological phase.

Another prime example is topological lasers in two dimensions~\cite{Bahari-17, Harari-18, *Bandres-18}, which are non-Hermitian extensions of the Chern insulator such as the Haldane model~\cite{Haldane-88} with energy gain. In Ref.~\cite{Bahari-17}, magneto-optic effects are used to break time-reversal symmetry, and a two-dimensional photonic crystal subject to an external magnetic field and uniform gain is realized. In Ref.~\cite{Harari-18, *Bandres-18}, on the other hand, a two-dimensional topological cavity array with optical gain is experimentally realized that does not require magnetic elements and has a larger photonic band gap. These topological lasers possess topologically protected chiral edge states even in the presence of non-Hermiticity, by which single-mode and high-efficiency lasers are realized due to the topological immunity against defects and disorder. The topological lasers do not rely on any symmetry and hence belong to AZ symmetry class A in two dimensions; the chiral edge states are attributed to the topological invariant characterized by an integer (Table~\ref{tab: complex AZ}). This $\mathbb{Z}$ topological invariant is given by the Chern number $C_{1}$ in Eq.~(\ref{eq: topological invariant - Chern - line})~\cite{Liang-13, Shen-18, Yao-18-SSH, *Yao-18-Chern, Kawabata-18-Chern}. Remarkably, only the edge resonators are selectively pumped, and non-Hermiticity is added only to the edges in Ref.~\cite{Bandres-18}; the bulk essentially remains the same as the Hermitian one. Nevertheless, it is nontrivial whether the chiral edge states in a Hermitian Chern insulator survive even in the presence of non-Hermiticity, and it is possible that they can be gapped out and disappear without closing a band gap, even if non-Hermiticity is added only at the edges. Importantly, our classification generally ensures that the topological edge states are immune to non-Hermiticity as long as it does not close a (real) line gap of the bulk. Here, again, the topological phase of this non-Hermitian system cannot be explained by the classification in Ref.~\cite{Gong-18, *Bandres-Segev-18}, which only considers point gaps.

\subsection{Symmetry-protected topological laser}
	\label{sec: SPT laser}

\begin{figure*}[t]
\centering
\includegraphics[width=144mm]{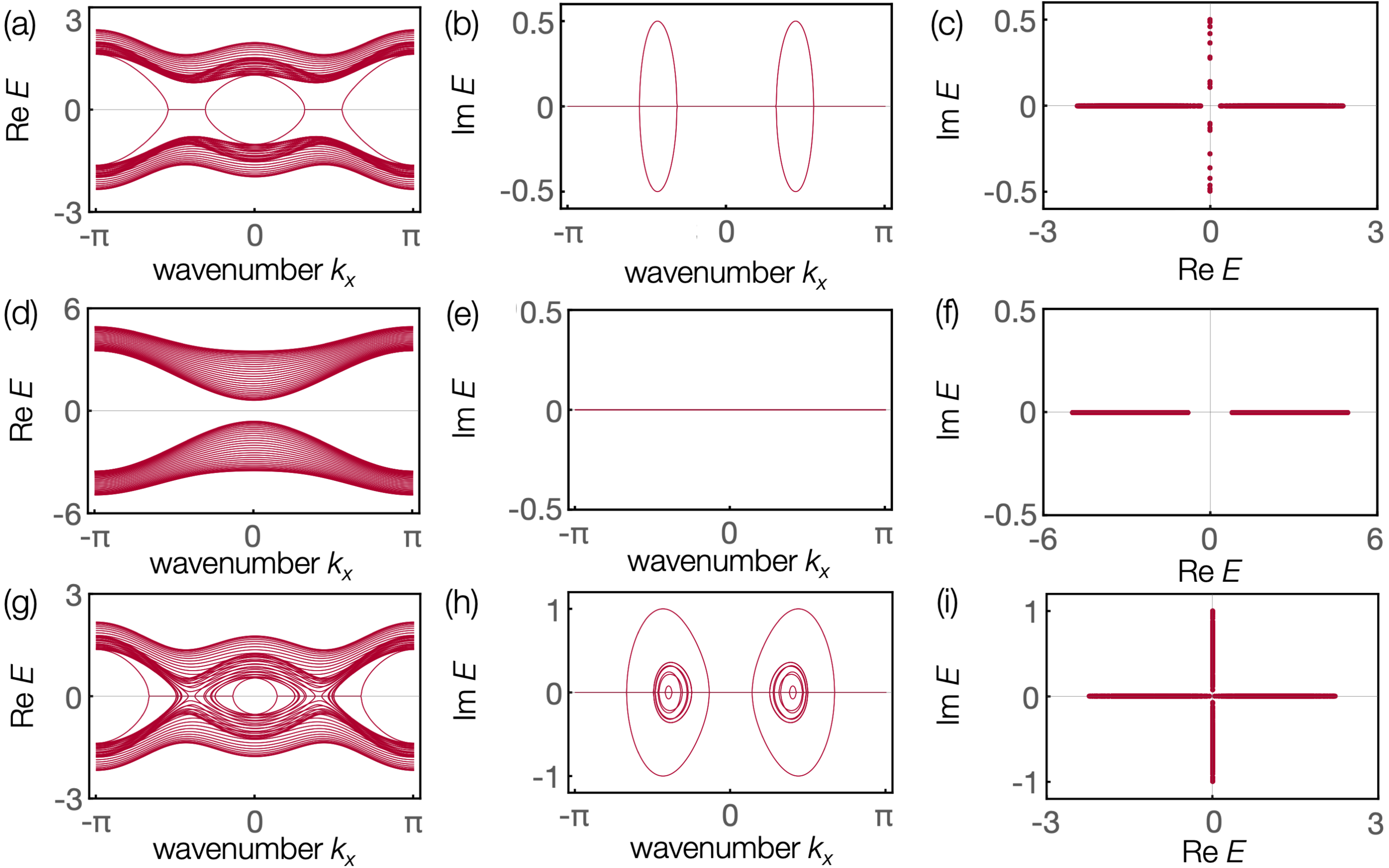} 
\caption{Complex spectrum of the symmetry-protected topological laser. In panels~(a, b, d, e, g, h), the open boundary condition is imposed for the $y$ direction ($30$ sites), whereas the periodic boundary condition is imposed for the $x$ direction, along which the wavenumber $k_{x}$ is defined. For panels~(c, f, i), the open boundary condition is imposed for both $x$ and $y$ directions ($30 \times 30$ sites). The complex spectrum is twofold degenerate as a result of the non-Hermitian Kramers theorem. (a-c)~Real-gapped phase with nontrivial topology ($t=1.0,\,m=-0.4,\,\gamma=0.5$). Whereas the bulk states have entirely real eigenenergies due to pseudo-Hermiticity, lasing helical edge states emerge with the nonzero imaginary part of the eigenenergies between the real-gapped bulk bands. The helical edge states form pairs of exceptional points. (d-f)~Real-gapped phase with trivial topology ($t=1.0,\,m=-3.0,\,\gamma=0.8$). The spectrum is entirely real, and no topologically protected edge states emerge. (g-i)~Real-gapless phase ($t=1.0,\,m=-0.4,\,\gamma=1.0$). The real gap is closed for the sufficiently strong non-Hermiticity, and even the bulk states can have the nonzero imaginary part of the eigenenergies due to spontaneous breaking of pseudo-Hermiticity.}
	\label{fig: SPT laser - spectrum}
\end{figure*}
	
Whereas the existing topological lasers in two dimensions~\cite{Bahari-17, Harari-18, *Bandres-18} do not rely on any symmetry, symmetry protection is needed for their systematic design. In fact, while a defining characteristic of topological lasers is the entire real spectra of bulk states and the complex spectra of topologically protected edge states, the entire reality is, in general, unfeasible without symmetry protection. Moreover, whereas the observed lasing edge states are chiral, certain symmetry is needed for a different type of lasing edge states such as helical ones. Our theory provides a general recipe for designing the symmetry-protected topological lasers. In particular, our framework incorporates pseudo-Hermiticity in Eq.~(\ref{eq: def pseudo-Hermiticity})~\cite{Mostafazadeh-02-1, *Mostafazadeh-02-2, *Mostafazadeh-02-3}, which leads to the reality of spectra in non-Hermitian systems as discussed in Sec.~\ref{sec : pH - topology}. Furthermore, the presence of a real gap is necessary for the real bulk spectrum, across which topologically protected edge states appear. There are a number of possible candidates for pseudo-Hermiticity-protected topological lasers in Tables~\ref{tab: complex AZ + pH} and \ref{tab: real AZ + pH}. It is also notable that all the bulk states are delocalized and no skin effects occur in the symmetry-protected topological lasers due to the real bulk spectrum ensured by pseudo-Hermiticity, as demonstrated in Sec.~\ref{sec: BEC - symmetry}.

\begin{figure}[t]
\centering
\includegraphics[width=86mm]{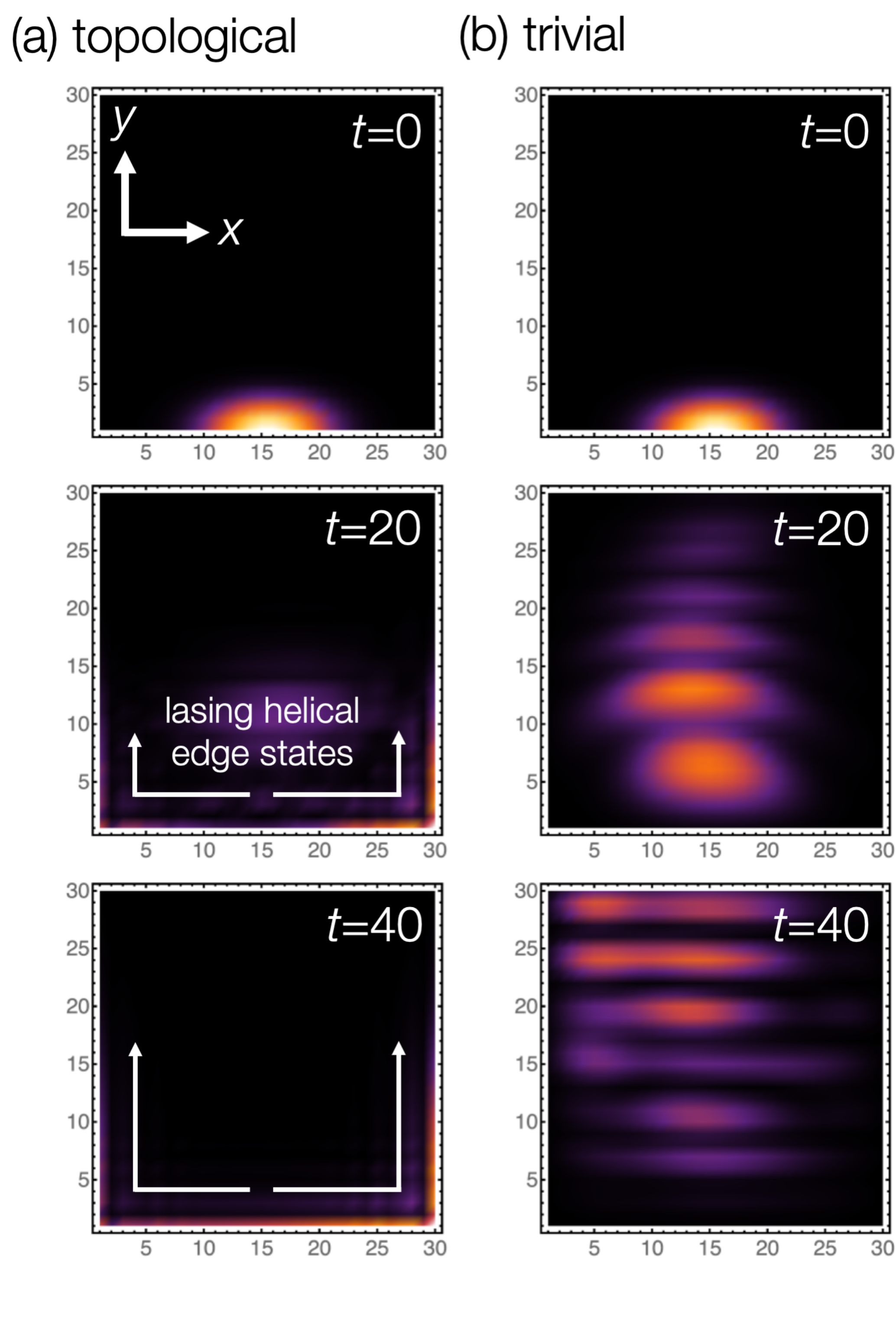} 
\caption{Dynamics of the symmetry-protected topological laser. The system has open boundaries along both directions with $30 \times 30$ sites. An initial state is a localized wave function $\ket{\psi_{0}} \propto \sum_{x, y} e^{-\left( x-15.5 \right)^{2}/40-\left( y-1 \right)^{2}/10}$, and the evolutions of the normalized intensity $\left| \braket{x,y | \psi_{t} }\right|^{2} / \braket{\psi_{t} | \psi_{t}}$ with $\ket{\psi_t} := e^{-\ii Ht} \ket{\psi_{0}}$ are shown for $t=0, 20, 40$. (a)~In the topological phase ($t=1.0, m=-0.4, \gamma =0.5$), the wavepacket remains localized and  moves bidirectionally along the edges due to the presence of the lasing helical edge states. (b)~In the trivial phase ($t=1.0, m=-3.0, \gamma =0.8$), the wavepacket quickly diffuses into the bulk due to the absence of the robust edge states.}
	\label{fig: SPT laser - dynamics}
\end{figure}

To illustrate how our theory systematically predicts symmetry-protected topological lasers, here we focus on the case in the presence of TRS in Eq.~(\ref{eq: TRS}) as well as pseudo-Hermiticity. Although two-dimensional topological phases are usually absent in the presence of TRS with $\mathcal{T}_{+} \mathcal{T}_{+}^{*} = + 1$, they can appear if TRS anticommutes with pseudo-Hermiticity as a unique feature of non-Hermitian symmetry~\cite{Esaki-11, *Sato-12} (see also Sec.~\ref{sec : pH - topology}). This situation corresponds to symmetry class AI with pseudo-Hermiticity $\eta_{-}$ in Table~\ref{tab: real AZ + pH}, which indeed hosts $\mathbb{Z}$ topological phases in two dimensions. Their topological invariants are given as the Chern number $C_{1}$ in Eq.~(\ref{eq: topological invariant - Chern - line}) with respect to $\eta H \left( {\bm k} \right)$, whereas the Chern number of the original Hamiltonian $H \left( {\bm k} \right)$ always vanishes due to the presence of TRS (see Appendix~\ref{appendix: pH - invariant} for details). We also note that $\mathrm{TRS}^{\dag}$ in Eq.~(\ref{eq: TRS-dag}) with $\mathcal{C}_{+} \mathcal{C}_{+}^{\dag} = -1$ is respected as a combination of TRS and pseudo-Hermiticity (i.e., $\mathcal{C}_{+} := \ii \eta \mathcal{T}_{+}$). As a result, the complex spectrum is twofold degenerate because of the Kramers theorem for non-Hermitian systems (see Sec.~\ref{sec: AZ-dag}), and thus lasing helical edge states emerge as a signature of the nontrivial topology.

A typical example of the pseudo-Hermiticity-protected topological lasers with TRS is given as
\begin{eqnarray}
H \left( {\bm k} \right)
&=& t \left( \cos k_{x} - \cos k_{y} \right) \tau_{x} +  t \left( \sin k_{x} \sin k_{y} \right) \sigma_{y} \tau_{y} \quad \nonumber \\ 
&~&\quad + \left( m+ t \cos k_{x} + t \cos k_{y} \right) \tau_{z} + \ii \gamma \sigma_{z} \tau_{y},
\end{eqnarray}
with $t, m, \gamma \in \mathbb{R}$. The system indeed respects both TRS and pseudo-Hermiticity:
\begin{equation}
H^{*} \left( {\bm k} \right) = H \left( - {\bm k} \right),\quad
\sigma_{y} H^{\dag} \left( {\bm k} \right) \sigma_{y}^{-1} = H \left( {\bm k} \right).
\end{equation}
As a combination of these symmetries, it also respects $\mathrm{TRS}^{\dag}$ with $\mathcal{C}_{+} \mathcal{C}_{+}^{\dag} = -1$, i.e.,
\begin{equation}
\sigma_{y} H^{T} \left( {\bm k} \right) \sigma_{y} = H \left( - {\bm k} \right),\quad
\sigma_{y} \sigma_{y}^{*} = -1.
\end{equation}
The presence of $\mathrm{TRS}^{\dag}$ leads to the Kramers degeneracy of the spectrum at the time-reversal-invariant momenta. Moreover, it respects parity (inversion) symmetry
\begin{equation}
H \left( {\bm k} \right) = H \left( - {\bm k} \right),
\end{equation}
and hence parity-time symmetry. The spectrum is obtained as 
\begin{eqnarray}
E \left( {\bm k} \right) 
&=& \pm \left[ t^{2} \left( \cos k_{x} - \cos k_{y} \right)^{2} + t^{2} \sin^{2} k_{x} \sin^{2} k_{y} \right. \nonumber \\
&~&\qquad\quad \left. \left( m+ t \cos k_{x} + t \cos k_{y} \right)^{2} - \gamma^{2} \right]^{1/2}.\qquad
\end{eqnarray}
If the non-Hermiticity $\gamma$ is sufficiently weak, a real gap remains to be open. In the real-gapped phase with the topologically nontrivial bulk, helical edge states emerge corresponding to the nonzero Chern number $C_{1} \neq 0$ for $\eta H \left( {\bm k} \right)$ [Fig.~\ref{fig: SPT laser - spectrum}\,(a-c)]. The bulk spectrum is entirely real and hence stable due to the presence of pseudo-Hermiticity and parity-time symmetry. Remarkably, the helical edge states form pairs of exceptional points in the complex spectrum and can have the nonzero imaginary parts of the eigenenergies, which leads to the amplification (lasing) of these helical edge states with time. In the real-gapped phase with the topologically trivial bulk, on the other hand, such lasing helical edge states do not appear, and the entire spectrum is real despite non-Hermiticity [Fig.~\ref{fig: SPT laser - spectrum}\,(d-f)]. Furthermore, the real gap is closed for sufficiently strong non-Hermiticity, and the bulk spectrum becomes complex and unstable due to spontaneous breaking of pseudo-Hermiticity and parity-time symmetry [Fig.~\ref{fig: SPT laser - spectrum}\,(g-i)]. It is also noteworthy that helical edge states are forbidden to have complex eigenenergies and cannot be lasing for $\left| C_{1} \right| = 1$. In fact, a pair of helical edge states may be degenerate only at the time-reversal-invariant momenta and the formation of exceptional points is impossible due to the Kramers theorem for $\left| C_{1} \right| = 1$. For Fig.~\ref{fig: SPT laser - spectrum}\,(a-c), on the other hand, two pairs of them emerge due to $\left| C_{1} \right| = 2$, which enables the degeneracy away from the time-reversal-invariant momenta and the formation of exceptional points.

A clear signature of the lasing helical edge states manifests itself in the dynamics. In particular, we consider the dynamics of a wavepacket whose initial state is prepared to be a localized wave function. Since the lasing helical edge states emerge in the topological phase, the wavepacket remains localized under the time evolution [Fig.~\ref{fig: SPT laser - dynamics}\,(a)]. Because the helical nature of the lasing edge states, they move bidirectionally along the edges, which is to be contrasted with the topological lasers that rely on no symmetry protection and support the lasing chiral edge states~\cite{Bahari-17, Harari-18, *Bandres-18}. Notably, the group velocity $\nabla_{\bm k} \left( \mathrm{Re}\,E \right)$ of the lasing edge states vanishes and the wavepacket moves slowly because $\mathrm{Re}\,E \left( {\bm k} \right)$ is flat for $\mathrm{Im}\,E \left( {\bm k} \right) \neq 0$ in this model [Fig.~\ref{fig: SPT laser - spectrum}\,(a-c)]. However, an arbitrary potential $\varepsilon \left( {\bm k}\right)$ is allowed in this symmetry class as long as it satisfies $\varepsilon \left( {\bm k}\right) = \varepsilon \left( -{\bm k}\right) = \varepsilon^{*} \left( {\bm k}\right)$, and hence the transport of the wavepacket can be controlled in a flexible manner. In the trivial phase, on the other hand, the wavepacket quickly diffuses into the bulk because of the absence of the robust edge states [Fig.~\ref{fig: SPT laser - dynamics}\,(b)]. Therefore, these dynamical features serve as a clear experimental signature of the lasing helical edge states. Importantly, whereas the specific symmetry class is considered here as an illustration, our classification tables~\ref{tab: complex AZ + pH} and \ref{tab: real AZ + pH} have the potential to bring about a number of different types of symmetry-protected topological lasers.

\subsection{Dissipative topological superconductor}
	\label{sec: NH TSC}
	
Non-Hermitian extensions of topological superconductors constitute another salient platform for our theoretical framework.	Remarkably, since the physical PHS defined by Eq.~(\ref{eq: PHS}) in terms of transposition has not been identified, the previous classification~\cite{Gong-18} does not encompass non-Hermitian superconductors. Here we consider non-Hermitian spinless superconductors, which intrinsically possess PHS with $\mathcal{C}_{-} \mathcal{C}_{-}^{*} = +1$ and belong to class D (Table~\ref{tab: real AZ}). Because of PHS in Eq.~(\ref{eq: PHS}), eigenenergies, in general, come in $\left( E, -E \right)$ pairs, and a simultaneous eigenstate of the Hamiltonian and the PHS operator possesses zero energy.

\begin{figure}[t]
\centering
\includegraphics[width=86mm]{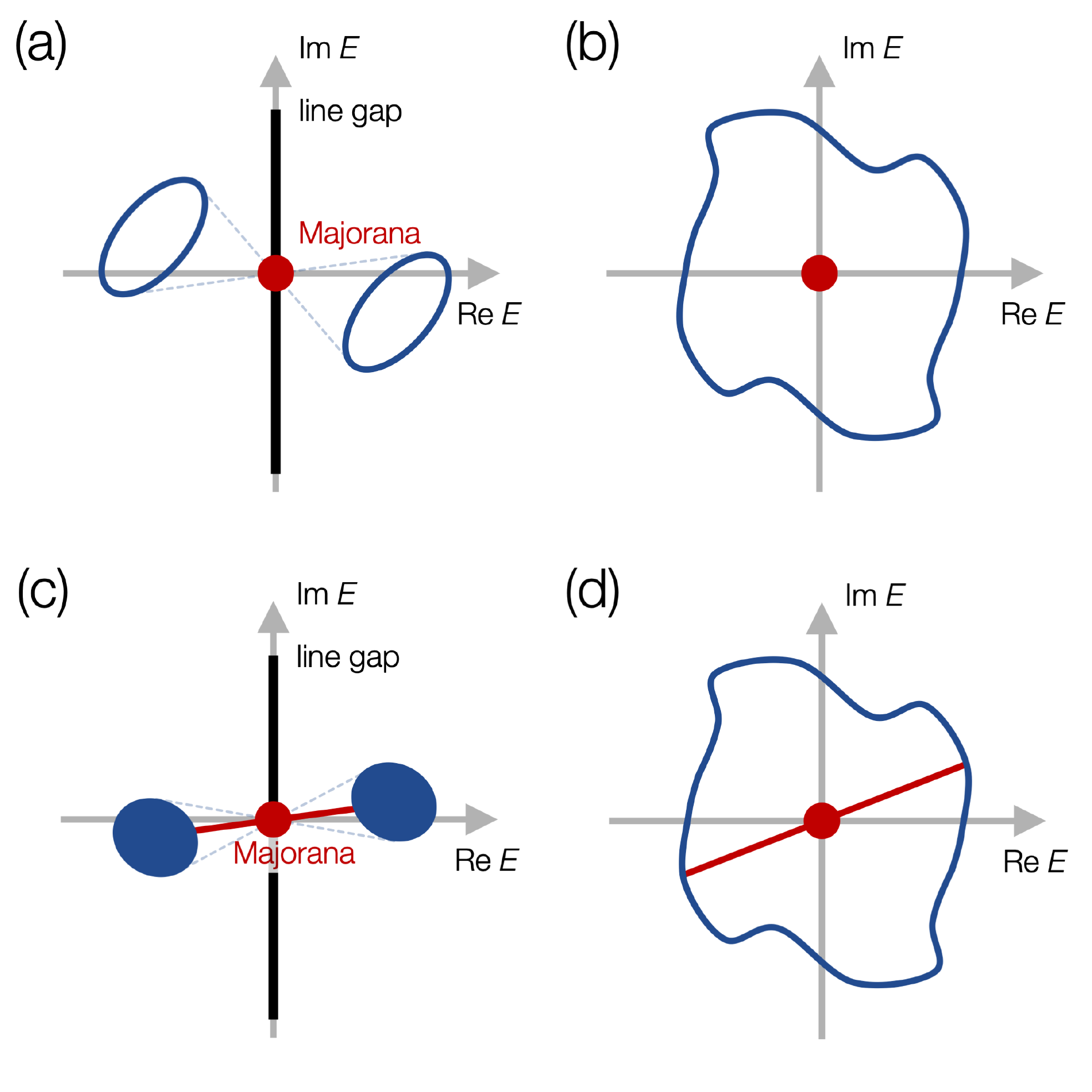} 
\caption{Complex spectra of non-Hermitian topological superconductors (class D). Complex eigenenergies (blue regions) come in $\left( E, -E \right)$ pairs due to particle-hole symmetry. In one dimension, the Majorana zero modes (red points) survive in the presence of (a)~a line gap and (b)~a point gap, which originates from the $\mathbb{Z}_{2}$ topology of the bulk. In two dimensions, the chiral Majorana edge states (red lines) survive in the presence of (c)~a line gap due to the $\mathbb{Z}$ topology (Chern number). Whereas their number is not invariant if the line gap is closed, their parity is invariant in the presence of (d)~a point gap due to the $\mathbb{Z}_{2}$ topology.}
	\label{fig: D}
\end{figure}

The salient feature of topological superconductors is the emergence of Majorana fermions at their boundaries~\cite{Alicea-review, Sato-review}. Because of the nontrivial $\mathbb{Z}_{2}$ topology in one dimension for a line gap, the Majorana zero modes are robust even in the presence of non-Hermiticity as long as a line gap is open [Fig.~\ref{fig: D}\,(a)]. The corresponding bulk topological invariant $\nu \in \{ 0, 1 \}$ is given by
\begin{equation} \begin{split}
\left( -1 \right)^{\nu} &:= \mathrm{sgn} \left\{
\frac{ \mathrm{Pf} \left[ H \left( \pi \right) \mathcal{C}_{-} \right] }{ \mathrm{Pf} \left[ H \left( 0 \right) \mathcal{C}_{-} \right] } \right. \\
&\quad\left. \times \exp \left[ 
-\frac{1}{2} \int_{k=0}^{k=\pi} d \log \det \left[ H \left( k \right) \mathcal{C}_{-} \right]
\right] \right\},
	\label{eq: 1D class D - Z2}
\end{split} \end{equation}
which is equivalent to the $\pi$-quantized Berry phase defined over the entire Brillouin zone (see Appendix~\ref{appendix: 1D class D - line} for details). As with the Hermitian case~\cite{Ryu-02}, PHS ensures that the Majorana zero modes possess zero energy even in non-Hermitian systems. In fact, if a Majorana zero mode $\hat{\Psi}_{\rm zero}$ localized at one edge is perturbed to have a nonzero energy $\delta E \neq 0$, there should exist the other mode $\hat{\cal C}\,\hat{\Psi}_{\rm zero}\,\hat{\cal C}^{-1}$ localized at the same edge with energy $-\delta E$, which is incompatible with the assumption that the number of topologically protected edge modes is at most one per one edge in the presence of a line gap. Furthermore, as a direct result of the nontrivial $\mathbb{Z}_{2}$ topology for a point gap, the Majorana zero modes survive even if the bulk complex spectrum encircles zero energy [Fig.~\ref{fig: D}\,(b)]. This remarkable immunity originates from the unique gap structure of non-Hermitian systems. The corresponding $\mathbb{Z}_{2}$ topological invariant is again given by Eq.~(\ref{eq: 1D class D - Z2}) (see Appendix~\ref{appendix: 1D class D - point} for details). 

A prime example of the non-Hermitian topological superconducting wire is a Kitaev chain~\cite{Kitaev-01} with the complex chemical potential:
\begin{eqnarray}
\hat{H}
&=& \sum_{j=1}^{L-1} \left( -t\,\hat{\psi}_{j}^{\dag} \hat{\psi}_{j+1} + \Delta\,\hat{\psi}_{j} \hat{\psi}_{j+1} + {\rm H.c.} \right) \nonumber \\
&~&\qquad\qquad\quad - \left( \mu + \ii \gamma \right) \sum_{j=1}^{L} \left( \hat{\psi}_{j}^{\dag} \hat{\psi}_{j} - \frac{1}{2} \right),
\end{eqnarray}
where $\hat{\psi}_{j}$ ($\hat{\psi}_{j}^{\dag}$) annihilates (creates) a fermion on site $j$, and $t$, $\Delta$, and $\mu$ are the hopping amplitude, the $p$-wave pairing gap, and the chemical potential, respectively. Moreover, $\gamma$ describes a degree of non-Hermiticity, which is attributed to the external gain and/or loss~\cite{Kawabata-18-Kitaev} or the finite lifetime of quasiparticles~\cite{Kozii-17}. In momentum space, the Hamiltonian reads
\begin{equation}
\hat{H} = - \frac{1}{2} \sum_{k} \begin{pmatrix}
	\hat{\psi}_{k}^{\dag} & \hat{\psi}_{-k} 
\end{pmatrix} H \left( k \right) \begin{pmatrix}
	\hat{\psi}_{k} \\ \hat{\psi}_{-k}^{\dag}
\end{pmatrix}
\end{equation}
with the BdG Hamiltonian
\begin{equation}
H \left( k \right) = \left( \begin{array}{@{\,}cc@{\,}} 
	\mu + \ii \gamma + 2t \cos k & 2\ii \Delta \sin k\\
	-2\ii \Delta^{*} \sin k & - \left( \mu + \ii \gamma + 2t \cos k \right)
	\end{array} \right).
\end{equation}
The bulk spectrum is obtained as
\begin{equation}
E \left( k \right) = \pm \sqrt{\left( \mu + \ii \gamma + 2t \cos k\right)^{2} + 4 \left| \Delta \right|^{2} \sin^{2} k},
\end{equation}
and both point and line gaps close for 
\begin{equation}
\left( \frac{\mu}{2t} \right)^{2} + \left( \frac{\gamma}{2 \left| \Delta \right|} \right)^{2} = 1.
\end{equation} 
The bulk Hamiltonian does not depend on the boundary conditions, and no skin effects occur in this system, which originates from parity symmetry in Eq.~(\ref{eq: def-parity}) with $\mathcal{P} := \sigma_{z}$:
\begin{equation}
\sigma_{z} H \left( k \right) \sigma_{z}^{-1} = H \left( - k \right).
\end{equation}

Remarkably, the non-Hermitian BdG Hamiltonian ($\gamma \neq 0$) indeed respects not $\mathrm{PHS}^{\dag}$ but PHS with $\mathcal{C}_{-} := \sigma_{x}$:
\begin{equation} \begin{split}
\sigma_{x}\,H^{T} \left( k \right) \sigma_{x} &= - H \left( -k \right),\\
\sigma_{x}\,H^{*} \left( k \right) \sigma_{x} &\neq - H \left( -k \right).
\end{split} \end{equation}
Because of PHS, the $\mathbb{Z}_{2}$ topological invariant $\nu$ in Eq.~(\ref{eq: 1D class D - Z2}) is well defined as long as complex-energy gaps are open; we have $\nu = 1$ ($\nu = 0$) for $\left( \mu/2t \right)^{2} + \left( \gamma/2 \left| \Delta \right| \right)^{2} < 1$ [$\left( \mu/2t \right)^{2} + \left( \gamma/2 \left| \Delta \right| \right)^{2} > 1$]. Corresponding to this $\mathbb{Z}_{2}$ topology of the bulk, a pair of Majorana edge states emerges, which satisfies $[ \hat{H}, \hat{\Psi}_{\rm zero} ] = 0$ ($L \to \infty$) and is given as
\begin{equation} \begin{split}
\hat{\Psi}_{\rm zero}^{(1)} 
&\propto \sum_{j=1}^{L} \left( - \frac{\mu+\ii \gamma}{2t} \right)^{j-1} \hat{a}_{j},\\
\hat{\Psi}_{\rm zero}^{(2)} 
&\propto \sum_{j=1}^{L} \left( - \frac{\mu+\ii \gamma}{2t} \right)^{j-1} \hat{b}_{L+1-j},
\end{split} \end{equation}
where $t=\Delta$ is assumed for simplicity, and $\hat{a}_{j} := \hat{\psi}_{j} + \hat{\psi}_{j}^{\dag}$ and $\hat{b}_{j} := (\hat{\psi}_{j} - \hat{\psi}_{j}^{\dag})/\ii$ are Majorana operators. Despite the persistence of the Majorana zero modes, non-Hermiticity alters their statistics into $\hat{\Psi}_{\rm zero}^{\dag} \neq \hat{\Psi}_{\rm zero}$ and
\begin{equation} \begin{split}
\{ \hat{\Psi}_{\rm zero}, \hat{\Psi}_{\rm zero} \}
&= 2 \left[ 1 - \frac{\left( 2\ii \mu - \gamma \right) \gamma}{4t^{2} - \left( \mu + \ii \gamma \right)^{2}} \right], \\
\{ \hat{\Psi}_{\rm zero}, \hat{\Psi}_{\rm zero}^{\dag} \}
&= 2 \left[ 1 - \frac{\gamma^{2}}{4t^{2} - \left( \mu^{2} + \gamma^{2} \right)} \right],
\end{split} \end{equation}
where the zero modes are normalized so that they satisfy the canonical anticommutation relations in the Hermitian limit ($\gamma = 0$). These anomalous statistics contrast with the conventional ones for Majorana fermions in the Hermitian counterpart, which originates from the distinction between right and left eigenstates in non-Hermitian systems~\cite{Kawabata-18-Kitaev}. In fact, the Majorana operators respect $\hat{\Psi}_{\rm zero}^{\dag} = \hat{\tilde{\Psi}}_{\rm zero}$ instead of $\hat{\Psi}_{\rm zero}^{\dag} = \hat{\Psi}_{\rm zero}$ with the Majorana zero modes $\hat{\tilde{\Psi}}_{\rm zero}$ for $\hat{H}^{\dag}$. Whereas a $p$-wave superconductor is considered here, similar Majorana zero modes can emerge also in non-Hermitian $s$-wave superconductors. Recently, Ref.~\cite{Okuma-19} has explicitly constructed a model of a non-Hermitian $s$-wave superconductor with a point gap that indeed supports Majorana zero modes.

In two dimensions, chiral Majorana modes emerge along the edges. Because of the nontrivial $\mathbb{Z}$ topology for a line gap, these chiral edge modes survive non-Hermiticity as long as a line gap is open [Fig.~\ref{fig: D}\,(c)]. The corresponding topological invariant is given as the Chern number $C_{1}$ in Eq.~(\ref{eq: topological invariant - Chern - line}). On the other hand, if the line gap is closed and only a point gap is open, $C_{1}$ is not well defined, and accordingly, the number of the chiral edge modes changes without closing the point gap. Nevertheless, only a pair of the chiral edge modes can be absorbed into the bulk and their parity is invariant as long as the point gap is open [Fig.~\ref{fig: D}\,(d)] because eigenenergies are paired by $\left( E \left( {\bm k} \right), - E \left( -{\bm k} \right) \right)$ due to PHS. This parity conservation of the chiral edge modes corresponds to the $\mathbb{Z}_{2}$ invariant $\nu \in \{ 0, 1\}$ defined by
\begin{equation} \begin{split}
( -1 &)^{\nu} := \prod_{\mathsf{X}=\mathrm{I, II}} \mathrm{sgn} \left\{
\frac{ \mathrm{Pf} \left[ H \left( {\bm k}_{\mathsf{X}+} \right) \mathcal{C}_{-} \right] }{ \mathrm{Pf} \left[ H \left( {\bm k}_{\mathsf{X}-} \right) \mathcal{C}_{-} \right] } \right. \\
&\left. \times \exp \left[ 
-\frac{1}{2} \int_{{\bm k} = {\bm k}_{\mathsf{X}-}}^{{\bm k} = {\bm k}_{\mathsf{X}+}} d \log \det \left[ H \left( {\bm k} \right) \mathcal{C}_{-} \right]
\right] \right\},
\end{split} \end{equation}
where (${\bm k}_{\mathrm{I}+}, {\bm k}_{\mathrm{I}-}$) and (${\bm k}_{\mathrm{II}+}, {\bm k}_{\mathrm{II}-}$) are two pairs of particle-hole-symmetric momenta (see Appendix~\ref{appendix: 2D class D - point} for details). Notably, $\nu$ is equivalent to the parity of $C_{1}$ [i.e., $C_{1}$ (mod $2$)] defined in the presence of the line gap.

\section{Free boson}
	\label{sec: free boson}
	
\subsection{Topological classification}

Whereas the topological classification of Hermitian free fermions was well established~\cite{Schnyder-08, Kitaev-09, Ryu-10}, its bosonic counterpart has been absent even in Hermitian systems. Our theory of non-Hermitian systems provides such topological classification of Hermitian and non-Hermitian free bosons. We consider a generic noninteracting (quadratic) bosonic system
\begin{equation}
\hat{H}
= \frac{1}{2} \left( \begin{array}{@{\,}cc@{\,}} 
	\hat{\bm a}^{\dag} & \hat{\bm a}
	\end{array} \right) H_{\rm BdG} \left( \begin{array}{@{\,}cc@{\,}} 
	\hat{\bm a} \\ \hat{\bm a}^{\dag}
	\end{array} \right),
\end{equation}
with a set of bosonic annihilation (creation) operators $\hat{\bm a} := ( \hat{a}_{1},\,\cdots,\,\hat{a}_{N} )$ [$\hat{\bm a}^{\dag} := ( \hat{a}_{1}^{\dag},\,\cdots,\,\hat{a}_{N}^{\dag} )$], which satisfies $[ a_{i},\,a_{j}^{\dag} ] = \delta_{ij}$, $[ a_{i},\,a_{j}] = [ a_{i}^{\dag},\,a_{j}^{\dag} ] = 0$. Here the non-Hermitian BdG Hamiltonian $H_{\rm BdG}$ is described by
\begin{equation}
H_{\rm BdG} = \left( \begin{array}{@{\,}cc@{\,}} 
	M & \Delta_{+} \\ \Delta_{-} & M^{T}
	\end{array} \right),
\end{equation}
where $M$ and $\Delta_{\pm}$ are $N \times N$ non-Hermitian matrices, and $\Delta_{\pm}$ are required to be symmetric (i.e., $\Delta_{\pm}^{T} = \Delta_{\pm}$) because of Bose statistics. In the presence of Hermiticity, $M$ becomes Hermitian and $\Delta_{\pm}$ satisfies $\Delta_{+}^{\dag} = \Delta_{-}$.  

In contrast to fermionic systems whose BdG Hamiltonians are diagonalized with unitary matrices, bosonic BdG Hamiltonians should be diagonalized with paraunitary matrices so that their quasiparticles fulfill Bose statistics~\cite{Kawaguchi-review}. In other words, we should diagonalize not the original BdG Hamiltonian $H_{\rm BdG}$ but the effective matrix
\begin{equation}
H_{\sigma {\rm BdG}}:=\sigma_{z} H_{\rm BdG} = \left( \begin{array}{@{\,}cc@{\,}} 
	M & \Delta_{+} \\ - \Delta_{-} & - M^{T}
	\end{array} \right).
\end{equation}
Here the effective matrix $H_{\sigma {\rm BdG}}$ is generally non-Hermitian even if the original BdG Hamiltonian $H_{\rm BdG}$ is Hermitian. Importantly, the non-Hermiticity results from Bose statistics, which may induce dynamical instability~\cite{Kawaguchi-review}.

To consider the topological phases of free bosons, symmetry imposed on the effective non-Hermitian matrix $H_{\sigma \rm BdG}$ is relevant. In general, owing to $\Delta^T_{\pm}=\Delta_{\pm}$, $H_{\sigma \rm BdG}$ respects PHS
\begin{equation}
{\cal C}_-^{-1}\,H_{\sigma\rm BdG}^{T}\,{\cal C}_- = - H_{\sigma\rm BdG}
	\label{eq: boson - PHS}
\end{equation}
with ${\cal C}_- := \sigma_y$, which reduces to Eq.~(\ref{eq: PHS}) in momentum space. Moreover, in the presence of Hermiticity for $H_{\rm BdG}$, $H_{\sigma\rm BdG}$ respects pseudo-Hermiticity
\begin{equation}
\eta^{-1}\,H_{\sigma\rm BdG}^{\dag}\,\eta =H_{\sigma\rm BdG},
	\label{eq: boson - pseudo-Hermiticity}
\end{equation}
with $\eta := \sigma_z$, which reduces to Eq.~(\ref{eq: def pseudo-Hermiticity}) in momentum space. Therefore, the topological classification of Hermitian and non-Hermitian free bosons reduces to that of the non-Hermitian matrix $H_{\sigma\rm BdG}$ that respects Eqs.~(\ref{eq: boson - PHS}) and/or (\ref{eq: boson - pseudo-Hermiticity}) in addition to some other symmetries, which is already obtained as the periodic tables.

PHS and pseudo-Hermiticity in Eqs.~(\ref{eq: boson - PHS}) and (\ref{eq: boson - pseudo-Hermiticity}) satisfy ${\cal C}_-{\cal C}_-^*=-1$ and $\{\eta, {\cal C}_-\}=0$. Therefore, in the absence of TRS and other additional symmetries, a noninteracting bosonic BdG system naturally belongs to class C (class C with $\eta_-$) for non-Hermitian (Hermitian) $H_{\rm BdG}$. On the other hand, in the presence of TRS, which usually obeys ${\cal T}_+{\cal T}_+^*=1$ for bosonic systems, the natural symmetry class is class CI (class CI with $\eta_{+-}$). To apply our classification to  bosonic systems, however, a more careful consideration for an energy gap is necessary. For Hermitian fermionic systems with PHS, we usually assume a gap at zero energy. In the case of Hermitian superconductors, for instance, we take a superconducting gap at zero energy since all states below the gap are fully occupied in the ground state at zero temperature, and the lowest excited state appears in the gap. For free bosons, on the other hand, this assumption is not obvious since any states are not fully occupied in the ground state because of Bose statistics. Thus, we can consider an energy gap away from zero energy. In this case, the choice does not respect PHS, and hence the topological classification effectively neglects PHS. Therefore, the relevant symmetry class is class A (class A+$\eta$) or class AI (class AI+$\eta_+$) for non-Hermitian (Hermitian) $H_{\rm BdG}$ instead.

Our topological classification describes topological phenomena of free bosons~\cite{Katsura-10, *Onose-10, Shindou-13, Barnett-13, *Galilo-15, Engelhardt-15, *Engelhardt-16, Peano-16-nc, *Peano-16-x, Lieu-18, Clerk-18, Kondo-18, Kawaguchi-review}.
In class A with $\eta$, Table~\ref{tab: complex AZ + pH} predicts the topological phase characterized by an integer in two dimensions in the presence of a real gap, which corroborates the magnon Hall effect~\cite{Katsura-10, *Onose-10} as a bosonic counterpart of the quantum Hall effect. 
It should be noted here that $H_{\rm BdG}$ in Ref.~\cite{Shindou-13} is positive definite as well as Hermitian, so that the ${\mathbb Z} \oplus {\mathbb Z}$ invariant reduces to the single Chern number (${\mathbb Z}$), as explained later.
Recently, a bosonic analogue of the $\mathbb{Z}_{2}$ topological insulator was also proposed in Ref.~\cite{Kondo-18}. In addition to Eqs.~(\ref{eq: boson - PHS}) and (\ref{eq: boson - pseudo-Hermiticity}), this system respects pseudo-time-reversal symmetry given by Eq.~(\ref{eq: TRS}) with $\Tp \Tp^{*} = -1$, which leads to symmetry class AII with pseudo-Hermiticity $\eta_{+}$. Table~\ref{tab: real AZ + pH} predicts the $\mathbb{Z}_{2}$ topological phase in two dimensions in the presence of a real gap, which is consistent with the $\mathbb{Z}_{2}$ topological invariant constructed in Ref.~\cite{Kondo-18}. Again, we note that the ${\mathbb Z}_2 \oplus {\mathbb Z}_2$ invariant reduces to the single ${\mathbb Z}_2$ number since $H_{\rm BdG}$ is positive definite. Remarkably, our topological classification not only justifies the known bosonic topological phenomena but also may lead to novel topological phases of free bosons.

\subsection{Pseudo-Hermiticity and paraunitary condition}

Because of pseudo-Hermiticity, $H_{\sigma \rm BdG}$ may host real eigenvalues despite non-Hermiticity and additional topological structures appear, as explained in Sec.~\ref{sec : pH - topology}. Meanwhile, as mentioned above, $H_{\sigma\rm BdG}$ is diagonalized by a paraunitary rather than unitary matrix. In fact, this unique feature of bosonic systems is a consequence of the real spectrum and pseudo-Hermiticity, as described in detail below.

Let $|u_n\rangle$ ($|u_n\rangle\!\rangle$) be a right (left) eigenstate of $H_{\sigma\rm BdG}$.
Using the same procedure as in Sec.~\ref{sec : pH - topology}, we take the biorthonormal basis $(|\phi^{\pm}_n\rangle, |\phi^{\pm}_n\rangle\!\rangle)$ in which $c_{mn} := \langle\!\langle u_m|\eta|u_n\rangle\!\rangle$ is diagonal.
When the eigenvalues of $H_{\sigma \rm BdG}$ are real, this basis also diagonalizes $H_{\sigma\rm BdG}$, and thus we have  
\begin{align}
H_{\sigma \rm BdG}\,|\phi^{\pm}_n\rangle
=E_n^{\pm}\,|\phi^{\pm}_n\rangle,\quad
\eta\,|\phi_n^{\pm}\rangle\!\rangle=\pm\,|\phi_n^{\pm}\rangle.
\end{align}
Therefore, introducing the following matrices, 
\begin{equation} \begin{split}
&R := (|\phi_1^{+}\rangle, \dots, |\phi_N^{+}\rangle, |\phi_1^{-}\rangle, \dots, |\phi_N^{-}\rangle),\\ 
&L := (|\phi_1^{+}\rangle\!\rangle, \dots, |\phi_N^{+}\rangle\!\rangle, |\phi_1^{-}\rangle\!\rangle, \dots, |\phi_N^{-}\rangle\!\rangle), 
\end{split} \end{equation}
we have 
\begin{align}
H_{\sigma\rm BdG}=R
\left(
\begin{array}{cc}
E^{+} & \\
&  E^{-} \\
\end{array}
\right) R^{-1},~
\eta L=R\,\sigma_z, 
	\label{eq : para}
\end{align}
with $E^{\pm} := {\rm diag} \left( E_1^{\pm}, \dots, E_N^{\pm} \right)$. Recalling $\eta=\sigma_z$ and the biorthonormal relation $LR^{\dagger}=1$,
the latter equation in Eq.~(\ref{eq : para}) yields nothing but the paraunitary condition given by
\begin{align}
R\,\sigma_z\,R^{\dagger} = \sigma_z. 
	\label{eq : paraunitary}
\end{align}

The original bosonic BdG Hamiltonian $H_{\rm BdG}$ is often supposed to be positive definite as well as Hermitian. 
In this case, we can construct $R$ as follows~\cite{Shindou-13}. From the Cholesky decomposition~\cite{MatrixAnalysis}, 
$H_{\rm BdG}$ is recast into the product of an invertible upper triangle matrix $K$ and its Hermitian conjugate $K^{\dagger}$ as
\begin{align}
H_{\rm BdG}=KK^{\dagger}.
\end{align}
Then, we introduce the Hermitian matrix $K\sigma_z K^{\dagger}$ and diagonalize it by a unitary matrix $U$,
\begin{align}
K\sigma_z K^{\dagger}=
U\left(
\begin{array}{cc}
\varepsilon & 0\\
0 & -\varepsilon
\end{array}
\right)U^{\dagger}
\label{eq : W}
\end{align}
where $\varepsilon := {\rm diag} \left( \varepsilon_1, \dots, \varepsilon_N \right)$ is a diagonal matrix consisting of positive eigenvalues of $K\sigma_z K^{\dagger}$.
From Sylvester's law of inertia~\cite{MatrixAnalysis}, the numbers of positive and negative eigenvalues of $K\sigma_z K^{\dagger}$ are the same, and thus $K\sigma_z K^{\dagger}$ can be diagonalized in the form of Eq.~(\ref{eq : W}). 
Rewriting the right-hand side of Eq.~(\ref{eq : W}) as
\begin{align}
U
\left(
\begin{array}{cc}
\varepsilon^{1/2} & 0 \\
0 & \varepsilon^{1/2}
\end{array}
\right) 
\left(
\begin{array}{cc}
\varepsilon & 0 \\
0 & -\varepsilon
\end{array}
\right) 
\left(
\begin{array}{cc}
\varepsilon^{-1/2} & 0 \\
0 & \varepsilon^{-1/2}
\end{array}
\right)U^{\dagger}, 
\end{align}
we obtain
\begin{align}
H_{\sigma \rm BdG}=\sigma_z K^{\dagger}K=R
\left(
\begin{array}{cc}
\varepsilon & 0\\
0 & -\varepsilon
\end{array}
\right) 
R^{-1},
\label{eq : para2}
\end{align}
where
\begin{align}
R := K^{-1}U\left(
\begin{array}{cc}
\varepsilon^{1/2} & 0 \\
0 & \varepsilon^{1/2}
\end{array}
\right)
\end{align}
satisfies the paraunitary condition in Eq.~(\ref{eq : paraunitary}).

From the above construction, we can see that the positive-definite Hermitian condition for $H_{\rm BdG}$ provides a strong constraint.
Comparing Eq.~(\ref{eq : para}) with Eq.~(\ref{eq : para2}), we have $E^{\pm}=\pm \varepsilon$. Therefore, a positive (negative) energy state $|u\rangle$ in Eq.~(\ref{eq : para2}) always satisfies $\eta\,|u\rangle\!\rangle=|u\rangle$ ($\eta\,|u\rangle\!\rangle=-|u\rangle)$. We can also show that positive-energy and negative-energy eigenstates are related to each other by PHS in Eq.~(\ref{eq: boson - PHS}). Thus, the sector with $\eta\,|u\rangle\!\rangle=|u\rangle$ and that with $\eta\,|u\rangle\!\rangle=-|u\rangle$ are not independent of each other, and thus they are characterized by the same topological invariant.
This constraint reduces possible independent topological invariants.

\section{Conclusion}
	\label{sec: conclusions}
	
We have clarified symmetry and complex-energy gaps in non-Hermitian physics and sorted out all the non-Hermitian topological phases. Whereas symmetries are unified in non-Hermitian physics~\cite{Kawabata-18}, they can also ramify due to the distinction between complex conjugation and transposition for non-Hermitian Hamiltonians. As a result, the non-Hermitian symmetry class is 38-fold beyond the celebrated 10-fold AZ symmetry class~\cite{AZ-97}, each of which describes intrinsic non-Hermitian topological phases as well as non-Hermitian random matrices. Moreover, a complex-energy gap can be either point-like (zero-dimensional) or line-like (one-dimensional) due to the complex nature of the energy spectrum, which enriches non-Hermitian topology. On the basis of these fundamental insights in non-Hermitian physics, we have classified all the non-Hermitian topological phases as summarized in the periodic tables~\ref{tab: complex AZ}-\ref{tab: real AZ + pH}. This classification corroborates the unique lasing and transport phenomena recently observed in experiments~\cite{Poli-15, Zeuner-15, Weimann-17, Obuse-17, *Xiao-17, St-Jean-17, Bahari-17, Zhao-18, Parto-18, Harari-18, *Bandres-18}. Although these experiments cannot be described by the previous classification provided in Ref.~\cite{Gong-18, *Bandres-Segev-18}, our work provides a more general and comprehensive framework, so that the book on non-Hermitian topological systems has now been closed.

The theoretical framework developed in the present work opens up new applications in non-Hermitian physics. One of the crucial ones is to find and design novel symmetry-protected topological lasers. Whereas a two-dimensional one has been discussed in this work, even three-dimensional ones can be systematically explored on the basis of our classification theory. Our framework can also be applied to find topological phases in non-Hermitian superconductors, which has the potential to be of use in topological quantum computation~\cite{Nayak-review}. We hope that our general theory of symmetry and topology in non-Hermitian physics will lead to such novel phenomena and functionalities that originate from the interplay of non-Hermiticity and topology.

\section*{Acknowledgements}
We thank Kenji Fukushima, Sho Higashikawa, Hosho Katsura, Max Lein, Nobuyuki Okuma, Tomoki Ozawa, Henning Schomerus, and Ryuichi Shindou for helpful discussions. In particular, K.K. appreciates stimulating discussions with Zongping Gong. This work was supported by a Grant-in-Aid for Scientific Research on Innovative Areas ``Topological Materials Science" (KAKENHI Grant No.~JP15H05855) from the Japan Society for the Promotion of Science (JSPS). K.K. was supported by the JSPS through Program for Leading Graduate Schools (ALPS) and KAKENHI Grant No.~JP19J21927. K.S. was supported by PRESTO, JST (JPMJPR18L4). M.U. was supported by KAKENHI Grant No.~JP18H01145 from the JSPS. M.S. was supported by KAKENHI Grant No.~JP17H02922 from the JSPS. K.K. and M.S. were supported in part by the International Centre for Theoretical Sciences (ICTS) during a visit for participating in the program - Non-Hermitian Physics - PHHQP XVIII (Code: ICTS/nhp2018/06). 

\bigskip
{\it Note added.\,---} After this work had been submitted, a related work by Zhou and Lee appeared~\cite{ZL-18}. Although Ref.~\cite{ZL-18} initially counted the number of symmetry classes as 42, Ref.~\cite{ZL-18} has corrected it as 38 after learning about our 38-fold symmetry classification~\cite{Lee-pc}.

\appendix

\section{Sublattice symmetry as an additional symmetry}
	\label{appendix: SLS}

\begin{table}[t]
	\centering
	\caption{Possible types [$t=0, 1$ (mod $2$)] of sublattice symmetry as an additional symmetry in the complex AZ symmetry class [$s=0, 1$ (mod $2$)]. The subscript of $\SLS_{\pm}$ specifies the commutation ($+$) or anticommutation ($-$) relation to chiral symmetry: $\CS \SLS_{\pm} = \pm \SLS_{\pm} \CS$. \\}
	\label{tab: symmetry - complex AZ + SLS}
	\small
     \begin{tabular}{cccc} \hline \hline
    ~$s$~ & ~AZ class~ & ~$t=0$~ & ~$t=1$~ \\ \hline
    $0$ & A & & $\SLS$ \\
    $1$ & AIII & $\SLS_{+}$ & $\SLS_{-}$ \\ \hline \hline
  \end{tabular}
\end{table}

\begin{table}[t]
	\centering
	\caption{Possible types [$t=0, 1, 2, 3$ (mod $4$)] of sublattice symmetry as an additional symmetry in the real AZ symmetry class [$s=0, 1, \cdots, 7$ (mod $8$)]. The subscript of $\SLS_{\pm}$ specifies the commutation ($+$) or anticommutation ($-$) relation between $\SLS_{\pm}$ and time-reversal symmetry (TRS) and/or particle-hole symmetry (PHS). For the symmetry classes that involve both TRS and PHS (BDI, DIII, CII, and CI), the first subscript specifies the relation to TRS and the second one to PHS. Classes AI with $\SLS_{-}$, BDI with $\SLS_{-+}$ or $\SLS_{--}$, and CII with $\SLS_{-+}$ or $\SLS_{--}$ are equivalent to classes AII with $\SLS_{-}$, DIII with $\SLS_{-+}$ or $\SLS_{--}$, and CI with $\SLS_{-+}$ or $\SLS_{--}$, respectively. \\}
	\label{tab: symmetry - real AZ + SLS}
	\small
     \begin{tabular}{cccccc} \hline \hline
    ~$s$~ & ~AZ class~ & ~$t=0$~ & ~$t=1$~ & ~$t=2$~ & ~$t=3$~ \\ \hline
    $0$ & AI &  & $\SLS_{-}$ &  & $\SLS_{+}$ \\
    $1$ & BDI & $\SLS_{++}$ & $\SLS_{-+}$ & $\SLS_{--}$ & $\SLS_{+-}$ \\
    $2$ & D &  & $\SLS_{+}$ &  & $\SLS_{-}$ \\
    $3$ & DIII & $\SLS_{--}$ & $\SLS_{-+}$ & $\SLS_{++}$ & $\SLS_{+-}$ \\
    $4$ & AII &  & $\SLS_{-}$ &  & $\SLS_{+}$ \\
    $5$ & CII & $\SLS_{++}$ & $\SLS_{-+}$ & $\SLS_{--}$ & $\SLS_{+-}$ \\
    $6$ & C &  & $\SLS_{+}$ &  & $\SLS_{-}$ \\
    $7$ & CI & $\SLS_{--}$ & $\SLS_{-+}$ & $\SLS_{++}$ & $\SLS_{+-}$ \\ \hline \hline
  \end{tabular}
\end{table}

\begin{table*}[t]
	\centering
	\caption{Equivalence between the real AZ symmetry class with sublattice symmetry (SLS) and the real $\text{AZ}^{\dag}$ symmetry class with SLS. The subscript of ${\cal S}_{\pm}$ specifies the commutation ($+$) or anticommutation ($-$) relation to TRS/$\text{TRS}^{\dag}$ and/or PHS/$\text{PHS}^{\dag}$. For the symmetry classes that involve both TRS/$\text{TRS}^{\dag}$ and PHS/$\text{PHS}^{\dag}$ (BDI, DIII, CII, and CI; $\text{BDI}^{\dag}$, $\text{DIII}^{\dag}$, $\text{CII}^{\dag}$, and $\text{CI}^{\dag}$), the first subscript specifies the relation to TRS/$\text{TRS}^{\dag}$, and the second one specifies the relation to PHS/$\text{PHS}^{\dag}$. \\}
	\label{tab: SLS - AZ & AZ-dag}
     \begin{tabular}{ccccccc} \hline \hline
    ~$\text{AZ}^{\dag}$ class~ & ~$\SLS_{+}$~ & ~$\SLS_{-}$~ & ~$\SLS_{++}$~ & ~$\SLS_{+-}$~ & ~$\SLS_{-+}$~ & ~$\SLS_{--}$~\\ \hline
    $\text{AI}^{\dag}$ & ~D + $\SLS_{+}$~ & ~C + $\SLS_{-}$~ & & & & \\
    $\text{BDI}^{\dag}$ & & & ~BDI + $\SLS_{++}$~ & ~DIII + $\SLS_{-+}$~ & ~CI + $\SLS_{+-}$~ & ~CII + $\SLS_{--}$~ \\
    $\text{D}^{\dag}$ & ~AI + $\SLS_{+}$~ & ~AII + $\SLS_{-}$~ & & & & \\
    $\text{DIII}^{\dag}$ & & & ~CI + $\SLS_{++}$~ & ~CII + $\SLS_{-+}$~ & ~BDI + $\SLS_{+-}$~ & ~DIII + $\SLS_{--}$~ \\
    $\text{AII}^{\dag}$ & ~C + $\SLS_{+}$~ & ~D + $\SLS_{-}$~ & & & & \\
    $\text{CII}^{\dag}$ & & & ~CII + $\SLS_{++}$~ & ~CI + $\SLS_{-+}$~ & ~DIII + $\SLS_{+-}$~ & ~BDI + $\SLS_{--}$~ \\
    $\text{C}^{\dag}$ & ~AII + $\SLS_{+}$~ & ~AI + $\SLS_{-}$~ & & & & \\
    $\text{CI}^{\dag}$ & & & ~DIII + $\SLS_{++}$~ & ~BDI + $\SLS_{-+}$~ & ~CII + $\SLS_{+-}$~ & ~CI + $\SLS_{--}$~ \\ \hline \hline
  \end{tabular}
\end{table*}

\begin{table*}[t]
	\centering
	\caption{Equivalence between pseudo-Hermiticity and sublattice symmetry as an additional symmetry in the AZ symmetry class. For the complex classes, the subscript of $\eta_{\pm}$ and $\SLS_{\pm}$ specifies the commutation ($+$) or anticommutation ($-$) relation to chiral symmetry. For the real classes, the subscript of $\eta_{\pm}$ and $\SLS_{\pm}$ specifies the commutation ($+$) or anticommutation ($-$) relation to time-reversal symmetry (TRS) and/or particle-hole symmetry (PHS). For the symmetry classes that involve both TRS and PHS (BDI, DIII, CII, and CI), the first subscript specifies the relation to TRS and the second one to PHS. \\}
	\label{tab: pH & SLS}
     \begin{tabular}{cccccccc} \hline \hline
    ~AZ class~ & ~$\eta$~ & ~$\eta_{+}$~ & ~$\eta_{-}$~ & ~$\eta_{++}$~ & ~$\eta_{+-}$~ & ~$\eta_{-+}$~ & ~$\eta_{--}$~\\ \hline
    A & ~AIII~ & & & & & & \\
    AIII & & ~AIII + $\SLS_{+}$~ & ~AIII + $\SLS_{-}$~ & & & & \\ \hline
    AI & & ~$\text{BDI}^{\dag}$~ & ~$\text{DIII}^{\dag}$~ & & & & \\
    BDI & & & & ~BDI + $\SLS_{++}$~ & ~BDI + $\SLS_{-+}$~ & ~BDI + $\SLS_{+-}$~ & ~BDI + $\SLS_{--}$~ \\
    D & & ~BDI~ & ~DIII~ & & & & \\
    DIII & & & & ~DIII + $\SLS_{--}$~ & ~DIII + $\SLS_{+-}$~ & ~DIII + $\SLS_{-+}$~ & ~DIII + $\SLS_{++}$~ \\
    AII & & ~$\text{CII}^{\dag}$~ & ~$\text{CI}^{\dag}$~ & & & & \\
    CII & & & & ~CII + $\SLS_{++}$~ & ~CII + $\SLS_{-+}$~ & ~CII + $\SLS_{+-}$~ & ~CII + $\SLS_{--}$~ \\
    C & & ~CII~ & ~CI~ & & & & \\
    CI & & & & ~CI + $\SLS_{--}$~ & ~CI + $\SLS_{+-}$~ & ~CI + $\SLS_{-+}$~ & ~CI + $\SLS_{++}$~ \\ \hline \hline
  \end{tabular}
\end{table*}

SLS can be considered an additional symmetry to the AZ symmetry~\cite{Shiozaki-Sato-14} as shown in Tables~\ref{tab: symmetry - complex AZ + SLS} and \ref{tab: symmetry - real AZ + SLS}. Moreover, Table~\ref{tab: SLS - AZ & AZ-dag} shows the equivalence between the real AZ symmetry class with SLS and the real $\text{AZ}^{\dag}$ symmetry class with SLS. Let us take class $\text{DIII}^{\dag}$ as an example. In this symmetry class, the Hamiltonian respects both $\text{TRS}^{\dag}$ and $\text{PHS}^{\dag}$:
\begin{equation} \begin{split}
\Cp H^{T} \left( {\bm k} \right) \Cp^{-1}
&= H \left( - {\bm k} \right),~~\Cp \Cp^{*} = -1; \\
\Tm H^{*} \left( {\bm k} \right) \Tm^{-1}
&= - H \left( - {\bm k} \right),~~\Tm \Tm^{*} = +1.
\end{split} \end{equation}
We consider adding SLS that satisfies
\begin{equation} \begin{split}
\SLS H \left( {\bm k} \right) \SLS^{-1}
= - H \left( {\bm k} \right),~~\SLS^{2} = 1; \\
\SLS \Cp = \epsilon_{c} \Cp \SLS^{*},~~
\SLS \Tm = \epsilon_{t} \Tm \SLS^{*},
\end{split} \end{equation}
with $\epsilon_{c}, \epsilon_{t} \in \{ \pm 1 \}$. Then, TRS can be constructed by combining $\text{PHS}^{\dag}$ and SLS as $\Tp := \SLS \Tm$, which satisfies
\begin{equation} \begin{split}
\Tp H^{*} \left( {\bm k} \right) \Tp^{-1}
&= H \left( - {\bm k} \right),~\Tp \Tp^{*} = \epsilon_{t}; \\
\SLS \Tp &= \epsilon_{t} \Tp \SLS^{*}.
\end{split} \end{equation}
Similarly, PHS can be constructed by combining $\text{TRS}^{\dag}$ and SLS as $\Cm := \SLS \Cp$ for $\epsilon_{c} \epsilon_{t} = +1$ and $\Cm := \ii \SLS \Cp$ for $\epsilon_{c} \epsilon_{t} = -1$, which satisfies
\begin{equation} \begin{split}
\Cm H^{T} \left( {\bm k} \right) \Cm^{-1}
&= - H \left( - {\bm k} \right),~\Cm \Cm^{*} = -\epsilon_{c}; \\
\SLS \Cm &= \epsilon_{c} \Cm \SLS^{*}.
\end{split} \end{equation}
Here $\Cm$ is chosen so that it commutes with $\Tp$:
\begin{equation}
\Tp \Cm^{*} = \Cm \Tp^{*}.
\end{equation}
Thus, class $\text{DIII}^{\dag}$ with SLS is equivalent to one of the AZ symmetry classes with SLS.

\section{Pseudo-Hermiticity as an additional symmetry}
	\label{appendix: pseudo-Hermiticity}

Table~\ref{tab: pH & SLS} shows the equivalence between pseudo-Hermiticity and SLS as an additional symmetry to the AZ symmetry. Let us consider class DIII as an example. In this symmetry class, the Hamiltonian respects both TRS and PHS:
\begin{equation} \begin{split}
\Tp H^{*} \left( {\bm k} \right) \Tp^{-1}
&= H \left( - {\bm k} \right),~~\Tp \Tp^{*} = -1; \\
\Cm H^{T} \left( {\bm k} \right) \Cm^{-1}
&= - H \left( - {\bm k} \right),~~\Cm \Cm^{*} = +1.
\end{split} \end{equation}
As a combination of TRS and PHS, the Hamiltonian also respects CS:
\begin{equation}
\CS H^{\dag} \left( {\bm k} \right) \CS^{-1}
= - H \left( {\bm k} \right),~~\CS^{2} = 1,
\end{equation}
with $\CS := \ii\,\Cm \Tp^{*}$. Then, we consider adding pseudo-Hermiticity that satisfies
\begin{equation} \begin{split}
\eta H^{\dag} \left( {\bm k} \right) \eta^{-1}
= H \left( {\bm k} \right),~~\eta^{2} = 1; \\
\eta \Tp = \epsilon_{t} \Tp \eta^{*},~~
\eta \Cm = \epsilon_{c} \Cm \eta^{*},
\end{split} \end{equation}
with $\epsilon_{t}, \epsilon_{c} \in \{ \pm 1 \}$. Here, SLS can be constructed by combining CS and pseudo-Hermiticity. In the case of $\epsilon_{t} \epsilon_{c} = +1$, SLS is defined as $\SLS := \eta \CS$ with $\SLS^{2} = 1$, which satisfies $\SLS \Tp = - \epsilon_{t} \Tp \SLS^{*}$ and $\SLS \Cm = - \epsilon_{c} \Cm \SLS^{*}$; in the case of $\epsilon_{t} \epsilon_{c} = -1$, SLS is defined as $\SLS := \ii \eta \CS$ with $\SLS^{2} = 1$, which satisfies $\SLS \Tp = \epsilon_{t} \Tp \SLS^{*}$ and $\SLS \Cm = \epsilon_{c} \Cm \SLS^{*}$.

\section{Proof of Theorem 1 (unitary flattening for point gaps)}
	\label{appendix: unitary flattening}
	
To prove Theorem 1, we introduce the following Hermitian Hamiltonian $\tilde{H}_{0} \left( {\bm k} \right)$ constructed from the non-Hermitian Hamiltonian $H \left( {\bm k} \right)$:
\begin{align}
\tilde{H}_{0} \left( {\bm k} \right) :=
\left( \begin{array}{cc}
	0 & H \left( {\bm k} \right) \\
	H^{\dagger} \left( {\bm k} \right) & 0
\end{array} \right), 
\end{align}  
which satisfies CS (SLS)
\begin{align}
\Sigma\,\tilde{H}_{0} \left( {\bm k} \right)\,\Sigma = - \tilde{H}_{0} \left( {\bm k} \right),
\quad
\Sigma :=
\left( \begin{array}{cc}
1 & 0 \\
0 & -1
\end{array} \right).
\end{align}
We note that $\tilde{H}_{0} \left( {\bm k} \right)$ is identical to $\tilde{H} \left( {\bm k} \right)$ in Eq.~(\ref{eq: def - tilde H}), except that the off-diagonal component $H \left( {\bm k} \right)$ is nonunitary ($\tilde{H}_{0}^{2} \left( {\bm k} \right) \neq 1$). For $H \left( {\bm k} \right)$ with certain symmetries, $\tilde{H}_0 \left( {\bm k} \right)$ has the corresponding symmetries defined by Eqs.~(\ref{eq: T - tilde})-(\ref{eq: pH - tilde}), where $\tilde{H} \left( {\bm k} \right)$ is replaced by $\tilde{H}_{0} \left( {\bm k} \right)$. Moreover, as long as $H \left( {\bm k} \right)$ retains a point gap, $\tilde{H}_{0} \left( {\bm k} \right)$ also has an energy gap, and vice versa~\cite{Gong-18, *Bandres-Segev-18, Roy-17}. Therefore, we can perform a continuous deformation of $H \left( {\bm k} \right)$ while maintaining its symmetries and point gap just by the corresponding deformation of $\tilde{H}_{0} \left( {\bm k} \right)$.

Now we show that we can continuously deform $\tilde{H}_{0} \left( {\bm k} \right)$ so that it satisfies $\tilde{H}^{2}_{0} \left( {\bm k} \right) = 1$, which immediately leads to Theorem 1. Since $\tilde{H}_{0} \left( {\bm k} \right)$ is Hermitian, it can be diagonalized as 
\begin{align}
\tilde{H}_{0} \left( {\bm k} \right) =
\Phi \left( {\bm k} \right)
\left( \begin{array}{ccc}
\varepsilon_{1} \left( {\bm k} \right) & & \\
& \varepsilon_{2} \left( {\bm k} \right)& \\
&&\ddots
\end{array}
\right) \Phi^{\dagger} \left( {\bm k} \right), 
\end{align}
where $\Phi \left( {\bm k} \right)$ is unitary and $\varepsilon_{i} \left( {\bm k} \right)$'s are real. We also have $\varepsilon_{i} \left( {\bm k} \right) \neq 0$ as $\tilde{H}_{0} \left( {\bm k} \right)$ is gapped. Using
\begin{align}
&\Omega_{H}^{-1/2} \left( {\bm k} \right) \nonumber \\
&:= \left[
	\tilde{H}_{0}^{\dagger} \left( {\bm k} \right) \tilde{H}_{0} \left( {\bm k} \right) 
\right]^{-1/4}
\nonumber\\
&=
\Phi \left( {\bm k} \right)
\left(
\begin{array}{ccc}
|\varepsilon_1 \left( {\bm k} \right)|^{-1/2} & & \\
& |\varepsilon_2 \left( {\bm k} \right)|^{-1/2}& \\
&&\ddots
\end{array}
\right) \Phi^{\dagger} \left( {\bm k} \right),
\end{align}
we introduce a one-parameter family of Hamiltonians, 
\begin{align}
&\tilde{H}_{\lambda} \left( {\bm k} \right) \nonumber \\
&:= \left[ \Omega_{H}^{-1/2} \left( {\bm k} \right) \lambda+1-\lambda \right]
\tilde{H}_{0} \left( {\bm k} \right) \left[ \Omega_{H}^{-1/2} \left( {\bm k} \right) \lambda+1-\lambda \right]
\end{align}
with $\lambda \in [0,1]$. Here, $\tilde{H}_{\lambda} \left( {\bm k} \right)$ is Hermitian and keeps a gap because $\Omega_{H}^{-1/2} \left( {\bm k} \right) \lambda+1-\lambda$ is positive definite. From $\tilde{H}_{1}^{2} \left( {\bm k} \right)=1$, $\tilde{H}_{\lambda} \left( {\bm k} \right)$ provides an adiabatic path with $\tilde{H}_{0}^{2} \left( {\bm k} \right) \to \tilde{H}_{1}^{2} \left( {\bm k} \right)=1$. Therefore, if $\tilde{H}_{\lambda} \left( {\bm k} \right)$ has the same symmetry as $\tilde{H}_{0} \left( {\bm k} \right)$, we have Theorem 1. 

In fact, $\tilde{H}_{\lambda} \left( {\bm k} \right)$ has the same symmetry. For instance, consider $\tilde{H}_{0} \left({\bm k} \right)$ with PHS,
\begin{align}
\tilde{\cal C}_{-} \tilde{H}_{0}^{*} \left( {\bm k} \right) \tilde{\cal C}_{-}^{-1}=-\tilde{H}_{0} \left( -{\bm k} \right).
\end{align}
In this case, we have
\begin{align}
\tilde{\cal C}_{-} \left[
\tilde{H}_{0}^{\dagger} \left( {\bm k} \right) \tilde{H}_{0} \left( {\bm k} \right)
\right]^{*}
\tilde{\cal C}_{-}^{-1} = \tilde{H}^{\dagger}_{0} \left( -{\bm k} \right) \tilde{H}_{0} \left( -{\bm k} \right),
\end{align}
which yields
\begin{align}
\tilde{\cal C}_{-} [\Omega^{-1/2}_{H} \left( {\bm k} \right)]^{*} \tilde{\cal C}_{-}^{-1} = \Omega_{H}^{-1/2} \left( -{\bm k} \right).
\end{align} 
As a result, we also have PHS for $\tilde{H}_{\lambda} \left( {\bm k} \right)$, 
\begin{align}
\tilde{\cal C}_{-} \tilde{H}_{\lambda}^{*} \left( {\bm k} \right) \tilde{\cal C}_{-}^{-1} = -\tilde{H}_{\lambda} \left( -{\bm k} \right).
\end{align}
In a similar manner, we can show that $\tilde{H}_{\lambda} \left( {\bm k} \right)$ has the same symmetry as $\tilde{H}_0({\bm k})$ for all the other symmetries.

\section{Proof of Theorem 2 (Hermitian flattening for line gaps)}
	\label{appendix: Hermitian flattening}

\subsection{Spectral flattening for line gaps}

Let us consider a non-Hermitian Hamiltonian $H({\bm k})$ with a line gap and denote the right and left eigenstates as
$|u_n({\bm k})\rangle$ and $|u_n({\bm k})\rangle\!\rangle$, respectively: 
\begin{equation} \begin{split}
&H({\bm k})\,|u_n({\bm k})\rangle=E_n({\bm k})\,|u_n({\bm k})\rangle,\\ 
&H^{\dagger}({\bm k})\,|u_n({\bm k})\rangle\!\rangle=E^*_n({\bm k})\,|u_n({\bm k})\rangle\!\rangle. 
\end{split} \end{equation}
For our purpose, it is sufficient to consider the case without exceptional points since they can be pair-annihilated without closing a line gap. Then, 
 $|u_n({\bm k})\rangle$ and $|u_n({\bm k})\rangle\!\rangle$ satisfy the biorthonormal condition~\cite{Brody-14}
\begin{eqnarray}
\langle\!\langle u_m({\bm k})|u_n({\bm k})\rangle 
=\langle u_m({\bm k})|u_n({\bm k})\rangle\!\rangle=\delta_{mn},
\end{eqnarray}
and the completeness condition
\begin{eqnarray}
\sum_n |u_n({\bm k})\rangle\langle\!\langle u_n({\bm k})|=
\sum _n |u_n({\bm k})\rangle\!\rangle \langle u_n({\bm k})|=1. 
\end{eqnarray}
For later convenience, we collect these eigenstates as row vectors $R({\bm k})$ and $L({\bm k})$: 
\begin{equation} \begin{split}
&R({\bm k}):=(|u_1({\bm k})\rangle, |u_2({\bm k})\rangle, \cdots ), \\ 
&L({\bm k}):=(|u_1({\bm k})\rangle\!\rangle, |u_2({\bm k})\rangle\!\rangle, \cdots ).
\end{split} \end{equation}
The above biorthonormal and completeness conditions are compactly written as
\begin{eqnarray}
L^{\dagger}({\bm k}) R({\bm k})= R^{\dagger}({\bm k}) L({\bm k})=
L({\bm k}) R^{\dagger}({\bm k})= R({\bm k}) L^{\dagger}({\bm k})=1,
\nonumber\\ 
\label{eq:RL}
\end{eqnarray}
and the right eigenequations are put together as
\begin{eqnarray}
H({\bm k}) R({\bm k})=R({\bm k})
\left(
\begin{array}{ccc}
E_1({\bm k}) & &  \\
& E_2({\bm k}) &  \\
& & \ddots
\end{array}
\right),  
\end{eqnarray}
which is recast into
\begin{eqnarray}
H({\bm k})=R({\bm k})
\left(
\begin{array}{ccc}
E_1({\bm k}) & &  \\
& E_2({\bm k}) &  \\
& & \ddots
\end{array}
\right)R^{-1}({\bm k}).  
\end{eqnarray}
Now we flatten the complex spectrum of $H({\bm k})$ without closing the line gap. As long as the system does not close the gap, it keeps the same topological structures, and hence this flattening procedure does not affect the classification of topological phases.

Importantly, the flattening process depends on the symmetry class. If any symmetry operation does not include complex or Hermitian conjugation, a line gap merely implies the presence of two disconnected parts of the band energies in the complex-energy plane. While keeping the line gap, we can continuously change one part of the spectrum into $+1$ and the other into $-1$ as
\begin{equation} \begin{split}
H({\bm k}) 
& \rightarrow
R({\bm k})
\left( \begin{array}{cc}
+1_{p\times p} & 0  \\
 0 & -1_{q\times q}  \\
\end{array} \right)
R^{-1}({\bm k}) \\
&=: R({\bm k})\,{\mathbb E}\,R^{-1}({\bm k}),  
	\label{eq:flattening}
\end{split} \end{equation}
where $p$ ($q$) is the number of bands contained in one (the other) part of the spectrum. After this flattening procedure, we obtain a non-Hermitian Hamiltonian $H({\bm k})$ with $H^2({\bm k})=1$. On the other hand, if a symmetry operation with complex or Hermitian conjugation is relevant, we have a real structure in the complex-energy spectrum. The real part of the spectrum can be distinguished from the imaginary one, and thus we have two distinct types of line gaps, i.e., a real gap and an imaginary gap, where a real (an imaginary) gap implies a gap in the real (imaginary) part of the complex spectrum. Correspondingly, there are two different flattening processes as follows. 
(i)~For a system with a real gap, one can continuously change the band energies with a larger (smaller) real part into $+1$ ($-1$) without closing the real gap. The resultant Hamiltonian has the same form as Eq.~(\ref{eq:flattening}).   
(ii)~For a system with an imaginary gap, one can continuously change the band energies with a larger (smaller) imaginary part into  $+\ii$ ($-\ii$) without closing the imaginary gap, 
\begin{eqnarray}
H({\bm k})
 \rightarrow
R({\bm k}) \left[
\ii\left(
\begin{array}{cc}
+1_{p\times p} & 0  \\
 0 & -1_{q\times q}  \\
\end{array}
\right) \right]
R^{-1}({\bm k}).
\label{eq:flattening 2}
\end{eqnarray}
Then, by multiplying $H({\bm k})$ by $-\ii$, the Hamiltonian takes the form of Eq.~(\ref{eq:flattening}) again~\cite{Kawabata-18}. However, this procedure gives an additional minus sign to the symmetry operations with complex or Hermitian conjugation. Therefore, after the flattening procedure, TRS (CS) becomes PHS$^{\dagger}$ (pseudo-Hermiticity), and vice versa.  

Thus, the classification problem reduces to the non-Hermitian Hamiltonian with the form 
\begin{eqnarray}
H({\bm k})=R({\bm k})\,{\mathbb E}\,R^{-1}({\bm k}), \quad {\mathbb E}^2=1
\label{eq:HRER}
\end{eqnarray} 
subject to proper symmetry constraints. Below, we show that the above non-Hermitian Hamiltonian can be deformed into a Hermitian one while keeping the symmetry constraints. 

\subsection{Symmetry constraints}

To fulfill the above purpose, we solve the symmetry constraints for $H({\bm k})$ in terms of  $R({\bm k})$ and $L({\bm k})$.

\subsubsection{PHS and TRS$^{\,\dagger}$}
First, we consider PHS in Eq.~(\ref{eq: PHS}). Taking complex conjugation of the Bloch-BdG equation, we have
\begin{eqnarray}
H^*({\bm k})\,|u^*_n({\bm k})\rangle=E^*_n({\bm k})\,|u^*_n({\bm k})\rangle,
\end{eqnarray}
so that the Hermitian conjugate of Eq.~(\ref{eq: PHS})
\begin{eqnarray}
{\cal C}_{-}\,H^{*}({\bm k})\,{\cal C}^{-1}_{-}=-H^{\dagger}(-{\bm k}), 
\end{eqnarray}
leads to
\begin{eqnarray}
H^{\dagger}\,({\bm k}) \left[ {\cal C}_{-}\,|u^*_n(-{\bm k})\rangle \right]=
- E_n^*(-{\bm k}) \left[ {\cal C}_{-}\,|u^*_n(-{\bm k})\rangle \right].~~~~~~
\end{eqnarray}
Therefore, ${\cal C}_{-}|u^*_n(-{\bm k})\rangle$ gives a left eigenstate of $H({\bm k})$. Since $|u_n({\bm k})\rangle\!\rangle$ forms a complete basis of $H({\bm k})$,
we have
\begin{eqnarray}
{\cal C}_{-}|u^*_n(-{\bm k})\rangle
= \sum_m |u_m({\bm k})\rangle\!\rangle\,[{\mathbb C}_{-}]_{mn}, 
\end{eqnarray}
with $[{\mathbb C}_-]_{mn} := \langle u_m({\bm k})|\,{\cal C}_{-}\,|u^*_n(-{\bm k})\rangle$.  Here, we choose a gauge of the biorthonormal basis such that ${\mathbb C}_-$ is a unitary matrix independent of ${\bm k}$ (such a gauge can be taken, at least locally). In terms of $R({\bm k})$ and $L({\bm k})$, the above relation is compactly summarized as
\begin{eqnarray}
{\cal C}_{-} R^*(-{\bm k}) = L({\bm k})\,{\mathbb C}_-.
\label{eq:cRL}
\end{eqnarray}
Multiplying Eq.~(\ref{eq:cRL}) by ${\cal C}_{-}^{\dagger}$ and ${\mathbb C}_-^{\dagger}$ from the left and right, respectively, we also have  
\begin{eqnarray}
R^{*}(-{\bm k})\,{\mathbb C}_{-}^{\dagger}={\cal C}_{-}^{\dagger}\,L({\bm k}).
\end{eqnarray}
Thus, Eq.~(\ref{eq:cRL}) is equivalent to 
\begin{eqnarray}
{\cal C}_{-}^T L^*(-{\bm k})=R({\bm k})\,{\mathbb C}_-^{T}.
\label{eq:cLR}
\end{eqnarray}

Using Eqs.~(\ref{eq:cRL}) and (\ref{eq:cLR}), we can rewrite PHS as a constraint on ${\mathbb C}_-$ and ${\mathbb E}$. 
It then follows from Eq.~(\ref{eq:RL}) that
\begin{equation} \begin{split}
&{\cal C}_{-}=L({\bm k})\,{\mathbb C}_{-}\,L^{T}(-{\bm k}),\\
&{\cal C}_{-}^T=R({\bm k})\,{\mathbb C}_-^{T}\,R^T(-{\bm k}).
\end{split} \end{equation}
The latter equation also implies 
\begin{eqnarray}
{\cal C}_-^{*} = R^{*}(-{\bm k})\,{\mathbb C}_{-}^{*}\,R^{\dagger}({\bm k}), 
\end{eqnarray}
so that we have
\begin{eqnarray}
{\cal C}_{-}{\cal C}_{-}^{*}
&=&L({\bm k})\,{\mathbb C}_{-}\,L^T(-{\bm k})\,R^*(-{\bm k})\,{\mathbb C}_{-}^{*}\,R^{\dagger}({\bm k})
\nonumber\\
&=&L({\bm k})\,{\mathbb C}_{-}\,{\mathbb C}_{-}^{*}\,R^{\dagger}({\bm k}).
\end{eqnarray}
Thus, the relation ${\cal C}_{-}{\cal C}_-^{*}=\pm 1$ of PHS reduces to 
\begin{eqnarray}
{\mathbb C}_-{\mathbb C}_-^{*}
=\pm 1. 
	\label{eq:cc}
\end{eqnarray}
In a similar manner, we can also show that PHS reduces to 
\begin{eqnarray}
{\mathbb C}_{-}\,{\mathbb E}^T=- {\mathbb E}\,{\mathbb C}_-. 
\label{eq:ce}
\end{eqnarray}

In the above, we derive the symmetry constraints on ${\mathbb C}_-$ and ${\mathbb E}$ [Eqs.~(\ref{eq:cc}) and (\ref{eq:ce})] from PHS for  $H({\bm k})$. Conversely, we can also show that $H({\bm k})$ in the form of Eq.~({\ref{eq:HRER}}) has PHS with ${\cal C}_-{\cal C}^*_-=\pm 1$ when $R({\bm k})$, $L({\bm k})$, ${\mathbb C}_-$, and ${\mathbb E}$ satisfy Eqs.~(\ref{eq:RL}), (\ref{eq:cRL}), (\ref{eq:cc}), and (\ref{eq:ce}). Therefore, when we keep a set of relations
\begin{equation} \begin{split}
&{\mathbb C}_{-}\,{\mathbb E}^T=- {\mathbb E}\,{\mathbb C}_-, 
\quad {\mathbb C}_{-}{\mathbb C}_-^{*}=\pm 1, 
\quad {\mathbb C}_-^{\dagger}{\mathbb C}_-=1, \\
&
{\cal C}_{-} R^{*}(-{\bm k})=L({\bm k})\,{\mathbb C}_-, 
\quad L^{\dagger}({\bm k})\,R({\bm k})=1,
\label{eq:solutionC}
\end{split} \end{equation}
the Hamiltonian given by
\begin{align}
H({\bm k})=R({\bm k})\,{\mathbb E}\,L^{\dagger}({\bm k}), 
\quad 
{\mathbb E}^2=1,
\quad {\mathbb E}={\mathbb E}^{\dagger} 
\label{eq:hRL},
\end{align}
retains PHS with ${\cal C}_{-} {\cal C}^{*}_{-}=\pm 1$.

In a similar manner, we can obtain the following relations 
\begin{equation} \begin{split}
&{\mathbb C}_+{\mathbb E}^T={\mathbb E}\,{\mathbb C}_+, 
\quad {\mathbb C}_+{\mathbb C}_+^{*}=\pm 1, 
\quad {\mathbb C}_+^{\dagger}{\mathbb C}_+=1, \\
&
{\cal C}_{+} R^{*}(-{\bm k})=L({\bm k})\,{\mathbb C}_+, 
\quad L^{\dagger}({\bm k})R({\bm k})=1
	\label{eq:solutionC2}
\end{split} \end{equation}
from TRS$^{\dagger}$ in Eq.~(\ref{eq: TRS-dag}). As long as we keep these relations, we can also retain TRS$^{\dagger}$ with ${\cal C}_+{\cal C}_+^{*}=\pm 1$ for $H({\bm k})$ in Eq.~(\ref{eq:hRL}).

\subsubsection{TRS and PHS$^{\,\dagger}$}
In a manner similar to the above argument, it can be shown that TRS in Eq.~(\ref{eq: TRS}) and PHS$^{\dagger}$ in Eq.~(\ref{eq: PHS-dag}) with ${\cal T}_{\pm}{\cal T}_{\pm}^*= \epsilon_{t} \in \{ \pm 1 \}$ can be obtained, provided that the following relations hold:
\begin{equation} \begin{split}
&{\mathbb T}_{\pm}\,{\mathbb E}^*=\pm {\mathbb E}\,{\mathbb T}_{\pm},
\quad {\mathbb T}_{\pm}{\mathbb T}_{\pm}^{*}=\epsilon_{t},
\quad {\mathbb T}_{\pm}^{\dagger}{\mathbb T}_{\pm}=1, \\
&
{\cal T}_{\pm}R^*(-{\bm k})=R({\bm k})\,{\mathbb T}_{\pm}, 
\quad L^{\dagger}({\bm k})R({\bm k})=1.
	\label{eq:solutionK}
\end{split} \end{equation}

\subsubsection{CS}
CS in Eq.~(\ref{eq: CS}) reduces to
\begin{equation} \begin{split}
&{\mathbb G}\,{\mathbb E}^{\dagger}=- {\mathbb E}\,{\mathbb G},
\quad {\mathbb G}^{\dagger}{\mathbb G}^{-1}=1,
\quad {\mathbb G}^{\dagger}{\mathbb G}=1, \\
&\Gamma R({\bm k})=L({\bm k})\,{\mathbb G}, 
\quad L^{\dagger}({\bm k})R({\bm k})=1.
	\label{eq:solutionQ}
\end{split} \end{equation}

\subsubsection{SLS}
SLS in Eq.~(\ref{eq: SLS}) reduces to
\begin{equation} \begin{split}
&{\mathbb S}\,{\mathbb E}=-{\mathbb E}\,{\mathbb S}, \quad {\mathbb S}^2=1, 
\quad {\mathbb S}^{\dagger}{\mathbb S}=1, \\
&
{\cal S}R({\bm k})=R({\bm k})\,{\mathbb S},
\quad L^{\dagger}({\bm k})R({\bm k})=1.
	\label{eq:solutionP}
\end{split} \end{equation}

\subsubsection{Pseudo-Hermiticity}

Pseudo-Hermiticity in Eq.~(\ref{eq: def pseudo-Hermiticity}) reduces to 
\begin{equation} \begin{split}
&{\mathbb H}\,{\mathbb E}^{\dagger}= {\mathbb E}\,{\mathbb H},
\quad {\mathbb H}^{\dagger}{\mathbb H}^{-1}=1,
\quad {\mathbb H}^{\dagger}{\mathbb H}=1, \\
&\eta R({\bm k})=L({\bm k})\,{\mathbb H}, 
\quad L^{\dagger}({\bm k})R({\bm k})=1.
	\label{eq:solutionQ2}
\end{split} \end{equation}

\subsection{Relations between symmetries}
When there are two or more symmetry operations, we have commutation or anticommutation relations between them. By choosing phases of operators, TRS and PHS (TRS$^{\dagger}$ and PHS$^{\dagger}$) can always be commutative:
\begin{eqnarray}
{\cal T}_{\pm}\,{\cal C}^*_{\mp}={\cal C}_{\mp}\,{\cal T}_{\pm}^*.
\end{eqnarray}
For SLS and pseudo-Hermiticity, we have 
\begin{align}
{\cal S}{\cal T}_+=\epsilon_t{\cal T}_+{\cal S}^*,
\quad  
{\cal S}{\cal C}_-=\epsilon_c{\cal C}_-{\cal S}^*,
\quad
{\cal S}\Gamma=\epsilon_{\Gamma}\Gamma{\cal S},
\end{align} 
and
\begin{align}
\eta\,{\cal T}_+=\epsilon_t{\cal T}_+\eta^*,
\quad  
\eta\,{\cal C}_-=\epsilon_c\,{\cal C}_-\eta^*,
\quad
\eta\,\Gamma=\epsilon_{\Gamma}\Gamma\,\eta,
\end{align} 
respectively. These relations can be satisfied when we have
\begin{equation} \begin{split}
&{\mathbb T}_{\pm}{\mathbb C}^*_{\mp}={\mathbb C}_{\mp}{\mathbb T}_{\pm}^*, \\
&{\mathbb S}{\mathbb T}_+=\epsilon_t{\mathbb T}_+{\mathbb S}^*,
\quad  
{\mathbb S}{\mathbb C}_-=\epsilon_c{\mathbb C}_-{\mathbb S}^*,
\quad
{\mathbb S}{\mathbb G}=\epsilon_\Gamma{\mathbb G}{\mathbb S}, \\
&{\mathbb H}{\mathbb T}_+=\epsilon_t{\mathbb T}_+{\mathbb H}^*,
\quad  
{\mathbb H}{\mathbb C}_-=\epsilon_c{\mathbb C}_-{\mathbb H}^*,
\quad
{\mathbb H}{\mathbb G}=\epsilon_{\Gamma}{\mathbb G}{\mathbb H}.
\end{split} \end{equation}

\subsection{Hermitianization}
	\label{appendix:Hermitianization}
Now we show that a non-Hermitian Hamiltonian in the form of Eq.~(\ref{eq:HRER}) can be continuously deformed into a Hermitian Hamiltonian while keeping symmetry constraints. For this purpose, we perform the polar decomposition of $R({\bm k})$:
\begin{eqnarray}
R({\bm k})={\Lambda}_R({\bm k})\,U_R({\bm k}).
\end{eqnarray}
Here $\Lambda_R({\bm k})$ is given as
\begin{eqnarray}
{\Lambda}_R({\bm k}) := \left[ R({\bm k}) R^{\dagger}({\bm k}) \right]^{1/2}, 
\end{eqnarray}
where the root of $R({\bm k}) R^{\dagger}({\bm k})$ is defined as follows. Since $R({\bm k})$ is invertible, $R({\bm k}) R^{\dagger}({\bm k})$ is a positive-definite Hermitian matrix, and thus $R({\bm k}) R^{\dagger}({\bm k})$ can be diagonalized as
\begin{align}
R({\bm k}) R^{\dagger}({\bm k}) = V({\bm k})
\left(
\begin{array}{ccc}
\lambda_1^2({\bm k})& & \\
&\lambda_2^2({\bm k})& \\ 
& & \ddots
\end{array}
\right)V^{\dagger}({\bm k}),
\end{align}
where $V({\bm k})$ is a unitary matrix and $\lambda_i({\bm k})$ is a positive number. Then, $\Lambda_R({\bm k}) := [ R({\bm k}) R^{\dagger}({\bm k}) ]^{1/2}$ is defined as
\begin{eqnarray}
[ R({\bm k}) R^{\dagger}({\bm k}) ]^{1/2}&=&V({\bm k})
\left(
\begin{array}{ccc}
\lambda_1({\bm k})& & \\
&\lambda_2({\bm k})& \\ 
& & \ddots
\end{array}
\right)V^{\dagger}({\bm k})
\nonumber\\
&=:&
V({\bm k})\,\lambda({\bm k})\,V^{\dagger}({\bm k}),
\end{eqnarray} 
with $\lambda({\bm k}) := {\rm diag}\,[\lambda_1({\bm k}), \lambda_2({\bm k}), \dots]$.
From this equation, we also have  
\begin{eqnarray}
\Lambda_R^{-1}({\bm k})=V({\bm k})\,
\lambda^{-1}({\bm k})\,
V^{\dagger}({\bm k}).
\end{eqnarray} 
Therefore, $U_R({\bm k})$ is uniquely determined as $U_R({\bm k})=\Lambda^{-1}_R({\bm k})\,R({\bm k})$, which is unitary:
\begin{eqnarray}
U_R({\bm k})\,U_R^{\dagger}({\bm k})
=\Lambda_R^{-1}({\bm k})R({\bm k})R^{\dagger}({\bm k})\Lambda_R^{-1}({\bm k})=1.~~~~
\end{eqnarray}

As easily seen, we can make a non-Hermitian Hamiltonian in the form of Eq.~(\ref{eq:HRER}) Hermitian by just deforming $\Lambda_R({\bm k})$ as $\Lambda_R({\bm k})\to 1$. This process also retains the line gap. However, we need to check whether this process can be done while keeping symmetry.

From the symmetry constraints for $R({\bm k})$ in Eqs.~(\ref{eq:solutionC}) and (\ref{eq:solutionC2})-(\ref{eq:solutionQ2}),
we have the following constraints on $R({\bm k}) R^{\dagger}({\bm k})$:
\begin{align}
&R({\bm k})\,R^{\dagger}({\bm k})\,{\cal C}_{\mp}\,
[R(-{\bm k})R^{\dagger}(-{\bm k})]^{*}={\cal C}_{\mp} \label{eq:RRcRR8} 
\end{align}
for PHS/TRS$^\dagger$,  
\begin{align}
&R({\bm k})\,R^{\dagger}({\bm k})\,{\cal T}_{\pm}\{[R(-{\bm k})R^{\dagger}(-{\bm k})]^{T}\}^{-1}
={\cal T}_{\pm}
\end{align} 
for TRS/PHS$^\dagger$, 
\begin{align}
R({\bm k})R^{\dagger}({\bm k})\,\Gamma\,R({\bm k})R^{\dagger}({\bm k})=\Gamma
\end{align}
for CS, 
\begin{align}
&R({\bm k})R^{\dagger}({\bm k})\,{\cal S}\,
[R({\bm k})R^{\dagger}({\bm k})]^{-1}={\cal S}
\end{align}
for SLS, and 
\begin{align}
&R({\bm k})R^{\dagger}({\bm k})\,\eta\,R({\bm k})R^{\dagger}({\bm k})=\eta
\end{align}
for pseudo-Hermiticity.
It can also be shown that these constraints are equivalent to the following ones:
\begin{equation} \begin{split}
&\Lambda_R({\bm k})\,{\cal C}_{\mp}\,\Lambda_R^*(-{\bm k})={\cal C}_{\mp},~  
\Lambda_R({\bm k})\,{\cal T}_{\pm}\,[\Lambda_R^T(-{\bm k})]^{-1}={\cal T}_{\pm}, \\
&\Lambda_R({\bm k})\,\Gamma\,\Lambda_R({\bm k})=\Gamma,~
\Lambda_R({\bm k})\,{\cal S}\,\Lambda_R^{-1}({\bm k})={\cal S}, \\
&\Lambda_R({\bm k})\,\eta\,\Lambda_R({\bm k})=\eta.
	\label{eq:LCL}
\end{split} \end{equation} 
For instance, the first equation in the above is derived as follows.
We first rewrite Eq.~(\ref{eq:RRcRR8}) as
\begin{align}
V({\bm k})\,\lambda^2({\bm k})\,V^{\dagger}({\bm k})\,{\cal C}_{\mp}\,
V^*(-{\bm k})\,\lambda^2(-{\bm k})\,V^T(-{\bm k})={\cal C}_{\mp},  
\end{align}
which leads to
\begin{align}
\lambda^2({\bm k})\,V^{\dagger}({\bm k})\,{\cal C}_{\mp}\,
V^*(-{\bm k})\,\lambda^2(-{\bm k})=V^{\dagger}({\bm k})\,{\cal C}_{\mp}\,V^*(-{\bm k}).
\end{align}
This equation is equivalent to 
\begin{align}
\lambda_n({\bm k})\,\lambda_m(-{\bm k})=1  
	\label{eq:lambda_constraints}
\end{align}
for 
$[V^{\dagger}({\bm k})\,{\cal C}_{\mp}\,V^*(-{\bm k})]_{mn}\neq 0$, which yields the first equation in Eq.~(\ref{eq:LCL}).
From Eq.~(\ref{eq:LCL}), we also have
\begin{equation} \begin{split}
&{\cal C}_{\mp}U_R^*(-{\bm k})=U_R^{\dagger}({\bm k}){\mathbb C}_{\mp},~
{\cal T}_{\pm}U^*_R(-{\bm k})=U_R({\bm k}){\mathbb T}_{\pm}, \\
&\Gamma U_R({\bm k})=U_R^{\dagger}({\bm k}){\mathbb G},~
{\cal S}{\mathbb U}_R({\bm k})=U_R({\bm k}){\mathbb S}, \\
&\eta U_R({\bm k})=U_R^{\dagger}({\bm k}){\mathbb H}.
	\label{eq:UCU}
\end{split} \end{equation}

Here it should be noted that $\lambda_n({\bm k})$ can be continuously deformed into $1$ while keeping Eq.~(\ref{eq:lambda_constraints}), and thus we can retain the first equation in Eq.~(\ref{eq:LCL}) during the process of 
$\Lambda_R({\bm k})\to 1$. Combining it with the first equation in Eq.~(\ref{eq:UCU}), we obtain the correct symmetry constraint on $R({\bm k})$ in Eq.~(\ref{eq:solutionC}) for PHS/TRS$^{\dagger}$. This means that we can deform a non-Hermitian Hamiltonian to a Hermitian one while keeping PHS/TRS$^\dagger$. In a similar manner, we can make a non-Hermitian Hamiltonian Hermitian while keeping any other symmetry.

\subsection{Patching different momentum regions}

To derive Eq.~(\ref{eq:cRL}), we have taken a special gauge in which ${\mathbb C}_{-}$ is independent of ${\bm k}$ (we have also chosen similar gauges for the other symmetries). Generally, such a gauge can be taken not globally but locally; 
if we take such a gauge globally, there arises a singularity in $R({\bm k})$ and $L({\bm k})$. To avoid this singularity, we divide the whole momentum space into several subregions and take a proper gauge in each region. Whereas the matrix ${\mathbb C}_{-}$ can take the same form in all regions, $R({\bm k})$ and $L({\bm k})$ are given locally so that they can be different in different regions. From the arguments in Appendix~\ref{appendix:Hermitianization}, we can deform in each region a non-Hermitian Hamiltonian into a Hermitian one while keeping the line gap and relevant symmetries. Now, we show that this Hermitianization process can be performed globally. For definiteness, we focus on the case with PHS below, but the generalization to the other cases is straightforward.

First, let us consider two regions I and II in
momentum space and denote $R({\bm k})$ in region I (II) as $R_{\rm I} ({\bm k})$ [$R_{\rm II}({\bm k})$]. Since both $R_{\rm I}({\bm k})$ and $R_{\rm II}({\bm k})$ are well defined on the boundary between regions I and II, they are related to each other by a gauge transformation
\begin{align}
R_{\rm I}({\bm k})=R_{\rm II}({\bm k})\,G({\bm k}),
\end{align}
with an invertible matrix $G({\bm k})$. Here, $G({\bm k})$ is unitary for Hermitian systems, but this is not necessarily so for non-Hermitian systems. Since both $R_{\rm I} ({\bm k})$ and $R_{\rm II}({\bm k})$ obey Eqs.~(\ref{eq:HRER}) and (\ref{eq:cRL}) on the boundary, $G({\bm k})$ is found to satisfy
\begin{align}
G({\bm k})\,{\mathbb E}\,G^{-1}({\bm k})={\mathbb E},~
G^{\dagger}({\bm k})\,{\mathbb C}_{-}\,G^*(-{\bm k})={\mathbb C}_-.
\label{eq:Ggauge}
\end{align}
We then perform the polar decomposition of $G({\bm k})$
\begin{align}
G({\bm k})=U_G({\bm k})\,\Omega_G({\bm k}),~
\Omega_G({\bm k}) := [G^{\dagger}({\bm k})G({\bm k})]^{1/2}
\end{align}
with a unitary matrix $U_G({\bm k})$. Here, $\Lambda_G({\bm k})$ is defined as follows. Since $G^{\dagger}({\bm k}) G({\bm k})$ is a positive-definite Hermitian matrix, it is diagonalized as
\begin{align}
G^{\dagger}({\bm k})G({\bm k})=
W({\bm k})
\left(
\begin{array}{ccc}
\omega_{1}^2({\bm k}) & & \\
& \omega^2_{2}({\bm k}) & \\
& & \ddots
\end{array}
\right)W^{\dagger}({\bm k}) 
\end{align}
with a unitary matrix $W({\bm k})$ and positive real numbers $\omega^2_{i}({\bm k})$'s. Then, $\Omega_G({\bm k})$ is defined as
\begin{align}
\Omega_G({\bm k}) :=
W({\bm k})
\left(
\begin{array}{ccc}
\omega_{1}({\bm k}) & & \\
& \omega_{2}({\bm k}) & \\
& & \ddots
\end{array}
\right)W^{\dagger}({\bm k})
\end{align}
with $\omega_{i}({\bm k})>0$. Using the polar decomposition of $G({\bm k})$, we recast Eq.~(\ref{eq:Ggauge}) into
\begin{align}
&\Omega_G({\bm k})\,{\mathbb E}\,\Omega_G^{-1}({\bm k})={\mathbb E},~
\Omega_G({\bm k})\,{\mathbb C}_{-}\,\Omega_G^T(-{\bm k})={\mathbb C}_-,
\label{eq:LG1}
\\
&U_G({\bm k})\,{\mathbb E}\,U_G^{\dagger}({\bm k})={\mathbb E},~
U_G({\bm k})\,{\mathbb C}_{-}\,U_G^T(-{\bm k})={\mathbb C}_-.
\label{eq:LG2}
\end{align}
Here, it should be noted that $\Omega_G({\bm k})$ can be extended to the whole region I without crossing a singularity. In fact, $\omega_{i}({\bm k})$ can be rewritten as $\omega_{i}({\bm k})=e^{-\rho_i({\bm k})}$ because of the positivity of $\omega_{i}({\bm k})$, and we can extrapolate $\rho_i({\bm k})$ as $\rho_i({\bm k})\to 0$ from the boundary to the center of region I while keeping Eq.~(\ref{eq:LG1}). As a result, we have a well defined $\Omega_G({\bm k})$ in the whole region I.

Using this $\Omega_G({\bm k})$, we can construct another well defined $R({\bm k})$ in region I, i.e., $R'_{\rm I}({\bm k}):= R_{\rm I} ({\bm k})\,\Omega_G^{-1}({\bm k})$. Although the new matrix $R'_{\rm I}({\bm k})$ satisfies Eqs.~(\ref{eq:HRER}) and (\ref{eq:cRL}) again, there is an important modification. Now, the gauge transformation between regions I and II becomes unitary,
\begin{align}
R'_{\rm I}({\bm k})=R_{\rm II}({\bm k})\,U_G({\bm k}), 
\end{align}
which yields 
\begin{align}
R'_{\rm I} ({\bm k})R'_{\rm I}{}^{\dagger}({\bm k})=R_{\rm II}({\bm k})R_{\rm II}^{\dagger}({\bm k}). 
\end{align}
Therefore, 
\begin{align}
\Lambda_R({\bm k}) :=
\left\{
\begin{array}{ll}
(R'_{\rm I}({\bm k})R'_{\rm I}{}^{\dagger}({\bm k}))^{1/2} & \mbox{for region I;}\\
(R_{\rm II}({\bm k})R_{\rm II}^{\dagger}({\bm k}))^{1/2}
&
\mbox{for region II} 
\end{array}
\right.
\end{align}
defines a continuous single-valued matrix function in the union of regions I and II. In a similar manner, all the gauge transformations between the different regions can be made unitary, indicating that $\Lambda_{R}({\bm k})$ can be defined 
continuously in the whole momentum space. This means that the Hermitianization process in Appendix~\ref{appendix:Hermitianization} can be performed globally.

\section{\textit{K}-theory classification based on unitary flattening for point gaps}
	\label{appendix: K-theory}

We provide the \textit{K}-theory classification of non-Hermitian Hamiltonians $H \left( \bm k \right)$ defined over the $d$-dimensional Brillouin-zone (BZ) torus $T^{d}$ based on unitary flattening for point gaps. In particular, we show that this classification is given by the twisted equivariant \textit{K}-group ${}^{\phi}K^{(\tau,c)-n-1}_G(T^d)$~\cite{Shiozaki-16, *Shiozaki-17, *Shiozaki-18, Freed-13, Gomi-17} with a shift of an integer degree, which coincides with the classification of adiabatic time evolutions with a certain period. 

We first formulate possible symmetries of non-Hermitian fermionic systems in the many-body Hilbert space. Let $G$ be a symmetry group and $\hat{\phi}: G \rightarrow \mathbb{Z}/2 = \{\pm 1\}$ be a homomorphism specifying whether $g \in G$ is unitary or antiunitary, i.e., $g \in G$ acts on the imaginary unit as 
\begin{equation}
g\,\ii\,g^{-1} = \hat{\phi}_{g}\,\ii.
\end{equation}
In addition, let $\hat{c}: G \rightarrow \mathbb{Z}/2 = \{\pm 1\}$ be a homomorphism specifying whether or not $g \in G$ is a particle-hole type, i.e., $g \in G$ acts on complex fermion operators as 
\begin{equation}
\left.\begin{array}{ll}
g\,\hat{\psi}^{\dag}_{\bm k}\,g^{-1} & ( \hat{c}_{g} = +1 ) \\
g\,\hat{\psi}_{\bm k}\,g^{-1} & ( \hat{c}_{g} = -1 ) \\
\end{array}\right\} = \hat{\psi}^{\dag}_{g{\bm k}} U_{g} \left( {\bm k} \right),
\end{equation}
where $\hat{\psi}_{\bm k}$ ($\hat{\psi}_{\bm k}^{\dag}$) is a complex fermion annihilation (creation) operator in the BZ, and $U_{g} \left( {\bm k} \right)$ is a unitary matrix. Based on $\hat{\phi}$ and $\hat{c}$, there are four types of symmetries:
\begin{enumerate}
\item Unitary symmetry $\hat{U}$: $\hat{\phi}_{g} = +1$ and $\hat{c}_{g} = +1$.
\item Time-reversal symmetry $\hat{\cal T}$: $\hat{\phi}_{g} = -1$ and $\hat{c}_{g} = +1$.
\item Particle-hole symmetry $\hat{\cal C}$: $\hat{\phi}_{g} = +1$ and $\hat{c}_{g} = -1$.
\item Chiral symmetry $\hat{\Gamma}$: $\hat{\phi}_{g} = -1$ and $\hat{c}_{g} = -1$.
\end{enumerate}
It is notable that particle-hole symmetry $\hat{\cal C}$ is unitary in the many-body Hilbert space. Furthermore, we fix the factor system of the symmetry $G$ that indicates a $U \left( 1 \right)$ phase among two symmetry actions $gh \in G$ and $hg \in G$ as
\begin{align}
\left.\begin{array}{ll}
U_{g} \left( h {\bm k} \right) U_{h} \left( {\bm k} \right) & ( \hat{\phi}_{g} \hat{c}_{g} = +1 ) \\
U_{g} \left( h {\bm k} \right) U_{h}^{*} \left( {\bm k} \right) & ( \hat{\phi}_{g} \hat{c}_{g} = -1 ) \\
\end{array} \right\}
= e^{\ii \tau_{g,h} \left( gh {\bm k} \right)} U_{gh} \left( {\bm k} \right), 
\end{align}
where the twist $\tau = \tau_{g, h} \left( \bm k \right)$ specifies the projective representation for internal degrees of freedom and nonprimitive lattice translations of space group symmetry~\cite{Freed-13}. For a free fermion Hamiltonian $\hat{H} = \sum_{\bm k} \hat{\psi}_{\bm k}^{\dag} H \left( \bm k \right) \hat{\psi}_{\bm k}$, the symmetry $g \hat{H} g^{-1} = \hat{H}$ is recast as
\begin{align}
\left.\begin{array}{lll}
\hat{U}: & U_{g}^{-1} \left( {\bm k} \right) H \left( {\bm k} \right) U_{g} \left( {\bm k} \right ) \\
\hat{\cal T}: & U_{g}^{-1} \left( {\bm k} \right) H^{*} \left( {\bm k} \right) U_{g} \left( {\bm k} \right) \\
\hat{\cal C}: & - U_{g}^{-1} \left( {\bm k} \right) H^{T} \left( {\bm k} \right) U_{g} \left( {\bm k} \right) \\
\hat{\Gamma}: & - U_{g}^{-1} \left( {\bm k} \right) H^{\dag} \left( {\bm k} \right) U_{g} \left( {\bm k} \right) \\
\end{array} \right\}
= H \left( g {\bm k} \right) 
\end{align}
for the single-particle Hamiltonian $H \left( \bm k \right)$. Here, we assume ${\rm tr} \left[ H \left( \bm k \right) \right] = 0$.

We now develop the \textit{K}-theory classification of $H \left( \bm k \right)$ based on the unitary flattening for point gaps [i.e., $\det H \left( \bm k \right) \neq 0$ for all ${\bm k}$]. Since $H \left( \bm k \right)$ can be assumed to be a unitary matrix due to Theorem 1 in Sec.~\ref{sec: classification (unitary flattening)}, $H \left( \bm k \right)$ is identified with an adiabatic time evolution of a Hermitian system with a certain period, which implies that the classification of non-Hermitian Hamiltonians $H \left( \bm k \right)$ under the unitary flattening is the same as that for unit adiabatic time evolutions. Here the unit adiabatic time evolutions are described by the \textit{K}-group with a shift of the integer degree $n$ by $+1$~\cite{Shiozaki-16, *Shiozaki-17, *Shiozaki-18}. We thus expect that the non-Hermitian Hamiltonians $H \left( \bm k \right)$ under the unitary flattening are classified by the \textit{K}-group ${}^{\phi} K^{(\tau, c)-1}_{G} ( T^{d} )$ with $\phi := \hat{\phi} \hat{c}$ and $c := \hat{c}$. In fact, for the extended Hermitian Hamiltonian $\tilde{H} \left( {\bm k} \right)$ with CS (SLS) $\Sigma$ [Eqs.~(\ref{eq: def - tilde H}) and (\ref{eq: sigma - tilde}) in Sec.~\ref{sec: classification (unitary flattening)}], symmetry $g \in G$ is represented as 
\begin{align}
\tilde{U}_{g} \left( {\bm k} \right) := 
\left\{\begin{array}{ll}
\left( \begin{array}{@{\,}cc@{\,}} 
	U_{g} \left( {\bm k} \right) & 0 \\
	0 & U_{g} \left( {\bm k} \right) \\ 
	\end{array} \right) & ( \hat{c}_{g} = +1 ); \\
\left( \begin{array}{@{\,}cc@{\,}} 
	0 & U_{g} \left( {\bm k} \right) \\
	U_{g} \left( {\bm k} \right) & 0 \\ 
	\end{array} \right) & ( \hat{c}_{g} = -1 ), \\
\end{array}\right.
\end{align}
so that the Hamiltonian satisfies
\begin{align}
&\left.\begin{array}{ll}
\tilde{U}_{g}^{-1} \left( {\bm k} \right) \tilde{H} \left( {\bm k} \right) \tilde{U}_{g} \left( {\bm k} \right) & (\hat{\phi}_{g} \hat{c}_{g} = +1) \\
\tilde{U}_{g}^{-1} \left( {\bm k} \right) \tilde{H}^{*} \left( {\bm k} \right) \tilde{U}_{g} \left( {\bm k} \right) & (\hat{\phi}_{g} \hat{c}_{g} = -1) \\
\end{array} \right\}
= \hat{c}_{g} \tilde{H} \left( {\bm k} \right),
\end{align}
\begin{align}
&\left.
\begin{array}{ll}
\tilde{U}_{g} \left( h{\bm k} \right) \tilde{U}_{h} \left( {\bm k} \right) & (\hat{\phi}_{g} \hat{c}_{g} = +1) \\
\tilde{U}_{g} \left( h{\bm k} \right) \tilde{U}_{h}^{*} \left( {\bm k} \right) & (\hat{\phi}_{g} \hat{c}_{g} = -1) \\
\end{array}\right\}
= e^{\ii \tau_{g,h} \left( gh{\bm k} \right)} \tilde{U}_{gh} \left( {\bm k} \right),
\end{align}
\begin{align}
&~~\tilde{U}_{g}^{-1}\,\left( {\bm k} \right)\,\Sigma\,\tilde{U}_{g} \left( {\bm k} \right) = \hat{c}_{g} \Sigma.
\end{align}
These conditions determine nothing but the symmetry class for Hermitian Hamiltonians with the integer grading $n=1$~\cite{Shiozaki-16, *Shiozaki-17, *Shiozaki-18}. Therefore, we conclude that non-Hermitian Hamiltonians $H \left( {\bm k} \right)$ under the unitary flattening are classified by the \textit{K}-group ${}^{\phi} K_{G}^{(\tau,c)-1} ( T^{d} )$. As a consequence, the periodic table for the AZ symmetry class is obtained as Tables~\ref{tab: complex AZ} and \ref{tab: real AZ}. 

Using the non-Hermitian Dirac matrices developed in Sec.~\ref{sec: Dirac matrix}, we can also define the symmetry class $(G,\phi,c,\tau,n)$ with the integer grading $n>0$~\cite{Shiozaki-16, *Shiozaki-17, *Shiozaki-18} for non-Hermitian Hamiltonians as follows. 
As in the Hermitian case, the shift of the integer grading is defined by adding CS for $\gamma_{i}$'s ($i=1,\cdots, n$) satisfying 
\begin{align}
&\gamma_i h^{\dag} (k) \gamma_i^{\dag} = - h(k), \qquad 
\gamma_i \gamma_j^{\dag} + \gamma_j \gamma_i^{\dag}  = 2 \delta_{ij}, \\
&\left.
\begin{array}{ll}
u_g(k) \gamma_i u_g^{\dag} (k) & (\hat \phi_g \hat c_g =1)\\
u_g(k) \gamma_i^* u_g^{\dag} (k) & (\hat \phi_g \hat c_g =-1)\\
\end{array}\right\}
= \hat c_g \gamma_i.
\end{align}
From the Hermitianization given by Eq.~(\ref{eq: def - tilde H}), the classification of non-Hermitian Hamiltonians with the symmetry class $(G,\phi,c,\tau,n)$ is given by ${}^{\phi}K_G^{(\tau,c)-n-1}(T^d)$, with $\phi=\hat \phi \hat c$ and $c = \hat c$. 

\section{Topological classification based on two antiunitary symmetries}
	\label{appendix: two antiunitary symmetries}

Topological classification based on two antiunitary symmetries $\Tp$, $\Tm$ and one unitary symmetry $\SLS$ was considered in Ref.~\cite{Gong-18, *Bandres-Segev-18}. This classification assumes point gaps, and the corresponding classification table for both complex-energy gaps is shown in Table~\ref{tab: two antiunitary symmetries}. Notably, two antiunitary symmetries $\Tp$ and $\Tm$ are topologically equivalent to each other for non-Hermitian Hamiltonians~\cite{Kawabata-18}, whereas they are clearly distinct for Hermitian Hamiltonians. As a result, some symmetry classes are equivalent to others in non-Hermitian physics. In fact, the symmetry class only having $\Tp$ with $\Tp \Tp^{*} = +1$ ($\Tp \Tp^{*} = -1$) [i.e., class AI (AII)] is equivalent to that only having $\Tm$ with $\Tm \Tm^{*} = +1$ ($\Tm \Tm^{*} = -1$) [i.e., class $\mathrm{D}^{\dag}$ ($\mathrm{C}^{\dag}$)], and the symmetry class having $\Tp$ with $\Tp \Tp^{*} = +1$ and $\Tm$ with $\Tm \Tm^{*} = -1$ is equivalent to that having $\Tp$ with $\Tp \Tp^{*} = -1$ and $\Tm$ with $\Tm \Tm^{*} = +1$.

\begin{table*}[t]
	\centering
	\caption{Topological classification table for non-Hermitian systems based on two antiunitary symmetries $\Tp$, $\Tm$ and unitary symmetry $\SLS$. Non-Hermitian topological phases are classified according to the symmetry, the spatial dimension $d$, and the definition of complex-energy point (P) or line (L) gaps. \\}
		\label{tab: two antiunitary symmetries}
	\small
     \begin{tabular}{cccccccccccc} \hline \hline
    ~$\left( \Tp,~\Tm \right)$~ & ~$\SLS$~ & ~Gap~ & ~Classifying space~ & ~$d=0$~ & ~$d=1$~ & ~$d=2$~ & ~$d=3$~ & ~$d=4$~ & ~$d=5$~ & ~$d=6$~ & ~$d=7$~ \\ \hline
    \multirow{2}{*}{$0$} & \multirow{2}{*}{$0$}
    & P & \Cb \\
    & & L & \Ca \\ \hline
    \multirow{2}{*}{$0$} & \multirow{2}{*}{$1$}
    & P & \Cbb \\
    & & L & \Cb \\ \hline \hline
    \multirow{3}{*}{$\left( +1, +1 \right)$} & \multirow{3}{*}{$1$} 
    & P & \Rbb \\
    & & \multirow{2}{*}{L} 
    & \Rc \\
    & & & \Rc \\ \hline
    \multirow{3}{*}{$+1$} & \multirow{3}{*}{$0$} 
    & P & \Rb \\
    & & \multirow{2}{*}{L} 
    & \Ra \\
    & & & \Rc \\ \hline
    \multirow{3}{*}{$\left( +1, -1 \right)$} & \multirow{3}{*}{$1$} 
    & P & \Cb \\
    & & \multirow{2}{*}{L} 
    & \Rh \\
    & & & \Rd \\ \hline
    \multirow{3}{*}{$-1$}& \multirow{3}{*}{$0$} 
    & P & \Rf \\
    & & \multirow{2}{*}{L} 
    & \Ree \\
    & & & \Rg \\ \hline
    \multirow{3}{*}{$\left( -1, -1 \right)$} & \multirow{3}{*}{$1$} 
    & P & \Rff \\
    & & \multirow{2}{*}{L} 
    & \Rf \\
    & & & \Rf \\ \hline \hline
  \end{tabular}
\end{table*}

\begin{table*}[t]
	\centering
	\caption{Topological invariants for point gaps in the AZ and $\mathrm{AZ}^{\dag}$ symmetry classes for the spatial dimension $d \leq 3$. The equation numbers or the section numbers of the corresponding topological invariants are shown for each topological phase. \\}
	\label{tab: AZ - invariants}
     \begin{tabular}{cccccc} \hline \hline
    ~Symmetry class~ & ~Classifying space~ & ~$d=0$~ & ~$d=1$~ & ~$d=2$~ & ~$d=3$~ \\ \hline
    A & $\mathcal{C}_{1}$ & $0$ & \Z~[Eq.~(\ref{appendix: topological invariant - winding})] & $0$ & \Z~[Eq.~(\ref{appendix: topological invariant - winding})] \\
    AIII & $\mathcal{C}_{0}$ & \Z~[Eq.~(\ref{eq: topological invariant - Chern})] & $0$ & \Z~[Eq.~(\ref{eq: topological invariant - Chern})] & $0$ \\ \hline
    AI & $\mathcal{R}_{1}$ & \Zt~[Eq.~(\ref{appendix: point - 0D AI})] & \Z~[Eq.~(\ref{appendix: topological invariant - winding})] & $0$ & $0$ \\
    BDI & $\mathcal{R}_{2}$ & \Zt~[Eq.~(\ref{appendix: point - 0D BDI})] & \Zt~[Eq.~(\ref{appendix: point - 1D BDI})] & \Z~[Eq.~(\ref{eq: topological invariant - Chern})] & $0$ \\
    D & $\mathcal{R}_{3}$ & 0 & \Zt~[Eq.~(\ref{appendix: 1D class D - point - Z2})] & \Zt~[Eq.~(\ref{appendix: point - 2D D})] & \Z~[Eq.~(\ref{appendix: topological invariant - winding})] \\
    DIII & $\mathcal{R}_{4}$ & $2$\Z~[Eq.~(\ref{eq: topological invariant - Chern})] & $0$ & \Zt~(Sec.~\ref{appendix: point - 2D DIII}) & \Zt~(Sec.~\ref{appendix: point - 3D DIII}) \\
    AII & $\mathcal{R}_{5}$ & $0$ & $2$\Z~[Eq.~(\ref{appendix: topological invariant - winding})] & $0$ & \Zt~(Sec.~\ref{appendix: point - 3D AII}) \\
    CII & $\mathcal{R}_{6}$ & $0$ & $0$ & $2$\Z~[Eq.~(\ref{eq: topological invariant - Chern})] & $0$ \\
    C & $\mathcal{R}_{7}$ & $0$ & $0$ & $0$ & $2$\Z~[Eq.~(\ref{appendix: topological invariant - winding})] \\
    CI & $\mathcal{R}_{0}$ & \Z~[Eq.~(\ref{eq: topological invariant - Chern})] & $0$ & $0$ & $0$ \\ \hline
    $\text{AI}^{\dag}$ & $\mathcal{R}_{7}$ & $0$ & $0$ & $0$ & $2$\Z~[Eq.~(\ref{appendix: topological invariant - winding})] \\
    $\text{BDI}^{\dag}$ & $\mathcal{R}_{0}$ & \Z~[Eq.~(\ref{eq: topological invariant - Chern})] & $0$ & $0$ & $0$ \\
    $\text{D}^{\dag}$ & $\mathcal{R}_{1}$ & \Zt~[Eq.~(\ref{appendix: point - 0D D-dag})] & \Z~[Eq.~(\ref{appendix: topological invariant - winding})] & $0$ & $0$ \\
    $\text{DIII}^{\dag}$ & $\mathcal{R}_{2}$ & \Zt~[Eq.~(\ref{appendix: point - 0D DIII-dag})] & \Zt~[Eq.~(\ref{appendix: point - 1D DIII-dag})] & \Z~[Eq.~(\ref{eq: topological invariant - Chern})] & $0$ \\
    $\text{AII}^{\dag}$ & $\mathcal{R}_{3}$ & 0 & \Zt~[Eq.~(\ref{appendix: point - 1D AII-dag})] & \Zt~[Eq.~(\ref{appendix: point - 2D AII-dag})] & \Z~[Eq.~(\ref{appendix: topological invariant - winding})] \\
    $\text{CII}^{\dag}$ & $\mathcal{R}_{4}$ & $2$\Z~[Eq.~(\ref{eq: topological invariant - Chern})] & $0$ & \Zt~(Sec.~\ref{appendix: point - 2D CII-dag}) & \Zt~(Sec.~\ref{appendix: point - 3D CII-dag}) \\
    $\text{C}^{\dag}$ & $\mathcal{R}_{5}$ & $0$ & $2$\Z~[Eq.~(\ref{appendix: topological invariant - winding})] & $0$ & \Zt~(Sec.~\ref{appendix: point - 3D C-dag}) \\
    $\text{CI}^{\dag}$ & $\mathcal{R}_{6}$ & $0$ & $0$ & $2$\Z~[Eq.~(\ref{eq: topological invariant - Chern})] & $0$ \\ \hline
    A + $\mathcal{S}$ ($\text{AIII}^{\dag}$)& $\mathcal{C}_{1} \times \mathcal{C}_{1}$ & $0$ & $\mathbb{Z} \oplus \mathbb{Z}$~(Sec.~\ref{appendix: point - A + SLS}) & $0$ & $\mathbb{Z} \oplus \mathbb{Z}$~(Sec.~\ref{appendix: point - A + SLS}) \\
    AIII + $\mathcal{S}_{+}$ & $\mathcal{C}_{1}$ & $0$ & \Z~[Eq.~(\ref{appendix: point - AIII + SLS+})] & $0$ & \Z~[Eq.~(\ref{appendix: point - AIII + SLS+})] \\
    AIII+ $\mathcal{S}_{-}$ & $\mathcal{C}_{0} \times \mathcal{C}_{0}$ & $\mathbb{Z} \oplus \mathbb{Z}$~(Sec.~\ref{appendix: point - AIII + SLS-}) & $0$ & $\mathbb{Z} \oplus \mathbb{Z}$~(Sec.~\ref{appendix: point - AIII + SLS-}) & $0$ \\ \hline \hline
  \end{tabular}
  	\label{tab: topological invariant - point}
\end{table*}

\section{Topological invariants for point gaps}
	\label{appendix: topological invariants - point}

We explicitly present topological invariants of non-Hermitian systems with point gaps. In particular, we focus on basic symmetry classes such as the AZ and $\mathrm{AZ}^{\dag}$ symmetry classes for spatial dimensions $d \leq 3$ (Table~\ref{tab: topological invariant - point}).

\subsection{$\mathbb{Z}$ invariants in even dimensions}

In the presence of CS defined by Eq.~(\ref{eq: CS}), we can define the $n$-th Chern number $C_{n}$ of the Hermitian Hamiltonian $\ii H \left( {\bm k} \right) \Gamma$ for even spatial dimensions $d=2n$. Introducing the Green function by
\begin{equation}
G^{-1} \left( \omega, {\bm k} \right) := 
\ii \omega - \ii H \left( {\bm k} \right) \Gamma,
\end{equation} 
we can define the $n$-th Chern number for $d=2n$ dimensions as~\cite{Ishikawa-Matsuyama, Volovik-textbook}
\begin{equation}
C_{n} = \frac{n!}{\left( 2\pi \ii \right)^{n+1} \left( 2n+1\right)!} \int_{\mathbb{R}_{\omega} \times \mathrm{BZ}^{d}} \mathrm{tr} \left( G dG^{-1} \right)^{2n+1}.\quad
	\label{eq: topological invariant - Chern}
\end{equation}
See also Appendix~\ref{appendix: line - Chern} for various expressions of the Chern number in Hermitian and non-Hermitian systems.

\subsection{$\mathbb{Z}$ invariants in odd dimensions}

The $\mathbb{Z}$ (and $2\mathbb{Z}$) topological invariants in odd spatial dimensions $d = 2n+1$ are given as the winding number $W_{2n+1}$ of the map $H = H \left( {\bm k} \right): \mathrm{BZ}^{d} \rightarrow \mathrm{GL}_{N} \left( \mathbb{C} \right)$, where $\mathrm{BZ}^{d}$ denotes the $d$-dimensional Brillouin zone and $\mathrm{GL}_{N} \left( \mathbb{C} \right)$ the general linear group of $N \times N$ invertible matrices (i.e., $\det H \neq 0$). The winding number $W_{2n+1}$ is explicitly given as 
\begin{equation}
W_{2n+1} := \frac{n!}{\left( 2\pi \ii \right)^{n+1} \left( 2n+1 \right)!} \int_{\mathrm{BZ}^{d}} \mathrm{tr} \left( H^{-1} dH \right)^{2n+1}.
	\label{appendix: topological invariant - winding}
\end{equation}

\subsection{0D class AI}

In class AI, we have TRS defined by Eq.~(\ref{eq: TRS}) with $\mathcal{T}_{+} \mathcal{T}_{+}^{*} = +1$, which leads to the reality of $\det H$. Thus, the $\mathbb{Z}_{2}$ topological invariant 
$\nu \in \{ 0, 1\}$ is given as
\begin{align}
\left( -1\right)^{\nu} := {\rm sgn} \det H. 
	\label{appendix: point - 0D AI}
\end{align}

\subsection{0D class BDI}

Because of the presence of TRS with $\mathcal{T}_{+}\mathcal{T}_{+}^{*} = +1$ and CS with $\Gamma = \mathcal{C}_{-} \mathcal{T}_{+}^{*}$ that commutes with TRS (i.e., $\mathcal{T}_{+} \Gamma^{*} = \Gamma \mathcal{T}_{+}$) in class BDI, we have
\begin{equation}
\mathcal{T}_{+} \left( \ii H \Gamma \right)^{*} \mathcal{T}_{+}^{-1} = - \ii H \Gamma.
\end{equation}
Hence the Hermitian matrix $\ii H \Gamma$ belongs to class D with the particle-symmetry operator $\mathcal{T}_{+}$. The $\mathbb{Z}_{2}$ topological invariant for the non-Hermitian matrix $H$ is thus given as the $\mathbb{Z}_{2}$ invariant for the Hermitian matrix $\ii H \Gamma$ in class D. In particular, a matrix $H\mathcal{C}_{-}$ is antisymmetric, i.e., 
\begin{equation}
\left( H\mathcal{C}_{-} \right)^{T} = - H\mathcal{C}_{-},
\end{equation}
and hence its Pfaffian is well defined. In addition, we have
\begin{equation}
\left[ \mathrm{Pf} \left( H \mathcal{C}_{-} \right) \right]^{*}
= \det\,( \mathcal{T}^{\dag}_{+} )\,\mathrm{Pf} \left( H \mathcal{C}_{-} \right),
	\label{eq: 0D BDI Pf}
\end{equation}
which leads to the reality of $\mathrm{Pf} \left( H \mathcal{C}_{-} \right)$ in an appropriate basis satisfying $\det\,( \mathcal{T}^{\dag}_{+} ) = 1$. In this basis, $H\mathcal{C}_{-}$ is a real antisymmetric matrix and thus the $\mathbb{Z}_{2}$ topological invariant $\nu \in \{ 0, 1\}$ is given as
\begin{equation}
\left( -1 \right)^{\nu} := \mathrm{sgn} \left[ \mathrm{Pf} \left( H\mathcal{C}_{-} \right) \right].
	\label{appendix: point - 0D BDI}
\end{equation}

\subsection{1D class BDI}

Since we have $\mathcal{T}_{+} \left[ \ii H \left( k \right) \Gamma \right]^{*} \mathcal{T}_{+}^{-1} = - \ii H \left( - k \right) \Gamma$ similarly to the zero-dimensional case, the matrix $H \left( k_{0} \right) \mathcal{C}_{-}$ is antisymmetric and satisfies Eq.~(\ref{eq: 0D BDI Pf}) at a particle-hole-symmetric momentum $k_{0} \in \{ 0, \pi \}$. The $\mathbb{Z}_{2}$ topological invariant $\nu \in \{ 0, 1\}$ is thus given as
\begin{equation}
\left( -1 \right)^{\nu} := \mathrm{sgn} \left[ \frac{\mathrm{Pf} \left( H \left( \pi \right) \mathcal{C}_{-} \right)}{\mathrm{Pf} \left( H \left( 0 \right) \mathcal{C}_{-} \right)} \right].
	\label{appendix: point - 1D BDI}
\end{equation}

\subsection{1D class D}
	\label{appendix: 1D class D - point}

In class D, we have PHS defined by Eq.~(\ref{eq: PHS}) with $\mathcal{C}_{-} \mathcal{C}_{-}^{*} = +1$, leading to 
\begin{equation}
\left[ H \left( k \right) \mathcal{C}_{-} \right]^{T}
= - H \left( -k \right) \mathcal{C}_{-}.
\end{equation}
Hence, $H \left( k \right) \mathcal{C}_{-}$ is antisymmetric, and its Pfaffian is well defined at a particle-hole-symmetric momentum $k_{0} \in \{ 0, \pi \}$. In a similar manner to 1D class DIII for the Hermitian case~\cite{QHZ-10, SR-11, Budich-13}, the $\mathbb{Z}_{2}$ invariant $\nu \in \{ 0, 1\}$ is given as 
\begin{equation} \begin{split}
\left( -1 \right)^{\nu} &:= \mathrm{sgn} \left\{
\frac{ \mathrm{Pf} \left[ H \left( \pi \right) \mathcal{C}_{-} \right] }{ \mathrm{Pf} \left[ H \left( 0 \right) \mathcal{C}_{-} \right] } \right. \\
&\quad\left. \times \exp \left[ 
-\frac{1}{2} \int_{k=0}^{k=\pi} d \log \det \left[ H \left( k \right) \mathcal{C}_{-} \right]
\right] \right\}.
	\label{appendix: 1D class D - point - Z2}
\end{split} \end{equation}

\subsection{2D class D}
	\label{appendix: 2D class D - point}

Since we have $\left[ H \left( {\bm k} \right) \mathcal{C}_{-} \right]^{T} = - H \left( -{\bm k} \right) \mathcal{C}_{-}$ similarly to the one-dimensional case, the $\mathbb{Z}_{2}$ invariant $\nu \in \{ 0, 1\}$ is given as
\begin{equation} \begin{split}
( -1 &)^{\nu} := \prod_{\mathsf{X}=\mathrm{I, II}} \mathrm{sgn} \left\{
\frac{ \mathrm{Pf} \left[ H \left( {\bm k}_{\mathsf{X}+} \right) \mathcal{C}_{-} \right] }{ \mathrm{Pf} \left[ H \left( {\bm k}_{\mathsf{X}-} \right) \mathcal{C}_{-} \right] } \right. \\
&\left. \times \exp \left[ 
-\frac{1}{2} \int_{{\bm k} = {\bm k}_{\mathsf{X}-}}^{{\bm k} = {\bm k}_{\mathsf{X}+}} d \log \det \left[ H \left( {\bm k} \right) \mathcal{C}_{-} \right]
\right] \right\},
	\label{appendix: point - 2D D}
\end{split} \end{equation}
where (${\bm k}_{\mathrm{I}+}, {\bm k}_{\mathrm{I}-}$) and (${\bm k}_{\mathrm{II}+}, {\bm k}_{\mathrm{II}-}$) are two pairs of particle-hole-symmetric momenta~\cite{SR-11}.

\subsection{2D class DIII}
	\label{appendix: point - 2D DIII}

Because of the presence of TRS with $\mathcal{T}_{+}\mathcal{T}_{+}^{*} = -1$ and CS with $\Gamma = \ii \mathcal{C}_{-} \mathcal{T}_{+}^{*}$ that anticommutes with TRS (i.e., $\mathcal{T}_{+} \Gamma^{*} = - \Gamma \mathcal{T}_{+}$) in class DIII, we have
\begin{equation}
\mathcal{T}_{+} \left[ \ii H \left( {\bm k} \right) \Gamma \right]^{*} \mathcal{T}_{+}^{-1} = \ii H \left( - {\bm k} \right) \Gamma.
\end{equation}
Hence, the Hermitian matrix $\ii H \left( {\bm k} \right) \Gamma$ belongs to class AII. The $\mathbb{Z}_{2}$ topological invariant for the non-Hermitian Hamiltonian $H$ is thus given by the Kane-Mele~\cite{Kane-Mele-05-Z2} or Fu-Kane~\cite{Fu-Kane-06} invariant for the Hermitian matrix $\ii H \left( {\bm k} \right) \Gamma$ in class AII.

\subsection{3D class DIII}
	\label{appendix: point - 3D DIII}

Since we have $\mathcal{T}_{+} \left[ \ii H \left( {\bm k} \right) \Gamma \right]^{*} \mathcal{T}_{+}^{-1} = \ii H \left( - {\bm k} \right) \Gamma$ similarly to the two-dimensional case, the Fu-Kane-Mele invariant~\cite{Fu-Kane-Mele-07} is defined for the Hermitian matrix $\ii H \left( {\bm k} \right) \Gamma$ in class AII. Alternatively, the integral of the Chern-Simons three form provides the same $\mathbb{Z}_{2}$ topological invariant~\cite{WQZ-10}.

\subsection{3D class AII}
	\label{appendix: point - 3D AII}

In class AII, we have TRS with $\mathcal{T}_{+} \mathcal{T}_{+}^{*} = -1$. The $\mathbb{Z}_{2}$ topological invariant of a non-Hermitian Hamiltonian in class AII thus reduces to that of a Hermitian Hamiltonian in class CII~\cite{Schnyder-Ryu-review}.

\subsection{0D class $\mathbf{D}^{\dag}$}

In class $\mathrm{D}^{\dag}$, we have $\mathrm{PHS}^{\dag}$ defined by Eq.~(\ref{eq: PHS-dag}), which leads to the reality of $\det \left( \ii H \right)$. Thus, the $\mathbb{Z}_{2}$ topological invariant 
$\nu \in \{ 0, 1\}$ is given as
\begin{align}
\left( -1\right)^{\nu} := {\rm sgn} \det \left( \ii H \right). 
	\label{appendix: point - 0D D-dag}
\end{align}

\subsection{0D class $\mathbf{DIII}^{\dag}$}

Because of the presence of $\mathrm{PHS}^{\dag}$ defined by Eq.~(\ref{eq: PHS-dag}) with $\mathcal{T}_{-}\mathcal{T}_{-}^{*} = +1$ and CS that anticommutes with $\mathrm{PHS}^{\dag}$ (i.e., $\mathcal{T}_{-} \Gamma^{*} = -\Gamma \mathcal{T}_{-}$) in class $\mathrm{DIII}^{\dag}$, we have
\begin{equation}
\mathcal{T}_{-} \left( \ii H \Gamma \right)^{*} \mathcal{T}_{-}^{-1} = - \ii H \Gamma.
\end{equation}
Hence, the Hermitian matrix $\ii H \Gamma$ belongs to class D with the particle-symmetry operator $\mathcal{T}_{-}$. Similarly to class BDI, the $\mathbb{Z}_{2}$ topological invariant for the non-Hermitian matrix $H$ in class $\mathrm{DIII}^{\dag}$ is given as the $\mathbb{Z}_{2}$ invariant for the Hermitian matrix $\ii H \Gamma$ in class D. In particular, the antisymmetric matrix $H\mathcal{C}_{+}$ satisfies
\begin{equation}
\left[ \mathrm{Pf} \left( H \mathcal{C}_{+} \right) \right]^{*}
= \det\,( \mathcal{T}^{\dag}_{-} )\,\mathrm{Pf} \left( H \mathcal{C}_{+} \right),
	\label{eq: 0D DIII-dag Pf}
\end{equation}
and thus the $\mathbb{Z}_{2}$ topological invariant $\nu \in \{ 0, 1\}$ is given as
\begin{equation}
\left( -1 \right)^{\nu} := \mathrm{sgn} \left[ \mathrm{Pf} \left( H\mathcal{C}_{+} \right) \right]
	\label{appendix: point - 0D DIII-dag}
\end{equation}
with an appropriate basis satisfying $\det\,(\mathcal{T}_{-}^{\dag}) = 1$.

\subsection{1D class $\mathbf{DIII}^{\dag}$}

Since we have $\mathcal{T}_{-} \left[ \ii H \left( k \right) \Gamma \right]^{*} \mathcal{T}_{-}^{-1} = - \ii H \left( - k \right) \Gamma$ similarly to the zero-dimensional case, the matrix $H \left( k_{0} \right) \mathcal{T}_{-}$ is antisymmetric and satisfies Eq.~(\ref{eq: 0D DIII-dag Pf}) at a particle-hole-symmetric momentum $k_{0} \in \{ 0, \pi \}$. The $\mathbb{Z}_{2}$ topological invariant $\nu \in \{ 0, 1\}$ is thus given as
\begin{equation}
\left( -1 \right)^{\nu} := \mathrm{sgn} \left[ \frac{\mathrm{Pf} \left( H \left( \pi \right) \mathcal{C}_{+} \right)}{\mathrm{Pf} \left( H \left( 0 \right) \mathcal{C}_{+} \right)} \right].
	\label{appendix: point - 1D DIII-dag}
\end{equation}

\subsection{1D class $\mathbf{AII}^{\dag}$}

In class $\mathrm{AII}^{\dag}$, we have $\mathrm{TRS}^{\dag}$ with $\mathcal{C}_{+} \mathcal{C}_{+}^{*} = -1$, leading to 
\begin{equation}
\left[ H \left( k \right) \mathcal{C}_{+} \right]^{T}
= - H \left( -k \right) \mathcal{C}_{+}.
\end{equation}
Hence, $H \left( k \right) \mathcal{C}_{+}$ is antisymmetric, and its Pfaffian is well defined at a time-reversal-invariant momentum $k_{0} \in \{ 0, \pi \}$. In similar to 1D class DIII for the Hermitian case~\cite{QHZ-10, SR-11, Budich-13}, the $\mathbb{Z}_{2}$ invariant $\nu \in \{ 0, 1\}$ is thus given as 
\begin{equation} \begin{split}
\left( -1 \right)^{\nu} &:= \mathrm{sgn} \left\{
\frac{ \mathrm{Pf} \left[ H \left( \pi \right) \mathcal{C}_{+} \right] }{ \mathrm{Pf} \left[ H \left( 0 \right) \mathcal{C}_{+} \right] } \right. \\
&~~\left. \times \exp \left[ 
-\frac{1}{2} \int_{k=0}^{k=\pi} d \log \det \left[ H \left( k \right) \mathcal{C}_{+} \right]
\right] \right\}.
	\label{appendix: point - 1D AII-dag}
\end{split} \end{equation}

\subsection{2D class $\mathbf{AII}^{\dag}$}

Since we have $\left[ H \left( {\bm k} \right) \mathcal{C}_{+} \right]^{T} = - H \left( -{\bm k} \right) \mathcal{C}_{+}$ similarly to the one-dimensional case, the $\mathbb{Z}_{2}$ invariant $\nu \in \{ 0, 1\}$ is given as 
\begin{equation} \begin{split}
( -1 &)^{\nu} := \prod_{\mathsf{X}=\mathrm{I, II}} \mathrm{sgn} \left\{
\frac{ \mathrm{Pf} \left[ H \left( {\bm k}_{\mathsf{X}+} \right) \mathcal{C}_{+} \right] }{ \mathrm{Pf} \left[ H \left( {\bm k}_{\mathsf{X}-} \right) \mathcal{C}_{+} \right] } \right. \\
&\left. \times \exp \left[ 
-\frac{1}{2} \int_{{\bm k} = {\bm k}_{\mathsf{X}-}}^{{\bm k} = {\bm k}_{\mathsf{X}+}} d \log \det \left[ H \left( {\bm k} \right) \mathcal{C}_{+} \right]
\right] \right\},
	\label{appendix: point - 2D AII-dag}
\end{split} \end{equation}
where (${\bm k}_{\mathrm{I}+}, {\bm k}_{\mathrm{I}-}$) and (${\bm k}_{\mathrm{II}+}, {\bm k}_{\mathrm{II}-}$) are two pairs of time-reversal-invariant momenta~\cite{SR-11}.

\subsection{2D class $\mathbf{CII}^{\dag}$}
	\label{appendix: point - 2D CII-dag}

Because of the presence of $\mathrm{PHS}^{\dag}$ with $\mathcal{T}_{-}\mathcal{T}_{-}^{*} = +1$ and CS that commutes with $\mathrm{PHS}^{\dag}$ (i.e., $\mathcal{T}_{-}\Gamma^{*} = \Gamma \mathcal{T}_{-}$) in class $\mathrm{CII}^{\dag}$, we have
\begin{equation}
\mathcal{T}_{-} \left[ \ii H \left( {\bm k} \right) \Gamma \right]^{*} \mathcal{T}_{-}^{-1} = \ii H \left( - {\bm k} \right) \Gamma.
\end{equation}
Hence the Hermitian matrix $\ii H \left( {\bm k} \right) \Gamma$ belongs to class AII. The $\mathbb{Z}_{2}$ topological invariant is thus given as the Kane-Mele~\cite{Kane-Mele-05-Z2} or Fu-Kane~\cite{Fu-Kane-06} invariant for this Hermitian matrix $\ii H \left( {\bm k} \right) \Gamma$ in class AII.

\subsection{3D class $\mathbf{CII}^{\dag}$}
	\label{appendix: point - 3D CII-dag}

Since we have $\mathcal{T}_{-} \left[ \ii H \left( {\bm k} \right) \Gamma \right]^{*} \mathcal{T}_{-}^{-1} = \ii H \left( - {\bm k} \right) \Gamma$ similarly to the two-dimensional case, the Fu-Kane-Mele~\cite{Fu-Kane-Mele-07} or the Chern-Simons invariant is defined for the Hermitian matrix $\ii H \left( {\bm k} \right) \Gamma$ in class AII~\cite{WQZ-10}.

\subsection{3D class $\mathbf{C}^{\dag}$}
	\label{appendix: point - 3D C-dag}

In class $\mathrm{C}^{\dag}$, we have $\mathrm{PHS}^{\dag}$ with $\mathcal{T}_{-} \mathcal{T}_{-}^{*} = -1$. The $\mathbb{Z}_{2}$ topological invariant of a non-Hermitian Hamiltonian in class $\mathrm{C}^{\dag}$ thus reduces to that of a Hermitian Hamiltonian in class CII~\cite{Schnyder-Ryu-review}.

\subsection{Class A with sublattice symmetry $\mathcal{S}$ (class $\text{AIII}^{\dag}$)}
	\label{appendix: point - A + SLS}

For gapped systems with SLS in Eq.~(\ref{eq: SLS}), any state $\ket{u_{n}}$ is independent of $\mathcal{S} \ket{u_{n}}$, from which we can construct two independent eigenstates of $\mathcal{S}$ with the opposite eigenvalues $\pm 1$. Thus, $\mathcal{S}$ can be chosen as a Pauli matrix $\mathcal{S} = \sigma_{z}$ without loss of generality. With this choice of $\mathcal{S}$, the Hamiltonian $H \left( {\bm k} \right)$ becomes
\begin{equation}
H = \begin{pmatrix}
0 & H_{+} \left( {\bm k} \right) \\ H_{-} \left( {\bm k} \right) & 0
\end{pmatrix}
	\label{eq: A+S - Hamiltonian}
\end{equation}
in this basis. Notably, $H_{+} \left( {\bm k} \right)$ and $H_{-} \left( {\bm k} \right)$ are independent of each other for generic non-Hermitian Hamiltonians, whereas $H_{+}^{\dag} \left( {\bm k} \right) = H_{-} \left( {\bm k} \right)$ is respected for Hermitian Hamiltonians. Moreover, we have $\det H_{+} \left( {\bm k} \right) \neq 0$ and $\det H_{-} \left( {\bm k} \right) \neq 0$ in the presence of a point gap (i.e., $\det H \left( {\bm k} \right) \neq 0$). Therefore, in odd dimensions $d=2n+1$, the two independent winding numbers $W^{\pm}_{2n+1}$ that are given by Eq.~(\ref{appendix: topological invariant - winding}) for $H_{\pm}\left( {\bm k} \right)$ constitute the $\mathbb{Z} \oplus \mathbb{Z}$ topological invariant.

\subsection{Class AIII with sublattice symmetry $\mathcal{S}_{+}$}

Because of the commutation relation between the chiral-symmetry operator and the sublattice-symmetry operator (i.e., $\Gamma \mathcal{S} = \mathcal{S} \Gamma$), we have
\begin{equation}
\mathcal{S} \left[ \ii H \left( {\bm k} \right) \Gamma \right] \mathcal{S}^{-1} = - \ii H \left( {\bm k} \right) \Gamma,
\end{equation}
which implies that the Hermitian matrix $\ii H \left( {\bm k} \right) \Gamma$ respects chiral symmetry and belongs to class AIII. Thus, in odd spatial dimensions $d=2n+1$, the $\mathbb{Z}$ winding number is given as
\begin{equation}
W_{2n+1} := \frac{N_{n}}{2} \int_{\mathrm{BZ}^{d}} \mathrm{tr} \left\{ \mathcal{S} \left[ \left( \ii H \Gamma \right)^{-1} d \left( \ii H \Gamma \right) \right]^{2n+1} \right\}
	\label{appendix: point - AIII + SLS+}
\end{equation}
with the coefficient $N_{n} := n!/\left( 2\pi \ii \right)^{n+1} \left( 2n+1 \right)!$. This formula reduces to
\begin{equation} \begin{split}
W_{2n+1} &= - \frac{N_{n}}{2} \int_{\mathrm{BZ}^{d}} \mathrm{tr} \left[ \mathcal{S} \left( H^{-1} dH \right)^{2n+1} \right] \\
&= \frac{W_{2n+1}^{+} - W_{2n+1}^{-} }{2},
\end{split} \end{equation}
where $W^{\pm}_{2n+1}$ is the winding number for $H^{\pm} \left( {\bm k} \right)$ in Eq.~(\ref{eq: A+S - Hamiltonian}).

\subsection{Class AIII with sublattice symmetry $\mathcal{S}_{-}$}
	\label{appendix: point - AIII + SLS-}

Because of the anticommutation relation between the chiral-symmetry operator and the sublattice-symmetry operator (i.e., $\Gamma \mathcal{S} = - \mathcal{S} \Gamma$), we have
\begin{equation}
\mathcal{S} \left[ \ii H \left( {\bm k} \right) \Gamma \right] \mathcal{S}^{-1} = \ii H \left( {\bm k} \right) \Gamma,\quad \mathcal{S}^{2} =1,
\end{equation}
which implies that the Hermitian matrix $\ii H \left( {\bm k} \right) \Gamma$ respects the $\mathbb{Z}_{2}$ unitary symmetry. Thus, $\ii H \left( {\bm k} \right) \Gamma$ can be block-diagonalized into the two Hermitian matrices $\left( \ii H \left( {\bm k} \right) \Gamma \right)_{\mathcal{S}=+1}$ and $\left( \ii H \left( {\bm k} \right) \Gamma \right)_{\mathcal{S}=-1}$ in the subspaces where the eigenvalues of $\mathcal{S}$ are $+1$ and $-1$, respectively. Therefore, in even spatial dimensions $d=2n$, the two independent $n$-th Chern numbers $C_{n}^{\pm}$ can be defined as the $\mathbb{Z} \oplus \mathbb{Z}$ topological invariant, each of which is given by Eq.~(\ref{eq: topological invariant - Chern}) for $\left( \ii H \left( {\bm k} \right) \Gamma \right)_{\mathcal{S}=\pm1}$.

\section{Topological invariants for line gaps}
	\label{appendix: topological invariants - line}

We present topological invariants of non-Hermitian systems with line gaps. We diagonalize the Hamiltonian as
\begin{equation}
H \left( {\bm k} \right) = \sum_{n} E_{n} \left( {\bm k} \right) \ket{ u_{n} \left( {\bm k} \right) } \langle\!\langle u_{n} \left( {\bm k} \right) |,
\end{equation}
where $E_{n} \left( {\bm k} \right) \in \mathbb{C}$ is an eigenenergy and $\ket{u_{n} \left( {\bm k} \right)}$ and $\langle\!\langle u_{n} \left( {\bm k} \right) |$ are the corresponding right and left eigenstates,
\begin{equation} \begin{split}
&H \left( {\bm k} \right) \ket{u_{n} \left( {\bm k} \right)} = E_{n} \left( {\bm k} \right) \ket{u_n \left( {\bm k} \right)},\\ 
&H^{\dag} \left( {\bm k} \right) | u_{n} \left( {\bm k} \right) \rangle\!\rangle = E_{n}^{*} \left( {\bm k} \right) | u_{n} \left( {\bm k} \right) \rangle\!\rangle, 
\end{split} \end{equation}
both of which are biorthogonal~\cite{Brody-14}
\begin{equation}
\langle u_{m} \left( {\bm k} \right) | u_{n} \left( {\bm k} \right) \rangle\!\rangle 
= \langle\!\langle u_{m} \left( {\bm k} \right) | u_{n} \left( {\bm k} \right) \rangle 
= \delta_{mn},
\end{equation}
and satisfy the completeness condition
\begin{eqnarray}
\sum_n |u_n({\bm k})\rangle\langle\!\langle u_n({\bm k})|=
\sum _n |u_n({\bm k})\rangle\!\rangle \langle u_n({\bm k})|=1.
\end{eqnarray}
Here we assume the presence of a real gap, i.e., ${\rm Re}\,E_{n} \left( {\bm k} \right) \neq 0$ for all $n$ and ${\bm k}$, and an eigenstate with a positive (negative) $n$ is chosen to satisfy ${\rm Re}\,E_{n} \left( {\bm k} \right) > 0$ (${\rm Re}\,E_{n} \left( {\bm k} \right) < 0$).

\subsection{$\mathbb{Z}$ invariants in even dimensions}
	\label{appendix: line - Chern}

The $\mathbb{Z}$ topological invariants in even spatial dimensions $d=2n$ are given as the $n$-th Chern number $C_{n}$. In non-Hermitian systems, a non-Abelian Berry connection is defined as
\begin{equation}
\mathcal{A}_{lm} := \langle\!\langle u_{l} | du_{m} \rangle
= \langle\!\langle u_{l} | \partial_{\bm k} u_{m} \rangle \cdot d{\bm k},
\end{equation}
and a Berry curvature is defined as
\begin{equation}
\mathcal{F} := d\mathcal{A} + \mathcal{A}^{2}.
\end{equation}
Using $\langle\!\langle du_{l} | u_{m} \rangle + \langle\!\langle u_{l} | du_{m} \rangle = d\,\langle\!\langle u_{l} | u_{m} \rangle = 0$, we have
\begin{eqnarray}
\mathcal{F}_{lm} &=& \langle\!\langle du_{l} | \left( 1 - \sum_{n<0} \ket{u_{n}} \langle\!\langle u_{n} | \right) |du_{m} \rangle.
\end{eqnarray}
The $n$-th Chern number $C_{n}$ is then given as
\begin{equation}
C_{n} := \frac{1}{n!} \left( \frac{\ii}{2\pi} \right)^{n} \int_{\mathrm{BZ}^{d}} \mathrm{tr}\,\mathcal{F}^{n},
\end{equation}
where the trace is taken over the occupied bands (complex bands with $n < 0$).

The Chern number $C_{n}$ can be expressed in terms of the $Q$ function. Here, we introduce the projector $\mathcal{P}_{\rm R}$ by
\begin{equation}
\mathcal{P}_{\rm R} := \sum_{n<0} \ket{u_{n}} \langle\!\langle u_{n} |,
\end{equation}
satisfying $\mathcal{P}_{\rm R}^{2} = \mathcal{P}_{\rm R}$. Then, the Berry curvature reduces to
\begin{eqnarray}
\mathcal{F}_{lm} &=& \langle\!\langle du_{l} | \left( 1 - \mathcal{P}_{\rm R} \right) |du_{m} \rangle
= \langle\!\langle u_{l} | \mathcal{P}_{\rm R} \left( d\mathcal{P}_{\rm R} \right)^{2} | u_{m} \rangle.\quad\quad
\end{eqnarray}
Defining the $Q$ matrix as
\begin{equation}
Q_{\rm R} := 1 - 2 \mathcal{P}_{\rm R},
\end{equation}
which is non-Hermitian but satisfies $Q_{\rm R}^{2} = 1$ and hence $Q_{\rm R} dQ_{\rm R} + dQ_{\rm R} Q_{\rm R} = 0$, we have
\begin{eqnarray}
C_{n} &=& \frac{1}{n!} \left( \frac{\ii}{2\pi} \right)^{n} \int_{\mathrm{BZ}^{d}} \mathrm{tr} \left[ \mathcal{P}_{\rm R} \left( d\mathcal{P}_{\rm R} \right)^{2} \right]^{n} \nonumber \\
&=& - \frac{1}{2^{2n+1}n!} \left( \frac{\ii}{2\pi} \right)^{n} \int_{\mathrm{BZ}^{d}} \mathrm{tr} \left[ Q_{\rm R} \left( dQ_{\rm R} \right)^{2n} \right].\quad\quad
	\label{eq: Chern - Q}
\end{eqnarray}

The Chern number can also be expressed in terms of the Green function defined by
\begin{equation}
G^{-1} \left( \omega, {\bm k} \right) := \ii \omega - H \left( {\bm k} \right).
\end{equation}
Since a real gap is assumed to be open, $G \left( \omega, {\bm k} \right)$ is invertible. The Green function is thus identified with a map 
\begin{equation}
\mathbb{R}_{\omega} \times \mathrm{BZ}^{d} \rightarrow \mathrm{GL}_{N} \left( \mathbb{C} \right),
\end{equation}
where $\mathrm{GL}_{N} \left( \mathbb{C} \right)$ is the general linear group of $N \times N$ invertible matrices. With this Green function, the $n$-th Chern number $C_{n}$ can be expressed as
\begin{equation}
C_{n} = \frac{n!}{\left( 2\pi \ii \right)^{n+1} \left( 2n+1\right)!} \int_{\mathbb{R}_{\omega} \times \mathrm{BZ}^{d}} \mathrm{tr} \left( G dG^{-1} \right)^{2n+1}.\quad
	\label{eq: Chern - Green}
\end{equation}
To derive this expression, we first show its topological invariance. In fact, a variation $G \rightarrow G + \delta G$ gives
\begin{align}
\delta\,&[ \mathrm{tr}\,( G dG^{-1} )^{2n+1} ] \nonumber \\
&= - \left( 2n+1 \right) d \left\{ \mathrm{tr} \left[ G^{-1} \delta G \left( dG^{-1} G\right)^{2n} \right] \right\},
\end{align}
which just gives a boundary term and vanishes after the integration. Since the right-hand side of Eq.~(\ref{eq: Chern - Green}) is topologically invariant, the Hamiltonian $H \left( {\bm k} \right)$ can be continuously deformed into the $Q$ function $Q_{\rm R} \left( {\bm k}\right)$. We denote $\tilde{G}^{-1} \left( \omega, {\bm k} \right) := \ii \omega - Q_{\rm R} \left( {\bm k} \right)$ as the Green function of the $Q$ function. Noticing
\begin{equation}
\tilde{G}\,(d_{\bm k}\tilde{G}^{-1})\,\tilde{G}\,(d_{\bm k}\tilde{G}^{-1})
= - \frac{\left( dQ_{\rm R} \right)^{2}}{1+\omega^{2}}
\end{equation}
because of $Q_{\rm R}dQ_{\rm R}+dQ_{\rm R}Q_{\rm R}=0$, we have
\begin{align}
\mathrm{tr}\,&( \tilde{G} d\tilde{G}^{-1} )^{2n+1} \nonumber \\
&= d\omega\frac{\ii \left( 2n+1 \right) \left( -1 \right)^{n+1}}{\left( 1+\omega^{2} \right)^{n+1}} \mathrm{tr}\,\left[ \left( \ii \omega + Q_{R} \right) (dQ_{R})^{2n} \right].
\end{align}
Integrating it with respect to $\omega \in \mathbb{R}$, we have
\begin{align}
\int_{\mathbb{R}_{\omega}} \mathrm{tr}\,&( \tilde{G} d\tilde{G}^{-1} )^{2n+1} \nonumber \\
&= \frac{\pi \ii \left( -1 \right)^{n+1} \left(2n+1\right)!}{2^{2n}\left( n!\right)^{2}}\,\mathrm{tr}\,[ Q_{R} (dQ_{R})^{2n} ],
\end{align}
which leads to the equivalence between Eqs.~(\ref{eq: Chern - Q}) and (\ref{eq: Chern - Green}).

\subsection{$\mathbb{Z}$ invariants in odd dimensions (real gap)}
	\label{appendix: CS - line - winding number}

In the presence of a real gap and CS defined by Eq.~(\ref{eq: CS}), we can define the winding number $W_{2n+1}$ in odd spatial dimensions $d=2n+1$. When $\ket{u_{n}}$ is a right eigenstate with $E_{n}$, $\Gamma \ket{u_{n}}$ is a left eigenstate with $-E_{n}^{*}$ due to CS. Therefore, eigenstates and eigenenergies can be expressed as
\begin{equation}
| u_{-n} \rangle\!\rangle = \Gamma\,| u_{n}\rangle,~~
E_{-n} = - E_{n}^{*},
	\label{eq: CS - energy and state}
\end{equation}
where an eigenstate with a positive (negative) $n$ is chosen to satisfy ${\rm Re}\,E_{n} > 0$ (${\rm Re}\,E_{n} < 0$). Here, we introduce the following projectors:
\begin{equation} \begin{split}
&{\cal P}_{\rm R} := \sum_{n<0} \ket{u_{n}} \langle\!\langle u_{n} |,\quad
{\cal P}_{\rm L} := \sum_{n<0} |u_{n} \rangle\!\rangle \bra{u_{n}},
\end{split} \end{equation}
which satisfy ${\cal P}_{\rm R/L}^{2} = {\cal P}_{\rm R/L}$ and ${\cal P}_{\rm R}^{\dag} = {\cal P}_{\rm L}$. Because of CS, we also have
\begin{eqnarray}
\Gamma\,{\cal P}_{\rm R} \Gamma
= 1 - {\cal P}_{\rm L},
	\label{eq: CS vp}
\end{eqnarray}
where Eq.~(\ref{eq: CS - energy and state}) and the completeness condition are used. We then define the $Q$ matrix by
\begin{equation}
Q := 1 - \left( {\cal P}_{\rm R} + {\cal P}_{\rm L} \right),
\end{equation}
which is Hermitian but $Q^{2} \neq 1$, and respects CS. The winding number $W_{2n+1}$ is thus defined as 
\begin{eqnarray} 
W_{2n+1} &:=& - \frac{n!}{2 \left( 2\pi \ii \right)^{n+1} \left( 2n+1 \right)!} \nonumber \\
&~&\qquad\times\int_{\mathrm{BZ}^{d}} \mathrm{tr} \left[ \Gamma \left( Q^{-1} dQ \right)^{2n+1} \right].\quad
\end{eqnarray}
Moreover, when the chiral-symmetry operator $\Gamma$ is diagonal (i.e., $\Gamma := \sigma_{z}$), the $Q$ matrix can be expressed as
\begin{equation}
Q = \left( \begin{array}{@{\,}cc@{\,}} 
	0 & q \\
	q^{\dag} & 0 \\ 
	\end{array} \right),
\end{equation}
where the off-diagonal part $q$ is an invertible matrix. Here $q$ is not necessarily unitary in contrast to the Hermitian case due to $Q^{2} \neq 1$. Nevertheless, $q$ is invertible and the winding number $W_{2n+1}$ is given as
\begin{equation}
W_{2n+1} = \frac{n!}{\left( 2\pi \ii \right)^{n+1} \left( 2n+1 \right)!} \int_{\mathrm{BZ}^{d}} \mathrm{tr} \left( q^{-1} dq \right)^{2n+1}.
\end{equation}
Remarkably, the winding number $W_{1}$ is constructed in a one-dimensional non-Hermitian system with CS in Ref.~\cite{Esaki-11, *Sato-12}.

\subsection{1D class D}
	\label{appendix: 1D class D - line}

The $\mathbb{Z}_{2}$ topological invariant in 1D class D is given as the quantized Berry phase. When $\ket{u_{n} \left( k \right)}$ is a right eigenstate with eigenenergy $E_{n} \left( k \right)$, we have
\begin{equation}
H^{\dag} \left( k \right) [ \mathcal{C}_{-} \ket{u_{n}^{*} \left( - k \right)} ]
= - E_{n}^{*} \left( -k \right) [ \mathcal{C}_{-} \ket{u_{n}^{*} \left( - k \right)} ]
\end{equation}
due to PHS defined by Eq.~(\ref{eq: PHS}), which implies that $\mathcal{C}_{-} \ket{u_{n}^{*} \left( - k \right)}$ is a left eigenstate of $H \left( k \right)$ with its eigenenergy $- E_{n} \left( -k \right)$. Therefore, eigenstates and eigenenergies can be expressed as
\begin{equation}
| u_{-n} \left( k \right) \rangle\!\rangle = \mathcal{C}_{-} \ket{u_{n}^{*} \left( - k \right)},~~
E_{-n} \left( k \right) = - E_{n} \left( - k \right),
\end{equation} 
where an eigenstate with a positive (negative) $n$ is chosen to satisfy ${\rm Re}\,E_{n} > 0$ (${\rm Re}\,E_{n} < 0$). As a result, the Berry connections for $n > 0$ and $n < 0$ are related to each other by
\begin{eqnarray}
\mathcal{A}_{-n} \left( k \right) &:=& \ii\,\langle\!\langle u_{-n} \left( k \right) | \partial_{k} u_{-n} \left( k \right) \rangle
= \mathcal{A}_{n} \left( -k \right).\quad\quad
\end{eqnarray}
Thus, the Berry phase $\varphi$ of all the occupied bands with $n<0$ accumulated over the entire Brillouin zone is given as
\begin{eqnarray}
\varphi &:=& \oint_{\rm BZ} \sum_{n<0} \mathcal{A}_{n} \left( k \right) dk \nonumber \\
&=& \frac{1}{2} \oint_{\rm BZ} \sum_{\mathrm{all}\,n} \mathcal{A}_{n} \left( k \right) dk \nonumber \\
&=& \frac{\ii}{2} \oint_{\rm BZ} \mathrm{tr} \left[ L^{\dag} \left( k \right) \partial_{k} R \left( k\right) \right] dk,
\end{eqnarray}
where $R$ and $L$ are given by collecting right and left eigenstates, respectively:
\begin{equation} \begin{split}
R \left( k \right) := \left( \ket{u_{1} \left( k\right)},\,\ket{u_{2} \left( k\right)}, \cdots \right), \\
L \left( k \right) := \left( | u_{1} \left( k\right) \rangle\!\rangle,\,| u_{2} \left( k\right) \rangle\!\rangle, \cdots \right). \\
\end{split} \end{equation}
Although $R$ and $L$ are nonunitary in the presence of non-Hermiticity, both of them are invertible and satisfy $R^{\dag} L = L^{\dag} R =1$ as long as a line gap is open. Consequently, we have $\varphi \equiv - \pi W_{1}$ (mod $2\pi$) with the phase winding $W_{1}$ of $\det R \left( k \right) \neq 0$:
\begin{eqnarray}
W_{1} := \oint_{\rm BZ} \frac{dk}{2\pi \ii} \mathrm{tr} \left( R^{-1}  \partial_{k} R \right)
= \oint_{\rm BZ} \frac{dk}{2\pi \ii} \partial_{k} \log \det R. \qquad\quad
\end{eqnarray}
The $\mathbb{Z}_{2}$ topological invariant is thus given as the $\pi$-quantized Berry phase $\varphi$, i.e., $W_{1}$ (mod $2$). Remarkably, a point gap is always open if a line gap is open. Consequently, the $\mathbb{Z}_{2}$ topological invariant for a point gap in Eq.~(\ref{appendix: 1D class D - point - Z2}) is also well defined, which is equivalent to the quantized Berry phase $\varphi$.

\subsection{$\mathbb{Z}$ invariants in pseudo-Hermitian systems}
	\label{appendix: pH - invariant}

In the presence of pseudo-Hermiticity defined by Eq.~(\ref{eq: def pseudo-Hermiticity}), the matrix $\eta H \left( {\bm k} \right)$ is Hermitian:
\begin{equation}
\left[ \eta H\left( {\bm k}\right) \right]^{\dag} = \eta H\left( {\bm k}\right).
\end{equation}
Moreover, eigenvalues of $\eta H \left( {\bm k} \right)$ coincide with singular values of $H \left( {\bm k} \right)$ due to
\begin{equation}
\left[ \eta H\left( {\bm k}\right) \right]^{\dag} \left[ \eta H\left( {\bm k}\right) \right]
= H^{\dag} \left( {\bm k}\right) H\left( {\bm k}\right),
\end{equation}
which implies that an energy gap of $\eta H\left( {\bm k}\right)$ is equivalent to a point gap of $H \left( {\bm k} \right)$. Consequently, in the presence of a real gap, we have the two independent Chern numbers $C_{n}$ in even spatial dimensions $d=2n$, one of which is defined for $H \left( {\bm k} \right)$ and the other of which is defined for $\eta H \left( {\bm k} \right)$. In the presence of additional TRS with $\mathcal{T}_{+} \mathcal{T}_{+}^{*} = +1$, the Chern number for $H \left( {\bm k} \right)$ vanishes. Furthermore, if the TRS commutes with pseudo-Hermiticity (i.e., $\mathcal{T}_{+} \eta_{+}^{*} = \eta_{+} \mathcal{T}_{+}$), $\eta H \left( {\bm k} \right)$ respects TRS and hence its Chern number also vanishes. If the TRS anticommutes with pseudo-Hermiticity (i.e., $\mathcal{T}_{+} \eta_{-}^{*} = -\eta_{-} \mathcal{T}_{+}$), on the other hand, the Chern number for $\eta H \left( {\bm k} \right)$ does not necessarily vanish and serves as the time-reversal-invariant Chern number in Sec.~\ref{sec : pH - topology}~\cite{Esaki-11, *Sato-12}. Whereas the time-reversal-invariant Chern number defined in Sec.~\ref{sec : pH - topology}~\cite{Esaki-11, *Sato-12} requires a real gap, the present one only requires a point gap.

\section{Exact solution to the non-Hermitian Su-Schrieffer-Heeger model with asymmetric hopping}
	\label{appendix: SSH}

We provide an exact solution to the non-Hermitian Su-Schrieffer-Heeger model defined by Eq.~(\ref{eq: SSH - asymmetric hopping}) with open boundaries. For simplicity, we assume $v \geq g \geq 0$, $w \geq 0$ for the sake of simplicity, but the discussion can be generalized straightforwardly. Our calculations are based on a systematic method of diagonalizing free fermions with generic boundaries described in Refs.~\cite{Kawabata-18-Kitaev, Katsura-15, *KKWK-17, Alase-16, *Alase-17, *Cobanera-18}, which is readily applied to generic non-Hermitian lattice models.
	
We denote an eigenenergy as $E \in \mathbb{C}$ and the corresponding right eigenstate as $\hat{\varphi} = \sum_{i=1}^{L} ( A_{i} \hat{a}_{i} + B_{i} \hat{b}_{i} )$ with coefficients $A_{i}, B_{i} \in \mathbb{C}$. The Schr\"odinger equation $[ \hat{H}, \hat{\varphi} ] = E\,\hat{\varphi}$ reads
\begin{equation} \begin{split}
w B_{i-1} + \left( v+g \right) B_{i} &= EA_{i}\quad\left( i = 2, 3, \cdots, L \right), \\
\left( v-g \right) A_{i} + w A_{i+1} &= EB_{i}\quad\left( i = 1, 2, \cdots, L-1 \right) \label{eq: SSH - bulk}
\end{split} \end{equation}
in the bulk, and
\begin{eqnarray}
\left( v+g \right) B_{1} = EA_{1},\quad
\left( v-g \right) A_{L} = EB_{L} \label{eq: SSH - edge}
\end{eqnarray}
at the edges. Defining $A_{L+1}$ and $B_{0}$ with Eq.~(\ref{eq: SSH - bulk}), the boundary equations~(\ref{eq: SSH - edge}) reduce to
\begin{equation}
A_{L+1} = B_{0} = 0.
	\label{eq: SSH - edge2}
\end{equation}

Here we take a plane-wave ansatz $A_{j} \sim A e^{\ii kj}, B_{j} \sim B e^{\ii kj}$ ($k \in \mathbb{C}$). Whereas the wavenumber is real ($k \in \mathbb{R}$) for the bulk states in Hermitian systems, it can be complex in non-Hermitian systems. The bulk equation~(\ref{eq: SSH - bulk}) reduces to 
\begin{equation}
\begin{pmatrix}
0 & \left( v+g\right) + we^{-\ii k} \\
\left( v-g\right) + we^{\ii k} & 0
\end{pmatrix} \begin{pmatrix}
A \\ B 
\end{pmatrix} = E \begin{pmatrix}
A \\ B 
\end{pmatrix}.
	\label{eq: SSH - bulk2}
\end{equation}
For a nontrivial solution, we have
\begin{equation}
E \left( k \right) = \pm \sqrt{v^{2} + w^{2} -g^{2} + \left( v+g\right) w e^{\ii k} + \left( v-g\right) w e^{-\ii k} }.
	\label{eq: SSH - spectrum}
\end{equation}
In the absence of the non-Hermiticity ($g=0$), if an eigenstate with wavenumber $k \in \mathbb{R}$ belongs to an eigenenergy $E \in \mathbb{R}$, another eigenstate with wavenumber $-k$ has the same eigenenergy $E$. In the presence of non-Hermiticity ($g \neq 0$), however, this fact breaks down; if two eigenstates with wavenumbers $k$ and $k'$ ($k \neq k'$) belong to the same eigenenergy $E$, Eqs.~(\ref{eq: SSH - bulk2}) and (\ref{eq: SSH - spectrum}) lead to $\left( v+g\right) e^{\ii k} + \left( v-g\right) e^{-\ii k} = \left( v+g\right) e^{\ii k'} + \left( v-g\right) e^{-\ii k'}$, i.e., 
\begin{equation}
e^{\ii k'} = \frac{v-g}{v+g} e^{-\ii k},
\end{equation}
which implies that the eigenstates are localized at either edge for $g \neq 0$ (non-Hermitian skin effect~\cite{Yao-18-SSH}). This equation naturally leads to the redefinition of the wavenumber as
\begin{equation}
e^{\ii k} =: \sqrt{\frac{v-g}{v+g}} e^{\ii q},
\end{equation}
where $q$ should be real for the bulk eigenmodes. Using this redefined wavenumber $q \in \mathbb{R}$, the bulk equation~(\ref{eq: SSH - bulk2}) reduces to
\begin{equation}
\begin{pmatrix}
0 & \sqrt{v^{2}-g^{2}} + we^{-\ii q} \\
\sqrt{v^{2}-g^{2}} + we^{\ii q} & 0
\end{pmatrix} \begin{pmatrix}
A \\ \tilde{B} 
\end{pmatrix} = E \begin{pmatrix}
A \\ \tilde{B} 
\end{pmatrix}
	\label{eq: SSH - bulk3}
\end{equation}
with $\tilde{B} := B \sqrt{\left( v+g \right)/\left( v-g \right)}$, and the eigenenergy in Eq.~(\ref{eq: SSH - spectrum}) reduces to
\begin{equation}
E \left( q \right) = \pm \sqrt{v^{2} + w^{2} - g^{2} + 2 \sqrt{v^{2}-g^{2}} w \cos q}.
\end{equation}
With the real wavenumber $q$, the bulk Hamiltonian under the open boundary condition is obtained as Eq.~(\ref{eq: SSH - H - OBC}). 

Now a generic eigenstate is described by
\begin{equation}
A_{j} = A_{+} e^{\ii qj} + A_{-} e^{-\ii qj},\quad
B_{j} = B_{+} e^{\ii qj} + B_{-} e^{-\ii qj}
\end{equation}
with $A_{+}, A_{-}, B_{+}, B_{-} \in \mathbb{C}$. The bulk eigenequation (\ref{eq: SSH - bulk3}) leads to
\begin{equation} \begin{split}
\frac{B_{+}}{A_{+}} 
&= \sqrt{\frac{v-g}{v+g}} \frac{\sqrt{v^{2}-g^{2}}+we^{\ii q}}{E \left( q \right)}
=: C \left( q \right),
\end{split} \end{equation}
and $B_{-}/A_{-} = C \left( -q \right)$. The boundary equation~(\ref{eq: SSH - edge2}) leads to the quantization of the wavenumber $q$. In fact, Eq.~(\ref{eq: SSH - edge2}) leads to
\begin{equation}
\begin{pmatrix}
e^{\ii q \left( L+1 \right)} & e^{- \ii q \left( L+1 \right)} \\
C \left( q \right) & C \left( - q \right)
\end{pmatrix} \begin{pmatrix}
A_{+} \\ A_{-} 
\end{pmatrix} = 0,
\end{equation}
which has a nontrivial solution if and only if the determinant of the coefficient matrix vanishes. After some calculations, we obtain
\begin{equation}
\frac{\sin q \left( L+1 \right)}{\sin qL} = - \frac{w}{\sqrt{v^{2}-g^{2}}}
\end{equation}
which quantizes $q$. In the trivial phase ($w < \sqrt{v^{2}-g^{2}}$), all the wavenumbers $q$ are real and thus all the eigenstates are localized at an edge with the same localization length for $L \to \infty$. In the topological phase ($w > \sqrt{v^{2}-g^{2}}$), on the other hand, some of the wavenumbers $q$ are complex even for $L \to \infty$, and those eigenmodes are topologically protected to have zero energy~\cite{Katsura-15, *KKWK-17}. Therefore, the winding number defined as Eq.~(\ref{eq: SSH - winding}) for $H_{\rm OBC}$ (Fig.~\ref{fig: SSH}) indeed predicts the emergence of the topologically protected edge states.

\bibliography{NH-top}

\begin{thebibliography}{288}%
\makeatletter
\providecommand \@ifxundefined [1]{%
 \@ifx{#1\undefined}
}%
\providecommand \@ifnum [1]{%
 \ifnum #1\expandafter \@firstoftwo
 \else \expandafter \@secondoftwo
 \fi
}%
\providecommand \@ifx [1]{%
 \ifx #1\expandafter \@firstoftwo
 \else \expandafter \@secondoftwo
 \fi
}%
\providecommand \natexlab [1]{#1}%
\providecommand \enquote  [1]{``#1''}%
\providecommand \bibnamefont  [1]{#1}%
\providecommand \bibfnamefont [1]{#1}%
\providecommand \citenamefont [1]{#1}%
\providecommand \href@noop [0]{\@secondoftwo}%
\providecommand \href [0]{\begingroup \@sanitize@url \@href}%
\providecommand \@href[1]{\@@startlink{#1}\@@href}%
\providecommand \@@href[1]{\endgroup#1\@@endlink}%
\providecommand \@sanitize@url [0]{\catcode `\\12\catcode `\$12\catcode
  `\&12\catcode `\#12\catcode `\^12\catcode `\_12\catcode `\%12\relax}%
\providecommand \@@startlink[1]{}%
\providecommand \@@endlink[0]{}%
\providecommand \url  [0]{\begingroup\@sanitize@url \@url }%
\providecommand \@url [1]{\endgroup\@href {#1}{\urlprefix }}%
\providecommand \urlprefix  [0]{URL }%
\providecommand \Eprint [0]{\href }%
\providecommand \doibase [0]{http://dx.doi.org/}%
\providecommand \selectlanguage [0]{\@gobble}%
\providecommand \bibinfo  [0]{\@secondoftwo}%
\providecommand \bibfield  [0]{\@secondoftwo}%
\providecommand \translation [1]{[#1]}%
\providecommand \BibitemOpen [0]{}%
\providecommand \bibitemStop [0]{}%
\providecommand \bibitemNoStop [0]{.\EOS\space}%
\providecommand \EOS [0]{\spacefactor3000\relax}%
\providecommand \BibitemShut  [1]{\csname bibitem#1\endcsname}%
\let\auto@bib@innerbib\@empty
\bibitem [{\citenamefont {Bender}\ and\ \citenamefont
  {Boettcher}(1998)}]{Bender-98}%
  \BibitemOpen
  \bibfield  {author} {\bibinfo {author} {\bibfnamefont {C.~M.}\ \bibnamefont
  {Bender}}\ and\ \bibinfo {author} {\bibfnamefont {S.}~\bibnamefont
  {Boettcher}},\ }\bibfield  {title} {\enquote {\bibinfo {title} {{R}eal
  {S}pectra in {N}on-{H}ermitian {H}amiltonians {H}aving $\mathcal{PT}$
  {S}ymmetry},}\ }\href@noop {} {\bibfield  {journal} {\bibinfo  {journal}
  {Phys. Rev. Lett.}\ }\textbf {\bibinfo {volume} {80}},\ \bibinfo {pages}
  {5243} (\bibinfo {year} {1998})}\BibitemShut {NoStop}%
\bibitem [{\citenamefont {Bender}\ \emph {et~al.}(2002)\citenamefont {Bender},
  \citenamefont {Brody},\ and\ \citenamefont {Jones}}]{Bender-02}%
  \BibitemOpen
  \bibfield  {author} {\bibinfo {author} {\bibfnamefont {C.~M.}\ \bibnamefont
  {Bender}}, \bibinfo {author} {\bibfnamefont {D.~C.}\ \bibnamefont {Brody}}, \
  and\ \bibinfo {author} {\bibfnamefont {H.~F.}\ \bibnamefont {Jones}},\
  }\bibfield  {title} {\enquote {\bibinfo {title} {{Complex Extension of
  Quantum Mechanics}},}\ }\href@noop {} {\bibfield  {journal} {\bibinfo
  {journal} {{Phys. Rev. Lett.}}\ }\textbf {\bibinfo {volume} {89}},\ \bibinfo
  {pages} {270401} (\bibinfo {year} {2002})}\BibitemShut {NoStop}%
\bibitem [{\citenamefont {Bender}(2007)}]{Bender-review}%
  \BibitemOpen
  \bibfield  {author} {\bibinfo {author} {\bibfnamefont {C.~M.}\ \bibnamefont
  {Bender}},\ }\bibfield  {title} {\enquote {\bibinfo {title} {{M}aking {S}ense
  of {N}on-{H}ermitian {H}amiltonians},}\ }\href@noop {} {\bibfield  {journal}
  {\bibinfo  {journal} {Rep. Prog. Phys.}\ }\textbf {\bibinfo {volume} {70}},\
  \bibinfo {pages} {947} (\bibinfo {year} {2007})}\BibitemShut {NoStop}%
\bibitem [{\citenamefont {Konotop}\ \emph {et~al.}(2016)\citenamefont
  {Konotop}, \citenamefont {Yang},\ and\ \citenamefont
  {Zezyulin}}]{Konotop-review}%
  \BibitemOpen
  \bibfield  {author} {\bibinfo {author} {\bibfnamefont {V.~V.}\ \bibnamefont
  {Konotop}}, \bibinfo {author} {\bibfnamefont {J.}~\bibnamefont {Yang}}, \
  and\ \bibinfo {author} {\bibfnamefont {D.~A.}\ \bibnamefont {Zezyulin}},\
  }\bibfield  {title} {\enquote {\bibinfo {title} {{N}onlinear {W}aves in
  $\mathcal{PT}$-{S}ymmetric {S}ystems},}\ }\href@noop {} {\bibfield  {journal}
  {\bibinfo  {journal} {Rev. Mod. Phys.}\ }\textbf {\bibinfo {volume} {88}},\
  \bibinfo {pages} {035002} (\bibinfo {year} {2016})}\BibitemShut {NoStop}%
\bibitem [{\citenamefont {Feng}\ \emph {et~al.}(2017)\citenamefont {Feng},
  \citenamefont {El-Ganainy},\ and\ \citenamefont {Ge}}]{Feng-review}%
  \BibitemOpen
  \bibfield  {author} {\bibinfo {author} {\bibfnamefont {L.}~\bibnamefont
  {Feng}}, \bibinfo {author} {\bibfnamefont {R.}~\bibnamefont {El-Ganainy}}, \
  and\ \bibinfo {author} {\bibfnamefont {L.}~\bibnamefont {Ge}},\ }\bibfield
  {title} {\enquote {\bibinfo {title} {{N}on-{H}ermitian {P}hotonics {B}ased on
  {P}arity-{T}ime {S}ymmetry},}\ }\href@noop {} {\bibfield  {journal} {\bibinfo
   {journal} {Nat. Photon.}\ }\textbf {\bibinfo {volume} {11}},\ \bibinfo
  {pages} {752} (\bibinfo {year} {2017})}\BibitemShut {NoStop}%
\bibitem [{\citenamefont {El-Ganainy}\ \emph {et~al.}(2018)\citenamefont
  {El-Ganainy}, \citenamefont {Makris}, \citenamefont {Khajavikhan},
  \citenamefont {Musslimani}, \citenamefont {Rotter},\ and\ \citenamefont
  {Christodoulides}}]{Christodoulides-review}%
  \BibitemOpen
  \bibfield  {author} {\bibinfo {author} {\bibfnamefont {R.}~\bibnamefont
  {El-Ganainy}}, \bibinfo {author} {\bibfnamefont {K.~G.}\ \bibnamefont
  {Makris}}, \bibinfo {author} {\bibfnamefont {M.}~\bibnamefont {Khajavikhan}},
  \bibinfo {author} {\bibfnamefont {Z.~H.}\ \bibnamefont {Musslimani}},
  \bibinfo {author} {\bibfnamefont {S.}~\bibnamefont {Rotter}}, \ and\ \bibinfo
  {author} {\bibfnamefont {D.~N.}\ \bibnamefont {Christodoulides}},\ }\bibfield
   {title} {\enquote {\bibinfo {title} {{N}on-{H}ermitian {P}hysics and {P}{T}
  {S}ymmetry},}\ }\href@noop {} {\bibfield  {journal} {\bibinfo  {journal}
  {Nat. Phys.}\ }\textbf {\bibinfo {volume} {14}},\ \bibinfo {pages} {11}
  (\bibinfo {year} {2018})}\BibitemShut {NoStop}%
\bibitem [{\citenamefont {Miri}\ and\ \citenamefont
  {Al\`u}(2019)}]{Alu-review}%
  \BibitemOpen
  \bibfield  {author} {\bibinfo {author} {\bibfnamefont {M.-A.}\ \bibnamefont
  {Miri}}\ and\ \bibinfo {author} {\bibfnamefont {A.}~\bibnamefont {Al\`u}},\
  }\bibfield  {title} {\enquote {\bibinfo {title} {{Exceptional Points in
  Optics and Photonics}},}\ }\href@noop {} {\bibfield  {journal} {\bibinfo
  {journal} {Science}\ }\textbf {\bibinfo {volume} {363}},\ \bibinfo {pages}
  {eaar7709} (\bibinfo {year} {2019})}\BibitemShut {NoStop}%
\bibitem [{\citenamefont {\c{S}. K.~\"Ozdemir}\ \emph
  {et~al.}(2019)\citenamefont {\c{S}. K.~\"Ozdemir}, \citenamefont {Rotter},
  \citenamefont {Nori},\ and\ \citenamefont {Yang}}]{Ozdemir-review}%
  \BibitemOpen
  \bibfield  {author} {\bibinfo {author} {\bibnamefont {\c{S}. K.~\"Ozdemir}},
  \bibinfo {author} {\bibfnamefont {S.}~\bibnamefont {Rotter}}, \bibinfo
  {author} {\bibfnamefont {F.}~\bibnamefont {Nori}}, \ and\ \bibinfo {author}
  {\bibfnamefont {L.}~\bibnamefont {Yang}},\ }\bibfield  {title} {\enquote
  {\bibinfo {title} {{Parity-Time Symmetry and Exceptional Points in
  Photonics}},}\ }\href@noop {} {\bibfield  {journal} {\bibinfo  {journal}
  {Nat. Mater.}\ }\textbf {\bibinfo {volume} {18}},\ \bibinfo {pages} {783}
  (\bibinfo {year} {2019})}\BibitemShut {NoStop}%
\bibitem [{\citenamefont {Kozii}\ and\ \citenamefont {Fu}()}]{Kozii-17}%
  \BibitemOpen
  \bibfield  {author} {\bibinfo {author} {\bibfnamefont {V.}~\bibnamefont
  {Kozii}}\ and\ \bibinfo {author} {\bibfnamefont {L.}~\bibnamefont {Fu}},\
  }\href@noop {} {\enquote {\bibinfo {title} {{N}on-{H}ermitian {T}opological
  {T}heory of {F}inite-{L}ifetime {Q}uasiparticles: {P}rediction of {B}ulk
  {F}ermi {A}rc due to {E}xceptional {P}oint},}\ }\bibinfo {note} {{a}rXiv:
  1708.05841}\BibitemShut {NoStop}%
\bibitem [{\citenamefont {Papaj}\ \emph {et~al.}(2019)\citenamefont {Papaj},
  \citenamefont {Isobe},\ and\ \citenamefont {Fu}}]{Papaj-18}%
  \BibitemOpen
  \bibfield  {author} {\bibinfo {author} {\bibfnamefont {M.}~\bibnamefont
  {Papaj}}, \bibinfo {author} {\bibfnamefont {H.}~\bibnamefont {Isobe}}, \ and\
  \bibinfo {author} {\bibfnamefont {L.}~\bibnamefont {Fu}},\ }\bibfield
  {title} {\enquote {\bibinfo {title} {{Nodal Arc of Disordered Dirac Fermions
  and Non-Hermitian Band Theory}},}\ }\href@noop {} {\bibfield  {journal}
  {\bibinfo  {journal} {Phys. Rev. B}\ }\textbf {\bibinfo {volume} {99}},\
  \bibinfo {pages} {201107(R)} (\bibinfo {year} {2019})}\BibitemShut {NoStop}%
\bibitem [{\citenamefont {Shen}\ and\ \citenamefont {Fu}(2018)}]{Shen-18-QO}%
  \BibitemOpen
  \bibfield  {author} {\bibinfo {author} {\bibfnamefont {H.}~\bibnamefont
  {Shen}}\ and\ \bibinfo {author} {\bibfnamefont {L.}~\bibnamefont {Fu}},\
  }\bibfield  {title} {\enquote {\bibinfo {title} {{Q}uantum {O}scillation from
  {I}n-{G}ap {S}tates and a {N}on-{H}ermitian {L}andau {L}evel {P}roblem},}\
  }\href@noop {} {\bibfield  {journal} {\bibinfo  {journal} {Phys. Rev. Lett.}\
  }\textbf {\bibinfo {volume} {121}},\ \bibinfo {pages} {026403} (\bibinfo
  {year} {2018})}\BibitemShut {NoStop}%
\bibitem [{\citenamefont {Kozii}\ and\ \citenamefont {Fu}(2018)}]{Kozii-18}%
  \BibitemOpen
  \bibfield  {author} {\bibinfo {author} {\bibfnamefont {V.}~\bibnamefont
  {Kozii}}\ and\ \bibinfo {author} {\bibfnamefont {L.}~\bibnamefont {Fu}},\
  }\bibfield  {title} {\enquote {\bibinfo {title} {{T}hermal {P}lasmon
  {R}esonantly {E}nhances {E}lectron {S}cattering in {D}irac/{W}eyl
  {S}emimetals},}\ }\href@noop {} {\bibfield  {journal} {\bibinfo  {journal}
  {Phys. Rev. B}\ }\textbf {\bibinfo {volume} {98}},\ \bibinfo {pages}
  {041109(R)} (\bibinfo {year} {2018})}\BibitemShut {NoStop}%
\bibitem [{\citenamefont {Zyuzin}\ and\ \citenamefont
  {Zyuzin}(2018)}]{Zyuzin-18}%
  \BibitemOpen
  \bibfield  {author} {\bibinfo {author} {\bibfnamefont {A.~A.}\ \bibnamefont
  {Zyuzin}}\ and\ \bibinfo {author} {\bibfnamefont {A.~Y.}\ \bibnamefont
  {Zyuzin}},\ }\bibfield  {title} {\enquote {\bibinfo {title} {{F}lat {B}and in
  {D}isorder-{D}riven {N}on-{H}ermitian {W}eyl {S}emimetals},}\ }\href@noop {}
  {\bibfield  {journal} {\bibinfo  {journal} {Phys. Rev. B}\ }\textbf {\bibinfo
  {volume} {97}},\ \bibinfo {pages} {041203(R)} (\bibinfo {year}
  {2018})}\BibitemShut {NoStop}%
\bibitem [{\citenamefont {Moors}\ \emph {et~al.}(2019)\citenamefont {Moors},
  \citenamefont {Zyuzin}, \citenamefont {Zyuzin}, \citenamefont {Tiwari},\ and\
  \citenamefont {Schmidt}}]{Moors-19}%
  \BibitemOpen
  \bibfield  {author} {\bibinfo {author} {\bibfnamefont {K.}~\bibnamefont
  {Moors}}, \bibinfo {author} {\bibfnamefont {A.~A.}\ \bibnamefont {Zyuzin}},
  \bibinfo {author} {\bibfnamefont {A.~Y.}\ \bibnamefont {Zyuzin}}, \bibinfo
  {author} {\bibfnamefont {R.~P.}\ \bibnamefont {Tiwari}}, \ and\ \bibinfo
  {author} {\bibfnamefont {T.~L.}\ \bibnamefont {Schmidt}},\ }\bibfield
  {title} {\enquote {\bibinfo {title} {{D}isorder-{D}riven {E}xceptional
  {L}ines and {F}ermi {R}ibbons in {T}ilted {N}odal-{L}ine {S}emimetals},}\
  }\href@noop {} {\bibfield  {journal} {\bibinfo  {journal} {{Phys. Rev. B}}\
  }\textbf {\bibinfo {volume} {99}},\ \bibinfo {pages} {041116(R)} (\bibinfo
  {year} {2019})}\BibitemShut {NoStop}%
\bibitem [{\citenamefont {Yoshida}\ \emph {et~al.}(2018)\citenamefont
  {Yoshida}, \citenamefont {Peters},\ and\ \citenamefont
  {Kawakami}}]{Yoshida-18}%
  \BibitemOpen
  \bibfield  {author} {\bibinfo {author} {\bibfnamefont {T.}~\bibnamefont
  {Yoshida}}, \bibinfo {author} {\bibfnamefont {R.}~\bibnamefont {Peters}}, \
  and\ \bibinfo {author} {\bibfnamefont {N.}~\bibnamefont {Kawakami}},\
  }\bibfield  {title} {\enquote {\bibinfo {title} {Non-{H}ermitian
  {P}erspective of the {B}and {S}tructure in {H}eavy-{F}ermion {S}ystems},}\
  }\href@noop {} {\bibfield  {journal} {\bibinfo  {journal} {Phys. Rev. B}\
  }\textbf {\bibinfo {volume} {98}},\ \bibinfo {pages} {035141} (\bibinfo
  {year} {2018})}\BibitemShut {NoStop}%
\bibitem [{\citenamefont {Yoshida}\ \emph {et~al.}(2019)\citenamefont
  {Yoshida}, \citenamefont {Peters}, \citenamefont {Kawakami},\ and\
  \citenamefont {Hatsugai}}]{Yoshida-19}%
  \BibitemOpen
  \bibfield  {author} {\bibinfo {author} {\bibfnamefont {T.}~\bibnamefont
  {Yoshida}}, \bibinfo {author} {\bibfnamefont {R.}~\bibnamefont {Peters}},
  \bibinfo {author} {\bibfnamefont {N.}~\bibnamefont {Kawakami}}, \ and\
  \bibinfo {author} {\bibfnamefont {Y.}~\bibnamefont {Hatsugai}},\ }\bibfield
  {title} {\enquote {\bibinfo {title} {{Symmetry-Protected Exceptional Rings in
  Two-Dimensional Correlated Systems with Chiral Symmetry}},}\ }\href@noop {}
  {\bibfield  {journal} {\bibinfo  {journal} {{Phys. Rev. B}}\ }\textbf
  {\bibinfo {volume} {99}},\ \bibinfo {pages} {121101(R)} (\bibinfo {year}
  {2019})}\BibitemShut {NoStop}%
\bibitem [{\citenamefont {Philip}\ \emph {et~al.}(2018)\citenamefont {Philip},
  \citenamefont {Hirsbrunner},\ and\ \citenamefont {Gilbert}}]{Philip-18}%
  \BibitemOpen
  \bibfield  {author} {\bibinfo {author} {\bibfnamefont {T.~M.}\ \bibnamefont
  {Philip}}, \bibinfo {author} {\bibfnamefont {M.~R.}\ \bibnamefont
  {Hirsbrunner}}, \ and\ \bibinfo {author} {\bibfnamefont {M.~J.}\ \bibnamefont
  {Gilbert}},\ }\bibfield  {title} {\enquote {\bibinfo {title} {{Loss of Hall
  Conductivity Quantization in a Non-Hermitian Quantum Anomalous Hall
  Insulator}},}\ }\href@noop {} {\bibfield  {journal} {\bibinfo  {journal}
  {Phys. Rev. B}\ }\textbf {\bibinfo {volume} {98}},\ \bibinfo {pages} {155430}
  (\bibinfo {year} {2018})}\BibitemShut {NoStop}%
\bibitem [{\citenamefont {Chen}\ and\ \citenamefont {Zhai}(2018)}]{Zhai-18}%
  \BibitemOpen
  \bibfield  {author} {\bibinfo {author} {\bibfnamefont {Y.}~\bibnamefont
  {Chen}}\ and\ \bibinfo {author} {\bibfnamefont {H.}~\bibnamefont {Zhai}},\
  }\bibfield  {title} {\enquote {\bibinfo {title} {{Hall Conductance of a
  Non-Hermitian Chern Insulator}},}\ }\href@noop {} {\bibfield  {journal}
  {\bibinfo  {journal} {Phys. Rev. B}\ }\textbf {\bibinfo {volume} {98}},\
  \bibinfo {pages} {245130} (\bibinfo {year} {2018})}\BibitemShut {NoStop}%
\bibitem [{\citenamefont {Bergholtz}\ and\ \citenamefont
  {Budich}()}]{Bergholtz-19}%
  \BibitemOpen
  \bibfield  {author} {\bibinfo {author} {\bibfnamefont {E.~J.}\ \bibnamefont
  {Bergholtz}}\ and\ \bibinfo {author} {\bibfnamefont {J.~C.}\ \bibnamefont
  {Budich}},\ }\href@noop {} {\enquote {\bibinfo {title} {{Non-Hermitian Weyl
  Physics in Topological Insulator Ferromagnet Junctions}},}\ }\bibinfo {note}
  {{a}rXiv: 1903.12187}\BibitemShut {NoStop}%
\bibitem [{\citenamefont {Hatano}\ and\ \citenamefont
  {Nelson}(1996)}]{Hatano-96}%
  \BibitemOpen
  \bibfield  {author} {\bibinfo {author} {\bibfnamefont {N.}~\bibnamefont
  {Hatano}}\ and\ \bibinfo {author} {\bibfnamefont {D.~R.}\ \bibnamefont
  {Nelson}},\ }\bibfield  {title} {\enquote {\bibinfo {title} {{L}ocalization
  {T}ransitions in {N}on-{H}ermitian {Q}uantum {M}echanics},}\ }\href@noop {}
  {\bibfield  {journal} {\bibinfo  {journal} {Phys. Rev. Lett.}\ }\textbf
  {\bibinfo {volume} {77}},\ \bibinfo {pages} {570} (\bibinfo {year}
  {1996})}\BibitemShut {NoStop}%
\bibitem [{\citenamefont {Hatano}\ and\ \citenamefont
  {Nelson}(1997)}]{Hatano-97}%
  \BibitemOpen
  \bibfield  {author} {\bibinfo {author} {\bibfnamefont {N.}~\bibnamefont
  {Hatano}}\ and\ \bibinfo {author} {\bibfnamefont {D.~R.}\ \bibnamefont
  {Nelson}},\ }\bibfield  {title} {\enquote {\bibinfo {title} {{V}ortex
  {P}inning and {N}on-{H}ermitian {Q}uantum {M}echanics},}\ }\href@noop {}
  {\bibfield  {journal} {\bibinfo  {journal} {Phys. Rev. B}\ }\textbf {\bibinfo
  {volume} {56}},\ \bibinfo {pages} {8651} (\bibinfo {year}
  {1997})}\BibitemShut {NoStop}%
\bibitem [{\citenamefont {Hatano}\ and\ \citenamefont
  {Nelson}(1998)}]{Hatano-98}%
  \BibitemOpen
  \bibfield  {author} {\bibinfo {author} {\bibfnamefont {N.}~\bibnamefont
  {Hatano}}\ and\ \bibinfo {author} {\bibfnamefont {D.~R.}\ \bibnamefont
  {Nelson}},\ }\bibfield  {title} {\enquote {\bibinfo {title}
  {{N}on-{H}ermitian {D}elocalization and {E}igenfunctions},}\ }\href@noop {}
  {\bibfield  {journal} {\bibinfo  {journal} {{Phys. Rev. B}}\ }\textbf
  {\bibinfo {volume} {58}},\ \bibinfo {pages} {8384} (\bibinfo {year}
  {1998})}\BibitemShut {NoStop}%
\bibitem [{\citenamefont {Efetov}(1997)}]{Efetov-97}%
  \BibitemOpen
  \bibfield  {author} {\bibinfo {author} {\bibfnamefont {K.~B.}\ \bibnamefont
  {Efetov}},\ }\bibfield  {title} {\enquote {\bibinfo {title} {{D}irected
  {Q}uantum {C}haos},}\ }\href@noop {} {\bibfield  {journal} {\bibinfo
  {journal} {Phys. Rev. Lett.}\ }\textbf {\bibinfo {volume} {79}},\ \bibinfo
  {pages} {491} (\bibinfo {year} {1997})}\BibitemShut {NoStop}%
\bibitem [{\citenamefont {Brouwer}\ \emph {et~al.}(1997)\citenamefont
  {Brouwer}, \citenamefont {Silvestrov},\ and\ \citenamefont
  {Beenakker}}]{Brouwer-97}%
  \BibitemOpen
  \bibfield  {author} {\bibinfo {author} {\bibfnamefont {P.~W.}\ \bibnamefont
  {Brouwer}}, \bibinfo {author} {\bibfnamefont {P.~G.}\ \bibnamefont
  {Silvestrov}}, \ and\ \bibinfo {author} {\bibfnamefont {C.~W.~J.}\
  \bibnamefont {Beenakker}},\ }\bibfield  {title} {\enquote {\bibinfo {title}
  {{T}heory of {D}irected {L}ocalization in {O}ne {D}imension},}\ }\href@noop
  {} {\bibfield  {journal} {\bibinfo  {journal} {Phys. Rev. B}\ }\textbf
  {\bibinfo {volume} {56}},\ \bibinfo {pages} {R4333(R)} (\bibinfo {year}
  {1997})}\BibitemShut {NoStop}%
\bibitem [{\citenamefont {Goldsheid}\ and\ \citenamefont
  {Khoruzhenko}(1998)}]{Goldsheid-98}%
  \BibitemOpen
  \bibfield  {author} {\bibinfo {author} {\bibfnamefont {I.~Y.}\ \bibnamefont
  {Goldsheid}}\ and\ \bibinfo {author} {\bibfnamefont {B.~A.}\ \bibnamefont
  {Khoruzhenko}},\ }\bibfield  {title} {\enquote {\bibinfo {title}
  {{D}istribution of {E}igenvalues in {N}on-{H}ermitian {A}nderson {M}odels},}\
  }\href@noop {} {\bibfield  {journal} {\bibinfo  {journal} {Phys. Rev. Lett.}\
  }\textbf {\bibinfo {volume} {80}},\ \bibinfo {pages} {2897} (\bibinfo {year}
  {1998})}\BibitemShut {NoStop}%
\bibitem [{\citenamefont {Mudry}\ \emph {et~al.}(1998)\citenamefont {Mudry},
  \citenamefont {Simons},\ and\ \citenamefont {Altland}}]{Mudry-98}%
  \BibitemOpen
  \bibfield  {author} {\bibinfo {author} {\bibfnamefont {C.}~\bibnamefont
  {Mudry}}, \bibinfo {author} {\bibfnamefont {B.~D.}\ \bibnamefont {Simons}}, \
  and\ \bibinfo {author} {\bibfnamefont {A.}~\bibnamefont {Altland}},\
  }\bibfield  {title} {\enquote {\bibinfo {title} {{Random Dirac Fermions and
  Non-Hermitian Quantum Mechanics}},}\ }\href@noop {} {\bibfield  {journal}
  {\bibinfo  {journal} {Phys. Rev. Lett.}\ }\textbf {\bibinfo {volume} {80}},\
  \bibinfo {pages} {4257} (\bibinfo {year} {1998})}\BibitemShut {NoStop}%
\bibitem [{\citenamefont {Shnerb}\ and\ \citenamefont
  {Nelson}(1998)}]{Shnerb-98}%
  \BibitemOpen
  \bibfield  {author} {\bibinfo {author} {\bibfnamefont {N.~M.}\ \bibnamefont
  {Shnerb}}\ and\ \bibinfo {author} {\bibfnamefont {D.~R.}\ \bibnamefont
  {Nelson}},\ }\bibfield  {title} {\enquote {\bibinfo {title} {{W}inding
  {N}umbers, {C}omplex {C}urrents, and {N}on-{H}ermitian {L}ocalization},}\
  }\href@noop {} {\bibfield  {journal} {\bibinfo  {journal} {Phys. Rev. Lett.}\
  }\textbf {\bibinfo {volume} {80}},\ \bibinfo {pages} {5172} (\bibinfo {year}
  {1998})}\BibitemShut {NoStop}%
\bibitem [{\citenamefont {Nelson}\ and\ \citenamefont
  {Shnerb}(1998)}]{Nelson-98}%
  \BibitemOpen
  \bibfield  {author} {\bibinfo {author} {\bibfnamefont {D.~R.}\ \bibnamefont
  {Nelson}}\ and\ \bibinfo {author} {\bibfnamefont {N.~M.}\ \bibnamefont
  {Shnerb}},\ }\bibfield  {title} {\enquote {\bibinfo {title} {{Non-Hermitian
  Localization and Population Biology}},}\ }\href@noop {} {\bibfield  {journal}
  {\bibinfo  {journal} {Phys. Rev. E}\ }\textbf {\bibinfo {volume} {58}},\
  \bibinfo {pages} {1383} (\bibinfo {year} {1998})}\BibitemShut {NoStop}%
\bibitem [{\citenamefont {Amir}\ \emph {et~al.}(2016)\citenamefont {Amir},
  \citenamefont {Hatano},\ and\ \citenamefont {Nelson}}]{Amir-16}%
  \BibitemOpen
  \bibfield  {author} {\bibinfo {author} {\bibfnamefont {A.}~\bibnamefont
  {Amir}}, \bibinfo {author} {\bibfnamefont {N.}~\bibnamefont {Hatano}}, \ and\
  \bibinfo {author} {\bibfnamefont {D.~R.}\ \bibnamefont {Nelson}},\ }\bibfield
   {title} {\enquote {\bibinfo {title} {{Non-Hermitian Localization in
  Biological Networks}},}\ }\href@noop {} {\bibfield  {journal} {\bibinfo
  {journal} {{Phys. Rev. E}}\ }\textbf {\bibinfo {volume} {93}},\ \bibinfo
  {pages} {042310} (\bibinfo {year} {2016})}\BibitemShut {NoStop}%
\bibitem [{\citenamefont {Fukui}\ and\ \citenamefont
  {Kawakami}(1998)}]{Fukui-98}%
  \BibitemOpen
  \bibfield  {author} {\bibinfo {author} {\bibfnamefont {T.}~\bibnamefont
  {Fukui}}\ and\ \bibinfo {author} {\bibfnamefont {N.}~\bibnamefont
  {Kawakami}},\ }\bibfield  {title} {\enquote {\bibinfo {title} {{Breakdown of
  the Mott Insulator: Exact Solution of an Asymmetric Hubbard Model}},}\
  }\href@noop {} {\bibfield  {journal} {\bibinfo  {journal} {{Phys. Rev. B}}\
  }\textbf {\bibinfo {volume} {58}},\ \bibinfo {pages} {16051} (\bibinfo {year}
  {1998})}\BibitemShut {NoStop}%
\bibitem [{\citenamefont {Feinberg}\ and\ \citenamefont
  {Zee}(1999)}]{Feinberg-99}%
  \BibitemOpen
  \bibfield  {author} {\bibinfo {author} {\bibfnamefont {J.}~\bibnamefont
  {Feinberg}}\ and\ \bibinfo {author} {\bibfnamefont {A.}~\bibnamefont {Zee}},\
  }\bibfield  {title} {\enquote {\bibinfo {title} {{N}on-{H}ermitian
  {L}ocalization and {D}elocalization},}\ }\href@noop {} {\bibfield  {journal}
  {\bibinfo  {journal} {Phys. Rev. E}\ }\textbf {\bibinfo {volume} {59}},\
  \bibinfo {pages} {6433} (\bibinfo {year} {1999})}\BibitemShut {NoStop}%
\bibitem [{\citenamefont {LeClair}(2000)}]{LeClair-00}%
  \BibitemOpen
  \bibfield  {author} {\bibinfo {author} {\bibfnamefont {A.}~\bibnamefont
  {LeClair}},\ }\bibfield  {title} {\enquote {\bibinfo {title} {{Relevance of
  Disorder for Dirac Fermions with Imaginary Vector Potentials}},}\ }\href@noop
  {} {\bibfield  {journal} {\bibinfo  {journal} {Phys. Rev. Lett.}\ }\textbf
  {\bibinfo {volume} {84}},\ \bibinfo {pages} {1292} (\bibinfo {year}
  {2000})}\BibitemShut {NoStop}%
\bibitem [{\citenamefont {Hamazaki}\ \emph {et~al.}()\citenamefont {Hamazaki},
  \citenamefont {Kawabata},\ and\ \citenamefont {Ueda}}]{Hamazaki-18}%
  \BibitemOpen
  \bibfield  {author} {\bibinfo {author} {\bibfnamefont {R.}~\bibnamefont
  {Hamazaki}}, \bibinfo {author} {\bibfnamefont {K.}~\bibnamefont {Kawabata}},
  \ and\ \bibinfo {author} {\bibfnamefont {M.}~\bibnamefont {Ueda}},\
  }\href@noop {} {\enquote {\bibinfo {title} {{Non-Hermitian Many-Body
  Localization}},}\ }\bibinfo {note} {{a}rXiv: 1811.11319}\BibitemShut
  {NoStop}%
\bibitem [{\citenamefont {Katsura}\ \emph {et~al.}(2010)\citenamefont
  {Katsura}, \citenamefont {Nagaosa},\ and\ \citenamefont {Lee}}]{Katsura-10}%
  \BibitemOpen
  \bibfield  {author} {\bibinfo {author} {\bibfnamefont {H.}~\bibnamefont
  {Katsura}}, \bibinfo {author} {\bibfnamefont {N.}~\bibnamefont {Nagaosa}}, \
  and\ \bibinfo {author} {\bibfnamefont {P.~A.}\ \bibnamefont {Lee}},\
  }\bibfield  {title} {\enquote {\bibinfo {title} {{T}heory of the {T}hermal
  {H}all {E}ffect in {Q}uantum {M}agnets},}\ }\href@noop {} {\bibfield
  {journal} {\bibinfo  {journal} {Phys. Rev. Lett.}\ }\textbf {\bibinfo
  {volume} {104}},\ \bibinfo {pages} {066403} (\bibinfo {year}
  {2010})}\BibitemShut {NoStop}%
\bibitem [{\citenamefont {Onose}\ \emph {et~al.}(2010)\citenamefont {Onose},
  \citenamefont {Ideue}, \citenamefont {Katsura}, \citenamefont {Shiomi},
  \citenamefont {Nagaosa},\ and\ \citenamefont {Tokura}}]{Onose-10}%
  \BibitemOpen
  \bibfield  {author} {\bibinfo {author} {\bibfnamefont {Y.}~\bibnamefont
  {Onose}}, \bibinfo {author} {\bibfnamefont {T.}~\bibnamefont {Ideue}},
  \bibinfo {author} {\bibfnamefont {H.}~\bibnamefont {Katsura}}, \bibinfo
  {author} {\bibfnamefont {Y.}~\bibnamefont {Shiomi}}, \bibinfo {author}
  {\bibfnamefont {N.}~\bibnamefont {Nagaosa}}, \ and\ \bibinfo {author}
  {\bibfnamefont {Y.}~\bibnamefont {Tokura}},\ }\bibfield  {title} {\enquote
  {\bibinfo {title} {{O}bservation of the {M}agnon {H}all {E}ffect},}\
  }\href@noop {} {\bibfield  {journal} {\bibinfo  {journal} {Science}\ }\textbf
  {\bibinfo {volume} {329}},\ \bibinfo {pages} {297} (\bibinfo {year}
  {2010})}\BibitemShut {NoStop}%
\bibitem [{\citenamefont {Shindou}\ \emph {et~al.}(2013)\citenamefont
  {Shindou}, \citenamefont {Matsumoto}, \citenamefont {Murakami},\ and\
  \citenamefont {Ohe}}]{Shindou-13}%
  \BibitemOpen
  \bibfield  {author} {\bibinfo {author} {\bibfnamefont {R.}~\bibnamefont
  {Shindou}}, \bibinfo {author} {\bibfnamefont {R.}~\bibnamefont {Matsumoto}},
  \bibinfo {author} {\bibfnamefont {S.}~\bibnamefont {Murakami}}, \ and\
  \bibinfo {author} {\bibfnamefont {J.}~\bibnamefont {Ohe}},\ }\bibfield
  {title} {\enquote {\bibinfo {title} {{T}opological {C}hiral {M}agnonic {E}dge
  {M}ode in a {M}agnonic {C}rystal},}\ }\href@noop {} {\bibfield  {journal}
  {\bibinfo  {journal} {Phys. Rev. B}\ }\textbf {\bibinfo {volume} {87}},\
  \bibinfo {pages} {174427} (\bibinfo {year} {2013})}\BibitemShut {NoStop}%
\bibitem [{\citenamefont {Barnett}(2013)}]{Barnett-13}%
  \BibitemOpen
  \bibfield  {author} {\bibinfo {author} {\bibfnamefont {R.}~\bibnamefont
  {Barnett}},\ }\bibfield  {title} {\enquote {\bibinfo {title} {{E}dge-{S}tate
  {I}nstabilities of {B}osons in a {T}opological {B}and},}\ }\href@noop {}
  {\bibfield  {journal} {\bibinfo  {journal} {Phys. Rev. A}\ }\textbf {\bibinfo
  {volume} {88}},\ \bibinfo {pages} {063631} (\bibinfo {year}
  {2013})}\BibitemShut {NoStop}%
\bibitem [{\citenamefont {Galilo}\ \emph {et~al.}(2015)\citenamefont {Galilo},
  \citenamefont {Lee},\ and\ \citenamefont {Barnett}}]{Galilo-15}%
  \BibitemOpen
  \bibfield  {author} {\bibinfo {author} {\bibfnamefont {B.}~\bibnamefont
  {Galilo}}, \bibinfo {author} {\bibfnamefont {D.~K.~K.}\ \bibnamefont {Lee}},
  \ and\ \bibinfo {author} {\bibfnamefont {R.}~\bibnamefont {Barnett}},\
  }\bibfield  {title} {\enquote {\bibinfo {title} {{S}elective {P}opulation of
  {E}dge {S}tates in a 2{D} {T}opological {B}and {S}ystem},}\ }\href@noop {}
  {\bibfield  {journal} {\bibinfo  {journal} {Phys. Rev. Lett.}\ }\textbf
  {\bibinfo {volume} {115}},\ \bibinfo {pages} {245302} (\bibinfo {year}
  {2015})}\BibitemShut {NoStop}%
\bibitem [{\citenamefont {Engelhardt}\ and\ \citenamefont
  {Brandes}(2015)}]{Engelhardt-15}%
  \BibitemOpen
  \bibfield  {author} {\bibinfo {author} {\bibfnamefont {G.}~\bibnamefont
  {Engelhardt}}\ and\ \bibinfo {author} {\bibfnamefont {T.}~\bibnamefont
  {Brandes}},\ }\bibfield  {title} {\enquote {\bibinfo {title} {{T}opological
  {B}ogoliubov {E}xcitations in {I}nversion-{S}ymmetric {S}ystems of
  {I}nteracting {B}osons},}\ }\href@noop {} {\bibfield  {journal} {\bibinfo
  {journal} {Phys. Rev. A}\ }\textbf {\bibinfo {volume} {91}},\ \bibinfo
  {pages} {053621} (\bibinfo {year} {2015})}\BibitemShut {NoStop}%
\bibitem [{\citenamefont {Engelhardt}\ \emph {et~al.}(2016)\citenamefont
  {Engelhardt}, \citenamefont {Benito}, \citenamefont {Platero},\ and\
  \citenamefont {Brandes}}]{Engelhardt-16}%
  \BibitemOpen
  \bibfield  {author} {\bibinfo {author} {\bibfnamefont {G.}~\bibnamefont
  {Engelhardt}}, \bibinfo {author} {\bibfnamefont {M.}~\bibnamefont {Benito}},
  \bibinfo {author} {\bibfnamefont {G.}~\bibnamefont {Platero}}, \ and\
  \bibinfo {author} {\bibfnamefont {T.}~\bibnamefont {Brandes}},\ }\bibfield
  {title} {\enquote {\bibinfo {title} {{T}opological {I}nstabilities in
  ac-{D}riven {B}osonic {S}ystems},}\ }\href@noop {} {\bibfield  {journal}
  {\bibinfo  {journal} {Phys. Rev. Lett.}\ }\textbf {\bibinfo {volume} {117}},\
  \bibinfo {pages} {045302} (\bibinfo {year} {2016})}\BibitemShut {NoStop}%
\bibitem [{\citenamefont {Peano}\ \emph
  {et~al.}(2016{\natexlab{a}})\citenamefont {Peano}, \citenamefont {Houde},
  \citenamefont {Brendel}, \citenamefont {Marquardt},\ and\ \citenamefont
  {Clerk}}]{Peano-16-nc}%
  \BibitemOpen
  \bibfield  {author} {\bibinfo {author} {\bibfnamefont {V.}~\bibnamefont
  {Peano}}, \bibinfo {author} {\bibfnamefont {M.}~\bibnamefont {Houde}},
  \bibinfo {author} {\bibfnamefont {C.}~\bibnamefont {Brendel}}, \bibinfo
  {author} {\bibfnamefont {F.}~\bibnamefont {Marquardt}}, \ and\ \bibinfo
  {author} {\bibfnamefont {A.~A.}\ \bibnamefont {Clerk}},\ }\bibfield  {title}
  {\enquote {\bibinfo {title} {{T}opological {P}hase {T}ransitions and {C}hiral
  {I}nelastic {T}ransport {I}nduced by the {S}queezing of {L}ight},}\
  }\href@noop {} {\bibfield  {journal} {\bibinfo  {journal} {Nat. Commun.}\
  }\textbf {\bibinfo {volume} {7}},\ \bibinfo {pages} {10779} (\bibinfo {year}
  {2016}{\natexlab{a}})}\BibitemShut {NoStop}%
\bibitem [{\citenamefont {Peano}\ \emph
  {et~al.}(2016{\natexlab{b}})\citenamefont {Peano}, \citenamefont {Houde},
  \citenamefont {Marquardt},\ and\ \citenamefont {Clerk}}]{Peano-16-x}%
  \BibitemOpen
  \bibfield  {author} {\bibinfo {author} {\bibfnamefont {V.}~\bibnamefont
  {Peano}}, \bibinfo {author} {\bibfnamefont {M.}~\bibnamefont {Houde}},
  \bibinfo {author} {\bibfnamefont {F.}~\bibnamefont {Marquardt}}, \ and\
  \bibinfo {author} {\bibfnamefont {A.~A.}\ \bibnamefont {Clerk}},\ }\bibfield
  {title} {\enquote {\bibinfo {title} {{T}opological {Q}uantum {F}luctuations
  and {T}raveling {W}ave {A}mplifiers},}\ }\href@noop {} {\bibfield  {journal}
  {\bibinfo  {journal} {Phys. Rev. X}\ }\textbf {\bibinfo {volume} {6}},\
  \bibinfo {pages} {041026} (\bibinfo {year} {2016}{\natexlab{b}})}\BibitemShut
  {NoStop}%
\bibitem [{\citenamefont {Lieu}(2018{\natexlab{a}})}]{Lieu-18}%
  \BibitemOpen
  \bibfield  {author} {\bibinfo {author} {\bibfnamefont {S.}~\bibnamefont
  {Lieu}},\ }\bibfield  {title} {\enquote {\bibinfo {title} {{T}opological
  {S}ymmetry {C}lasses for {N}on-{H}ermitian {M}odels and {C}onnections to the
  {B}osonic {B}ogoliubov-de {G}ennes {E}quation},}\ }\href@noop {} {\bibfield
  {journal} {\bibinfo  {journal} {Phys. Rev. B}\ }\textbf {\bibinfo {volume}
  {98}},\ \bibinfo {pages} {115135} (\bibinfo {year}
  {2018}{\natexlab{a}})}\BibitemShut {NoStop}%
\bibitem [{\citenamefont {McDonald}\ \emph {et~al.}(2018)\citenamefont
  {McDonald}, \citenamefont {Pereg-Barnea},\ and\ \citenamefont
  {Clerk}}]{Clerk-18}%
  \BibitemOpen
  \bibfield  {author} {\bibinfo {author} {\bibfnamefont {A.}~\bibnamefont
  {McDonald}}, \bibinfo {author} {\bibfnamefont {T.}~\bibnamefont
  {Pereg-Barnea}}, \ and\ \bibinfo {author} {\bibfnamefont {A.~A.}\
  \bibnamefont {Clerk}},\ }\bibfield  {title} {\enquote {\bibinfo {title}
  {{Phase-Dependent Chiral Transport and Effective Non-Hermitian Dynamics in a
  Bosonic Kitaev-Majorana Chain}},}\ }\href@noop {} {\bibfield  {journal}
  {\bibinfo  {journal} {Phys. Rev. X}\ }\textbf {\bibinfo {volume} {8}},\
  \bibinfo {pages} {041031} (\bibinfo {year} {2018})}\BibitemShut {NoStop}%
\bibitem [{\citenamefont {Kondo}\ \emph {et~al.}(2019)\citenamefont {Kondo},
  \citenamefont {Akagi},\ and\ \citenamefont {Katsura}}]{Kondo-18}%
  \BibitemOpen
  \bibfield  {author} {\bibinfo {author} {\bibfnamefont {H.}~\bibnamefont
  {Kondo}}, \bibinfo {author} {\bibfnamefont {Y.}~\bibnamefont {Akagi}}, \ and\
  \bibinfo {author} {\bibfnamefont {H.}~\bibnamefont {Katsura}},\ }\bibfield
  {title} {\enquote {\bibinfo {title} {{$\mathbb{Z}_{2}$ Topological Invariant
  for Magnon Spin Hall Systems}},}\ }\href@noop {} {\bibfield  {journal}
  {\bibinfo  {journal} {Phys. Rev. B}\ }\textbf {\bibinfo {volume} {99}},\
  \bibinfo {pages} {041110(R)} (\bibinfo {year} {2019})}\BibitemShut {NoStop}%
\bibitem [{\citenamefont {Kawaguchi}\ and\ \citenamefont
  {Ueda}(2012)}]{Kawaguchi-review}%
  \BibitemOpen
  \bibfield  {author} {\bibinfo {author} {\bibfnamefont {Y.}~\bibnamefont
  {Kawaguchi}}\ and\ \bibinfo {author} {\bibfnamefont {M.}~\bibnamefont
  {Ueda}},\ }\bibfield  {title} {\enquote {\bibinfo {title} {{S}pinor
  {B}ose-{E}instein {C}ondensates},}\ }\href@noop {} {\bibfield  {journal}
  {\bibinfo  {journal} {Phys. Rep.}\ }\textbf {\bibinfo {volume} {520}},\
  \bibinfo {pages} {253} (\bibinfo {year} {2012})}\BibitemShut {NoStop}%
\bibitem [{\citenamefont {Kane}\ and\ \citenamefont
  {Lubensky}(2014)}]{Kane-Lubensky-14}%
  \BibitemOpen
  \bibfield  {author} {\bibinfo {author} {\bibfnamefont {C.~L.}\ \bibnamefont
  {Kane}}\ and\ \bibinfo {author} {\bibfnamefont {T.~C.}\ \bibnamefont
  {Lubensky}},\ }\bibfield  {title} {\enquote {\bibinfo {title} {{T}opological
  {B}oundary {M}odes in {I}sostatic {L}attices},}\ }\href@noop {} {\bibfield
  {journal} {\bibinfo  {journal} {Nat. Phys.}\ }\textbf {\bibinfo {volume}
  {10}},\ \bibinfo {pages} {39} (\bibinfo {year} {2014})}\BibitemShut {NoStop}%
\bibitem [{\citenamefont {Roychowdhury}\ \emph {et~al.}(2018)\citenamefont
  {Roychowdhury}, \citenamefont {Rocklin},\ and\ \citenamefont
  {Lawler}}]{Roychowdhury-18-L}%
  \BibitemOpen
  \bibfield  {author} {\bibinfo {author} {\bibfnamefont {K.}~\bibnamefont
  {Roychowdhury}}, \bibinfo {author} {\bibfnamefont {D.~Z.}\ \bibnamefont
  {Rocklin}}, \ and\ \bibinfo {author} {\bibfnamefont {M.~J.}\ \bibnamefont
  {Lawler}},\ }\bibfield  {title} {\enquote {\bibinfo {title} {{T}opology and
  {G}eometry of {S}pin {O}rigami},}\ }\href@noop {} {\bibfield  {journal}
  {\bibinfo  {journal} {Phys. Rev. Lett.}\ }\textbf {\bibinfo {volume} {121}},\
  \bibinfo {pages} {177201} (\bibinfo {year} {2018})}\BibitemShut {NoStop}%
\bibitem [{\citenamefont {Roychowdhury}\ and\ \citenamefont
  {Lawler}(2018)}]{Roychowdhury-18-B}%
  \BibitemOpen
  \bibfield  {author} {\bibinfo {author} {\bibfnamefont {K.}~\bibnamefont
  {Roychowdhury}}\ and\ \bibinfo {author} {\bibfnamefont {M.~J.}\ \bibnamefont
  {Lawler}},\ }\bibfield  {title} {\enquote {\bibinfo {title} {{C}lassification
  of {M}agnetic {F}rustration and {M}etamaterials from {T}opology},}\
  }\href@noop {} {\bibfield  {journal} {\bibinfo  {journal} {Phys. Rev. B}\
  }\textbf {\bibinfo {volume} {98}},\ \bibinfo {pages} {094432} (\bibinfo
  {year} {2018})}\BibitemShut {NoStop}%
\bibitem [{\citenamefont {Moiseyev}(2011)}]{Moiseyev-11}%
  \BibitemOpen
  \bibfield  {author} {\bibinfo {author} {\bibfnamefont {N.}~\bibnamefont
  {Moiseyev}},\ }\href@noop {} {\emph {\bibinfo {title} {{N}on-{H}ermitian
  {Q}uantum {M}echanics}}}\ (\bibinfo  {publisher} {Cambridge University Press,
  Cambridge, England},\ \bibinfo {year} {2011})\BibitemShut {NoStop}%
\bibitem [{\citenamefont {Brody}(2014)}]{Brody-14}%
  \BibitemOpen
  \bibfield  {author} {\bibinfo {author} {\bibfnamefont {D.~C.}\ \bibnamefont
  {Brody}},\ }\bibfield  {title} {\enquote {\bibinfo {title} {{B}iorthogonal
  {Q}uantum {M}echanics},}\ }\href@noop {} {\bibfield  {journal} {\bibinfo
  {journal} {J. Phys. A}\ }\textbf {\bibinfo {volume} {47}},\ \bibinfo {pages}
  {035305} (\bibinfo {year} {2014})}\BibitemShut {NoStop}%
\bibitem [{\citenamefont {Berry}(2004)}]{Berry-04}%
  \BibitemOpen
  \bibfield  {author} {\bibinfo {author} {\bibfnamefont {M.~V.}\ \bibnamefont
  {Berry}},\ }\bibfield  {title} {\enquote {\bibinfo {title} {{P}hysics of
  {N}onhermitian {D}egeneracies},}\ }\href@noop {} {\bibfield  {journal}
  {\bibinfo  {journal} {Czech. J. Phys.}\ }\textbf {\bibinfo {volume} {54}},\
  \bibinfo {pages} {1039} (\bibinfo {year} {2004})}\BibitemShut {NoStop}%
\bibitem [{\citenamefont {Heiss}(2012)}]{Heiss-12}%
  \BibitemOpen
  \bibfield  {author} {\bibinfo {author} {\bibfnamefont {W.~D.}\ \bibnamefont
  {Heiss}},\ }\bibfield  {title} {\enquote {\bibinfo {title} {{T}he {P}hysics
  of {E}xceptional {P}oints},}\ }\href@noop {} {\bibfield  {journal} {\bibinfo
  {journal} {J. Phys. A}\ }\textbf {\bibinfo {volume} {45}},\ \bibinfo {pages}
  {444016} (\bibinfo {year} {2012})}\BibitemShut {NoStop}%
\bibitem [{\citenamefont {Bender}\ \emph {et~al.}(2007)\citenamefont {Bender},
  \citenamefont {Brody}, \citenamefont {Jones},\ and\ \citenamefont
  {Meister}}]{Bender-07}%
  \BibitemOpen
  \bibfield  {author} {\bibinfo {author} {\bibfnamefont {C.~M.}\ \bibnamefont
  {Bender}}, \bibinfo {author} {\bibfnamefont {D.~C.}\ \bibnamefont {Brody}},
  \bibinfo {author} {\bibfnamefont {H.~F.}\ \bibnamefont {Jones}}, \ and\
  \bibinfo {author} {\bibfnamefont {B.~K.}\ \bibnamefont {Meister}},\
  }\bibfield  {title} {\enquote {\bibinfo {title} {{F}aster than {H}ermitian
  {Q}uantum {M}echanics},}\ }\href@noop {} {\bibfield  {journal} {\bibinfo
  {journal} {Phys. Rev. Lett.}\ }\textbf {\bibinfo {volume} {98}},\ \bibinfo
  {pages} {040403} (\bibinfo {year} {2007})}\BibitemShut {NoStop}%
\bibitem [{\citenamefont {Musslimani}\ \emph {et~al.}(2008)\citenamefont
  {Musslimani}, \citenamefont {Makris}, \citenamefont {El-Ganainy},\ and\
  \citenamefont {Christodoulides}}]{Musslimani-08}%
  \BibitemOpen
  \bibfield  {author} {\bibinfo {author} {\bibfnamefont {Z.~H.}\ \bibnamefont
  {Musslimani}}, \bibinfo {author} {\bibfnamefont {K.~G.}\ \bibnamefont
  {Makris}}, \bibinfo {author} {\bibfnamefont {R.}~\bibnamefont {El-Ganainy}},
  \ and\ \bibinfo {author} {\bibfnamefont {D.~N.}\ \bibnamefont
  {Christodoulides}},\ }\bibfield  {title} {\enquote {\bibinfo {title}
  {{O}ptical {S}olitons in $\mathcal{PT}$ {P}eriodic {P}otentials},}\
  }\href@noop {} {\bibfield  {journal} {\bibinfo  {journal} {Phys. Rev. Lett.}\
  }\textbf {\bibinfo {volume} {100}},\ \bibinfo {pages} {030402} (\bibinfo
  {year} {2008})}\BibitemShut {NoStop}%
\bibitem [{\citenamefont {Makris}\ \emph {et~al.}(2008)\citenamefont {Makris},
  \citenamefont {El-Ganainy}, \citenamefont {Christodoulides},\ and\
  \citenamefont {Musslimani}}]{Makris-08}%
  \BibitemOpen
  \bibfield  {author} {\bibinfo {author} {\bibfnamefont {K.~G.}\ \bibnamefont
  {Makris}}, \bibinfo {author} {\bibfnamefont {R.}~\bibnamefont {El-Ganainy}},
  \bibinfo {author} {\bibfnamefont {D.~N.}\ \bibnamefont {Christodoulides}}, \
  and\ \bibinfo {author} {\bibfnamefont {Z.~H.}\ \bibnamefont {Musslimani}},\
  }\bibfield  {title} {\enquote {\bibinfo {title} {{B}eam {D}ynamics in
  $\mathcal{PT}$ {S}ymmetric {O}ptical {L}attices},}\ }\href@noop {} {\bibfield
   {journal} {\bibinfo  {journal} {{Phys. Rev. Lett.}}\ }\textbf {\bibinfo
  {volume} {100}},\ \bibinfo {pages} {103904} (\bibinfo {year}
  {2008})}\BibitemShut {NoStop}%
\bibitem [{\citenamefont {Klaiman}\ \emph {et~al.}(2008)\citenamefont
  {Klaiman}, \citenamefont {G\"unther},\ and\ \citenamefont
  {Moiseyev}}]{Klaiman-08}%
  \BibitemOpen
  \bibfield  {author} {\bibinfo {author} {\bibfnamefont {S.}~\bibnamefont
  {Klaiman}}, \bibinfo {author} {\bibfnamefont {U.}~\bibnamefont {G\"unther}},
  \ and\ \bibinfo {author} {\bibfnamefont {N.}~\bibnamefont {Moiseyev}},\
  }\bibfield  {title} {\enquote {\bibinfo {title} {{V}isualization of {B}ranch
  {P}oints in $\mathcal{PT}$-{S}ymmetric {W}aveguides},}\ }\href@noop {}
  {\bibfield  {journal} {\bibinfo  {journal} {Phys. Rev. Lett.}\ }\textbf
  {\bibinfo {volume} {101}},\ \bibinfo {pages} {080402} (\bibinfo {year}
  {2008})}\BibitemShut {NoStop}%
\bibitem [{\citenamefont {Graefe}\ \emph {et~al.}(2008)\citenamefont {Graefe},
  \citenamefont {Korsch},\ and\ \citenamefont {Niederle}}]{Graefe-08}%
  \BibitemOpen
  \bibfield  {author} {\bibinfo {author} {\bibfnamefont {E.-M.}\ \bibnamefont
  {Graefe}}, \bibinfo {author} {\bibfnamefont {H.~J.}\ \bibnamefont {Korsch}},
  \ and\ \bibinfo {author} {\bibfnamefont {A.~E.}\ \bibnamefont {Niederle}},\
  }\bibfield  {title} {\enquote {\bibinfo {title} {{M}ean-{F}ield {D}ynamics of
  a {N}on-{H}ermitian {B}ose-{H}ubbard {D}imer},}\ }\href@noop {} {\bibfield
  {journal} {\bibinfo  {journal} {Phys. Rev. Lett.}\ }\textbf {\bibinfo
  {volume} {101}},\ \bibinfo {pages} {150408} (\bibinfo {year}
  {2008})}\BibitemShut {NoStop}%
\bibitem [{\citenamefont {G\"unther}\ and\ \citenamefont
  {Samsonov}(2008)}]{Gunther-08}%
  \BibitemOpen
  \bibfield  {author} {\bibinfo {author} {\bibfnamefont {U.}~\bibnamefont
  {G\"unther}}\ and\ \bibinfo {author} {\bibfnamefont {B.~F.}\ \bibnamefont
  {Samsonov}},\ }\bibfield  {title} {\enquote {\bibinfo {title}
  {{N}aimark-{D}ilated $\mathcal{PT}$-{S}ymmetric {B}rachistochrone},}\
  }\href@noop {} {\bibfield  {journal} {\bibinfo  {journal} {Phys. Rev. Lett.}\
  }\textbf {\bibinfo {volume} {101}},\ \bibinfo {pages} {230404} (\bibinfo
  {year} {2008})}\BibitemShut {NoStop}%
\bibitem [{\citenamefont {Mostafazadeh}(2009)}]{Mostafazadeh-09}%
  \BibitemOpen
  \bibfield  {author} {\bibinfo {author} {\bibfnamefont {A.}~\bibnamefont
  {Mostafazadeh}},\ }\bibfield  {title} {\enquote {\bibinfo {title} {{S}pectral
  {S}ingularities of {C}omplex {S}cattering {P}otentials and {I}nfinite
  {R}eflection and {T}ransmission {C}oefficients at {R}eal {E}nergies},}\
  }\href@noop {} {\bibfield  {journal} {\bibinfo  {journal} {Phys. Rev. Lett.}\
  }\textbf {\bibinfo {volume} {102}},\ \bibinfo {pages} {220402} (\bibinfo
  {year} {2009})}\BibitemShut {NoStop}%
\bibitem [{\citenamefont {Longhi}(2009)}]{Longhi-09}%
  \BibitemOpen
  \bibfield  {author} {\bibinfo {author} {\bibfnamefont {S.}~\bibnamefont
  {Longhi}},\ }\bibfield  {title} {\enquote {\bibinfo {title} {{B}loch
  {O}scillations in {C}omplex {C}rystals with $\mathcal{PT}$ {S}ymmetry},}\
  }\href@noop {} {\bibfield  {journal} {\bibinfo  {journal} {Phys. Rev. Lett.}\
  }\textbf {\bibinfo {volume} {103}},\ \bibinfo {pages} {123601} (\bibinfo
  {year} {2009})}\BibitemShut {NoStop}%
\bibitem [{\citenamefont {Longhi}(2010)}]{Longhi-10}%
  \BibitemOpen
  \bibfield  {author} {\bibinfo {author} {\bibfnamefont {S.}~\bibnamefont
  {Longhi}},\ }\bibfield  {title} {\enquote {\bibinfo {title}
  {$\mathcal{PT}$-{S}ymmetric {L}aser {A}bsorber},}\ }\href@noop {} {\bibfield
  {journal} {\bibinfo  {journal} {Phys. Rev. A}\ }\textbf {\bibinfo {volume}
  {82}},\ \bibinfo {pages} {031801(R)} (\bibinfo {year} {2010})}\BibitemShut
  {NoStop}%
\bibitem [{\citenamefont {Chong}\ \emph {et~al.}(2011)\citenamefont {Chong},
  \citenamefont {Ge},\ and\ \citenamefont {Stone}}]{Chong-11}%
  \BibitemOpen
  \bibfield  {author} {\bibinfo {author} {\bibfnamefont {Y.~D.}\ \bibnamefont
  {Chong}}, \bibinfo {author} {\bibfnamefont {L.}~\bibnamefont {Ge}}, \ and\
  \bibinfo {author} {\bibfnamefont {A.~D.}\ \bibnamefont {Stone}},\ }\bibfield
  {title} {\enquote {\bibinfo {title} {$\mathcal{PT}$-{S}ymmetry {B}reaking and
  {L}aser-{A}bsorber {M}odes in {O}ptical {S}cattering {S}ystems},}\
  }\href@noop {} {\bibfield  {journal} {\bibinfo  {journal} {Phys. Rev. Lett.}\
  }\textbf {\bibinfo {volume} {106}},\ \bibinfo {pages} {093902} (\bibinfo
  {year} {2011})}\BibitemShut {NoStop}%
\bibitem [{\citenamefont {Lin}\ \emph {et~al.}(2011)\citenamefont {Lin},
  \citenamefont {Ramezani}, \citenamefont {Eichelkraut}, \citenamefont
  {Kottos}, \citenamefont {Cao},\ and\ \citenamefont
  {Christodoulides}}]{Lin-11}%
  \BibitemOpen
  \bibfield  {author} {\bibinfo {author} {\bibfnamefont {Z.}~\bibnamefont
  {Lin}}, \bibinfo {author} {\bibfnamefont {H.}~\bibnamefont {Ramezani}},
  \bibinfo {author} {\bibfnamefont {T.}~\bibnamefont {Eichelkraut}}, \bibinfo
  {author} {\bibfnamefont {T.}~\bibnamefont {Kottos}}, \bibinfo {author}
  {\bibfnamefont {H.}~\bibnamefont {Cao}}, \ and\ \bibinfo {author}
  {\bibfnamefont {D.~N.}\ \bibnamefont {Christodoulides}},\ }\bibfield  {title}
  {\enquote {\bibinfo {title} {{U}nidirectional {I}nvisibility {I}nduced by
  $\mathcal{PT}$-{S}ymmetric {P}eriodic {S}tructures},}\ }\href@noop {}
  {\bibfield  {journal} {\bibinfo  {journal} {Phys. Rev. Lett.}\ }\textbf
  {\bibinfo {volume} {106}},\ \bibinfo {pages} {213901} (\bibinfo {year}
  {2011})}\BibitemShut {NoStop}%
\bibitem [{\citenamefont {Brody}\ and\ \citenamefont
  {Graefe}(2012)}]{Brody-12}%
  \BibitemOpen
  \bibfield  {author} {\bibinfo {author} {\bibfnamefont {D.~C.}\ \bibnamefont
  {Brody}}\ and\ \bibinfo {author} {\bibfnamefont {E.-M.}\ \bibnamefont
  {Graefe}},\ }\bibfield  {title} {\enquote {\bibinfo {title} {{M}ixed-{S}tate
  {E}volution in the {P}resence of {G}ain and {L}oss},}\ }\href@noop {}
  {\bibfield  {journal} {\bibinfo  {journal} {Phys. Rev. Lett.}\ }\textbf
  {\bibinfo {volume} {109}},\ \bibinfo {pages} {230405} (\bibinfo {year}
  {2012})}\BibitemShut {NoStop}%
\bibitem [{\citenamefont {Wiersig}(2014)}]{Wiersig-14}%
  \BibitemOpen
  \bibfield  {author} {\bibinfo {author} {\bibfnamefont {J.}~\bibnamefont
  {Wiersig}},\ }\bibfield  {title} {\enquote {\bibinfo {title} {{E}nhancing the
  {S}ensitivity of {F}requency and {E}nergy {S}plitting {D}etection by {U}sing
  {E}xceptional {P}oints: {A}pplication to {M}icrocavity {S}ensors for
  {S}ingle-{P}article {D}etection},}\ }\href@noop {} {\bibfield  {journal}
  {\bibinfo  {journal} {Phys. Rev. Lett.}\ }\textbf {\bibinfo {volume} {112}},\
  \bibinfo {pages} {203901} (\bibinfo {year} {2014})}\BibitemShut {NoStop}%
\bibitem [{\citenamefont {Jing}\ \emph {et~al.}(2014)\citenamefont {Jing},
  \citenamefont {\c{S}. K.~\"Ozdemir}, \citenamefont {L\"u}, \citenamefont
  {Zhang}, \citenamefont {Yang},\ and\ \citenamefont {Nori}}]{Jing-14}%
  \BibitemOpen
  \bibfield  {author} {\bibinfo {author} {\bibfnamefont {H.}~\bibnamefont
  {Jing}}, \bibinfo {author} {\bibnamefont {\c{S}. K.~\"Ozdemir}}, \bibinfo
  {author} {\bibfnamefont {X.-Y.}\ \bibnamefont {L\"u}}, \bibinfo {author}
  {\bibfnamefont {J.}~\bibnamefont {Zhang}}, \bibinfo {author} {\bibfnamefont
  {L.}~\bibnamefont {Yang}}, \ and\ \bibinfo {author} {\bibfnamefont
  {F.}~\bibnamefont {Nori}},\ }\bibfield  {title} {\enquote {\bibinfo {title}
  {$\mathcal{PT}$-{S}ymmetric {P}honon {L}aser},}\ }\href@noop {} {\bibfield
  {journal} {\bibinfo  {journal} {Phys. Rev. Lett.}\ }\textbf {\bibinfo
  {volume} {113}},\ \bibinfo {pages} {053604} (\bibinfo {year}
  {2014})}\BibitemShut {NoStop}%
\bibitem [{\citenamefont {Zhu}\ \emph {et~al.}(2014)\citenamefont {Zhu},
  \citenamefont {Ramezani}, \citenamefont {Shi}, \citenamefont {Zhu},\ and\
  \citenamefont {Zhang}}]{Zhu-14}%
  \BibitemOpen
  \bibfield  {author} {\bibinfo {author} {\bibfnamefont {X.}~\bibnamefont
  {Zhu}}, \bibinfo {author} {\bibfnamefont {H.}~\bibnamefont {Ramezani}},
  \bibinfo {author} {\bibfnamefont {C.}~\bibnamefont {Shi}}, \bibinfo {author}
  {\bibfnamefont {J.}~\bibnamefont {Zhu}}, \ and\ \bibinfo {author}
  {\bibfnamefont {X.}~\bibnamefont {Zhang}},\ }\bibfield  {title} {\enquote
  {\bibinfo {title} {$\mathcal{PT}$-{S}ymmetric {A}coustics},}\ }\href@noop {}
  {\bibfield  {journal} {\bibinfo  {journal} {Phys. Rev. X}\ }\textbf {\bibinfo
  {volume} {4}},\ \bibinfo {pages} {031042} (\bibinfo {year}
  {2014})}\BibitemShut {NoStop}%
\bibitem [{\citenamefont {Lee}\ and\ \citenamefont {Chan}(2014)}]{Lee-14X}%
  \BibitemOpen
  \bibfield  {author} {\bibinfo {author} {\bibfnamefont {T.~E.}\ \bibnamefont
  {Lee}}\ and\ \bibinfo {author} {\bibfnamefont {C.-K.}\ \bibnamefont {Chan}},\
  }\bibfield  {title} {\enquote {\bibinfo {title} {{H}eralded {M}agnetism in
  {N}on-{H}ermitian {A}tomic {S}ystems},}\ }\href@noop {} {\bibfield  {journal}
  {\bibinfo  {journal} {Phys. Rev. X}\ }\textbf {\bibinfo {volume} {4}},\
  \bibinfo {pages} {041001} (\bibinfo {year} {2014})}\BibitemShut {NoStop}%
\bibitem [{\citenamefont {Lee}\ \emph {et~al.}(2014)\citenamefont {Lee},
  \citenamefont {Reiter},\ and\ \citenamefont {Moiseyev}}]{Lee-14L}%
  \BibitemOpen
  \bibfield  {author} {\bibinfo {author} {\bibfnamefont {T.~E.}\ \bibnamefont
  {Lee}}, \bibinfo {author} {\bibfnamefont {F.}~\bibnamefont {Reiter}}, \ and\
  \bibinfo {author} {\bibfnamefont {N.}~\bibnamefont {Moiseyev}},\ }\bibfield
  {title} {\enquote {\bibinfo {title} {{E}ntanglement and {S}pin {S}queezing in
  {N}on-{H}ermitian {P}hase {T}ransitions},}\ }\href@noop {} {\bibfield
  {journal} {\bibinfo  {journal} {Phys. Rev. Lett.}\ }\textbf {\bibinfo
  {volume} {113}},\ \bibinfo {pages} {250401} (\bibinfo {year}
  {2014})}\BibitemShut {NoStop}%
\bibitem [{\citenamefont {Dana}\ \emph {et~al.}(2015)\citenamefont {Dana},
  \citenamefont {Bahabad},\ and\ \citenamefont {Malomed}}]{Dana-15}%
  \BibitemOpen
  \bibfield  {author} {\bibinfo {author} {\bibfnamefont {B.}~\bibnamefont
  {Dana}}, \bibinfo {author} {\bibfnamefont {A.}~\bibnamefont {Bahabad}}, \
  and\ \bibinfo {author} {\bibfnamefont {B.~A.}\ \bibnamefont {Malomed}},\
  }\bibfield  {title} {\enquote {\bibinfo {title} {{$\mathcal{CP}$ Symmetry in
  Optical Systems}},}\ }\href@noop {} {\bibfield  {journal} {\bibinfo
  {journal} {Phys. Rev. A}\ }\textbf {\bibinfo {volume} {91}},\ \bibinfo
  {pages} {043808} (\bibinfo {year} {2015})}\BibitemShut {NoStop}%
\bibitem [{\citenamefont {Kirikchi}\ \emph {et~al.}(2018)\citenamefont
  {Kirikchi}, \citenamefont {Malomed}, \citenamefont {Karjannto}, \citenamefont
  {Kusdiantara},\ and\ \citenamefont {Susanto}}]{Kirikchi-18}%
  \BibitemOpen
  \bibfield  {author} {\bibinfo {author} {\bibfnamefont {O.~B.}\ \bibnamefont
  {Kirikchi}}, \bibinfo {author} {\bibfnamefont {B.~A.}\ \bibnamefont
  {Malomed}}, \bibinfo {author} {\bibfnamefont {N.}~\bibnamefont {Karjannto}},
  \bibinfo {author} {\bibfnamefont {R.}~\bibnamefont {Kusdiantara}}, \ and\
  \bibinfo {author} {\bibfnamefont {H.}~\bibnamefont {Susanto}},\ }\bibfield
  {title} {\enquote {\bibinfo {title} {{Solitons in a Chain of
  Charge-Parity-Symmetric Dimers}},}\ }\href@noop {} {\bibfield  {journal}
  {\bibinfo  {journal} {{Phys. Rev. A}}\ }\textbf {\bibinfo {volume} {98}},\
  \bibinfo {pages} {063841} (\bibinfo {year} {2018})}\BibitemShut {NoStop}%
\bibitem [{\citenamefont {Longhi}\ \emph
  {et~al.}(2015{\natexlab{a}})\citenamefont {Longhi}, \citenamefont {Gatti},\
  and\ \citenamefont {Valle}}]{Longhi-15-SR}%
  \BibitemOpen
  \bibfield  {author} {\bibinfo {author} {\bibfnamefont {S.}~\bibnamefont
  {Longhi}}, \bibinfo {author} {\bibfnamefont {D.}~\bibnamefont {Gatti}}, \
  and\ \bibinfo {author} {\bibfnamefont {G.~Della}\ \bibnamefont {Valle}},\
  }\bibfield  {title} {\enquote {\bibinfo {title} {{R}obust {L}ight {T}ransport
  in {N}on-{H}ermitian {P}hotonic {L}attices},}\ }\href@noop {} {\bibfield
  {journal} {\bibinfo  {journal} {Sci. Rep.}\ }\textbf {\bibinfo {volume}
  {5}},\ \bibinfo {pages} {13376} (\bibinfo {year}
  {2015}{\natexlab{a}})}\BibitemShut {NoStop}%
\bibitem [{\citenamefont {Longhi}\ \emph
  {et~al.}(2015{\natexlab{b}})\citenamefont {Longhi}, \citenamefont {Gatti},\
  and\ \citenamefont {Valle}}]{Longhi-15-B}%
  \BibitemOpen
  \bibfield  {author} {\bibinfo {author} {\bibfnamefont {S.}~\bibnamefont
  {Longhi}}, \bibinfo {author} {\bibfnamefont {D.}~\bibnamefont {Gatti}}, \
  and\ \bibinfo {author} {\bibfnamefont {G.~Della}\ \bibnamefont {Valle}},\
  }\bibfield  {title} {\enquote {\bibinfo {title} {{N}on-{H}ermitian
  {T}ransparency and {O}ne-{W}ay {T}ransport in {L}ow-{D}imensional {L}attices
  by an {I}maginary {G}auge {F}ield},}\ }\href@noop {} {\bibfield  {journal}
  {\bibinfo  {journal} {Phys. Rev. B}\ }\textbf {\bibinfo {volume} {92}},\
  \bibinfo {pages} {094204} (\bibinfo {year} {2015}{\natexlab{b}})}\BibitemShut
  {NoStop}%
\bibitem [{\citenamefont {Kominis}(2015)}]{Kominis-15}%
  \BibitemOpen
  \bibfield  {author} {\bibinfo {author} {\bibfnamefont {Y.}~\bibnamefont
  {Kominis}},\ }\bibfield  {title} {\enquote {\bibinfo {title} {{Dynamic Power
  Balance for Nonlinear Waves in Unbalanced Gain and Loss Landscapes}},}\
  }\href@noop {} {\bibfield  {journal} {\bibinfo  {journal} {Phys. Rev. A}\
  }\textbf {\bibinfo {volume} {92}},\ \bibinfo {pages} {063849} (\bibinfo
  {year} {2015})}\BibitemShut {NoStop}%
\bibitem [{\citenamefont {Kominis}\ \emph {et~al.}(2016)\citenamefont
  {Kominis}, \citenamefont {Bountis},\ and\ \citenamefont
  {Flach}}]{Kominis-16}%
  \BibitemOpen
  \bibfield  {author} {\bibinfo {author} {\bibfnamefont {Y.}~\bibnamefont
  {Kominis}}, \bibinfo {author} {\bibfnamefont {T.}~\bibnamefont {Bountis}}, \
  and\ \bibinfo {author} {\bibfnamefont {S.}~\bibnamefont {Flach}},\ }\bibfield
   {title} {\enquote {\bibinfo {title} {{The Asymmetric Active Coupler: Stable
  Nonlinear Supermodes and Directed Transport}},}\ }\href@noop {} {\bibfield
  {journal} {\bibinfo  {journal} {Sci. Rep.}\ }\textbf {\bibinfo {volume}
  {6}},\ \bibinfo {pages} {33699} (\bibinfo {year} {2016})}\BibitemShut
  {NoStop}%
\bibitem [{\citenamefont {Liu}\ \emph {et~al.}(2016)\citenamefont {Liu},
  \citenamefont {Zhang}, \citenamefont {\c{S}. K.~\"Ozdemir}, \citenamefont
  {Peng}, \citenamefont {Jing}, \citenamefont {L\"u}, \citenamefont {Li},
  \citenamefont {Yang}, \citenamefont {Nori},\ and\ \citenamefont
  {x.~Liu}}]{Liu-16}%
  \BibitemOpen
  \bibfield  {author} {\bibinfo {author} {\bibfnamefont {Z.-P.}\ \bibnamefont
  {Liu}}, \bibinfo {author} {\bibfnamefont {J.}~\bibnamefont {Zhang}}, \bibinfo
  {author} {\bibnamefont {\c{S}. K.~\"Ozdemir}}, \bibinfo {author}
  {\bibfnamefont {B.}~\bibnamefont {Peng}}, \bibinfo {author} {\bibfnamefont
  {H.}~\bibnamefont {Jing}}, \bibinfo {author} {\bibfnamefont {X.-Y.}\
  \bibnamefont {L\"u}}, \bibinfo {author} {\bibfnamefont {C.-W.}\ \bibnamefont
  {Li}}, \bibinfo {author} {\bibfnamefont {L.}~\bibnamefont {Yang}}, \bibinfo
  {author} {\bibfnamefont {F.}~\bibnamefont {Nori}}, \ and\ \bibinfo {author}
  {\bibfnamefont {Y.}~\bibnamefont {x.~Liu}},\ }\bibfield  {title} {\enquote
  {\bibinfo {title} {{M}etrology with $\mathcal{PT}$-{S}ymmetric {C}avities:
  {E}nhanced {S}ensitivity near the $\mathcal{PT}$-{P}hase {T}ransition},}\
  }\href@noop {} {\bibfield  {journal} {\bibinfo  {journal} {Phys. Rev. Lett.}\
  }\textbf {\bibinfo {volume} {117}},\ \bibinfo {pages} {110802} (\bibinfo
  {year} {2016})}\BibitemShut {NoStop}%
\bibitem [{\citenamefont {Ge}(2017)}]{Ge-17}%
  \BibitemOpen
  \bibfield  {author} {\bibinfo {author} {\bibfnamefont {L.}~\bibnamefont
  {Ge}},\ }\bibfield  {title} {\enquote {\bibinfo {title}
  {{S}ymmetry-{P}rotected {Z}ero-{M}ode {L}aser with a {T}unable {S}patial
  {P}rofile},}\ }\href@noop {} {\bibfield  {journal} {\bibinfo  {journal}
  {Phys. Rev. A}\ }\textbf {\bibinfo {volume} {95}},\ \bibinfo {pages} {023812}
  (\bibinfo {year} {2017})}\BibitemShut {NoStop}%
\bibitem [{\citenamefont {Ashida}\ \emph {et~al.}(2017)\citenamefont {Ashida},
  \citenamefont {Furukawa},\ and\ \citenamefont {Ueda}}]{Ashida-17}%
  \BibitemOpen
  \bibfield  {author} {\bibinfo {author} {\bibfnamefont {Y.}~\bibnamefont
  {Ashida}}, \bibinfo {author} {\bibfnamefont {S.}~\bibnamefont {Furukawa}}, \
  and\ \bibinfo {author} {\bibfnamefont {M.}~\bibnamefont {Ueda}},\ }\bibfield
  {title} {\enquote {\bibinfo {title} {{P}arity-{T}ime-{S}ymmetric {Q}uantum
  {C}ritical {P}henomena},}\ }\href@noop {} {\bibfield  {journal} {\bibinfo
  {journal} {Nat. Commun.}\ }\textbf {\bibinfo {volume} {8}},\ \bibinfo {pages}
  {15791} (\bibinfo {year} {2017})}\BibitemShut {NoStop}%
\bibitem [{\citenamefont {Kawabata}\ \emph
  {et~al.}(2017{\natexlab{a}})\citenamefont {Kawabata}, \citenamefont
  {Ashida},\ and\ \citenamefont {Ueda}}]{KK-QI-17}%
  \BibitemOpen
  \bibfield  {author} {\bibinfo {author} {\bibfnamefont {K.}~\bibnamefont
  {Kawabata}}, \bibinfo {author} {\bibfnamefont {Y.}~\bibnamefont {Ashida}}, \
  and\ \bibinfo {author} {\bibfnamefont {M.}~\bibnamefont {Ueda}},\ }\bibfield
  {title} {\enquote {\bibinfo {title} {{I}nformation {R}etrieval and
  {C}riticality in {P}arity-{T}ime-{S}ymmetric {S}ystems},}\ }\href@noop {}
  {\bibfield  {journal} {\bibinfo  {journal} {Phys. Rev. Lett.}\ }\textbf
  {\bibinfo {volume} {119}},\ \bibinfo {pages} {190401} (\bibinfo {year}
  {2017}{\natexlab{a}})}\BibitemShut {NoStop}%
\bibitem [{\citenamefont {Ghatak}\ and\ \citenamefont {Das}(2018)}]{Ghatak-18}%
  \BibitemOpen
  \bibfield  {author} {\bibinfo {author} {\bibfnamefont {A.}~\bibnamefont
  {Ghatak}}\ and\ \bibinfo {author} {\bibfnamefont {T.}~\bibnamefont {Das}},\
  }\bibfield  {title} {\enquote {\bibinfo {title} {{T}heory of
  {S}uperconductivity with {N}on-{H}ermitian and {P}arity-{T}ime-{R}eversal
  {S}ymmetric {C}ooper {P}airing {S}ymmetry},}\ }\href@noop {} {\bibfield
  {journal} {\bibinfo  {journal} {Phys. Rev. B}\ }\textbf {\bibinfo {volume}
  {97}},\ \bibinfo {pages} {014512} (\bibinfo {year} {2018})}\BibitemShut
  {NoStop}%
\bibitem [{\citenamefont {Qi}\ \emph {et~al.}(2018)\citenamefont {Qi},
  \citenamefont {Zhang},\ and\ \citenamefont {Ge}}]{Qi-18}%
  \BibitemOpen
  \bibfield  {author} {\bibinfo {author} {\bibfnamefont {B.}~\bibnamefont
  {Qi}}, \bibinfo {author} {\bibfnamefont {L.}~\bibnamefont {Zhang}}, \ and\
  \bibinfo {author} {\bibfnamefont {L.}~\bibnamefont {Ge}},\ }\bibfield
  {title} {\enquote {\bibinfo {title} {{D}efect {S}tates {E}merging from a
  {N}on-{H}ermitian {F}latband of {P}hotonic {Z}ero {M}odes},}\ }\href@noop {}
  {\bibfield  {journal} {\bibinfo  {journal} {Phys. Rev. Lett.}\ }\textbf
  {\bibinfo {volume} {120}},\ \bibinfo {pages} {093901} (\bibinfo {year}
  {2018})}\BibitemShut {NoStop}%
\bibitem [{\citenamefont {Konotop}\ and\ \citenamefont
  {Zezyulin}(2018)}]{Konotop-18}%
  \BibitemOpen
  \bibfield  {author} {\bibinfo {author} {\bibfnamefont {V.~V.}\ \bibnamefont
  {Konotop}}\ and\ \bibinfo {author} {\bibfnamefont {D.~A.}\ \bibnamefont
  {Zezyulin}},\ }\bibfield  {title} {\enquote {\bibinfo {title} {{O}dd-{T}ime
  {R}eversal $\mathcal{PT}$ {S}ymmetry {I}nduced by an
  {A}nti-$\mathcal{PT}$-{S}ymmetric {M}edium},}\ }\href@noop {} {\bibfield
  {journal} {\bibinfo  {journal} {Phys. Rev. Lett.}\ }\textbf {\bibinfo
  {volume} {120}},\ \bibinfo {pages} {123902} (\bibinfo {year}
  {2018})}\BibitemShut {NoStop}%
\bibitem [{\citenamefont {Quijandr\'ia}\ \emph {et~al.}(2018)\citenamefont
  {Quijandr\'ia}, \citenamefont {Naether}, \citenamefont {\c{S}. K.~\"Ozdemir},
  \citenamefont {Nori},\ and\ \citenamefont {Zueco}}]{Quijandria-18}%
  \BibitemOpen
  \bibfield  {author} {\bibinfo {author} {\bibfnamefont {F.}~\bibnamefont
  {Quijandr\'ia}}, \bibinfo {author} {\bibfnamefont {U.}~\bibnamefont
  {Naether}}, \bibinfo {author} {\bibnamefont {\c{S}. K.~\"Ozdemir}}, \bibinfo
  {author} {\bibfnamefont {F.}~\bibnamefont {Nori}}, \ and\ \bibinfo {author}
  {\bibfnamefont {D.}~\bibnamefont {Zueco}},\ }\bibfield  {title} {\enquote
  {\bibinfo {title} {$\mathcal{PT}$-{S}ymmetric {C}ircuit-{QED}},}\ }\href@noop
  {} {\bibfield  {journal} {\bibinfo  {journal} {Phys. Rev. A}\ }\textbf
  {\bibinfo {volume} {97}},\ \bibinfo {pages} {053846} (\bibinfo {year}
  {2018})}\BibitemShut {NoStop}%
\bibitem [{\citenamefont {Louren\c{c}o}\ \emph {et~al.}(2018)\citenamefont
  {Louren\c{c}o}, \citenamefont {Eneias},\ and\ \citenamefont
  {Pereira}}]{Lourenco-18}%
  \BibitemOpen
  \bibfield  {author} {\bibinfo {author} {\bibfnamefont {J.~A.~S.}\
  \bibnamefont {Louren\c{c}o}}, \bibinfo {author} {\bibfnamefont {R.~L.}\
  \bibnamefont {Eneias}}, \ and\ \bibinfo {author} {\bibfnamefont {R.~G.}\
  \bibnamefont {Pereira}},\ }\bibfield  {title} {\enquote {\bibinfo {title}
  {{K}ondo {E}ffect in a $\mathcal{PT}$-{S}ymmetric {N}on-{H}ermitian
  {H}amiltonian},}\ }\href@noop {} {\bibfield  {journal} {\bibinfo  {journal}
  {Phys. Rev. B}\ }\textbf {\bibinfo {volume} {98}},\ \bibinfo {pages} {085126}
  (\bibinfo {year} {2018})}\BibitemShut {NoStop}%
\bibitem [{\citenamefont {Lau}\ and\ \citenamefont {Clerk}(2018)}]{Lau-18}%
  \BibitemOpen
  \bibfield  {author} {\bibinfo {author} {\bibfnamefont {H.-K.}\ \bibnamefont
  {Lau}}\ and\ \bibinfo {author} {\bibfnamefont {A.~A.}\ \bibnamefont
  {Clerk}},\ }\bibfield  {title} {\enquote {\bibinfo {title} {{F}undamental
  {L}imits and {N}on-{R}eciprocal {A}pproaches in {N}on-{H}ermitian {Q}uantum
  {S}ensing},}\ }\href@noop {} {\bibfield  {journal} {\bibinfo  {journal} {Nat.
  Commun.}\ }\textbf {\bibinfo {volume} {9}},\ \bibinfo {pages} {4320}
  (\bibinfo {year} {2018})}\BibitemShut {NoStop}%
\bibitem [{\citenamefont {Nakagawa}\ \emph {et~al.}(2018)\citenamefont
  {Nakagawa}, \citenamefont {Kawakami},\ and\ \citenamefont
  {Ueda}}]{Nakagawa-18}%
  \BibitemOpen
  \bibfield  {author} {\bibinfo {author} {\bibfnamefont {M.}~\bibnamefont
  {Nakagawa}}, \bibinfo {author} {\bibfnamefont {N.}~\bibnamefont {Kawakami}},
  \ and\ \bibinfo {author} {\bibfnamefont {M.}~\bibnamefont {Ueda}},\
  }\bibfield  {title} {\enquote {\bibinfo {title} {{Non-Hermitian Kondo Effect
  in Ultracold Alkaline-Earth Atoms}},}\ }\href@noop {} {\bibfield  {journal}
  {\bibinfo  {journal} {Phys. Rev. Lett.}\ }\textbf {\bibinfo {volume} {121}},\
  \bibinfo {pages} {203001} (\bibinfo {year} {2018})}\BibitemShut {NoStop}%
\bibitem [{\citenamefont {D\'ora}\ \emph {et~al.}(2019)\citenamefont {D\'ora},
  \citenamefont {Heyl},\ and\ \citenamefont {Moessner}}]{Dora-19}%
  \BibitemOpen
  \bibfield  {author} {\bibinfo {author} {\bibfnamefont {B.}~\bibnamefont
  {D\'ora}}, \bibinfo {author} {\bibfnamefont {M.}~\bibnamefont {Heyl}}, \ and\
  \bibinfo {author} {\bibfnamefont {R.}~\bibnamefont {Moessner}},\ }\bibfield
  {title} {\enquote {\bibinfo {title} {{The Kibble-Zurek Mechanism at
  Exceptional Points}},}\ }\href@noop {} {\bibfield  {journal} {\bibinfo
  {journal} {Nat. Commun.}\ }\textbf {\bibinfo {volume} {10}},\ \bibinfo
  {pages} {2254} (\bibinfo {year} {2019})}\BibitemShut {NoStop}%
\bibitem [{\citenamefont {Shibata}\ and\ \citenamefont
  {Katsura}(2019{\natexlab{a}})}]{Shibata-19-Kitaev}%
  \BibitemOpen
  \bibfield  {author} {\bibinfo {author} {\bibfnamefont {N.}~\bibnamefont
  {Shibata}}\ and\ \bibinfo {author} {\bibfnamefont {H.}~\bibnamefont
  {Katsura}},\ }\bibfield  {title} {\enquote {\bibinfo {title} {{Dissipative
  Spin Chain as a Non-Hermitian Kitaev Ladder}},}\ }\href@noop {} {\bibfield
  {journal} {\bibinfo  {journal} {Phys. Rev. B}\ }\textbf {\bibinfo {volume}
  {99}},\ \bibinfo {pages} {174303} (\bibinfo {year}
  {2019}{\natexlab{a}})}\BibitemShut {NoStop}%
\bibitem [{\citenamefont {Shibata}\ and\ \citenamefont
  {Katsura}(2019{\natexlab{b}})}]{Shibata-19-AT}%
  \BibitemOpen
  \bibfield  {author} {\bibinfo {author} {\bibfnamefont {N.}~\bibnamefont
  {Shibata}}\ and\ \bibinfo {author} {\bibfnamefont {H.}~\bibnamefont
  {Katsura}},\ }\bibfield  {title} {\enquote {\bibinfo {title} {{Dissipative
  Quantum Ising Chain as a Non-Hermitian Ashkin-Teller Model}},}\ }\href@noop
  {} {\bibfield  {journal} {\bibinfo  {journal} {{Phys. Rev. B}}\ }\textbf
  {\bibinfo {volume} {99}},\ \bibinfo {pages} {224432} (\bibinfo {year}
  {2019}{\natexlab{b}})}\BibitemShut {NoStop}%
\bibitem [{\citenamefont {Guo}\ \emph {et~al.}(2009)\citenamefont {Guo},
  \citenamefont {Salamo}, \citenamefont {Duchesne}, \citenamefont {Morandotti},
  \citenamefont {Volatier-Ravat}, \citenamefont {Aimez}, \citenamefont
  {Siviloglou},\ and\ \citenamefont {Christodoulides}}]{Guo-09}%
  \BibitemOpen
  \bibfield  {author} {\bibinfo {author} {\bibfnamefont {A.}~\bibnamefont
  {Guo}}, \bibinfo {author} {\bibfnamefont {G.~J.}\ \bibnamefont {Salamo}},
  \bibinfo {author} {\bibfnamefont {D.}~\bibnamefont {Duchesne}}, \bibinfo
  {author} {\bibfnamefont {R.}~\bibnamefont {Morandotti}}, \bibinfo {author}
  {\bibfnamefont {M.}~\bibnamefont {Volatier-Ravat}}, \bibinfo {author}
  {\bibfnamefont {V.}~\bibnamefont {Aimez}}, \bibinfo {author} {\bibfnamefont
  {G.~A.}\ \bibnamefont {Siviloglou}}, \ and\ \bibinfo {author} {\bibfnamefont
  {D.~N.}\ \bibnamefont {Christodoulides}},\ }\bibfield  {title} {\enquote
  {\bibinfo {title} {{O}bservation of $\mathcal{PT}$-{S}ymmetry {B}reaking in
  {C}omplex {O}ptical {P}otentials},}\ }\href@noop {} {\bibfield  {journal}
  {\bibinfo  {journal} {Phys. Rev. Lett.}\ }\textbf {\bibinfo {volume} {103}},\
  \bibinfo {pages} {093902} (\bibinfo {year} {2009})}\BibitemShut {NoStop}%
\bibitem [{\citenamefont {R\"uter}\ \emph {et~al.}(2010)\citenamefont
  {R\"uter}, \citenamefont {Makris}, \citenamefont {El-Ganainy}, \citenamefont
  {Christodoulides}, \citenamefont {Segev},\ and\ \citenamefont
  {Kip}}]{Ruter-10}%
  \BibitemOpen
  \bibfield  {author} {\bibinfo {author} {\bibfnamefont {C.~E.}\ \bibnamefont
  {R\"uter}}, \bibinfo {author} {\bibfnamefont {K.~G.}\ \bibnamefont {Makris}},
  \bibinfo {author} {\bibfnamefont {R.}~\bibnamefont {El-Ganainy}}, \bibinfo
  {author} {\bibfnamefont {D.~N.}\ \bibnamefont {Christodoulides}}, \bibinfo
  {author} {\bibfnamefont {M.}~\bibnamefont {Segev}}, \ and\ \bibinfo {author}
  {\bibfnamefont {D.}~\bibnamefont {Kip}},\ }\bibfield  {title} {\enquote
  {\bibinfo {title} {{O}bservation of {P}arity-{T}ime {S}ymmetry in
  {O}ptics},}\ }\href@noop {} {\bibfield  {journal} {\bibinfo  {journal} {Nat.
  Phys.}\ }\textbf {\bibinfo {volume} {6}},\ \bibinfo {pages} {192} (\bibinfo
  {year} {2010})}\BibitemShut {NoStop}%
\bibitem [{\citenamefont {Schindler}\ \emph {et~al.}(2011)\citenamefont
  {Schindler}, \citenamefont {Li}, \citenamefont {Zheng}, \citenamefont
  {Ellis},\ and\ \citenamefont {Kottos}}]{Schindler-11}%
  \BibitemOpen
  \bibfield  {author} {\bibinfo {author} {\bibfnamefont {J.}~\bibnamefont
  {Schindler}}, \bibinfo {author} {\bibfnamefont {A.}~\bibnamefont {Li}},
  \bibinfo {author} {\bibfnamefont {M.~C.}\ \bibnamefont {Zheng}}, \bibinfo
  {author} {\bibfnamefont {F.~M.}\ \bibnamefont {Ellis}}, \ and\ \bibinfo
  {author} {\bibfnamefont {T.}~\bibnamefont {Kottos}},\ }\bibfield  {title}
  {\enquote {\bibinfo {title} {{Experimental Study of Active \textit{LRC}
  Circuits with $\mathcal{PT}$ Symmetries}},}\ }\href@noop {} {\bibfield
  {journal} {\bibinfo  {journal} {Phys. Rev. A}\ }\textbf {\bibinfo {volume}
  {84}},\ \bibinfo {pages} {040101(R)} (\bibinfo {year} {2011})}\BibitemShut
  {NoStop}%
\bibitem [{\citenamefont {Regensburger}\ \emph {et~al.}(2012)\citenamefont
  {Regensburger}, \citenamefont {Bersch}, \citenamefont {Miri}, \citenamefont
  {Onishchukov}, \citenamefont {Christodoulides},\ and\ \citenamefont
  {Peschel}}]{Regensburger-12}%
  \BibitemOpen
  \bibfield  {author} {\bibinfo {author} {\bibfnamefont {A.}~\bibnamefont
  {Regensburger}}, \bibinfo {author} {\bibfnamefont {C.}~\bibnamefont
  {Bersch}}, \bibinfo {author} {\bibfnamefont {M.-A.}\ \bibnamefont {Miri}},
  \bibinfo {author} {\bibfnamefont {G.}~\bibnamefont {Onishchukov}}, \bibinfo
  {author} {\bibfnamefont {D.~N.}\ \bibnamefont {Christodoulides}}, \ and\
  \bibinfo {author} {\bibfnamefont {U.}~\bibnamefont {Peschel}},\ }\bibfield
  {title} {\enquote {\bibinfo {title} {{P}arity-{T}ime {S}ynthetic {P}hotonic
  {L}attices},}\ }\href@noop {} {\bibfield  {journal} {\bibinfo  {journal}
  {Nature}\ }\textbf {\bibinfo {volume} {488}},\ \bibinfo {pages} {167}
  (\bibinfo {year} {2012})}\BibitemShut {NoStop}%
\bibitem [{\citenamefont {Feng}\ \emph {et~al.}(2013)\citenamefont {Feng},
  \citenamefont {Xu}, \citenamefont {Fegadolli}, \citenamefont {Lu},
  \citenamefont {Oliveira}, \citenamefont {Almeida}, \citenamefont {Chen},\
  and\ \citenamefont {Scherer}}]{Feng-13}%
  \BibitemOpen
  \bibfield  {author} {\bibinfo {author} {\bibfnamefont {L.}~\bibnamefont
  {Feng}}, \bibinfo {author} {\bibfnamefont {Y.-L.}\ \bibnamefont {Xu}},
  \bibinfo {author} {\bibfnamefont {W.~S.}\ \bibnamefont {Fegadolli}}, \bibinfo
  {author} {\bibfnamefont {M.-H.}\ \bibnamefont {Lu}}, \bibinfo {author}
  {\bibfnamefont {J.~E.~B.}\ \bibnamefont {Oliveira}}, \bibinfo {author}
  {\bibfnamefont {V.~R.}\ \bibnamefont {Almeida}}, \bibinfo {author}
  {\bibfnamefont {Y.-F.}\ \bibnamefont {Chen}}, \ and\ \bibinfo {author}
  {\bibfnamefont {A.}~\bibnamefont {Scherer}},\ }\bibfield  {title} {\enquote
  {\bibinfo {title} {{E}xperimental {D}emonstration of a {U}nidirectional
  {R}eflectionless {P}arity-{T}ime {M}etamaterial at {O}ptical
  {F}requencies},}\ }\href@noop {} {\bibfield  {journal} {\bibinfo  {journal}
  {Nat. Mater.}\ }\textbf {\bibinfo {volume} {12}},\ \bibinfo {pages} {108}
  (\bibinfo {year} {2013})}\BibitemShut {NoStop}%
\bibitem [{\citenamefont {Bender}\ \emph {et~al.}(2013)\citenamefont {Bender},
  \citenamefont {Berntson}, \citenamefont {Parker},\ and\ \citenamefont
  {Samuel}}]{Bender-13}%
  \BibitemOpen
  \bibfield  {author} {\bibinfo {author} {\bibfnamefont {C.~M.}\ \bibnamefont
  {Bender}}, \bibinfo {author} {\bibfnamefont {B.~K.}\ \bibnamefont
  {Berntson}}, \bibinfo {author} {\bibfnamefont {D.}~\bibnamefont {Parker}}, \
  and\ \bibinfo {author} {\bibfnamefont {E.}~\bibnamefont {Samuel}},\
  }\bibfield  {title} {\enquote {\bibinfo {title} {{O}bservation of
  $\mathcal{PT}$ {P}hase {T}ransition in a {S}imple {M}echanical {S}ystem},}\
  }\href@noop {} {\bibfield  {journal} {\bibinfo  {journal} {Am. J. Phys.}\
  }\textbf {\bibinfo {volume} {81}},\ \bibinfo {pages} {173} (\bibinfo {year}
  {2013})}\BibitemShut {NoStop}%
\bibitem [{\citenamefont {Peng}\ \emph
  {et~al.}(2014{\natexlab{a}})\citenamefont {Peng}, \citenamefont {\c{S}.
  K.~\"Ozdemir}, \citenamefont {Lei}, \citenamefont {Monifi}, \citenamefont
  {Gianfreda}, \citenamefont {Long}, \citenamefont {Fan}, \citenamefont {Nori},
  \citenamefont {Bender},\ and\ \citenamefont {Yang}}]{Peng-14-NP}%
  \BibitemOpen
  \bibfield  {author} {\bibinfo {author} {\bibfnamefont {B.}~\bibnamefont
  {Peng}}, \bibinfo {author} {\bibnamefont {\c{S}. K.~\"Ozdemir}}, \bibinfo
  {author} {\bibfnamefont {F.}~\bibnamefont {Lei}}, \bibinfo {author}
  {\bibfnamefont {F.}~\bibnamefont {Monifi}}, \bibinfo {author} {\bibfnamefont
  {M.}~\bibnamefont {Gianfreda}}, \bibinfo {author} {\bibfnamefont {G.~L.}\
  \bibnamefont {Long}}, \bibinfo {author} {\bibfnamefont {S.}~\bibnamefont
  {Fan}}, \bibinfo {author} {\bibfnamefont {F.}~\bibnamefont {Nori}}, \bibinfo
  {author} {\bibfnamefont {C.~M.}\ \bibnamefont {Bender}}, \ and\ \bibinfo
  {author} {\bibfnamefont {L.}~\bibnamefont {Yang}},\ }\bibfield  {title}
  {\enquote {\bibinfo {title} {{P}arity-{T}ime-{S}ymmetric
  {W}hispering-{G}allery {M}icrocavities},}\ }\href@noop {} {\bibfield
  {journal} {\bibinfo  {journal} {Nat. Phys.}\ }\textbf {\bibinfo {volume}
  {10}},\ \bibinfo {pages} {394} (\bibinfo {year}
  {2014}{\natexlab{a}})}\BibitemShut {NoStop}%
\bibitem [{\citenamefont {Peng}\ \emph
  {et~al.}(2014{\natexlab{b}})\citenamefont {Peng}, \citenamefont {\c{S}.
  K.~\"Ozdemir}, \citenamefont {Rotter}, \citenamefont {Yilmaz}, \citenamefont
  {Liertzer}, \citenamefont {Monifi}, \citenamefont {Bender}, \citenamefont
  {Nori},\ and\ \citenamefont {Yang}}]{Peng-14-S}%
  \BibitemOpen
  \bibfield  {author} {\bibinfo {author} {\bibfnamefont {B.}~\bibnamefont
  {Peng}}, \bibinfo {author} {\bibnamefont {\c{S}. K.~\"Ozdemir}}, \bibinfo
  {author} {\bibfnamefont {S.}~\bibnamefont {Rotter}}, \bibinfo {author}
  {\bibfnamefont {H.}~\bibnamefont {Yilmaz}}, \bibinfo {author} {\bibfnamefont
  {M.}~\bibnamefont {Liertzer}}, \bibinfo {author} {\bibfnamefont
  {F.}~\bibnamefont {Monifi}}, \bibinfo {author} {\bibfnamefont {C.~M.}\
  \bibnamefont {Bender}}, \bibinfo {author} {\bibfnamefont {F.}~\bibnamefont
  {Nori}}, \ and\ \bibinfo {author} {\bibfnamefont {L.}~\bibnamefont {Yang}},\
  }\bibfield  {title} {\enquote {\bibinfo {title} {{L}oss-{I}nduced
  {S}uppression and {R}evival of {L}asing},}\ }\href@noop {} {\bibfield
  {journal} {\bibinfo  {journal} {Science}\ }\textbf {\bibinfo {volume}
  {346}},\ \bibinfo {pages} {328} (\bibinfo {year}
  {2014}{\natexlab{b}})}\BibitemShut {NoStop}%
\bibitem [{\citenamefont {Feng}\ \emph {et~al.}(2014)\citenamefont {Feng},
  \citenamefont {Wong}, \citenamefont {Ma}, \citenamefont {Wang},\ and\
  \citenamefont {Zhang}}]{Feng-14}%
  \BibitemOpen
  \bibfield  {author} {\bibinfo {author} {\bibfnamefont {L.}~\bibnamefont
  {Feng}}, \bibinfo {author} {\bibfnamefont {Z.~J.}\ \bibnamefont {Wong}},
  \bibinfo {author} {\bibfnamefont {R.-M.}\ \bibnamefont {Ma}}, \bibinfo
  {author} {\bibfnamefont {Y.}~\bibnamefont {Wang}}, \ and\ \bibinfo {author}
  {\bibfnamefont {X.}~\bibnamefont {Zhang}},\ }\bibfield  {title} {\enquote
  {\bibinfo {title} {{S}ingle-{M}ode {L}aser by {P}arity-{T}ime {S}ymmetry
  {B}reaking},}\ }\href@noop {} {\bibfield  {journal} {\bibinfo  {journal}
  {Science}\ }\textbf {\bibinfo {volume} {346}},\ \bibinfo {pages} {972}
  (\bibinfo {year} {2014})}\BibitemShut {NoStop}%
\bibitem [{\citenamefont {Hodaei}\ \emph {et~al.}(2014)\citenamefont {Hodaei},
  \citenamefont {Miri}, \citenamefont {Heinrich}, \citenamefont
  {Christodoulides},\ and\ \citenamefont {Khajavikhan}}]{Hodaei-14}%
  \BibitemOpen
  \bibfield  {author} {\bibinfo {author} {\bibfnamefont {H.}~\bibnamefont
  {Hodaei}}, \bibinfo {author} {\bibfnamefont {M.-A.}\ \bibnamefont {Miri}},
  \bibinfo {author} {\bibfnamefont {M.}~\bibnamefont {Heinrich}}, \bibinfo
  {author} {\bibfnamefont {D.~N.}\ \bibnamefont {Christodoulides}}, \ and\
  \bibinfo {author} {\bibfnamefont {M.}~\bibnamefont {Khajavikhan}},\
  }\bibfield  {title} {\enquote {\bibinfo {title} {{P}arity-{T}ime-{S}ymmetric
  {M}icroring {L}asers},}\ }\href@noop {} {\bibfield  {journal} {\bibinfo
  {journal} {Science}\ }\textbf {\bibinfo {volume} {346}},\ \bibinfo {pages}
  {975} (\bibinfo {year} {2014})}\BibitemShut {NoStop}%
\bibitem [{\citenamefont {Fleury}\ \emph {et~al.}(2015)\citenamefont {Fleury},
  \citenamefont {Sounas},\ and\ \citenamefont {Al\`u}}]{Fleury-15}%
  \BibitemOpen
  \bibfield  {author} {\bibinfo {author} {\bibfnamefont {R.}~\bibnamefont
  {Fleury}}, \bibinfo {author} {\bibfnamefont {D.}~\bibnamefont {Sounas}}, \
  and\ \bibinfo {author} {\bibfnamefont {A.}~\bibnamefont {Al\`u}},\ }\bibfield
   {title} {\enquote {\bibinfo {title} {{A}n {I}nvisible {A}coustic {S}ensor
  {B}ased on {P}arity-{T}ime-{S}ymmetry},}\ }\href@noop {} {\bibfield
  {journal} {\bibinfo  {journal} {Nat. Commun.}\ }\textbf {\bibinfo {volume}
  {6}},\ \bibinfo {pages} {5905} (\bibinfo {year} {2015})}\BibitemShut
  {NoStop}%
\bibitem [{\citenamefont {Gao}\ \emph {et~al.}(2015)\citenamefont {Gao},
  \citenamefont {Estrecho}, \citenamefont {Bliokh}, \citenamefont {Liew},
  \citenamefont {Fraser}, \citenamefont {Brodbeck}, \citenamefont {Kemp},
  \citenamefont {Schneider}, \citenamefont {H\"ofling}, \citenamefont
  {Yamamoto}, \citenamefont {Nori}, \citenamefont {Kivshar}, \citenamefont
  {Truscott}, \citenamefont {Dall},\ and\ \citenamefont
  {Ostrovskaya}}]{Gao-15}%
  \BibitemOpen
  \bibfield  {author} {\bibinfo {author} {\bibfnamefont {T.}~\bibnamefont
  {Gao}}, \bibinfo {author} {\bibfnamefont {E.}~\bibnamefont {Estrecho}},
  \bibinfo {author} {\bibfnamefont {K.~Y.}\ \bibnamefont {Bliokh}}, \bibinfo
  {author} {\bibfnamefont {T.~C.~H.}\ \bibnamefont {Liew}}, \bibinfo {author}
  {\bibfnamefont {M.~D.}\ \bibnamefont {Fraser}}, \bibinfo {author}
  {\bibfnamefont {S.}~\bibnamefont {Brodbeck}}, \bibinfo {author}
  {\bibfnamefont {M.}~\bibnamefont {Kemp}}, \bibinfo {author} {\bibfnamefont
  {C.}~\bibnamefont {Schneider}}, \bibinfo {author} {\bibfnamefont
  {S.}~\bibnamefont {H\"ofling}}, \bibinfo {author} {\bibfnamefont
  {Y.}~\bibnamefont {Yamamoto}}, \bibinfo {author} {\bibfnamefont
  {F.}~\bibnamefont {Nori}}, \bibinfo {author} {\bibfnamefont {Y.~S.}\
  \bibnamefont {Kivshar}}, \bibinfo {author} {\bibfnamefont {A.~G.}\
  \bibnamefont {Truscott}}, \bibinfo {author} {\bibfnamefont {R.~G.}\
  \bibnamefont {Dall}}, \ and\ \bibinfo {author} {\bibfnamefont {E.~A.}\
  \bibnamefont {Ostrovskaya}},\ }\bibfield  {title} {\enquote {\bibinfo {title}
  {{O}bservation of {N}on-{H}ermitian {D}egeneracies in a {C}haotic
  {E}xciton-{P}olariton {B}illiard},}\ }\href@noop {} {\bibfield  {journal}
  {\bibinfo  {journal} {Nature}\ }\textbf {\bibinfo {volume} {526}},\ \bibinfo
  {pages} {554} (\bibinfo {year} {2015})}\BibitemShut {NoStop}%
\bibitem [{\citenamefont {Peng}\ \emph
  {et~al.}(2016{\natexlab{a}})\citenamefont {Peng}, \citenamefont
  {{\"O}zdemir}, \citenamefont {Liertzer}, \citenamefont {Chen}, \citenamefont
  {Kramer}, \citenamefont {Y{\i}lmaz}, \citenamefont {Wiersig}, \citenamefont
  {Rotter},\ and\ \citenamefont {Yang}}]{Peng-16-PNAS}%
  \BibitemOpen
  \bibfield  {author} {\bibinfo {author} {\bibfnamefont {B.}~\bibnamefont
  {Peng}}, \bibinfo {author} {\bibfnamefont {{\c S}.~K.}\ \bibnamefont
  {{\"O}zdemir}}, \bibinfo {author} {\bibfnamefont {M.}~\bibnamefont
  {Liertzer}}, \bibinfo {author} {\bibfnamefont {W.}~\bibnamefont {Chen}},
  \bibinfo {author} {\bibfnamefont {J.}~\bibnamefont {Kramer}}, \bibinfo
  {author} {\bibfnamefont {H.}~\bibnamefont {Y{\i}lmaz}}, \bibinfo {author}
  {\bibfnamefont {J.}~\bibnamefont {Wiersig}}, \bibinfo {author} {\bibfnamefont
  {S.}~\bibnamefont {Rotter}}, \ and\ \bibinfo {author} {\bibfnamefont
  {L.}~\bibnamefont {Yang}},\ }\bibfield  {title} {\enquote {\bibinfo {title}
  {{Chiral Modes and Directional Lasing at Exceptional Points}},}\ }\href@noop
  {} {\bibfield  {journal} {\bibinfo  {journal} {Proc. Natl. Acad. Sci.
  U.S.A.}\ }\textbf {\bibinfo {volume} {113}},\ \bibinfo {pages} {6845}
  (\bibinfo {year} {2016}{\natexlab{a}})}\BibitemShut {NoStop}%
\bibitem [{\citenamefont {Miao}\ \emph {et~al.}(2016)\citenamefont {Miao},
  \citenamefont {Zhang}, \citenamefont {Sun}, \citenamefont {Walasik},
  \citenamefont {Longhi}, \citenamefont {Litchinitser},\ and\ \citenamefont
  {Feng}}]{Miao-16}%
  \BibitemOpen
  \bibfield  {author} {\bibinfo {author} {\bibfnamefont {P.}~\bibnamefont
  {Miao}}, \bibinfo {author} {\bibfnamefont {Z.}~\bibnamefont {Zhang}},
  \bibinfo {author} {\bibfnamefont {J.}~\bibnamefont {Sun}}, \bibinfo {author}
  {\bibfnamefont {W.}~\bibnamefont {Walasik}}, \bibinfo {author} {\bibfnamefont
  {S.}~\bibnamefont {Longhi}}, \bibinfo {author} {\bibfnamefont {N.~M.}\
  \bibnamefont {Litchinitser}}, \ and\ \bibinfo {author} {\bibfnamefont
  {L.}~\bibnamefont {Feng}},\ }\bibfield  {title} {\enquote {\bibinfo {title}
  {{O}rbital {A}ngular {M}omentum {M}icrolaser},}\ }\href@noop {} {\bibfield
  {journal} {\bibinfo  {journal} {Science}\ }\textbf {\bibinfo {volume}
  {353}},\ \bibinfo {pages} {464} (\bibinfo {year} {2016})}\BibitemShut
  {NoStop}%
\bibitem [{\citenamefont {Peng}\ \emph
  {et~al.}(2016{\natexlab{b}})\citenamefont {Peng}, \citenamefont {Cao},
  \citenamefont {Shen}, \citenamefont {Qu}, \citenamefont {Wen}, \citenamefont
  {Jiang},\ and\ \citenamefont {Xiao}}]{Peng-16}%
  \BibitemOpen
  \bibfield  {author} {\bibinfo {author} {\bibfnamefont {P.}~\bibnamefont
  {Peng}}, \bibinfo {author} {\bibfnamefont {W.}~\bibnamefont {Cao}}, \bibinfo
  {author} {\bibfnamefont {C.}~\bibnamefont {Shen}}, \bibinfo {author}
  {\bibfnamefont {W.}~\bibnamefont {Qu}}, \bibinfo {author} {\bibfnamefont
  {J.}~\bibnamefont {Wen}}, \bibinfo {author} {\bibfnamefont {L.}~\bibnamefont
  {Jiang}}, \ and\ \bibinfo {author} {\bibfnamefont {Y.}~\bibnamefont {Xiao}},\
  }\bibfield  {title} {\enquote {\bibinfo {title} {{A}nti-{P}arity-{T}ime
  {S}ymmetry with {F}lying {A}toms},}\ }\href@noop {} {\bibfield  {journal}
  {\bibinfo  {journal} {Nat. Phys.}\ }\textbf {\bibinfo {volume} {12}},\
  \bibinfo {pages} {1139} (\bibinfo {year} {2016}{\natexlab{b}})}\BibitemShut
  {NoStop}%
\bibitem [{\citenamefont {Doppler}\ \emph {et~al.}(2016)\citenamefont
  {Doppler}, \citenamefont {Mailybaev}, \citenamefont {B\"ohm}, \citenamefont
  {Kuhl}, \citenamefont {Girschikm}, \citenamefont {Libisch}, \citenamefont
  {Milburn}, \citenamefont {Rabl}, \citenamefont {Moiseyev},\ and\
  \citenamefont {Rotter}}]{Doppler-16}%
  \BibitemOpen
  \bibfield  {author} {\bibinfo {author} {\bibfnamefont {J.}~\bibnamefont
  {Doppler}}, \bibinfo {author} {\bibfnamefont {A.~A.}\ \bibnamefont
  {Mailybaev}}, \bibinfo {author} {\bibfnamefont {J.}~\bibnamefont {B\"ohm}},
  \bibinfo {author} {\bibfnamefont {U.}~\bibnamefont {Kuhl}}, \bibinfo {author}
  {\bibfnamefont {A.}~\bibnamefont {Girschikm}}, \bibinfo {author}
  {\bibfnamefont {F.}~\bibnamefont {Libisch}}, \bibinfo {author} {\bibfnamefont
  {T.~J.}\ \bibnamefont {Milburn}}, \bibinfo {author} {\bibfnamefont
  {P.}~\bibnamefont {Rabl}}, \bibinfo {author} {\bibfnamefont {N.}~\bibnamefont
  {Moiseyev}}, \ and\ \bibinfo {author} {\bibfnamefont {S.}~\bibnamefont
  {Rotter}},\ }\bibfield  {title} {\enquote {\bibinfo {title} {{D}ynamically
  {E}ncircling an {E}xceptional {P}oint for {A}symmetric {M}ode {S}witching},}\
  }\href@noop {} {\bibfield  {journal} {\bibinfo  {journal} {Nature}\ }\textbf
  {\bibinfo {volume} {537}},\ \bibinfo {pages} {76} (\bibinfo {year}
  {2016})}\BibitemShut {NoStop}%
\bibitem [{\citenamefont {Xu}\ \emph {et~al.}(2016)\citenamefont {Xu},
  \citenamefont {Mason}, \citenamefont {Jiang},\ and\ \citenamefont
  {Harris}}]{Xu-16}%
  \BibitemOpen
  \bibfield  {author} {\bibinfo {author} {\bibfnamefont {H.}~\bibnamefont
  {Xu}}, \bibinfo {author} {\bibfnamefont {D.}~\bibnamefont {Mason}}, \bibinfo
  {author} {\bibfnamefont {L.}~\bibnamefont {Jiang}}, \ and\ \bibinfo {author}
  {\bibfnamefont {J.~G.~E.}\ \bibnamefont {Harris}},\ }\bibfield  {title}
  {\enquote {\bibinfo {title} {{T}opological {E}nergy {T}ransfer in an
  {O}ptomechanical {S}ystem with {E}xceptional {P}oints},}\ }\href@noop {}
  {\bibfield  {journal} {\bibinfo  {journal} {Nature}\ }\textbf {\bibinfo
  {volume} {537}},\ \bibinfo {pages} {80} (\bibinfo {year} {2016})}\BibitemShut
  {NoStop}%
\bibitem [{\citenamefont {Zhang}\ \emph {et~al.}(2016)\citenamefont {Zhang},
  \citenamefont {Zhang}, \citenamefont {Sheng}, \citenamefont {Yang},
  \citenamefont {Miri}, \citenamefont {Christodoulides}, \citenamefont {He},
  \citenamefont {Zhang},\ and\ \citenamefont {Xiao}}]{Zhang-16}%
  \BibitemOpen
  \bibfield  {author} {\bibinfo {author} {\bibfnamefont {Z.}~\bibnamefont
  {Zhang}}, \bibinfo {author} {\bibfnamefont {Y.}~\bibnamefont {Zhang}},
  \bibinfo {author} {\bibfnamefont {J.}~\bibnamefont {Sheng}}, \bibinfo
  {author} {\bibfnamefont {L.}~\bibnamefont {Yang}}, \bibinfo {author}
  {\bibfnamefont {M.-A.}\ \bibnamefont {Miri}}, \bibinfo {author}
  {\bibfnamefont {D.~N.}\ \bibnamefont {Christodoulides}}, \bibinfo {author}
  {\bibfnamefont {B.}~\bibnamefont {He}}, \bibinfo {author} {\bibfnamefont
  {Y.}~\bibnamefont {Zhang}}, \ and\ \bibinfo {author} {\bibfnamefont
  {M.}~\bibnamefont {Xiao}},\ }\bibfield  {title} {\enquote {\bibinfo {title}
  {{O}bservation of {P}arity-{T}ime {S}ymmetry in {O}ptically {I}nduced
  {A}tomic {L}attices},}\ }\href@noop {} {\bibfield  {journal} {\bibinfo
  {journal} {Phys. Rev. Lett.}\ }\textbf {\bibinfo {volume} {117}},\ \bibinfo
  {pages} {123601} (\bibinfo {year} {2016})}\BibitemShut {NoStop}%
\bibitem [{\citenamefont {Assawaworrarit}\ \emph {et~al.}(2017)\citenamefont
  {Assawaworrarit}, \citenamefont {Yu},\ and\ \citenamefont
  {Fan}}]{Assawaworrarit-17}%
  \BibitemOpen
  \bibfield  {author} {\bibinfo {author} {\bibfnamefont {S.}~\bibnamefont
  {Assawaworrarit}}, \bibinfo {author} {\bibfnamefont {X.}~\bibnamefont {Yu}},
  \ and\ \bibinfo {author} {\bibfnamefont {S.}~\bibnamefont {Fan}},\ }\bibfield
   {title} {\enquote {\bibinfo {title} {{R}obust {W}ireless {P}ower {T}ransfer
  {U}sing a {N}onlinear {P}arity-{T}ime-{S}ymmetric {C}ircuit},}\ }\href@noop
  {} {\bibfield  {journal} {\bibinfo  {journal} {Nature}\ }\textbf {\bibinfo
  {volume} {546}},\ \bibinfo {pages} {387} (\bibinfo {year}
  {2017})}\BibitemShut {NoStop}%
\bibitem [{\citenamefont {Hodaei}\ \emph {et~al.}(2017)\citenamefont {Hodaei},
  \citenamefont {Hassan}, \citenamefont {Wittek}, \citenamefont
  {Garcia-Gracia}, \citenamefont {El-Ganainy}, \citenamefont
  {Christodoulides},\ and\ \citenamefont {Khajavikhan}}]{Hodaei-17}%
  \BibitemOpen
  \bibfield  {author} {\bibinfo {author} {\bibfnamefont {H.}~\bibnamefont
  {Hodaei}}, \bibinfo {author} {\bibfnamefont {A.~U.}\ \bibnamefont {Hassan}},
  \bibinfo {author} {\bibfnamefont {S.}~\bibnamefont {Wittek}}, \bibinfo
  {author} {\bibfnamefont {H.}~\bibnamefont {Garcia-Gracia}}, \bibinfo {author}
  {\bibfnamefont {R.}~\bibnamefont {El-Ganainy}}, \bibinfo {author}
  {\bibfnamefont {D.~N.}\ \bibnamefont {Christodoulides}}, \ and\ \bibinfo
  {author} {\bibfnamefont {M.}~\bibnamefont {Khajavikhan}},\ }\bibfield
  {title} {\enquote {\bibinfo {title} {{E}nhanced {S}ensitivity at
  {H}igher-{O}rder {E}xceptional {P}oints},}\ }\href@noop {} {\bibfield
  {journal} {\bibinfo  {journal} {Nature}\ }\textbf {\bibinfo {volume} {548}},\
  \bibinfo {pages} {187} (\bibinfo {year} {2017})}\BibitemShut {NoStop}%
\bibitem [{\citenamefont {Chen}\ \emph {et~al.}(2017)\citenamefont {Chen},
  \citenamefont {\c{S}. K.~\"Ozdemir}, \citenamefont {Zhao}, \citenamefont
  {Wiersig},\ and\ \citenamefont {Yang}}]{Chen-17}%
  \BibitemOpen
  \bibfield  {author} {\bibinfo {author} {\bibfnamefont {W.}~\bibnamefont
  {Chen}}, \bibinfo {author} {\bibnamefont {\c{S}. K.~\"Ozdemir}}, \bibinfo
  {author} {\bibfnamefont {G.}~\bibnamefont {Zhao}}, \bibinfo {author}
  {\bibfnamefont {J.}~\bibnamefont {Wiersig}}, \ and\ \bibinfo {author}
  {\bibfnamefont {L.}~\bibnamefont {Yang}},\ }\bibfield  {title} {\enquote
  {\bibinfo {title} {{E}xceptional {P}oints {E}nhance {S}ensing in an {O}ptical
  {M}icrocavity},}\ }\href@noop {} {\bibfield  {journal} {\bibinfo  {journal}
  {Nature}\ }\textbf {\bibinfo {volume} {548}},\ \bibinfo {pages} {192}
  (\bibinfo {year} {2017})}\BibitemShut {NoStop}%
\bibitem [{\citenamefont {Louren\c{c}o-Martins}\ \emph
  {et~al.}(2018)\citenamefont {Louren\c{c}o-Martins}, \citenamefont {Das},
  \citenamefont {Tizei}, \citenamefont {Weil},\ and\ \citenamefont
  {Kociak}}]{Martins-18}%
  \BibitemOpen
  \bibfield  {author} {\bibinfo {author} {\bibfnamefont {H.}~\bibnamefont
  {Louren\c{c}o-Martins}}, \bibinfo {author} {\bibfnamefont {P.}~\bibnamefont
  {Das}}, \bibinfo {author} {\bibfnamefont {L.~H.~G.}\ \bibnamefont {Tizei}},
  \bibinfo {author} {\bibfnamefont {R.}~\bibnamefont {Weil}}, \ and\ \bibinfo
  {author} {\bibfnamefont {M.}~\bibnamefont {Kociak}},\ }\bibfield  {title}
  {\enquote {\bibinfo {title} {{Self-Hybridization within Non-Hermitian
  Localized Plasmonic System}},}\ }\href@noop {} {\bibfield  {journal}
  {\bibinfo  {journal} {Nat. Phys.}\ }\textbf {\bibinfo {volume} {14}},\
  \bibinfo {pages} {360} (\bibinfo {year} {2018})}\BibitemShut {NoStop}%
\bibitem [{\citenamefont {Rivet}\ \emph {et~al.}(2018)\citenamefont {Rivet},
  \citenamefont {Brandst\"otter}, \citenamefont {Makris}, \citenamefont
  {Lissek}, \citenamefont {Rotter},\ and\ \citenamefont {Fleury}}]{Rivet-18}%
  \BibitemOpen
  \bibfield  {author} {\bibinfo {author} {\bibfnamefont {E.}~\bibnamefont
  {Rivet}}, \bibinfo {author} {\bibfnamefont {A.}~\bibnamefont
  {Brandst\"otter}}, \bibinfo {author} {\bibfnamefont {K.~G.}\ \bibnamefont
  {Makris}}, \bibinfo {author} {\bibfnamefont {H.}~\bibnamefont {Lissek}},
  \bibinfo {author} {\bibfnamefont {S.}~\bibnamefont {Rotter}}, \ and\ \bibinfo
  {author} {\bibfnamefont {R.}~\bibnamefont {Fleury}},\ }\bibfield  {title}
  {\enquote {\bibinfo {title} {{C}onstant-{P}ressure {S}ound {W}aves in
  {N}on-{H}ermitian {D}isordered {M}edia},}\ }\href@noop {} {\bibfield
  {journal} {\bibinfo  {journal} {Nat. Phys.}\ }\textbf {\bibinfo {volume}
  {14}},\ \bibinfo {pages} {942} (\bibinfo {year} {2018})}\BibitemShut
  {NoStop}%
\bibitem [{\citenamefont {M\"ullers}\ \emph {et~al.}(2018)\citenamefont
  {M\"ullers}, \citenamefont {Santra}, \citenamefont {Baals}, \citenamefont
  {Jiang}, \citenamefont {Benary}, \citenamefont {Labouvie}, \citenamefont
  {Zezyulin}, \citenamefont {Konotop},\ and\ \citenamefont {Ott}}]{Mullers-18}%
  \BibitemOpen
  \bibfield  {author} {\bibinfo {author} {\bibfnamefont {A.}~\bibnamefont
  {M\"ullers}}, \bibinfo {author} {\bibfnamefont {B.}~\bibnamefont {Santra}},
  \bibinfo {author} {\bibfnamefont {C.}~\bibnamefont {Baals}}, \bibinfo
  {author} {\bibfnamefont {J.}~\bibnamefont {Jiang}}, \bibinfo {author}
  {\bibfnamefont {J.}~\bibnamefont {Benary}}, \bibinfo {author} {\bibfnamefont
  {R.}~\bibnamefont {Labouvie}}, \bibinfo {author} {\bibfnamefont {D.~A.}\
  \bibnamefont {Zezyulin}}, \bibinfo {author} {\bibfnamefont {V.~V.}\
  \bibnamefont {Konotop}}, \ and\ \bibinfo {author} {\bibfnamefont
  {H.}~\bibnamefont {Ott}},\ }\bibfield  {title} {\enquote {\bibinfo {title}
  {{C}oherent {P}erfect {A}bsorption of {N}onlinear {M}atter {W}aves},}\
  }\href@noop {} {\bibfield  {journal} {\bibinfo  {journal} {Sci. Adv.}\
  }\textbf {\bibinfo {volume} {4}},\ \bibinfo {pages} {eaat6539} (\bibinfo
  {year} {2018})}\BibitemShut {NoStop}%
\bibitem [{\citenamefont {Yoon}\ \emph {et~al.}(2018)\citenamefont {Yoon},
  \citenamefont {Choi}, \citenamefont {Hahn}, \citenamefont {Kim},
  \citenamefont {Song}, \citenamefont {Yang}, \citenamefont {Lee},
  \citenamefont {Kim}, \citenamefont {Lee}, \citenamefont {Shin}, \citenamefont
  {Lee},\ and\ \citenamefont {Berini}}]{Yoon-18}%
  \BibitemOpen
  \bibfield  {author} {\bibinfo {author} {\bibfnamefont {J.~W.}\ \bibnamefont
  {Yoon}}, \bibinfo {author} {\bibfnamefont {Y.}~\bibnamefont {Choi}}, \bibinfo
  {author} {\bibfnamefont {C.}~\bibnamefont {Hahn}}, \bibinfo {author}
  {\bibfnamefont {G.}~\bibnamefont {Kim}}, \bibinfo {author} {\bibfnamefont
  {S.~H.}\ \bibnamefont {Song}}, \bibinfo {author} {\bibfnamefont {K.-Y.}\
  \bibnamefont {Yang}}, \bibinfo {author} {\bibfnamefont {J.~Y.}\ \bibnamefont
  {Lee}}, \bibinfo {author} {\bibfnamefont {Y.}~\bibnamefont {Kim}}, \bibinfo
  {author} {\bibfnamefont {C.~S.}\ \bibnamefont {Lee}}, \bibinfo {author}
  {\bibfnamefont {J.~K.}\ \bibnamefont {Shin}}, \bibinfo {author}
  {\bibfnamefont {H.-S.}\ \bibnamefont {Lee}}, \ and\ \bibinfo {author}
  {\bibfnamefont {P.}~\bibnamefont {Berini}},\ }\bibfield  {title} {\enquote
  {\bibinfo {title} {{T}ime-{A}symmetric {L}oop around an {E}xceptional {P}oint
  over the {F}ull {O}ptical {C}ommunications {B}and},}\ }\href@noop {}
  {\bibfield  {journal} {\bibinfo  {journal} {Nature}\ }\textbf {\bibinfo
  {volume} {562}},\ \bibinfo {pages} {86} (\bibinfo {year} {2018})}\BibitemShut
  {NoStop}%
\bibitem [{\citenamefont {Li}\ \emph {et~al.}(2019{\natexlab{a}})\citenamefont
  {Li}, \citenamefont {Harter}, \citenamefont {Liu}, \citenamefont {de~Melo},
  \citenamefont {Joglekar},\ and\ \citenamefont {Luo}}]{Joglekar-Luo-19}%
  \BibitemOpen
  \bibfield  {author} {\bibinfo {author} {\bibfnamefont {J.}~\bibnamefont
  {Li}}, \bibinfo {author} {\bibfnamefont {A.~K.}\ \bibnamefont {Harter}},
  \bibinfo {author} {\bibfnamefont {J.}~\bibnamefont {Liu}}, \bibinfo {author}
  {\bibfnamefont {L.}~\bibnamefont {de~Melo}}, \bibinfo {author} {\bibfnamefont
  {Y.~N.}\ \bibnamefont {Joglekar}}, \ and\ \bibinfo {author} {\bibfnamefont
  {L.}~\bibnamefont {Luo}},\ }\bibfield  {title} {\enquote {\bibinfo {title}
  {{Observation of Parity-Time Symmetry Breaking Transitions in a Dissipative
  Floquet System of Ultracold Atoms}},}\ }\href@noop {} {\bibfield  {journal}
  {\bibinfo  {journal} {Nat. Commun.}\ }\textbf {\bibinfo {volume} {10}},\
  \bibinfo {pages} {855} (\bibinfo {year} {2019}{\natexlab{a}})}\BibitemShut
  {NoStop}%
\bibitem [{\citenamefont {Li}\ \emph {et~al.}(2019{\natexlab{b}})\citenamefont
  {Li}, \citenamefont {Peng}, \citenamefont {Han}, \citenamefont {Miri},
  \citenamefont {Li}, \citenamefont {Xiao}, \citenamefont {Zhu}, \citenamefont
  {Zhao}, \citenamefont {Al\`u}, \citenamefont {Fan},\ and\ \citenamefont
  {Qiu}}]{Li-19}%
  \BibitemOpen
  \bibfield  {author} {\bibinfo {author} {\bibfnamefont {Y.}~\bibnamefont
  {Li}}, \bibinfo {author} {\bibfnamefont {Y.-G.}\ \bibnamefont {Peng}},
  \bibinfo {author} {\bibfnamefont {L.}~\bibnamefont {Han}}, \bibinfo {author}
  {\bibfnamefont {M.-A.}\ \bibnamefont {Miri}}, \bibinfo {author}
  {\bibfnamefont {W.}~\bibnamefont {Li}}, \bibinfo {author} {\bibfnamefont
  {M.}~\bibnamefont {Xiao}}, \bibinfo {author} {\bibfnamefont {X.-F.}\
  \bibnamefont {Zhu}}, \bibinfo {author} {\bibfnamefont {J.}~\bibnamefont
  {Zhao}}, \bibinfo {author} {\bibfnamefont {A.}~\bibnamefont {Al\`u}},
  \bibinfo {author} {\bibfnamefont {S.}~\bibnamefont {Fan}}, \ and\ \bibinfo
  {author} {\bibfnamefont {C.-W.}\ \bibnamefont {Qiu}},\ }\bibfield  {title}
  {\enquote {\bibinfo {title} {{Anti-Parity-Time Symmetry in Diffusive
  Systems}},}\ }\href@noop {} {\bibfield  {journal} {\bibinfo  {journal}
  {Science}\ }\textbf {\bibinfo {volume} {364}},\ \bibinfo {pages} {170}
  (\bibinfo {year} {2019}{\natexlab{b}})}\BibitemShut {NoStop}%
\bibitem [{\citenamefont {Wu}\ \emph {et~al.}(2019)\citenamefont {Wu},
  \citenamefont {Liu}, \citenamefont {Geng}, \citenamefont {Song},
  \citenamefont {Ye}, \citenamefont {Duan}, \citenamefont {Rong},\ and\
  \citenamefont {Du}}]{Wu-19}%
  \BibitemOpen
  \bibfield  {author} {\bibinfo {author} {\bibfnamefont {Y.}~\bibnamefont
  {Wu}}, \bibinfo {author} {\bibfnamefont {W.}~\bibnamefont {Liu}}, \bibinfo
  {author} {\bibfnamefont {J.}~\bibnamefont {Geng}}, \bibinfo {author}
  {\bibfnamefont {X.}~\bibnamefont {Song}}, \bibinfo {author} {\bibfnamefont
  {X.}~\bibnamefont {Ye}}, \bibinfo {author} {\bibfnamefont {C.-K.}\
  \bibnamefont {Duan}}, \bibinfo {author} {\bibfnamefont {X.}~\bibnamefont
  {Rong}}, \ and\ \bibinfo {author} {\bibfnamefont {J.}~\bibnamefont {Du}},\
  }\bibfield  {title} {\enquote {\bibinfo {title} {{Observation of Parity-Time
  Symmetry Breaking in a Single-Spin System}},}\ }\href@noop {} {\bibfield
  {journal} {\bibinfo  {journal} {Science}\ }\textbf {\bibinfo {volume}
  {364}},\ \bibinfo {pages} {878} (\bibinfo {year} {2019})}\BibitemShut
  {NoStop}%
\bibitem [{\citenamefont {Xiao}\ \emph {et~al.}()\citenamefont {Xiao},
  \citenamefont {Wang}, \citenamefont {Zhan}, \citenamefont {Bian},
  \citenamefont {Kawabata}, \citenamefont {Ueda}, \citenamefont {Yi},\ and\
  \citenamefont {Xue}}]{Xiao-18}%
  \BibitemOpen
  \bibfield  {author} {\bibinfo {author} {\bibfnamefont {L.}~\bibnamefont
  {Xiao}}, \bibinfo {author} {\bibfnamefont {K.}~\bibnamefont {Wang}}, \bibinfo
  {author} {\bibfnamefont {X.}~\bibnamefont {Zhan}}, \bibinfo {author}
  {\bibfnamefont {Z.}~\bibnamefont {Bian}}, \bibinfo {author} {\bibfnamefont
  {K.}~\bibnamefont {Kawabata}}, \bibinfo {author} {\bibfnamefont
  {M.}~\bibnamefont {Ueda}}, \bibinfo {author} {\bibfnamefont {W.}~\bibnamefont
  {Yi}}, \ and\ \bibinfo {author} {\bibfnamefont {P.}~\bibnamefont {Xue}},\
  }\href@noop {} {\enquote {\bibinfo {title} {{Observation of Critical
  Phenomena in Parity-Time-Symmetric Quantum Dynamics}},}\ }\bibinfo {note}
  {{arXiv: 1812.01213}}\BibitemShut {NoStop}%
\bibitem [{\citenamefont {Rudner}\ and\ \citenamefont
  {Levitov}(2009)}]{Rudner-09}%
  \BibitemOpen
  \bibfield  {author} {\bibinfo {author} {\bibfnamefont {M.~S.}\ \bibnamefont
  {Rudner}}\ and\ \bibinfo {author} {\bibfnamefont {L.~S.}\ \bibnamefont
  {Levitov}},\ }\bibfield  {title} {\enquote {\bibinfo {title} {{T}opological
  {T}ransition in a {N}on-{H}ermitian {Q}uantum {W}alk},}\ }\href@noop {}
  {\bibfield  {journal} {\bibinfo  {journal} {Phys. Rev. Lett.}\ }\textbf
  {\bibinfo {volume} {102}},\ \bibinfo {pages} {065703} (\bibinfo {year}
  {2009})}\BibitemShut {NoStop}%
\bibitem [{\citenamefont {Rudner}\ and\ \citenamefont
  {Levitov}(2010)}]{Rudner-10}%
  \BibitemOpen
  \bibfield  {author} {\bibinfo {author} {\bibfnamefont {M.~S.}\ \bibnamefont
  {Rudner}}\ and\ \bibinfo {author} {\bibfnamefont {L.~S.}\ \bibnamefont
  {Levitov}},\ }\bibfield  {title} {\enquote {\bibinfo {title} {{Phase
  Transitions in Dissipative Quantum Transport and Mesoscopic Nuclear Spin
  Pumping}},}\ }\href@noop {} {\bibfield  {journal} {\bibinfo  {journal} {Phys.
  Rev. B}\ }\textbf {\bibinfo {volume} {82}},\ \bibinfo {pages} {155418}
  (\bibinfo {year} {2010})}\BibitemShut {NoStop}%
\bibitem [{\citenamefont {Hu}\ and\ \citenamefont {Hughes}(2011)}]{Hu-11}%
  \BibitemOpen
  \bibfield  {author} {\bibinfo {author} {\bibfnamefont {Y.~C.}\ \bibnamefont
  {Hu}}\ and\ \bibinfo {author} {\bibfnamefont {T.~L.}\ \bibnamefont
  {Hughes}},\ }\bibfield  {title} {\enquote {\bibinfo {title} {{Absence of
  Topological Insulator Phases in Non-Hermitian \textit{PT}-Symmetric
  Hamiltonians}},}\ }\href@noop {} {\bibfield  {journal} {\bibinfo  {journal}
  {Phys. Rev. B}\ }\textbf {\bibinfo {volume} {84}},\ \bibinfo {pages} {153101}
  (\bibinfo {year} {2011})}\BibitemShut {NoStop}%
\bibitem [{\citenamefont {Esaki}\ \emph {et~al.}(2011)\citenamefont {Esaki},
  \citenamefont {Sato}, \citenamefont {Hasebe},\ and\ \citenamefont
  {Kohmoto}}]{Esaki-11}%
  \BibitemOpen
  \bibfield  {author} {\bibinfo {author} {\bibfnamefont {K.}~\bibnamefont
  {Esaki}}, \bibinfo {author} {\bibfnamefont {M.}~\bibnamefont {Sato}},
  \bibinfo {author} {\bibfnamefont {K.}~\bibnamefont {Hasebe}}, \ and\ \bibinfo
  {author} {\bibfnamefont {M.}~\bibnamefont {Kohmoto}},\ }\bibfield  {title}
  {\enquote {\bibinfo {title} {{E}dge {S}tates and {T}opological {P}hases in
  {N}on-{H}ermitian {S}ystems},}\ }\href@noop {} {\bibfield  {journal}
  {\bibinfo  {journal} {Phys. Rev. B}\ }\textbf {\bibinfo {volume} {84}},\
  \bibinfo {pages} {205128} (\bibinfo {year} {2011})}\BibitemShut {NoStop}%
\bibitem [{\citenamefont {Sato}\ \emph {et~al.}(2012)\citenamefont {Sato},
  \citenamefont {Hasebe}, \citenamefont {Esaki},\ and\ \citenamefont
  {Kohmoto}}]{Sato-12}%
  \BibitemOpen
  \bibfield  {author} {\bibinfo {author} {\bibfnamefont {M.}~\bibnamefont
  {Sato}}, \bibinfo {author} {\bibfnamefont {K.}~\bibnamefont {Hasebe}},
  \bibinfo {author} {\bibfnamefont {K.}~\bibnamefont {Esaki}}, \ and\ \bibinfo
  {author} {\bibfnamefont {M.}~\bibnamefont {Kohmoto}},\ }\bibfield  {title}
  {\enquote {\bibinfo {title} {{T}ime-{R}eversal {S}ymmetry in
  {N}on-{H}ermitian {S}ystems},}\ }\href@noop {} {\bibfield  {journal}
  {\bibinfo  {journal} {Prog. Theor. Phys.}\ }\textbf {\bibinfo {volume}
  {127}},\ \bibinfo {pages} {937} (\bibinfo {year} {2012})}\BibitemShut
  {NoStop}%
\bibitem [{\citenamefont {Pikulin}\ and\ \citenamefont
  {Nazarov}(2012)}]{Pikulin-12}%
  \BibitemOpen
  \bibfield  {author} {\bibinfo {author} {\bibfnamefont {D.~I.}\ \bibnamefont
  {Pikulin}}\ and\ \bibinfo {author} {\bibfnamefont {Y.~V.}\ \bibnamefont
  {Nazarov}},\ }\bibfield  {title} {\enquote {\bibinfo {title} {{T}opological
  {P}roperties of {S}uperconducting {J}unctions},}\ }\href@noop {} {\bibfield
  {journal} {\bibinfo  {journal} {JETP Lett.}\ }\textbf {\bibinfo {volume}
  {94}},\ \bibinfo {pages} {693} (\bibinfo {year} {2012})}\BibitemShut
  {NoStop}%
\bibitem [{\citenamefont {Pikulin}\ and\ \citenamefont
  {Nazarov}(2013)}]{Pikulin-13}%
  \BibitemOpen
  \bibfield  {author} {\bibinfo {author} {\bibfnamefont {D.~I.}\ \bibnamefont
  {Pikulin}}\ and\ \bibinfo {author} {\bibfnamefont {Y.~V.}\ \bibnamefont
  {Nazarov}},\ }\bibfield  {title} {\enquote {\bibinfo {title} {{T}wo {T}ypes
  of {T}opological {T}ransitions in {F}inite {M}ajorana {W}ires},}\ }\href@noop
  {} {\bibfield  {journal} {\bibinfo  {journal} {Phys. Rev. B}\ }\textbf
  {\bibinfo {volume} {87}},\ \bibinfo {pages} {235421} (\bibinfo {year}
  {2013})}\BibitemShut {NoStop}%
\bibitem [{\citenamefont {Liang}\ and\ \citenamefont {Huang}(2013)}]{Liang-13}%
  \BibitemOpen
  \bibfield  {author} {\bibinfo {author} {\bibfnamefont {S.-D.}\ \bibnamefont
  {Liang}}\ and\ \bibinfo {author} {\bibfnamefont {G.-Y.}\ \bibnamefont
  {Huang}},\ }\bibfield  {title} {\enquote {\bibinfo {title} {{T}opological
  {I}nvariance and {G}lobal {B}erry {P}hase in {N}on-{H}ermitian {S}ystems},}\
  }\href@noop {} {\bibfield  {journal} {\bibinfo  {journal} {Phys. Rev. A}\
  }\textbf {\bibinfo {volume} {87}},\ \bibinfo {pages} {012118} (\bibinfo
  {year} {2013})}\BibitemShut {NoStop}%
\bibitem [{\citenamefont {Schomerus}(2013)}]{Schomerus-13}%
  \BibitemOpen
  \bibfield  {author} {\bibinfo {author} {\bibfnamefont {H.}~\bibnamefont
  {Schomerus}},\ }\bibfield  {title} {\enquote {\bibinfo {title}
  {{T}opologically {P}rotected {M}idgap {S}tates in {C}omplex {P}hotonic
  {L}attices},}\ }\href@noop {} {\bibfield  {journal} {\bibinfo  {journal}
  {Opt. Lett.}\ }\textbf {\bibinfo {volume} {38}},\ \bibinfo {pages} {1912}
  (\bibinfo {year} {2013})}\BibitemShut {NoStop}%
\bibitem [{\citenamefont {Malzard}\ \emph {et~al.}(2015)\citenamefont
  {Malzard}, \citenamefont {Poli},\ and\ \citenamefont
  {Schomerus}}]{Malzard-15}%
  \BibitemOpen
  \bibfield  {author} {\bibinfo {author} {\bibfnamefont {S.}~\bibnamefont
  {Malzard}}, \bibinfo {author} {\bibfnamefont {C.}~\bibnamefont {Poli}}, \
  and\ \bibinfo {author} {\bibfnamefont {H.}~\bibnamefont {Schomerus}},\
  }\bibfield  {title} {\enquote {\bibinfo {title} {{T}opologically {P}rotected
  {D}efect {S}tates in {O}pen {P}hotonic {S}ystems with {N}on-{H}ermitian
  {C}harge-{C}onjugation and {P}arity-{T}ime {S}ymmetry},}\ }\href@noop {}
  {\bibfield  {journal} {\bibinfo  {journal} {Phys. Rev. Lett.}\ }\textbf
  {\bibinfo {volume} {115}},\ \bibinfo {pages} {200402} (\bibinfo {year}
  {2015})}\BibitemShut {NoStop}%
\bibitem [{\citenamefont {San-Jose}\ \emph {et~al.}(2016)\citenamefont
  {San-Jose}, \citenamefont {Cayao}, \citenamefont {Prada},\ and\ \citenamefont
  {Aguado}}]{SanJose-16}%
  \BibitemOpen
  \bibfield  {author} {\bibinfo {author} {\bibfnamefont {P.}~\bibnamefont
  {San-Jose}}, \bibinfo {author} {\bibfnamefont {J.}~\bibnamefont {Cayao}},
  \bibinfo {author} {\bibfnamefont {E.}~\bibnamefont {Prada}}, \ and\ \bibinfo
  {author} {\bibfnamefont {R.}~\bibnamefont {Aguado}},\ }\bibfield  {title}
  {\enquote {\bibinfo {title} {{M}ajorana {B}ound {S}tates from {E}xceptional
  {P}oints in {N}on-{T}opological {S}uperconductors},}\ }\href@noop {}
  {\bibfield  {journal} {\bibinfo  {journal} {Sci. Rep.}\ }\textbf {\bibinfo
  {volume} {6}},\ \bibinfo {pages} {21427} (\bibinfo {year}
  {2016})}\BibitemShut {NoStop}%
\bibitem [{\citenamefont {Lee}(2016)}]{Lee-16}%
  \BibitemOpen
  \bibfield  {author} {\bibinfo {author} {\bibfnamefont {T.~E.}\ \bibnamefont
  {Lee}},\ }\bibfield  {title} {\enquote {\bibinfo {title} {{A}nomalous {E}dge
  {S}tate in a {N}on-{H}ermitian {L}attice},}\ }\href@noop {} {\bibfield
  {journal} {\bibinfo  {journal} {Phys. Rev. Lett.}\ }\textbf {\bibinfo
  {volume} {116}},\ \bibinfo {pages} {133903} (\bibinfo {year}
  {2016})}\BibitemShut {NoStop}%
\bibitem [{\citenamefont {Gonz\'alez}\ and\ \citenamefont
  {Molina}(2016)}]{Gonzalez-16}%
  \BibitemOpen
  \bibfield  {author} {\bibinfo {author} {\bibfnamefont {J.}~\bibnamefont
  {Gonz\'alez}}\ and\ \bibinfo {author} {\bibfnamefont {R.~A.}\ \bibnamefont
  {Molina}},\ }\bibfield  {title} {\enquote {\bibinfo {title} {{M}acroscopic
  {D}egeneracy of {Z}ero-{M}ode {R}otating {S}urface {S}tates in 3{D} {D}irac
  and {W}eyl {S}emimetals under {R}adiation},}\ }\href@noop {} {\bibfield
  {journal} {\bibinfo  {journal} {Phys. Rev. Lett.}\ }\textbf {\bibinfo
  {volume} {116}},\ \bibinfo {pages} {156803} (\bibinfo {year}
  {2016})}\BibitemShut {NoStop}%
\bibitem [{\citenamefont {Gonz\'alez}\ and\ \citenamefont
  {Molina}(2017)}]{Gonzalez-17}%
  \BibitemOpen
  \bibfield  {author} {\bibinfo {author} {\bibfnamefont {J.}~\bibnamefont
  {Gonz\'alez}}\ and\ \bibinfo {author} {\bibfnamefont {R.~A.}\ \bibnamefont
  {Molina}},\ }\bibfield  {title} {\enquote {\bibinfo {title} {{T}opological
  {P}rotection from {E}xceptional {P}oints in {W}eyl and {N}odal-{L}ine
  {S}emimetals},}\ }\href@noop {} {\bibfield  {journal} {\bibinfo  {journal}
  {Phys. Rev. B}\ }\textbf {\bibinfo {volume} {96}},\ \bibinfo {pages} {045437}
  (\bibinfo {year} {2017})}\BibitemShut {NoStop}%
\bibitem [{\citenamefont {Molina}\ and\ \citenamefont
  {Gonz\'alez}(2018)}]{Molina-18}%
  \BibitemOpen
  \bibfield  {author} {\bibinfo {author} {\bibfnamefont {R.~A.}\ \bibnamefont
  {Molina}}\ and\ \bibinfo {author} {\bibfnamefont {J.}~\bibnamefont
  {Gonz\'alez}},\ }\bibfield  {title} {\enquote {\bibinfo {title} {{S}urface
  and 3{D} {Q}uantum {H}all {E}ffects from {E}ngineering of {E}xceptional
  {P}oints in {N}odal-{L}ine {S}emimetals},}\ }\href@noop {} {\bibfield
  {journal} {\bibinfo  {journal} {Phys. Rev. Lett.}\ }\textbf {\bibinfo
  {volume} {120}},\ \bibinfo {pages} {146601} (\bibinfo {year}
  {2018})}\BibitemShut {NoStop}%
\bibitem [{\citenamefont {Harter}\ \emph {et~al.}(2016)\citenamefont {Harter},
  \citenamefont {Lee},\ and\ \citenamefont {Joglekar}}]{Harter-16}%
  \BibitemOpen
  \bibfield  {author} {\bibinfo {author} {\bibfnamefont {A.~K.}\ \bibnamefont
  {Harter}}, \bibinfo {author} {\bibfnamefont {T.~E.}\ \bibnamefont {Lee}}, \
  and\ \bibinfo {author} {\bibfnamefont {Y.~N.}\ \bibnamefont {Joglekar}},\
  }\bibfield  {title} {\enquote {\bibinfo {title} {$\mathcal{PT}$-{B}reaking
  {T}hreshold in {S}patially {A}symmetric {A}ubry-{A}ndr\'e and {H}arper
  {M}odels: {H}idden {S}ymmetry and {T}opological {S}tates},}\ }\href@noop {}
  {\bibfield  {journal} {\bibinfo  {journal} {Phys. Rev. A}\ }\textbf {\bibinfo
  {volume} {93}},\ \bibinfo {pages} {062101} (\bibinfo {year}
  {2016})}\BibitemShut {NoStop}%
\bibitem [{\citenamefont {Leykam}\ \emph {et~al.}(2017)\citenamefont {Leykam},
  \citenamefont {Bliokh}, \citenamefont {Huang}, \citenamefont {Chong},\ and\
  \citenamefont {Nori}}]{Leykam-17}%
  \BibitemOpen
  \bibfield  {author} {\bibinfo {author} {\bibfnamefont {D.}~\bibnamefont
  {Leykam}}, \bibinfo {author} {\bibfnamefont {K.~Y.}\ \bibnamefont {Bliokh}},
  \bibinfo {author} {\bibfnamefont {C.}~\bibnamefont {Huang}}, \bibinfo
  {author} {\bibfnamefont {Y.~D.}\ \bibnamefont {Chong}}, \ and\ \bibinfo
  {author} {\bibfnamefont {F.}~\bibnamefont {Nori}},\ }\bibfield  {title}
  {\enquote {\bibinfo {title} {{E}dge {M}odes, {D}egeneracies, and
  {T}opological {N}umbers in {N}on-{H}ermitian {S}ystems},}\ }\href@noop {}
  {\bibfield  {journal} {\bibinfo  {journal} {Phys. Rev. Lett.}\ }\textbf
  {\bibinfo {volume} {118}},\ \bibinfo {pages} {040401} (\bibinfo {year}
  {2017})}\BibitemShut {NoStop}%
\bibitem [{\citenamefont {Xu}\ \emph {et~al.}(2017)\citenamefont {Xu},
  \citenamefont {Wang},\ and\ \citenamefont {Duan}}]{Xu-17}%
  \BibitemOpen
  \bibfield  {author} {\bibinfo {author} {\bibfnamefont {Y.}~\bibnamefont
  {Xu}}, \bibinfo {author} {\bibfnamefont {S.-T.}\ \bibnamefont {Wang}}, \ and\
  \bibinfo {author} {\bibfnamefont {L.-M.}\ \bibnamefont {Duan}},\ }\bibfield
  {title} {\enquote {\bibinfo {title} {{W}eyl {E}xceptional {R}ings in a
  {T}hree-{D}imensional {D}issipative {C}old {A}tomic {G}as},}\ }\href@noop {}
  {\bibfield  {journal} {\bibinfo  {journal} {Phys. Rev. Lett.}\ }\textbf
  {\bibinfo {volume} {118}},\ \bibinfo {pages} {045701} (\bibinfo {year}
  {2017})}\BibitemShut {NoStop}%
\bibitem [{\citenamefont {Menke}\ and\ \citenamefont
  {Hirschmann}(2017)}]{Menke-17}%
  \BibitemOpen
  \bibfield  {author} {\bibinfo {author} {\bibfnamefont {H.}~\bibnamefont
  {Menke}}\ and\ \bibinfo {author} {\bibfnamefont {M.}~\bibnamefont
  {Hirschmann}},\ }\bibfield  {title} {\enquote {\bibinfo {title}
  {{T}opological {Q}uantum {W}ires with {B}alanced {G}ain and {L}oss},}\
  }\href@noop {} {\bibfield  {journal} {\bibinfo  {journal} {Phys. Rev. B}\
  }\textbf {\bibinfo {volume} {95}},\ \bibinfo {pages} {174506} (\bibinfo
  {year} {2017})}\BibitemShut {NoStop}%
\bibitem [{\citenamefont {Lieu}(2018{\natexlab{b}})}]{Lieu-18-SSH}%
  \BibitemOpen
  \bibfield  {author} {\bibinfo {author} {\bibfnamefont {S.}~\bibnamefont
  {Lieu}},\ }\bibfield  {title} {\enquote {\bibinfo {title} {{T}opological
  {P}hases in the {N}on-{H}ermitian {S}u-{S}chrieffer-{H}eeger {M}odel},}\
  }\href@noop {} {\bibfield  {journal} {\bibinfo  {journal} {Phys. Rev. B}\
  }\textbf {\bibinfo {volume} {97}},\ \bibinfo {pages} {045106} (\bibinfo
  {year} {2018}{\natexlab{b}})}\BibitemShut {NoStop}%
\bibitem [{\citenamefont {Cerjan}\ \emph {et~al.}(2018)\citenamefont {Cerjan},
  \citenamefont {Xiao}, \citenamefont {Yuan},\ and\ \citenamefont
  {Fan}}]{Cerjan-18}%
  \BibitemOpen
  \bibfield  {author} {\bibinfo {author} {\bibfnamefont {A.}~\bibnamefont
  {Cerjan}}, \bibinfo {author} {\bibfnamefont {M.}~\bibnamefont {Xiao}},
  \bibinfo {author} {\bibfnamefont {L.}~\bibnamefont {Yuan}}, \ and\ \bibinfo
  {author} {\bibfnamefont {S.}~\bibnamefont {Fan}},\ }\bibfield  {title}
  {\enquote {\bibinfo {title} {{E}ffects of {N}on-{H}ermitian {P}erturbations
  on {W}eyl {H}amiltonians with {A}rbitrary {T}opological {C}harges},}\
  }\href@noop {} {\bibfield  {journal} {\bibinfo  {journal} {Phys. Rev. B}\
  }\textbf {\bibinfo {volume} {97}},\ \bibinfo {pages} {075128} (\bibinfo
  {year} {2018})}\BibitemShut {NoStop}%
\bibitem [{\citenamefont {Alvarez}\ \emph {et~al.}(2018)\citenamefont
  {Alvarez}, \citenamefont {Vargas},\ and\ \citenamefont
  {Torres}}]{MartinezAlvarez-18}%
  \BibitemOpen
  \bibfield  {author} {\bibinfo {author} {\bibfnamefont {V.~M.~Martinez}\
  \bibnamefont {Alvarez}}, \bibinfo {author} {\bibfnamefont {J.~E.~Barrios}\
  \bibnamefont {Vargas}}, \ and\ \bibinfo {author} {\bibfnamefont {L.~E.
  F.~Foa}\ \bibnamefont {Torres}},\ }\bibfield  {title} {\enquote {\bibinfo
  {title} {{N}on-{H}ermitian {R}obust {E}dge {S}tates in {O}ne {D}imension:
  {A}nomalous {L}ocalization and {E}igenspace {C}ondensation at {E}xceptional
  {P}oints},}\ }\href@noop {} {\bibfield  {journal} {\bibinfo  {journal} {Phys.
  Rev. B}\ }\textbf {\bibinfo {volume} {97}},\ \bibinfo {pages} {121401(R)}
  (\bibinfo {year} {2018})}\BibitemShut {NoStop}%
\bibitem [{\citenamefont {Shen}\ \emph {et~al.}(2018)\citenamefont {Shen},
  \citenamefont {Zhen},\ and\ \citenamefont {Fu}}]{Shen-18}%
  \BibitemOpen
  \bibfield  {author} {\bibinfo {author} {\bibfnamefont {H.}~\bibnamefont
  {Shen}}, \bibinfo {author} {\bibfnamefont {B.}~\bibnamefont {Zhen}}, \ and\
  \bibinfo {author} {\bibfnamefont {L.}~\bibnamefont {Fu}},\ }\bibfield
  {title} {\enquote {\bibinfo {title} {{T}opological {B}and {T}heory for
  {N}on-{H}ermitian {H}amiltonians},}\ }\href@noop {} {\bibfield  {journal}
  {\bibinfo  {journal} {Phys. Rev. Lett.}\ }\textbf {\bibinfo {volume} {120}},\
  \bibinfo {pages} {146402} (\bibinfo {year} {2018})}\BibitemShut {NoStop}%
\bibitem [{\citenamefont {Yin}\ \emph {et~al.}(2018)\citenamefont {Yin},
  \citenamefont {Jiang}, \citenamefont {Li}, \citenamefont {L\"u},\ and\
  \citenamefont {Chen}}]{Yin-18}%
  \BibitemOpen
  \bibfield  {author} {\bibinfo {author} {\bibfnamefont {C.}~\bibnamefont
  {Yin}}, \bibinfo {author} {\bibfnamefont {H.}~\bibnamefont {Jiang}}, \bibinfo
  {author} {\bibfnamefont {L.}~\bibnamefont {Li}}, \bibinfo {author}
  {\bibfnamefont {R.}~\bibnamefont {L\"u}}, \ and\ \bibinfo {author}
  {\bibfnamefont {S.}~\bibnamefont {Chen}},\ }\bibfield  {title} {\enquote
  {\bibinfo {title} {{G}eometrical {M}eaning of {W}inding {N}umber and {I}ts
  {C}haracterization of {T}opological {P}hases in {O}ne-{D}imensional {C}hiral
  {N}on-{H}ermitian {S}ystems},}\ }\href@noop {} {\bibfield  {journal}
  {\bibinfo  {journal} {Phys. Rev. A}\ }\textbf {\bibinfo {volume} {97}},\
  \bibinfo {pages} {052115} (\bibinfo {year} {2018})}\BibitemShut {NoStop}%
\bibitem [{\citenamefont {Kunst}\ \emph {et~al.}(2018)\citenamefont {Kunst},
  \citenamefont {Edvardsson}, \citenamefont {Budich},\ and\ \citenamefont
  {Bergholtz}}]{Kunst-18}%
  \BibitemOpen
  \bibfield  {author} {\bibinfo {author} {\bibfnamefont {F.~K.}\ \bibnamefont
  {Kunst}}, \bibinfo {author} {\bibfnamefont {E.}~\bibnamefont {Edvardsson}},
  \bibinfo {author} {\bibfnamefont {J.~C.}\ \bibnamefont {Budich}}, \ and\
  \bibinfo {author} {\bibfnamefont {E.~J.}\ \bibnamefont {Bergholtz}},\
  }\bibfield  {title} {\enquote {\bibinfo {title} {{Biorthogonal Bulk-Boundary
  Correspondence in Non-Hermitian Systems}},}\ }\href@noop {} {\bibfield
  {journal} {\bibinfo  {journal} {Phys. Rev. Lett.}\ }\textbf {\bibinfo
  {volume} {121}},\ \bibinfo {pages} {026808} (\bibinfo {year}
  {2018})}\BibitemShut {NoStop}%
\bibitem [{\citenamefont {Kawabata}\ \emph
  {et~al.}(2018{\natexlab{a}})\citenamefont {Kawabata}, \citenamefont {Ashida},
  \citenamefont {Katsura},\ and\ \citenamefont {Ueda}}]{Kawabata-18-Kitaev}%
  \BibitemOpen
  \bibfield  {author} {\bibinfo {author} {\bibfnamefont {K.}~\bibnamefont
  {Kawabata}}, \bibinfo {author} {\bibfnamefont {Y.}~\bibnamefont {Ashida}},
  \bibinfo {author} {\bibfnamefont {H.}~\bibnamefont {Katsura}}, \ and\
  \bibinfo {author} {\bibfnamefont {M.}~\bibnamefont {Ueda}},\ }\bibfield
  {title} {\enquote {\bibinfo {title} {{P}arity-{T}ime-{S}ymmetric
  {T}opological {S}uperconductor},}\ }\href@noop {} {\bibfield  {journal}
  {\bibinfo  {journal} {Phys. Rev. B}\ }\textbf {\bibinfo {volume} {98}},\
  \bibinfo {pages} {085116} (\bibinfo {year} {2018}{\natexlab{a}})}\BibitemShut
  {NoStop}%
\bibitem [{\citenamefont {Yao}\ and\ \citenamefont {Wang}(2018)}]{Yao-18-SSH}%
  \BibitemOpen
  \bibfield  {author} {\bibinfo {author} {\bibfnamefont {S.}~\bibnamefont
  {Yao}}\ and\ \bibinfo {author} {\bibfnamefont {Z.}~\bibnamefont {Wang}},\
  }\bibfield  {title} {\enquote {\bibinfo {title} {{E}dge {S}tates and
  {T}opological {I}nvariants of {N}on-{H}ermitian {S}ystems},}\ }\href@noop {}
  {\bibfield  {journal} {\bibinfo  {journal} {Phys. Rev. Lett.}\ }\textbf
  {\bibinfo {volume} {121}},\ \bibinfo {pages} {086803} (\bibinfo {year}
  {2018})}\BibitemShut {NoStop}%
\bibitem [{\citenamefont {Yao}\ \emph {et~al.}(2018)\citenamefont {Yao},
  \citenamefont {Song},\ and\ \citenamefont {Wang}}]{Yao-18-Chern}%
  \BibitemOpen
  \bibfield  {author} {\bibinfo {author} {\bibfnamefont {S.}~\bibnamefont
  {Yao}}, \bibinfo {author} {\bibfnamefont {F.}~\bibnamefont {Song}}, \ and\
  \bibinfo {author} {\bibfnamefont {Z.}~\bibnamefont {Wang}},\ }\bibfield
  {title} {\enquote {\bibinfo {title} {{N}on-{H}ermitian {C}hern {B}ands},}\
  }\href@noop {} {\bibfield  {journal} {\bibinfo  {journal} {{Phys. Rev.
  Lett.}}\ }\textbf {\bibinfo {volume} {121}},\ \bibinfo {pages} {136802}
  (\bibinfo {year} {2018})}\BibitemShut {NoStop}%
\bibitem [{\citenamefont {Gong}\ \emph {et~al.}(2018)\citenamefont {Gong},
  \citenamefont {Ashida}, \citenamefont {Kawabata}, \citenamefont {Takasan},
  \citenamefont {Higashikawa},\ and\ \citenamefont {Ueda}}]{Gong-18}%
  \BibitemOpen
  \bibfield  {author} {\bibinfo {author} {\bibfnamefont {Z.}~\bibnamefont
  {Gong}}, \bibinfo {author} {\bibfnamefont {Y.}~\bibnamefont {Ashida}},
  \bibinfo {author} {\bibfnamefont {K.}~\bibnamefont {Kawabata}}, \bibinfo
  {author} {\bibfnamefont {K.}~\bibnamefont {Takasan}}, \bibinfo {author}
  {\bibfnamefont {S.}~\bibnamefont {Higashikawa}}, \ and\ \bibinfo {author}
  {\bibfnamefont {M.}~\bibnamefont {Ueda}},\ }\bibfield  {title} {\enquote
  {\bibinfo {title} {{T}opological {P}hases of {N}on-{H}ermitian {S}ystems},}\
  }\href@noop {} {\bibfield  {journal} {\bibinfo  {journal} {Phys. Rev. X}\
  }\textbf {\bibinfo {volume} {8}},\ \bibinfo {pages} {031079} (\bibinfo {year}
  {2018})}\BibitemShut {NoStop}%
\bibitem [{\citenamefont {Bandres}\ and\ \citenamefont
  {Segev}(2018)}]{Bandres-Segev-18}%
  \BibitemOpen
  \bibfield  {author} {\bibinfo {author} {\bibfnamefont {M.~A.}\ \bibnamefont
  {Bandres}}\ and\ \bibinfo {author} {\bibfnamefont {M.}~\bibnamefont
  {Segev}},\ }\bibfield  {title} {\enquote {\bibinfo {title} {{V}iewpoint:
  {N}on-{H}ermitian {T}opological {S}ystems},}\ }\href@noop {} {\bibfield
  {journal} {\bibinfo  {journal} {Physics}\ }\textbf {\bibinfo {volume} {11}},\
  \bibinfo {pages} {96} (\bibinfo {year} {2018})}\BibitemShut {NoStop}%
\bibitem [{\citenamefont {Carlstr\"om}\ and\ \citenamefont
  {Bergholtz}(2018)}]{Carlstrom-18}%
  \BibitemOpen
  \bibfield  {author} {\bibinfo {author} {\bibfnamefont {J.}~\bibnamefont
  {Carlstr\"om}}\ and\ \bibinfo {author} {\bibfnamefont {E.~J.}\ \bibnamefont
  {Bergholtz}},\ }\bibfield  {title} {\enquote {\bibinfo {title} {{E}xceptional
  {L}inks and {T}wisted {F}ermi {R}ibbons in {N}on-{H}ermitian {S}ystems},}\
  }\href@noop {} {\bibfield  {journal} {\bibinfo  {journal} {Phys. Rev. A}\
  }\textbf {\bibinfo {volume} {98}},\ \bibinfo {pages} {042114} (\bibinfo
  {year} {2018})}\BibitemShut {NoStop}%
\bibitem [{\citenamefont {Kawabata}\ \emph
  {et~al.}(2018{\natexlab{b}})\citenamefont {Kawabata}, \citenamefont
  {Shiozaki},\ and\ \citenamefont {Ueda}}]{Kawabata-18-Chern}%
  \BibitemOpen
  \bibfield  {author} {\bibinfo {author} {\bibfnamefont {K.}~\bibnamefont
  {Kawabata}}, \bibinfo {author} {\bibfnamefont {K.}~\bibnamefont {Shiozaki}},
  \ and\ \bibinfo {author} {\bibfnamefont {M.}~\bibnamefont {Ueda}},\
  }\bibfield  {title} {\enquote {\bibinfo {title} {{A}nomalous {H}elical {E}dge
  {S}tates in a {N}on-{H}ermitian {C}hern {I}nsulator},}\ }\href@noop {}
  {\bibfield  {journal} {\bibinfo  {journal} {Phys. Rev. B}\ }\textbf {\bibinfo
  {volume} {98}},\ \bibinfo {pages} {165148} (\bibinfo {year}
  {2018}{\natexlab{b}})}\BibitemShut {NoStop}%
\bibitem [{\citenamefont {Takata}\ and\ \citenamefont
  {Notomi}(2018)}]{Takata-18}%
  \BibitemOpen
  \bibfield  {author} {\bibinfo {author} {\bibfnamefont {K.}~\bibnamefont
  {Takata}}\ and\ \bibinfo {author} {\bibfnamefont {M.}~\bibnamefont
  {Notomi}},\ }\bibfield  {title} {\enquote {\bibinfo {title} {{Photonic
  Topological Insulating Phase Induced Solely by Gain and Loss}},}\ }\href@noop
  {} {\bibfield  {journal} {\bibinfo  {journal} {Phys. Rev. Lett.}\ }\textbf
  {\bibinfo {volume} {121}},\ \bibinfo {pages} {213902} (\bibinfo {year}
  {2018})}\BibitemShut {NoStop}%
\bibitem [{\citenamefont {Kawabata}\ \emph {et~al.}(2019)\citenamefont
  {Kawabata}, \citenamefont {Higashikawa}, \citenamefont {Gong}, \citenamefont
  {Ashida},\ and\ \citenamefont {Ueda}}]{Kawabata-18}%
  \BibitemOpen
  \bibfield  {author} {\bibinfo {author} {\bibfnamefont {K.}~\bibnamefont
  {Kawabata}}, \bibinfo {author} {\bibfnamefont {S.}~\bibnamefont
  {Higashikawa}}, \bibinfo {author} {\bibfnamefont {Z.}~\bibnamefont {Gong}},
  \bibinfo {author} {\bibfnamefont {Y.}~\bibnamefont {Ashida}}, \ and\ \bibinfo
  {author} {\bibfnamefont {M.}~\bibnamefont {Ueda}},\ }\bibfield  {title}
  {\enquote {\bibinfo {title} {{Topological Unification of Time-Reversal and
  Particle-Hole Symmetries in Non-Hermitian Physics}},}\ }\href@noop {}
  {\bibfield  {journal} {\bibinfo  {journal} {Nat. Commun.}\ }\textbf {\bibinfo
  {volume} {10}},\ \bibinfo {pages} {297} (\bibinfo {year} {2019})}\BibitemShut
  {NoStop}%
\bibitem [{\citenamefont {Okugawa}\ and\ \citenamefont
  {Yokoyama}(2019)}]{Okugawa-19}%
  \BibitemOpen
  \bibfield  {author} {\bibinfo {author} {\bibfnamefont {R.}~\bibnamefont
  {Okugawa}}\ and\ \bibinfo {author} {\bibfnamefont {T.}~\bibnamefont
  {Yokoyama}},\ }\bibfield  {title} {\enquote {\bibinfo {title} {{Topological
  Exceptional Surfaces in Non-Hermitian Systems with Parity-Time and
  Parity-Particle-Hole Symmetries}},}\ }\href@noop {} {\bibfield  {journal}
  {\bibinfo  {journal} {Phys. Rev. B}\ }\textbf {\bibinfo {volume} {99}},\
  \bibinfo {pages} {041202(R)} (\bibinfo {year} {2019})}\BibitemShut {NoStop}%
\bibitem [{\citenamefont {Budich}\ \emph {et~al.}(2019)\citenamefont {Budich},
  \citenamefont {Carlstr\"om}, \citenamefont {Kunst},\ and\ \citenamefont
  {Bergholtz}}]{Budich-19}%
  \BibitemOpen
  \bibfield  {author} {\bibinfo {author} {\bibfnamefont {J.~C.}\ \bibnamefont
  {Budich}}, \bibinfo {author} {\bibfnamefont {J.}~\bibnamefont {Carlstr\"om}},
  \bibinfo {author} {\bibfnamefont {F.~K.}\ \bibnamefont {Kunst}}, \ and\
  \bibinfo {author} {\bibfnamefont {E.~J.}\ \bibnamefont {Bergholtz}},\
  }\bibfield  {title} {\enquote {\bibinfo {title} {{Symmetry-Protected Nodal
  Phases in Non-Hermitian Systems}},}\ }\href@noop {} {\bibfield  {journal}
  {\bibinfo  {journal} {Phys. Rev. B}\ }\textbf {\bibinfo {volume} {99}},\
  \bibinfo {pages} {041406(R)} (\bibinfo {year} {2019})}\BibitemShut {NoStop}%
\bibitem [{\citenamefont {Yang}\ and\ \citenamefont {Hu}(2019)}]{Yang-19}%
  \BibitemOpen
  \bibfield  {author} {\bibinfo {author} {\bibfnamefont {Z.}~\bibnamefont
  {Yang}}\ and\ \bibinfo {author} {\bibfnamefont {J.}~\bibnamefont {Hu}},\
  }\bibfield  {title} {\enquote {\bibinfo {title} {{Non-Hermitian Hopf-Link
  Exceptional Line Semimetals}},}\ }\href@noop {} {\bibfield  {journal}
  {\bibinfo  {journal} {Phys. Rev. B}\ }\textbf {\bibinfo {volume} {99}},\
  \bibinfo {pages} {081102(R)} (\bibinfo {year} {2019})}\BibitemShut {NoStop}%
\bibitem [{\citenamefont {Jin}\ and\ \citenamefont {Song}(2019)}]{Jin-19}%
  \BibitemOpen
  \bibfield  {author} {\bibinfo {author} {\bibfnamefont {L.}~\bibnamefont
  {Jin}}\ and\ \bibinfo {author} {\bibfnamefont {Z.}~\bibnamefont {Song}},\
  }\bibfield  {title} {\enquote {\bibinfo {title} {{Bulk-Boundary
  Correspondence in a Non-Hermitian System in One Dimension with Chiral
  Inversion Symmetry}},}\ }\href@noop {} {\bibfield  {journal} {\bibinfo
  {journal} {Phys. Rev. B}\ }\textbf {\bibinfo {volume} {99}},\ \bibinfo
  {pages} {081103(R)} (\bibinfo {year} {2019})}\BibitemShut {NoStop}%
\bibitem [{\citenamefont {Zhou}\ \emph {et~al.}(2019)\citenamefont {Zhou},
  \citenamefont {Lee}, \citenamefont {Liu},\ and\ \citenamefont
  {Zhen}}]{Zhou-19}%
  \BibitemOpen
  \bibfield  {author} {\bibinfo {author} {\bibfnamefont {H.}~\bibnamefont
  {Zhou}}, \bibinfo {author} {\bibfnamefont {J.~Y.}\ \bibnamefont {Lee}},
  \bibinfo {author} {\bibfnamefont {S.}~\bibnamefont {Liu}}, \ and\ \bibinfo
  {author} {\bibfnamefont {B.}~\bibnamefont {Zhen}},\ }\bibfield  {title}
  {\enquote {\bibinfo {title} {{Exceptional Surfaces in
  $\mathcal{PT}$-Symmetric Non-Hermitian Photonic Systems}},}\ }\href@noop {}
  {\bibfield  {journal} {\bibinfo  {journal} {Optica}\ }\textbf {\bibinfo
  {volume} {6}},\ \bibinfo {pages} {190} (\bibinfo {year} {2019})}\BibitemShut
  {NoStop}%
\bibitem [{\citenamefont {Wang}\ \emph
  {et~al.}(2019{\natexlab{a}})\citenamefont {Wang}, \citenamefont {Ruan},\ and\
  \citenamefont {Zhang}}]{Wang-19}%
  \BibitemOpen
  \bibfield  {author} {\bibinfo {author} {\bibfnamefont {H.}~\bibnamefont
  {Wang}}, \bibinfo {author} {\bibfnamefont {J.}~\bibnamefont {Ruan}}, \ and\
  \bibinfo {author} {\bibfnamefont {H.}~\bibnamefont {Zhang}},\ }\bibfield
  {title} {\enquote {\bibinfo {title} {{Non-Hermitian Nodal-Line Semimetals
  with an Anomalous Bulk-Boundary Correspondence}},}\ }\href@noop {} {\bibfield
   {journal} {\bibinfo  {journal} {Phys. Rev. B}\ }\textbf {\bibinfo {volume}
  {99}},\ \bibinfo {pages} {075130} (\bibinfo {year}
  {2019}{\natexlab{a}})}\BibitemShut {NoStop}%
\bibitem [{\citenamefont {Liu}\ \emph {et~al.}(2019)\citenamefont {Liu},
  \citenamefont {Zhang}, \citenamefont {Ai}, \citenamefont {Gong},
  \citenamefont {Kawabata}, \citenamefont {Ueda},\ and\ \citenamefont
  {Nori}}]{Liu-19}%
  \BibitemOpen
  \bibfield  {author} {\bibinfo {author} {\bibfnamefont {T.}~\bibnamefont
  {Liu}}, \bibinfo {author} {\bibfnamefont {Y.-R.}\ \bibnamefont {Zhang}},
  \bibinfo {author} {\bibfnamefont {Q.}~\bibnamefont {Ai}}, \bibinfo {author}
  {\bibfnamefont {Z.}~\bibnamefont {Gong}}, \bibinfo {author} {\bibfnamefont
  {K.}~\bibnamefont {Kawabata}}, \bibinfo {author} {\bibfnamefont
  {M.}~\bibnamefont {Ueda}}, \ and\ \bibinfo {author} {\bibfnamefont
  {F.}~\bibnamefont {Nori}},\ }\bibfield  {title} {\enquote {\bibinfo {title}
  {{Second-Order Topological Phases in Non-Hermitian Systems}},}\ }\href@noop
  {} {\bibfield  {journal} {\bibinfo  {journal} {Phys. Rev. Lett.}\ }\textbf
  {\bibinfo {volume} {122}},\ \bibinfo {pages} {076801} (\bibinfo {year}
  {2019})}\BibitemShut {NoStop}%
\bibitem [{\citenamefont {Edvardsson}\ \emph {et~al.}(2019)\citenamefont
  {Edvardsson}, \citenamefont {Kunst},\ and\ \citenamefont
  {Bergholtz}}]{Edvardsson-19}%
  \BibitemOpen
  \bibfield  {author} {\bibinfo {author} {\bibfnamefont {E.}~\bibnamefont
  {Edvardsson}}, \bibinfo {author} {\bibfnamefont {F.~K.}\ \bibnamefont
  {Kunst}}, \ and\ \bibinfo {author} {\bibfnamefont {E.~J.}\ \bibnamefont
  {Bergholtz}},\ }\bibfield  {title} {\enquote {\bibinfo {title}
  {{Non-Hermitian Extensions of Higher-Order Topological Phases and Their
  Biorthogonal Bulk-Boundary Correspondence}},}\ }\href@noop {} {\bibfield
  {journal} {\bibinfo  {journal} {Phys. Rev. B}\ }\textbf {\bibinfo {volume}
  {99}},\ \bibinfo {pages} {081302(R)} (\bibinfo {year} {2019})}\BibitemShut
  {NoStop}%
\bibitem [{\citenamefont {Carlstr\"om}\ \emph {et~al.}(2019)\citenamefont
  {Carlstr\"om}, \citenamefont {St\r{a}lhammar}, \citenamefont {Budich},\ and\
  \citenamefont {Bergholtz}}]{Carlstrom-19}%
  \BibitemOpen
  \bibfield  {author} {\bibinfo {author} {\bibfnamefont {J.}~\bibnamefont
  {Carlstr\"om}}, \bibinfo {author} {\bibfnamefont {M.}~\bibnamefont
  {St\r{a}lhammar}}, \bibinfo {author} {\bibfnamefont {J.~C.}\ \bibnamefont
  {Budich}}, \ and\ \bibinfo {author} {\bibfnamefont {E.~J.}\ \bibnamefont
  {Bergholtz}},\ }\bibfield  {title} {\enquote {\bibinfo {title} {{Knotted
  Non-Hermitian Metals}},}\ }\href@noop {} {\bibfield  {journal} {\bibinfo
  {journal} {Phys. Rev. B}\ }\textbf {\bibinfo {volume} {99}},\ \bibinfo
  {pages} {161115(R)} (\bibinfo {year} {2019})}\BibitemShut {NoStop}%
\bibitem [{\citenamefont {Lee}\ and\ \citenamefont {Thomale}(2019)}]{Lee-19}%
  \BibitemOpen
  \bibfield  {author} {\bibinfo {author} {\bibfnamefont {C.~H.}\ \bibnamefont
  {Lee}}\ and\ \bibinfo {author} {\bibfnamefont {R.}~\bibnamefont {Thomale}},\
  }\bibfield  {title} {\enquote {\bibinfo {title} {{Anatomy of Skin Modes and
  Topology in Non-Hermitian Systems}},}\ }\href@noop {} {\bibfield  {journal}
  {\bibinfo  {journal} {Phys. Rev. B}\ }\textbf {\bibinfo {volume} {99}},\
  \bibinfo {pages} {201103(R)} (\bibinfo {year} {2019})}\BibitemShut {NoStop}%
\bibitem [{\citenamefont {Kunst}\ and\ \citenamefont
  {Dwivedi}(2019)}]{Kunst-19}%
  \BibitemOpen
  \bibfield  {author} {\bibinfo {author} {\bibfnamefont {F.~K.}\ \bibnamefont
  {Kunst}}\ and\ \bibinfo {author} {\bibfnamefont {V.}~\bibnamefont
  {Dwivedi}},\ }\bibfield  {title} {\enquote {\bibinfo {title} {{Non-Hermitian
  Systems and Topology: A Transfer-Matrix Perspective}},}\ }\href@noop {}
  {\bibfield  {journal} {\bibinfo  {journal} {Phys. Rev. B}\ }\textbf {\bibinfo
  {volume} {99}},\ \bibinfo {pages} {245116} (\bibinfo {year}
  {2019})}\BibitemShut {NoStop}%
\bibitem [{\citenamefont {Longhi}(2019)}]{Longhi-19}%
  \BibitemOpen
  \bibfield  {author} {\bibinfo {author} {\bibfnamefont {S.}~\bibnamefont
  {Longhi}},\ }\bibfield  {title} {\enquote {\bibinfo {title} {{Topological
  Phase Transition in non-Hermitian Quasicrystals}},}\ }\href@noop {}
  {\bibfield  {journal} {\bibinfo  {journal} {Phys. Rev. Lett.}\ }\textbf
  {\bibinfo {volume} {122}},\ \bibinfo {pages} {237601} (\bibinfo {year}
  {2019})}\BibitemShut {NoStop}%
\bibitem [{\citenamefont {Lee}\ \emph {et~al.}(2019)\citenamefont {Lee},
  \citenamefont {Li},\ and\ \citenamefont {Gong}}]{Lee-Li-19}%
  \BibitemOpen
  \bibfield  {author} {\bibinfo {author} {\bibfnamefont {C.~H.}\ \bibnamefont
  {Lee}}, \bibinfo {author} {\bibfnamefont {L.}~\bibnamefont {Li}}, \ and\
  \bibinfo {author} {\bibfnamefont {J.}~\bibnamefont {Gong}},\ }\bibfield
  {title} {\enquote {\bibinfo {title} {{Hybrid Higher-Order Skin-Topological
  Modes in Nonreciprocal Systems}},}\ }\href@noop {} {\bibfield  {journal}
  {\bibinfo  {journal} {Phys. Rev. Lett.}\ }\textbf {\bibinfo {volume} {123}},\
  \bibinfo {pages} {016805} (\bibinfo {year} {2019})}\BibitemShut {NoStop}%
\bibitem [{\citenamefont {Rudner}\ \emph {et~al.}()\citenamefont {Rudner},
  \citenamefont {Levin},\ and\ \citenamefont {Levitov}}]{Rudner-16}%
  \BibitemOpen
  \bibfield  {author} {\bibinfo {author} {\bibfnamefont {M.~S.}\ \bibnamefont
  {Rudner}}, \bibinfo {author} {\bibfnamefont {M.}~\bibnamefont {Levin}}, \
  and\ \bibinfo {author} {\bibfnamefont {L.~S.}\ \bibnamefont {Levitov}},\
  }\href@noop {} {\enquote {\bibinfo {title} {{S}urvival, {D}ecay, and
  {T}opological {P}rotection in {N}on-{H}ermitian {Q}uantum {T}ransport},}\
  }\bibinfo {note} {{a}rXiv: 1605.07652}\BibitemShut {NoStop}%
\bibitem [{\citenamefont {Qiu}\ \emph {et~al.}()\citenamefont {Qiu},
  \citenamefont {Deng}, \citenamefont {Hu}, \citenamefont {Xue},\ and\
  \citenamefont {Yi}}]{Qiu-18}%
  \BibitemOpen
  \bibfield  {author} {\bibinfo {author} {\bibfnamefont {X.}~\bibnamefont
  {Qiu}}, \bibinfo {author} {\bibfnamefont {T.-S.}\ \bibnamefont {Deng}},
  \bibinfo {author} {\bibfnamefont {Y.}~\bibnamefont {Hu}}, \bibinfo {author}
  {\bibfnamefont {P.}~\bibnamefont {Xue}}, \ and\ \bibinfo {author}
  {\bibfnamefont {W.}~\bibnamefont {Yi}},\ }\href@noop {} {\enquote {\bibinfo
  {title} {{F}ixed {P}oints and {E}mergent {T}opological {P}henomena in a
  {P}arity-{T}ime-{S}ymmetric {Q}uantum {Q}uench},}\ }\bibinfo {note} {{a}rXiv:
  1806.10268}\BibitemShut {NoStop}%
\bibitem [{\citenamefont {Zeng}\ \emph {et~al.}()\citenamefont {Zeng},
  \citenamefont {Yang},\ and\ \citenamefont {Xu}}]{Zeng-19}%
  \BibitemOpen
  \bibfield  {author} {\bibinfo {author} {\bibfnamefont {Q.-B.}\ \bibnamefont
  {Zeng}}, \bibinfo {author} {\bibfnamefont {Y.-B.}\ \bibnamefont {Yang}}, \
  and\ \bibinfo {author} {\bibfnamefont {Y.}~\bibnamefont {Xu}},\ }\href@noop
  {} {\enquote {\bibinfo {title} {{Topological Phases in Non-Hermitian
  Aubry-Andr\'e-Harper Models}},}\ }\bibinfo {note} {{arXiv:
  1901.08060}}\BibitemShut {NoStop}%
\bibitem [{\citenamefont {Yokomizo}\ and\ \citenamefont
  {Murakami}()}]{Yokomizo-19}%
  \BibitemOpen
  \bibfield  {author} {\bibinfo {author} {\bibfnamefont {K.}~\bibnamefont
  {Yokomizo}}\ and\ \bibinfo {author} {\bibfnamefont {S.}~\bibnamefont
  {Murakami}},\ }\href@noop {} {\enquote {\bibinfo {title} {{Bloch Band Theory
  for Non-Hermitian Systems}},}\ }\bibinfo {note} {{a}rXiv:
  1902.10958}\BibitemShut {NoStop}%
\bibitem [{\citenamefont {Okuma}\ and\ \citenamefont {Sato}()}]{Okuma-19}%
  \BibitemOpen
  \bibfield  {author} {\bibinfo {author} {\bibfnamefont {N.}~\bibnamefont
  {Okuma}}\ and\ \bibinfo {author} {\bibfnamefont {M.}~\bibnamefont {Sato}},\
  }\href@noop {} {\enquote {\bibinfo {title} {{Topological Phase Transition
  Driven by Infinitesimal Instability: Majorana Fermions in Non-Hermitian
  Spintronics}},}\ }\bibinfo {note} {{a}rXiv: 1904.06355}\BibitemShut {NoStop}%
\bibitem [{\citenamefont {Poli}\ \emph {et~al.}(2015)\citenamefont {Poli},
  \citenamefont {Bellec}, \citenamefont {Kuhl}, \citenamefont {Mortessagne},\
  and\ \citenamefont {Schomerus}}]{Poli-15}%
  \BibitemOpen
  \bibfield  {author} {\bibinfo {author} {\bibfnamefont {C.}~\bibnamefont
  {Poli}}, \bibinfo {author} {\bibfnamefont {M.}~\bibnamefont {Bellec}},
  \bibinfo {author} {\bibfnamefont {U.}~\bibnamefont {Kuhl}}, \bibinfo {author}
  {\bibfnamefont {F.}~\bibnamefont {Mortessagne}}, \ and\ \bibinfo {author}
  {\bibfnamefont {H.}~\bibnamefont {Schomerus}},\ }\bibfield  {title} {\enquote
  {\bibinfo {title} {{S}elective {E}nhancement of {T}opologically {I}nduced
  {I}nterface {S}tates in a {D}ielectric {R}esonator {C}hain},}\ }\href@noop {}
  {\bibfield  {journal} {\bibinfo  {journal} {Nat. Commun.}\ }\textbf {\bibinfo
  {volume} {6}},\ \bibinfo {pages} {6710} (\bibinfo {year} {2015})}\BibitemShut
  {NoStop}%
\bibitem [{\citenamefont {Zeuner}\ \emph {et~al.}(2015)\citenamefont {Zeuner},
  \citenamefont {Rechtsman}, \citenamefont {Plotnik}, \citenamefont {Lumer},
  \citenamefont {Nolte}, \citenamefont {Rudner}, \citenamefont {Segev},\ and\
  \citenamefont {Szameit}}]{Zeuner-15}%
  \BibitemOpen
  \bibfield  {author} {\bibinfo {author} {\bibfnamefont {J.~M.}\ \bibnamefont
  {Zeuner}}, \bibinfo {author} {\bibfnamefont {M.~C.}\ \bibnamefont
  {Rechtsman}}, \bibinfo {author} {\bibfnamefont {Y.}~\bibnamefont {Plotnik}},
  \bibinfo {author} {\bibfnamefont {Y.}~\bibnamefont {Lumer}}, \bibinfo
  {author} {\bibfnamefont {S.}~\bibnamefont {Nolte}}, \bibinfo {author}
  {\bibfnamefont {M.~S.}\ \bibnamefont {Rudner}}, \bibinfo {author}
  {\bibfnamefont {M.}~\bibnamefont {Segev}}, \ and\ \bibinfo {author}
  {\bibfnamefont {A.}~\bibnamefont {Szameit}},\ }\bibfield  {title} {\enquote
  {\bibinfo {title} {{O}bservation of a {T}opological {T}ransition in the
  {B}ulk of a {N}on-{H}ermitian {S}ystem},}\ }\href@noop {} {\bibfield
  {journal} {\bibinfo  {journal} {Phys. Rev. Lett.}\ }\textbf {\bibinfo
  {volume} {115}},\ \bibinfo {pages} {040402} (\bibinfo {year}
  {2015})}\BibitemShut {NoStop}%
\bibitem [{\citenamefont {Zhen}\ \emph {et~al.}(2015)\citenamefont {Zhen},
  \citenamefont {Hsu}, \citenamefont {Igarashi}, \citenamefont {Lu},
  \citenamefont {Kaminer}, \citenamefont {Pick}, \citenamefont {Chua},
  \citenamefont {Joannopoulos},\ and\ \citenamefont
  {Solja\u{c}i\'c}}]{Zhen-15}%
  \BibitemOpen
  \bibfield  {author} {\bibinfo {author} {\bibfnamefont {B.}~\bibnamefont
  {Zhen}}, \bibinfo {author} {\bibfnamefont {C.~W.}\ \bibnamefont {Hsu}},
  \bibinfo {author} {\bibfnamefont {Y.}~\bibnamefont {Igarashi}}, \bibinfo
  {author} {\bibfnamefont {L.}~\bibnamefont {Lu}}, \bibinfo {author}
  {\bibfnamefont {I.}~\bibnamefont {Kaminer}}, \bibinfo {author} {\bibfnamefont
  {A.}~\bibnamefont {Pick}}, \bibinfo {author} {\bibfnamefont {S.-L.}\
  \bibnamefont {Chua}}, \bibinfo {author} {\bibfnamefont {J.~D.}\ \bibnamefont
  {Joannopoulos}}, \ and\ \bibinfo {author} {\bibfnamefont {M.}~\bibnamefont
  {Solja\u{c}i\'c}},\ }\bibfield  {title} {\enquote {\bibinfo {title}
  {{S}pawning {R}ings of {E}xceptional {P}oints out of {D}irac {C}ones},}\
  }\href@noop {} {\bibfield  {journal} {\bibinfo  {journal} {Nature}\ }\textbf
  {\bibinfo {volume} {525}},\ \bibinfo {pages} {354} (\bibinfo {year}
  {2015})}\BibitemShut {NoStop}%
\bibitem [{\citenamefont {Weimann}\ \emph {et~al.}(2017)\citenamefont
  {Weimann}, \citenamefont {Kremer}, \citenamefont {Plotnik}, \citenamefont
  {Lumer}, \citenamefont {Nolte}, \citenamefont {Makris}, \citenamefont
  {Segev}, \citenamefont {Rechtsman},\ and\ \citenamefont
  {Szameit}}]{Weimann-17}%
  \BibitemOpen
  \bibfield  {author} {\bibinfo {author} {\bibfnamefont {S.}~\bibnamefont
  {Weimann}}, \bibinfo {author} {\bibfnamefont {M.}~\bibnamefont {Kremer}},
  \bibinfo {author} {\bibfnamefont {Y.}~\bibnamefont {Plotnik}}, \bibinfo
  {author} {\bibfnamefont {Y.}~\bibnamefont {Lumer}}, \bibinfo {author}
  {\bibfnamefont {S.}~\bibnamefont {Nolte}}, \bibinfo {author} {\bibfnamefont
  {K.~G.}\ \bibnamefont {Makris}}, \bibinfo {author} {\bibfnamefont
  {M.}~\bibnamefont {Segev}}, \bibinfo {author} {\bibfnamefont {M.~C.}\
  \bibnamefont {Rechtsman}}, \ and\ \bibinfo {author} {\bibfnamefont
  {A.}~\bibnamefont {Szameit}},\ }\bibfield  {title} {\enquote {\bibinfo
  {title} {{T}opologically {P}rotected {B}ound {S}tates in {P}hotonic
  {P}arity-{T}ime-{S}ymmetric {C}rystals},}\ }\href@noop {} {\bibfield
  {journal} {\bibinfo  {journal} {Nat. Mater.}\ }\textbf {\bibinfo {volume}
  {16}},\ \bibinfo {pages} {433} (\bibinfo {year} {2017})}\BibitemShut
  {NoStop}%
\bibitem [{\citenamefont {Kim}\ \emph {et~al.}()\citenamefont {Kim},
  \citenamefont {Mochizuki}, \citenamefont {Kawakami},\ and\ \citenamefont
  {Obuse}}]{Obuse-17}%
  \BibitemOpen
  \bibfield  {author} {\bibinfo {author} {\bibfnamefont {D.}~\bibnamefont
  {Kim}}, \bibinfo {author} {\bibfnamefont {K.}~\bibnamefont {Mochizuki}},
  \bibinfo {author} {\bibfnamefont {N.}~\bibnamefont {Kawakami}}, \ and\
  \bibinfo {author} {\bibfnamefont {H.}~\bibnamefont {Obuse}},\ }\href@noop {}
  {\enquote {\bibinfo {title} {{F}loquet {T}opological {P}hases {D}riven by
  $\mathcal{PT}$ {S}ymmetric {N}onunitary {T}ime {E}volution},}\ }\bibinfo
  {note} {{a}rXiv: 1609.09650}\BibitemShut {NoStop}%
\bibitem [{\citenamefont {Xiao}\ \emph {et~al.}(2017)\citenamefont {Xiao},
  \citenamefont {Zhan}, \citenamefont {Bian}, \citenamefont {Wang},
  \citenamefont {Zhang}, \citenamefont {Wang}, \citenamefont {Li},
  \citenamefont {Mochizuki}, \citenamefont {Kim}, \citenamefont {Kawakami},
  \citenamefont {Yi}, \citenamefont {Obuse}, \citenamefont {Sanders},\ and\
  \citenamefont {Xue}}]{Xiao-17}%
  \BibitemOpen
  \bibfield  {author} {\bibinfo {author} {\bibfnamefont {L.}~\bibnamefont
  {Xiao}}, \bibinfo {author} {\bibfnamefont {X.}~\bibnamefont {Zhan}}, \bibinfo
  {author} {\bibfnamefont {Z.~H.}\ \bibnamefont {Bian}}, \bibinfo {author}
  {\bibfnamefont {K.~K.}\ \bibnamefont {Wang}}, \bibinfo {author}
  {\bibfnamefont {X.}~\bibnamefont {Zhang}}, \bibinfo {author} {\bibfnamefont
  {X.~P.}\ \bibnamefont {Wang}}, \bibinfo {author} {\bibfnamefont
  {J.}~\bibnamefont {Li}}, \bibinfo {author} {\bibfnamefont {K.}~\bibnamefont
  {Mochizuki}}, \bibinfo {author} {\bibfnamefont {D.}~\bibnamefont {Kim}},
  \bibinfo {author} {\bibfnamefont {N.}~\bibnamefont {Kawakami}}, \bibinfo
  {author} {\bibfnamefont {W.}~\bibnamefont {Yi}}, \bibinfo {author}
  {\bibfnamefont {H.}~\bibnamefont {Obuse}}, \bibinfo {author} {\bibfnamefont
  {B.~C.}\ \bibnamefont {Sanders}}, \ and\ \bibinfo {author} {\bibfnamefont
  {P.}~\bibnamefont {Xue}},\ }\bibfield  {title} {\enquote {\bibinfo {title}
  {{O}bservation of {T}opological {E}dge {S}tates in
  {P}arity-{T}ime-{S}ymmetric {Q}uantum {W}alks},}\ }\href@noop {} {\bibfield
  {journal} {\bibinfo  {journal} {Nat. Phys.}\ }\textbf {\bibinfo {volume}
  {13}},\ \bibinfo {pages} {1117} (\bibinfo {year} {2017})}\BibitemShut
  {NoStop}%
\bibitem [{\citenamefont {St-Jean}\ \emph {et~al.}(2017)\citenamefont
  {St-Jean}, \citenamefont {Goblot}, \citenamefont {Galopin}, \citenamefont
  {Lema\^itre}, \citenamefont {Ozawa}, \citenamefont {Gratiet}, \citenamefont
  {Sagnes}, \citenamefont {Bloch},\ and\ \citenamefont {Amo}}]{St-Jean-17}%
  \BibitemOpen
  \bibfield  {author} {\bibinfo {author} {\bibfnamefont {P.}~\bibnamefont
  {St-Jean}}, \bibinfo {author} {\bibfnamefont {V.}~\bibnamefont {Goblot}},
  \bibinfo {author} {\bibfnamefont {E.}~\bibnamefont {Galopin}}, \bibinfo
  {author} {\bibfnamefont {A.}~\bibnamefont {Lema\^itre}}, \bibinfo {author}
  {\bibfnamefont {T.}~\bibnamefont {Ozawa}}, \bibinfo {author} {\bibfnamefont
  {L.~Le}\ \bibnamefont {Gratiet}}, \bibinfo {author} {\bibfnamefont
  {I.}~\bibnamefont {Sagnes}}, \bibinfo {author} {\bibfnamefont
  {J.}~\bibnamefont {Bloch}}, \ and\ \bibinfo {author} {\bibfnamefont
  {A.}~\bibnamefont {Amo}},\ }\bibfield  {title} {\enquote {\bibinfo {title}
  {{L}asing in {T}opological {E}dge {S}tates of a {O}ne-{D}imensional
  {L}attice},}\ }\href@noop {} {\bibfield  {journal} {\bibinfo  {journal} {Nat.
  Photon.}\ }\textbf {\bibinfo {volume} {11}},\ \bibinfo {pages} {651}
  (\bibinfo {year} {2017})}\BibitemShut {NoStop}%
\bibitem [{\citenamefont {Bahari}\ \emph {et~al.}(2017)\citenamefont {Bahari},
  \citenamefont {Ndao}, \citenamefont {Vallini}, \citenamefont {Amili},
  \citenamefont {Fainman},\ and\ \citenamefont {Kant\'e}}]{Bahari-17}%
  \BibitemOpen
  \bibfield  {author} {\bibinfo {author} {\bibfnamefont {B.}~\bibnamefont
  {Bahari}}, \bibinfo {author} {\bibfnamefont {A.}~\bibnamefont {Ndao}},
  \bibinfo {author} {\bibfnamefont {F.}~\bibnamefont {Vallini}}, \bibinfo
  {author} {\bibfnamefont {A.~El}\ \bibnamefont {Amili}}, \bibinfo {author}
  {\bibfnamefont {Y.}~\bibnamefont {Fainman}}, \ and\ \bibinfo {author}
  {\bibfnamefont {B.}~\bibnamefont {Kant\'e}},\ }\bibfield  {title} {\enquote
  {\bibinfo {title} {{Nonreciprocal Lasing in Topological Cavities of Arbitrary
  Geometries}},}\ }\href@noop {} {\bibfield  {journal} {\bibinfo  {journal}
  {Science}\ }\textbf {\bibinfo {volume} {358}},\ \bibinfo {pages} {636}
  (\bibinfo {year} {2017})}\BibitemShut {NoStop}%
\bibitem [{\citenamefont {Zhou}\ \emph {et~al.}(2018)\citenamefont {Zhou},
  \citenamefont {Peng}, \citenamefont {Yoon}, \citenamefont {Hsu},
  \citenamefont {Nelson}, \citenamefont {Fu}, \citenamefont {Joannopoulos},
  \citenamefont {Solja\u{c}i\'c},\ and\ \citenamefont {Zhen}}]{Zhou-18-exp}%
  \BibitemOpen
  \bibfield  {author} {\bibinfo {author} {\bibfnamefont {H.}~\bibnamefont
  {Zhou}}, \bibinfo {author} {\bibfnamefont {C.}~\bibnamefont {Peng}}, \bibinfo
  {author} {\bibfnamefont {Y.}~\bibnamefont {Yoon}}, \bibinfo {author}
  {\bibfnamefont {C.~W.}\ \bibnamefont {Hsu}}, \bibinfo {author} {\bibfnamefont
  {K.~A.}\ \bibnamefont {Nelson}}, \bibinfo {author} {\bibfnamefont
  {L.}~\bibnamefont {Fu}}, \bibinfo {author} {\bibfnamefont {J.~D.}\
  \bibnamefont {Joannopoulos}}, \bibinfo {author} {\bibfnamefont
  {M.}~\bibnamefont {Solja\u{c}i\'c}}, \ and\ \bibinfo {author} {\bibfnamefont
  {B.}~\bibnamefont {Zhen}},\ }\bibfield  {title} {\enquote {\bibinfo {title}
  {{O}bservation of {B}ulk {F}ermi {A}rc and {P}olarization {H}alf {C}harge
  from {P}aired {E}xceptional {P}oints},}\ }\href@noop {} {\bibfield  {journal}
  {\bibinfo  {journal} {Science}\ }\textbf {\bibinfo {volume} {359}},\ \bibinfo
  {pages} {1009} (\bibinfo {year} {2018})}\BibitemShut {NoStop}%
\bibitem [{\citenamefont {Zhao}\ \emph {et~al.}(2018)\citenamefont {Zhao},
  \citenamefont {Miao}, \citenamefont {Teimourpour}, \citenamefont {Malzard},
  \citenamefont {El-Ganainy}, \citenamefont {Schomerus},\ and\ \citenamefont
  {Feng}}]{Zhao-18}%
  \BibitemOpen
  \bibfield  {author} {\bibinfo {author} {\bibfnamefont {H.}~\bibnamefont
  {Zhao}}, \bibinfo {author} {\bibfnamefont {P.}~\bibnamefont {Miao}}, \bibinfo
  {author} {\bibfnamefont {M.~H.}\ \bibnamefont {Teimourpour}}, \bibinfo
  {author} {\bibfnamefont {S.}~\bibnamefont {Malzard}}, \bibinfo {author}
  {\bibfnamefont {R.}~\bibnamefont {El-Ganainy}}, \bibinfo {author}
  {\bibfnamefont {H.}~\bibnamefont {Schomerus}}, \ and\ \bibinfo {author}
  {\bibfnamefont {L.}~\bibnamefont {Feng}},\ }\bibfield  {title} {\enquote
  {\bibinfo {title} {{T}opological {H}ybrid {S}ilicon {M}icrolasers},}\
  }\href@noop {} {\bibfield  {journal} {\bibinfo  {journal} {Nat. Commun.}\
  }\textbf {\bibinfo {volume} {9}},\ \bibinfo {pages} {981} (\bibinfo {year}
  {2018})}\BibitemShut {NoStop}%
\bibitem [{\citenamefont {Parto}\ \emph {et~al.}(2018)\citenamefont {Parto},
  \citenamefont {Wittek}, \citenamefont {Hodaei}, \citenamefont {Harari},
  \citenamefont {Bandres}, \citenamefont {Ren}, \citenamefont {Rechtsman},
  \citenamefont {Segev}, \citenamefont {Christodoulides},\ and\ \citenamefont
  {Khajavikhan}}]{Parto-18}%
  \BibitemOpen
  \bibfield  {author} {\bibinfo {author} {\bibfnamefont {M.}~\bibnamefont
  {Parto}}, \bibinfo {author} {\bibfnamefont {S.}~\bibnamefont {Wittek}},
  \bibinfo {author} {\bibfnamefont {H.}~\bibnamefont {Hodaei}}, \bibinfo
  {author} {\bibfnamefont {G.}~\bibnamefont {Harari}}, \bibinfo {author}
  {\bibfnamefont {M.~A.}\ \bibnamefont {Bandres}}, \bibinfo {author}
  {\bibfnamefont {J.}~\bibnamefont {Ren}}, \bibinfo {author} {\bibfnamefont
  {M.~C.}\ \bibnamefont {Rechtsman}}, \bibinfo {author} {\bibfnamefont
  {M.}~\bibnamefont {Segev}}, \bibinfo {author} {\bibfnamefont {D.~N.}\
  \bibnamefont {Christodoulides}}, \ and\ \bibinfo {author} {\bibfnamefont
  {M.}~\bibnamefont {Khajavikhan}},\ }\bibfield  {title} {\enquote {\bibinfo
  {title} {{Edge-Mode Lasing in 1D Topological Active Arrays}},}\ }\href@noop
  {} {\bibfield  {journal} {\bibinfo  {journal} {Phys. Rev. Lett.}\ }\textbf
  {\bibinfo {volume} {120}},\ \bibinfo {pages} {113901} (\bibinfo {year}
  {2018})}\BibitemShut {NoStop}%
\bibitem [{\citenamefont {Harari}\ \emph {et~al.}(2018)\citenamefont {Harari},
  \citenamefont {Bandres}, \citenamefont {Lumer}, \citenamefont {Rechtsman},
  \citenamefont {Chong}, \citenamefont {Khajavikhan}, \citenamefont
  {Christodoulides},\ and\ \citenamefont {Segev}}]{Harari-18}%
  \BibitemOpen
  \bibfield  {author} {\bibinfo {author} {\bibfnamefont {G.}~\bibnamefont
  {Harari}}, \bibinfo {author} {\bibfnamefont {M.~A.}\ \bibnamefont {Bandres}},
  \bibinfo {author} {\bibfnamefont {Y.}~\bibnamefont {Lumer}}, \bibinfo
  {author} {\bibfnamefont {M.~C.}\ \bibnamefont {Rechtsman}}, \bibinfo {author}
  {\bibfnamefont {Y.~D.}\ \bibnamefont {Chong}}, \bibinfo {author}
  {\bibfnamefont {M.}~\bibnamefont {Khajavikhan}}, \bibinfo {author}
  {\bibfnamefont {D.~N.}\ \bibnamefont {Christodoulides}}, \ and\ \bibinfo
  {author} {\bibfnamefont {M.}~\bibnamefont {Segev}},\ }\bibfield  {title}
  {\enquote {\bibinfo {title} {{T}opological {I}nsulator {L}aser: {T}heory},}\
  }\href@noop {} {\bibfield  {journal} {\bibinfo  {journal} {Science}\ }\textbf
  {\bibinfo {volume} {359}},\ \bibinfo {pages} {eaar4003} (\bibinfo {year}
  {2018})}\BibitemShut {NoStop}%
\bibitem [{\citenamefont {Bandres}\ \emph {et~al.}(2018)\citenamefont
  {Bandres}, \citenamefont {Wittek}, \citenamefont {Harari}, \citenamefont
  {Parto}, \citenamefont {Ren}, \citenamefont {Segev}, \citenamefont
  {Christodoulides},\ and\ \citenamefont {Khajavikhan}}]{Bandres-18}%
  \BibitemOpen
  \bibfield  {author} {\bibinfo {author} {\bibfnamefont {M.~A.}\ \bibnamefont
  {Bandres}}, \bibinfo {author} {\bibfnamefont {S.}~\bibnamefont {Wittek}},
  \bibinfo {author} {\bibfnamefont {G.}~\bibnamefont {Harari}}, \bibinfo
  {author} {\bibfnamefont {M.}~\bibnamefont {Parto}}, \bibinfo {author}
  {\bibfnamefont {J.}~\bibnamefont {Ren}}, \bibinfo {author} {\bibfnamefont
  {M.}~\bibnamefont {Segev}}, \bibinfo {author} {\bibfnamefont
  {D.}~\bibnamefont {Christodoulides}}, \ and\ \bibinfo {author} {\bibfnamefont
  {M.}~\bibnamefont {Khajavikhan}},\ }\bibfield  {title} {\enquote {\bibinfo
  {title} {{T}opological {I}nsulator {L}aser: {E}xperiments},}\ }\href@noop {}
  {\bibfield  {journal} {\bibinfo  {journal} {{Science}}\ }\textbf {\bibinfo
  {volume} {359}},\ \bibinfo {pages} {eaar4005} (\bibinfo {year}
  {2018})}\BibitemShut {NoStop}%
\bibitem [{\citenamefont {Wang}\ \emph
  {et~al.}(2019{\natexlab{b}})\citenamefont {Wang}, \citenamefont {Qiu},
  \citenamefont {Xiao}, \citenamefont {Zhan}, \citenamefont {Bian},
  \citenamefont {Sanders}, \citenamefont {Yi},\ and\ \citenamefont
  {Xue}}]{Wang-18}%
  \BibitemOpen
  \bibfield  {author} {\bibinfo {author} {\bibfnamefont {K.}~\bibnamefont
  {Wang}}, \bibinfo {author} {\bibfnamefont {X.}~\bibnamefont {Qiu}}, \bibinfo
  {author} {\bibfnamefont {L.}~\bibnamefont {Xiao}}, \bibinfo {author}
  {\bibfnamefont {X.}~\bibnamefont {Zhan}}, \bibinfo {author} {\bibfnamefont
  {Z.}~\bibnamefont {Bian}}, \bibinfo {author} {\bibfnamefont {B.~C.}\
  \bibnamefont {Sanders}}, \bibinfo {author} {\bibfnamefont {W.}~\bibnamefont
  {Yi}}, \ and\ \bibinfo {author} {\bibfnamefont {P.}~\bibnamefont {Xue}},\
  }\bibfield  {title} {\enquote {\bibinfo {title} {{Observation of Emergent
  Momentum-Time Skyrmions in Parity-Time-Symmetric Non-Unitary Quench
  Dynamics}},}\ }\href@noop {} {\bibfield  {journal} {\bibinfo  {journal} {Nat.
  Commun.}\ }\textbf {\bibinfo {volume} {10}},\ \bibinfo {pages} {2293}
  (\bibinfo {year} {2019}{\natexlab{b}})}\BibitemShut {NoStop}%
\bibitem [{\citenamefont {Cerjan}\ \emph {et~al.}(2019)\citenamefont {Cerjan},
  \citenamefont {Huang}, \citenamefont {Chen}, \citenamefont {Chong},\ and\
  \citenamefont {Rechtsman}}]{Cerjan-18-exp}%
  \BibitemOpen
  \bibfield  {author} {\bibinfo {author} {\bibfnamefont {A.}~\bibnamefont
  {Cerjan}}, \bibinfo {author} {\bibfnamefont {S.}~\bibnamefont {Huang}},
  \bibinfo {author} {\bibfnamefont {K.~P.}\ \bibnamefont {Chen}}, \bibinfo
  {author} {\bibfnamefont {Y.}~\bibnamefont {Chong}}, \ and\ \bibinfo {author}
  {\bibfnamefont {M.~C.}\ \bibnamefont {Rechtsman}},\ }\bibfield  {title}
  {\enquote {\bibinfo {title} {{Experimental Realization of a Weyl Exceptional
  Ring}},}\ }\href@noop {} {\bibfield  {journal} {\bibinfo  {journal} {Nat.
  Photon.}\ }\textbf {\bibinfo {volume} {13}},\ \bibinfo {pages} {623}
  (\bibinfo {year} {2019})}\BibitemShut {NoStop}%
\bibitem [{\citenamefont {Su}\ \emph {et~al.}(1979)\citenamefont {Su},
  \citenamefont {Schrieffer},\ and\ \citenamefont {Heeger}}]{SSH-79}%
  \BibitemOpen
  \bibfield  {author} {\bibinfo {author} {\bibfnamefont {W.~P.}\ \bibnamefont
  {Su}}, \bibinfo {author} {\bibfnamefont {J.~R.}\ \bibnamefont {Schrieffer}},
  \ and\ \bibinfo {author} {\bibfnamefont {A.~J.}\ \bibnamefont {Heeger}},\
  }\bibfield  {title} {\enquote {\bibinfo {title} {{S}olitons in
  {P}olyacetylene},}\ }\href@noop {} {\bibfield  {journal} {\bibinfo  {journal}
  {Phys. Rev. Lett.}\ }\textbf {\bibinfo {volume} {42}},\ \bibinfo {pages}
  {1698} (\bibinfo {year} {1979})}\BibitemShut {NoStop}%
\bibitem [{\citenamefont {Thouless}\ \emph {et~al.}(1982)\citenamefont
  {Thouless}, \citenamefont {Kohmoto}, \citenamefont {Nightingale},\ and\
  \citenamefont {den Nijs}}]{TKNN-82}%
  \BibitemOpen
  \bibfield  {author} {\bibinfo {author} {\bibfnamefont {D.~J.}\ \bibnamefont
  {Thouless}}, \bibinfo {author} {\bibfnamefont {M.}~\bibnamefont {Kohmoto}},
  \bibinfo {author} {\bibfnamefont {M.~P.}\ \bibnamefont {Nightingale}}, \ and\
  \bibinfo {author} {\bibfnamefont {M.}~\bibnamefont {den Nijs}},\ }\bibfield
  {title} {\enquote {\bibinfo {title} {{Q}uantized {H}all {C}onductance in a
  {T}wo-{D}imensional {P}eriodic {P}otential},}\ }\href@noop {} {\bibfield
  {journal} {\bibinfo  {journal} {Phys. Rev. Lett.}\ }\textbf {\bibinfo
  {volume} {49}},\ \bibinfo {pages} {405} (\bibinfo {year} {1982})}\BibitemShut
  {NoStop}%
\bibitem [{\citenamefont {Kohmoto}(1985)}]{Kohmoto-85}%
  \BibitemOpen
  \bibfield  {author} {\bibinfo {author} {\bibfnamefont {M.}~\bibnamefont
  {Kohmoto}},\ }\bibfield  {title} {\enquote {\bibinfo {title} {{T}opological
  {I}nvariant and the {Q}uantization of the {H}all {C}onductance},}\
  }\href@noop {} {\bibfield  {journal} {\bibinfo  {journal} {Ann. Phys.}\
  }\textbf {\bibinfo {volume} {160}},\ \bibinfo {pages} {343} (\bibinfo {year}
  {1985})}\BibitemShut {NoStop}%
\bibitem [{\citenamefont {Haldane}(1988)}]{Haldane-88}%
  \BibitemOpen
  \bibfield  {author} {\bibinfo {author} {\bibfnamefont {F.~D.~M.}\
  \bibnamefont {Haldane}},\ }\bibfield  {title} {\enquote {\bibinfo {title}
  {{M}odel for a {Q}uantum {H}all {E}ffect without {L}andau {L}evels:
  {C}ondensed-{M}atter {R}ealization of the ``{P}arity {A}nomaly''},}\
  }\href@noop {} {\bibfield  {journal} {\bibinfo  {journal} {Phys. Rev. Lett.}\
  }\textbf {\bibinfo {volume} {61}},\ \bibinfo {pages} {2015} (\bibinfo {year}
  {1988})}\BibitemShut {NoStop}%
\bibitem [{\citenamefont {Kane}\ and\ \citenamefont
  {Mele}(2005{\natexlab{a}})}]{Kane-Mele-05-Z2}%
  \BibitemOpen
  \bibfield  {author} {\bibinfo {author} {\bibfnamefont {C.~L.}\ \bibnamefont
  {Kane}}\ and\ \bibinfo {author} {\bibfnamefont {E.~J.}\ \bibnamefont
  {Mele}},\ }\bibfield  {title} {\enquote {\bibinfo {title} {{$Z_{\rm 2}$}
  {T}opological {O}rder and the {Q}uantum {S}pin {H}all {E}ffect},}\
  }\href@noop {} {\bibfield  {journal} {\bibinfo  {journal} {Phys. Rev. Lett.}\
  }\textbf {\bibinfo {volume} {95}},\ \bibinfo {pages} {146802} (\bibinfo
  {year} {2005}{\natexlab{a}})}\BibitemShut {NoStop}%
\bibitem [{\citenamefont {Kane}\ and\ \citenamefont
  {Mele}(2005{\natexlab{b}})}]{Kane-Mele-05-QSH}%
  \BibitemOpen
  \bibfield  {author} {\bibinfo {author} {\bibfnamefont {C.~L.}\ \bibnamefont
  {Kane}}\ and\ \bibinfo {author} {\bibfnamefont {E.~J.}\ \bibnamefont
  {Mele}},\ }\bibfield  {title} {\enquote {\bibinfo {title} {{Q}uantum {S}pin
  {H}all {E}ffect in {G}raphene},}\ }\href@noop {} {\bibfield  {journal}
  {\bibinfo  {journal} {{Phys. Rev. Lett.}}\ }\textbf {\bibinfo {volume}
  {95}},\ \bibinfo {pages} {226801} (\bibinfo {year}
  {2005}{\natexlab{b}})}\BibitemShut {NoStop}%
\bibitem [{\citenamefont {Fu}\ and\ \citenamefont {Kane}(2006)}]{Fu-Kane-06}%
  \BibitemOpen
  \bibfield  {author} {\bibinfo {author} {\bibfnamefont {L.}~\bibnamefont
  {Fu}}\ and\ \bibinfo {author} {\bibfnamefont {C.~L.}\ \bibnamefont {Kane}},\
  }\bibfield  {title} {\enquote {\bibinfo {title} {{T}ime {R}eversal
  {P}olarization and a {$Z_{2}$} {A}diabatic {S}pin {P}ump},}\ }\href@noop {}
  {\bibfield  {journal} {\bibinfo  {journal} {Phys. Rev. B}\ }\textbf {\bibinfo
  {volume} {74}},\ \bibinfo {pages} {195312} (\bibinfo {year}
  {2006})}\BibitemShut {NoStop}%
\bibitem [{\citenamefont {Bernevig}\ \emph {et~al.}(2006)\citenamefont
  {Bernevig}, \citenamefont {Hughes},\ and\ \citenamefont {Zhang}}]{BHZ-06}%
  \BibitemOpen
  \bibfield  {author} {\bibinfo {author} {\bibfnamefont {B.~A.}\ \bibnamefont
  {Bernevig}}, \bibinfo {author} {\bibfnamefont {T.~L.}\ \bibnamefont
  {Hughes}}, \ and\ \bibinfo {author} {\bibfnamefont {S.-C.}\ \bibnamefont
  {Zhang}},\ }\bibfield  {title} {\enquote {\bibinfo {title} {{Q}uantum {S}pin
  {H}all {E}ffect and {T}opological {P}hase {T}ransition in {H}g{T}e {Q}uantum
  {W}ells},}\ }\href@noop {} {\bibfield  {journal} {\bibinfo  {journal}
  {Science}\ }\textbf {\bibinfo {volume} {314}},\ \bibinfo {pages} {1757}
  (\bibinfo {year} {2006})}\BibitemShut {NoStop}%
\bibitem [{\citenamefont {Fu}\ \emph {et~al.}(2007)\citenamefont {Fu},
  \citenamefont {Kane},\ and\ \citenamefont {Mele}}]{Fu-Kane-Mele-07}%
  \BibitemOpen
  \bibfield  {author} {\bibinfo {author} {\bibfnamefont {L.}~\bibnamefont
  {Fu}}, \bibinfo {author} {\bibfnamefont {C.~L.}\ \bibnamefont {Kane}}, \ and\
  \bibinfo {author} {\bibfnamefont {E.~J.}\ \bibnamefont {Mele}},\ }\bibfield
  {title} {\enquote {\bibinfo {title} {{T}opological {I}nsulators in {T}hree
  {D}imensions},}\ }\href@noop {} {\bibfield  {journal} {\bibinfo  {journal}
  {Phys. Rev. Lett.}\ }\textbf {\bibinfo {volume} {98}},\ \bibinfo {pages}
  {106803} (\bibinfo {year} {2007})}\BibitemShut {NoStop}%
\bibitem [{\citenamefont {Moore}\ and\ \citenamefont
  {Balents}(2007)}]{Moore-Balents-07}%
  \BibitemOpen
  \bibfield  {author} {\bibinfo {author} {\bibfnamefont {J.~E.}\ \bibnamefont
  {Moore}}\ and\ \bibinfo {author} {\bibfnamefont {L.}~\bibnamefont
  {Balents}},\ }\bibfield  {title} {\enquote {\bibinfo {title} {{Topological
  Invariants of Time-Reversal-Invariant Band Structures}},}\ }\href@noop {}
  {\bibfield  {journal} {\bibinfo  {journal} {Phys. Rev. B}\ }\textbf {\bibinfo
  {volume} {75}},\ \bibinfo {pages} {121306(R)} (\bibinfo {year}
  {2007})}\BibitemShut {NoStop}%
\bibitem [{\citenamefont {Fu}\ and\ \citenamefont {Kane}(2007)}]{Fu-Kane-07}%
  \BibitemOpen
  \bibfield  {author} {\bibinfo {author} {\bibfnamefont {L.}~\bibnamefont
  {Fu}}\ and\ \bibinfo {author} {\bibfnamefont {C.~L.}\ \bibnamefont {Kane}},\
  }\bibfield  {title} {\enquote {\bibinfo {title} {{Topological Insulators with
  Inversion Symmetry}},}\ }\href@noop {} {\bibfield  {journal} {\bibinfo
  {journal} {Phys. Rev. B}\ }\textbf {\bibinfo {volume} {76}},\ \bibinfo
  {pages} {045302} (\bibinfo {year} {2007})}\BibitemShut {NoStop}%
\bibitem [{\citenamefont {Qi}\ \emph {et~al.}(2008)\citenamefont {Qi},
  \citenamefont {Hughes},\ and\ \citenamefont {Zhang}}]{Qi-08}%
  \BibitemOpen
  \bibfield  {author} {\bibinfo {author} {\bibfnamefont {X.-L.}\ \bibnamefont
  {Qi}}, \bibinfo {author} {\bibfnamefont {T.~L.}\ \bibnamefont {Hughes}}, \
  and\ \bibinfo {author} {\bibfnamefont {S.-C.}\ \bibnamefont {Zhang}},\
  }\bibfield  {title} {\enquote {\bibinfo {title} {{Topological Field Theory of
  Time-Reversal Invariant Insulators}},}\ }\href@noop {} {\bibfield  {journal}
  {\bibinfo  {journal} {Phys. Rev. B}\ }\textbf {\bibinfo {volume} {78}},\
  \bibinfo {pages} {195424} (\bibinfo {year} {2008})}\BibitemShut {NoStop}%
\bibitem [{\citenamefont {Roy}(2009)}]{Roy-09}%
  \BibitemOpen
  \bibfield  {author} {\bibinfo {author} {\bibfnamefont {R.}~\bibnamefont
  {Roy}},\ }\bibfield  {title} {\enquote {\bibinfo {title} {{Topological Phases
  and the Quantum Spin Hall Effect in Three Dimensions}},}\ }\href@noop {}
  {\bibfield  {journal} {\bibinfo  {journal} {Phys. Rev. B}\ }\textbf {\bibinfo
  {volume} {79}},\ \bibinfo {pages} {195322} (\bibinfo {year}
  {2009})}\BibitemShut {NoStop}%
\bibitem [{\citenamefont {K\"onig}\ \emph {et~al.}(2007)\citenamefont
  {K\"onig}, \citenamefont {Wiedmann}, \citenamefont {Br\"une}, \citenamefont
  {Roth}, \citenamefont {Buhmann}, \citenamefont {Molenkamp}, \citenamefont
  {Qi},\ and\ \citenamefont {Zhang}}]{Konig-07}%
  \BibitemOpen
  \bibfield  {author} {\bibinfo {author} {\bibfnamefont {M.}~\bibnamefont
  {K\"onig}}, \bibinfo {author} {\bibfnamefont {S.}~\bibnamefont {Wiedmann}},
  \bibinfo {author} {\bibfnamefont {C.}~\bibnamefont {Br\"une}}, \bibinfo
  {author} {\bibfnamefont {A.}~\bibnamefont {Roth}}, \bibinfo {author}
  {\bibfnamefont {H.}~\bibnamefont {Buhmann}}, \bibinfo {author} {\bibfnamefont
  {L.}~\bibnamefont {Molenkamp}}, \bibinfo {author} {\bibfnamefont {X.-L.}\
  \bibnamefont {Qi}}, \ and\ \bibinfo {author} {\bibfnamefont {S.-C.}\
  \bibnamefont {Zhang}},\ }\bibfield  {title} {\enquote {\bibinfo {title}
  {{Quantum Spin Hall Insulator State in HgTe Quantum Wells}},}\ }\href@noop {}
  {\bibfield  {journal} {\bibinfo  {journal} {Science}\ }\textbf {\bibinfo
  {volume} {318}},\ \bibinfo {pages} {766} (\bibinfo {year}
  {2007})}\BibitemShut {NoStop}%
\bibitem [{\citenamefont {Hsieh}\ \emph {et~al.}(2008)\citenamefont {Hsieh},
  \citenamefont {Qian}, \citenamefont {Wray}, \citenamefont {Xia},
  \citenamefont {Hor}, \citenamefont {Cava},\ and\ \citenamefont
  {Hasan}}]{Hsieh-08}%
  \BibitemOpen
  \bibfield  {author} {\bibinfo {author} {\bibfnamefont {D.}~\bibnamefont
  {Hsieh}}, \bibinfo {author} {\bibfnamefont {D.}~\bibnamefont {Qian}},
  \bibinfo {author} {\bibfnamefont {L.}~\bibnamefont {Wray}}, \bibinfo {author}
  {\bibfnamefont {Y.}~\bibnamefont {Xia}}, \bibinfo {author} {\bibfnamefont
  {Y.~S.}\ \bibnamefont {Hor}}, \bibinfo {author} {\bibfnamefont {R.~J.}\
  \bibnamefont {Cava}}, \ and\ \bibinfo {author} {\bibfnamefont {M.~Z.}\
  \bibnamefont {Hasan}},\ }\bibfield  {title} {\enquote {\bibinfo {title} {{A
  Topological Dirac Insulator in a Quantum Spin Hall Phase}},}\ }\href@noop {}
  {\bibfield  {journal} {\bibinfo  {journal} {Nature}\ }\textbf {\bibinfo
  {volume} {452}},\ \bibinfo {pages} {970} (\bibinfo {year}
  {2008})}\BibitemShut {NoStop}%
\bibitem [{\citenamefont {Hasan}\ and\ \citenamefont
  {Kane}(2010)}]{Kane-review}%
  \BibitemOpen
  \bibfield  {author} {\bibinfo {author} {\bibfnamefont {M.~Z.}\ \bibnamefont
  {Hasan}}\ and\ \bibinfo {author} {\bibfnamefont {C.~L.}\ \bibnamefont
  {Kane}},\ }\bibfield  {title} {\enquote {\bibinfo {title} {{C}olloquium:
  {T}opological {I}nsulators},}\ }\href@noop {} {\bibfield  {journal} {\bibinfo
   {journal} {Rev. Mod. Phys.}\ }\textbf {\bibinfo {volume} {82}},\ \bibinfo
  {pages} {3045} (\bibinfo {year} {2010})}\BibitemShut {NoStop}%
\bibitem [{\citenamefont {Qi}\ and\ \citenamefont
  {Zhang}(2011)}]{Zhang-review}%
  \BibitemOpen
  \bibfield  {author} {\bibinfo {author} {\bibfnamefont {X.-L.}\ \bibnamefont
  {Qi}}\ and\ \bibinfo {author} {\bibfnamefont {S.-C.}\ \bibnamefont {Zhang}},\
  }\bibfield  {title} {\enquote {\bibinfo {title} {{T}opological {I}nsulators
  and {S}uperconductors},}\ }\href@noop {} {\bibfield  {journal} {\bibinfo
  {journal} {Rev. Mod. Phys.}\ }\textbf {\bibinfo {volume} {83}},\ \bibinfo
  {pages} {1057} (\bibinfo {year} {2011})}\BibitemShut {NoStop}%
\bibitem [{\citenamefont {Read}\ and\ \citenamefont {Green}(2000)}]{Read-00}%
  \BibitemOpen
  \bibfield  {author} {\bibinfo {author} {\bibfnamefont {N.}~\bibnamefont
  {Read}}\ and\ \bibinfo {author} {\bibfnamefont {D.}~\bibnamefont {Green}},\
  }\bibfield  {title} {\enquote {\bibinfo {title} {{Paired States of Fermions
  in Two Dimensions with Breaking of Parity and Time-Reversal Symmetries and
  the Fractional Quantum Hall Effect}},}\ }\href@noop {} {\bibfield  {journal}
  {\bibinfo  {journal} {Phys. Rev. B}\ }\textbf {\bibinfo {volume} {61}},\
  \bibinfo {pages} {10267} (\bibinfo {year} {2000})}\BibitemShut {NoStop}%
\bibitem [{\citenamefont {Kitaev}(2001)}]{Kitaev-01}%
  \BibitemOpen
  \bibfield  {author} {\bibinfo {author} {\bibfnamefont {A.~Y.}\ \bibnamefont
  {Kitaev}},\ }\bibfield  {title} {\enquote {\bibinfo {title} {{Unpaired
  Majorana Fermions in Quantum Wires}},}\ }\href@noop {} {\bibfield  {journal}
  {\bibinfo  {journal} {Phys.-Usp.}\ }\textbf {\bibinfo {volume} {44}},\
  \bibinfo {pages} {131} (\bibinfo {year} {2001})}\BibitemShut {NoStop}%
\bibitem [{\citenamefont {Ivanov}(2001)}]{Ivanov-01}%
  \BibitemOpen
  \bibfield  {author} {\bibinfo {author} {\bibfnamefont {D.~A.}\ \bibnamefont
  {Ivanov}},\ }\bibfield  {title} {\enquote {\bibinfo {title} {{Non-Abelian
  Statistics of Half-Quantum Vortices in $p$-Wave Superconductors}},}\
  }\href@noop {} {\bibfield  {journal} {\bibinfo  {journal} {Phys. Rev. Lett.}\
  }\textbf {\bibinfo {volume} {86}},\ \bibinfo {pages} {268} (\bibinfo {year}
  {2001})}\BibitemShut {NoStop}%
\bibitem [{\citenamefont {Fu}\ and\ \citenamefont {Kane}(2008)}]{Fu-Kane-08}%
  \BibitemOpen
  \bibfield  {author} {\bibinfo {author} {\bibfnamefont {L.}~\bibnamefont
  {Fu}}\ and\ \bibinfo {author} {\bibfnamefont {C.~L.}\ \bibnamefont {Kane}},\
  }\bibfield  {title} {\enquote {\bibinfo {title} {{Superconducting Proximity
  Effect and Majorana Fermions at the Surface of a Topological Insulator}},}\
  }\href@noop {} {\bibfield  {journal} {\bibinfo  {journal} {Phys. Rev. Lett.}\
  }\textbf {\bibinfo {volume} {100}},\ \bibinfo {pages} {096407} (\bibinfo
  {year} {2008})}\BibitemShut {NoStop}%
\bibitem [{\citenamefont {Sato}\ \emph {et~al.}(2009)\citenamefont {Sato},
  \citenamefont {Takahashi},\ and\ \citenamefont {Fujimoto}}]{Sato-09}%
  \BibitemOpen
  \bibfield  {author} {\bibinfo {author} {\bibfnamefont {M.}~\bibnamefont
  {Sato}}, \bibinfo {author} {\bibfnamefont {Y.}~\bibnamefont {Takahashi}}, \
  and\ \bibinfo {author} {\bibfnamefont {S.}~\bibnamefont {Fujimoto}},\
  }\bibfield  {title} {\enquote {\bibinfo {title} {{Non-Abelian Topological
  Order in $s$-Wave Superfluids of Ultracold Fermionic Atoms}},}\ }\href@noop
  {} {\bibfield  {journal} {\bibinfo  {journal} {Phys. Rev. Lett.}\ }\textbf
  {\bibinfo {volume} {103}},\ \bibinfo {pages} {020401} (\bibinfo {year}
  {2009})}\BibitemShut {NoStop}%
\bibitem [{\citenamefont {Sato}\ \emph {et~al.}(2010)\citenamefont {Sato},
  \citenamefont {Takahashi},\ and\ \citenamefont {Fujimoto}}]{Sato-10}%
  \BibitemOpen
  \bibfield  {author} {\bibinfo {author} {\bibfnamefont {M.}~\bibnamefont
  {Sato}}, \bibinfo {author} {\bibfnamefont {Y.}~\bibnamefont {Takahashi}}, \
  and\ \bibinfo {author} {\bibfnamefont {S.}~\bibnamefont {Fujimoto}},\
  }\bibfield  {title} {\enquote {\bibinfo {title} {{Non-Abelian Topological
  Orders and Majorana Fermions in Spin-Singlet Superconductors}},}\ }\href@noop
  {} {\bibfield  {journal} {\bibinfo  {journal} {Phys. Rev. B}\ }\textbf
  {\bibinfo {volume} {82}},\ \bibinfo {pages} {134521} (\bibinfo {year}
  {2010})}\BibitemShut {NoStop}%
\bibitem [{\citenamefont {Sau}\ \emph {et~al.}(2010)\citenamefont {Sau},
  \citenamefont {R.~M.~Lutchyn},\ and\ \citenamefont {Sarma}}]{Sau-10}%
  \BibitemOpen
  \bibfield  {author} {\bibinfo {author} {\bibfnamefont {J.~D.}\ \bibnamefont
  {Sau}}, \bibinfo {author} {\bibfnamefont {S.~Tewari}\ \bibnamefont
  {R.~M.~Lutchyn}}, \ and\ \bibinfo {author} {\bibfnamefont {S.~Das}\
  \bibnamefont {Sarma}},\ }\bibfield  {title} {\enquote {\bibinfo {title}
  {{Generic New Platform for Topological Quantum Computation Using
  Semiconductor Heterostructures}},}\ }\href@noop {} {\bibfield  {journal}
  {\bibinfo  {journal} {Phys. Rev. Lett.}\ }\textbf {\bibinfo {volume} {104}},\
  \bibinfo {pages} {040502} (\bibinfo {year} {2010})}\BibitemShut {NoStop}%
\bibitem [{\citenamefont {Lutchyn}\ \emph {et~al.}(2010)\citenamefont
  {Lutchyn}, \citenamefont {Sau},\ and\ \citenamefont {Sarma}}]{Lutchyn-10}%
  \BibitemOpen
  \bibfield  {author} {\bibinfo {author} {\bibfnamefont {R.~M.}\ \bibnamefont
  {Lutchyn}}, \bibinfo {author} {\bibfnamefont {J.~D.}\ \bibnamefont {Sau}}, \
  and\ \bibinfo {author} {\bibfnamefont {S.~Das}\ \bibnamefont {Sarma}},\
  }\bibfield  {title} {\enquote {\bibinfo {title} {{Majorana Fermions and a
  Topological Phase Transition in Semiconducotr-Superconductor
  Heterostructures}},}\ }\href@noop {} {\bibfield  {journal} {\bibinfo
  {journal} {{Phys. Rev. Lett.}}\ }\textbf {\bibinfo {volume} {105}},\ \bibinfo
  {pages} {077001} (\bibinfo {year} {2010})}\BibitemShut {NoStop}%
\bibitem [{\citenamefont {Oreg}\ \emph {et~al.}(2010)\citenamefont {Oreg},
  \citenamefont {Refael},\ and\ \citenamefont {von Oppen}}]{Oreg-10}%
  \BibitemOpen
  \bibfield  {author} {\bibinfo {author} {\bibfnamefont {Y.}~\bibnamefont
  {Oreg}}, \bibinfo {author} {\bibfnamefont {G.}~\bibnamefont {Refael}}, \ and\
  \bibinfo {author} {\bibfnamefont {F.}~\bibnamefont {von Oppen}},\ }\bibfield
  {title} {\enquote {\bibinfo {title} {{Helical Liquids and Majorana Bound
  States in Quantum Wires}},}\ }\href@noop {} {\bibfield  {journal} {\bibinfo
  {journal} {Phys. Rev. Lett.}\ }\textbf {\bibinfo {volume} {105}},\ \bibinfo
  {pages} {177002} (\bibinfo {year} {2010})}\BibitemShut {NoStop}%
\bibitem [{\citenamefont {Alicea}\ \emph {et~al.}(2011)\citenamefont {Alicea},
  \citenamefont {Oreg}, \citenamefont {Refael}, \citenamefont {von Oppen},\
  and\ \citenamefont {Fisher}}]{Alicea-11}%
  \BibitemOpen
  \bibfield  {author} {\bibinfo {author} {\bibfnamefont {J.}~\bibnamefont
  {Alicea}}, \bibinfo {author} {\bibfnamefont {Y.}~\bibnamefont {Oreg}},
  \bibinfo {author} {\bibfnamefont {G.}~\bibnamefont {Refael}}, \bibinfo
  {author} {\bibfnamefont {F.}~\bibnamefont {von Oppen}}, \ and\ \bibinfo
  {author} {\bibfnamefont {M.~P.~A.}\ \bibnamefont {Fisher}},\ }\bibfield
  {title} {\enquote {\bibinfo {title} {{Non-Abelian Statistics and Topological
  Quantum Information Processing in 1D Wire Networks}},}\ }\href@noop {}
  {\bibfield  {journal} {\bibinfo  {journal} {Nat. Phys.}\ }\textbf {\bibinfo
  {volume} {7}},\ \bibinfo {pages} {412} (\bibinfo {year} {2011})}\BibitemShut
  {NoStop}%
\bibitem [{\citenamefont {Alicea}(2012)}]{Alicea-review}%
  \BibitemOpen
  \bibfield  {author} {\bibinfo {author} {\bibfnamefont {J.}~\bibnamefont
  {Alicea}},\ }\bibfield  {title} {\enquote {\bibinfo {title} {{New Directions
  in the Pursuit of Majorana Fermions in Solid State Systems}},}\ }\href@noop
  {} {\bibfield  {journal} {\bibinfo  {journal} {Rep. Prog. Phys.}\ }\textbf
  {\bibinfo {volume} {75}},\ \bibinfo {pages} {076501} (\bibinfo {year}
  {2012})}\BibitemShut {NoStop}%
\bibitem [{\citenamefont {Sato}\ and\ \citenamefont
  {Ando}(2017)}]{Sato-review}%
  \BibitemOpen
  \bibfield  {author} {\bibinfo {author} {\bibfnamefont {M.}~\bibnamefont
  {Sato}}\ and\ \bibinfo {author} {\bibfnamefont {Y.}~\bibnamefont {Ando}},\
  }\bibfield  {title} {\enquote {\bibinfo {title} {{Topological
  Superconductors: A Review}},}\ }\href@noop {} {\bibfield  {journal} {\bibinfo
   {journal} {Rep. Prog. Phys.}\ }\textbf {\bibinfo {volume} {80}},\ \bibinfo
  {pages} {076501} (\bibinfo {year} {2017})}\BibitemShut {NoStop}%
\bibitem [{\citenamefont {Haldane}\ and\ \citenamefont
  {Raghu}(2008)}]{Haldane-08}%
  \BibitemOpen
  \bibfield  {author} {\bibinfo {author} {\bibfnamefont {F.~D.~M.}\
  \bibnamefont {Haldane}}\ and\ \bibinfo {author} {\bibfnamefont
  {S.}~\bibnamefont {Raghu}},\ }\bibfield  {title} {\enquote {\bibinfo {title}
  {{Possible Realization of Directional Optical Waveguides in Photonic Crystals
  with Broken Time-Reversal Symmetry}},}\ }\href@noop {} {\bibfield  {journal}
  {\bibinfo  {journal} {Phys. Rev. Lett.}\ }\textbf {\bibinfo {volume} {100}},\
  \bibinfo {pages} {013904} (\bibinfo {year} {2008})}\BibitemShut {NoStop}%
\bibitem [{\citenamefont {Raghu}\ and\ \citenamefont
  {Haldane}(2008)}]{Raghu-08}%
  \BibitemOpen
  \bibfield  {author} {\bibinfo {author} {\bibfnamefont {S.}~\bibnamefont
  {Raghu}}\ and\ \bibinfo {author} {\bibfnamefont {F.~D.~M.}\ \bibnamefont
  {Haldane}},\ }\bibfield  {title} {\enquote {\bibinfo {title} {{Analogs of
  Quantum-Hall-Effect Edge States in Photonic Crystals}},}\ }\href@noop {}
  {\bibfield  {journal} {\bibinfo  {journal} {Phys. Rev. A}\ }\textbf {\bibinfo
  {volume} {78}},\ \bibinfo {pages} {033834} (\bibinfo {year}
  {2008})}\BibitemShut {NoStop}%
\bibitem [{\citenamefont {Wang}\ \emph {et~al.}(2008)\citenamefont {Wang},
  \citenamefont {Chong}, \citenamefont {Joannopoulos},\ and\ \citenamefont
  {Solja\u{c}i\'c}}]{Wang-08}%
  \BibitemOpen
  \bibfield  {author} {\bibinfo {author} {\bibfnamefont {Z.}~\bibnamefont
  {Wang}}, \bibinfo {author} {\bibfnamefont {Y.~D.}\ \bibnamefont {Chong}},
  \bibinfo {author} {\bibfnamefont {J.~D.}\ \bibnamefont {Joannopoulos}}, \
  and\ \bibinfo {author} {\bibfnamefont {M.}~\bibnamefont {Solja\u{c}i\'c}},\
  }\bibfield  {title} {\enquote {\bibinfo {title} {{Reflection-Free One-Way
  Edge Modes in a Gyromagnetic Photonic Crystal}},}\ }\href@noop {} {\bibfield
  {journal} {\bibinfo  {journal} {Phys. Rev. Lett.}\ }\textbf {\bibinfo
  {volume} {100}},\ \bibinfo {pages} {013905} (\bibinfo {year}
  {2008})}\BibitemShut {NoStop}%
\bibitem [{\citenamefont {Wang}\ \emph {et~al.}(2009)\citenamefont {Wang},
  \citenamefont {Chong}, \citenamefont {Joannopoulos},\ and\ \citenamefont
  {Solja\u{c}i\'c}}]{Wang-09}%
  \BibitemOpen
  \bibfield  {author} {\bibinfo {author} {\bibfnamefont {Z.}~\bibnamefont
  {Wang}}, \bibinfo {author} {\bibfnamefont {Y.~D.}\ \bibnamefont {Chong}},
  \bibinfo {author} {\bibfnamefont {J.~D.}\ \bibnamefont {Joannopoulos}}, \
  and\ \bibinfo {author} {\bibfnamefont {M.}~\bibnamefont {Solja\u{c}i\'c}},\
  }\bibfield  {title} {\enquote {\bibinfo {title} {{Observation of
  Unidirectional Backscattering-Immune Topological Electromagnetic States}},}\
  }\href@noop {} {\bibfield  {journal} {\bibinfo  {journal} {Nature}\ }\textbf
  {\bibinfo {volume} {461}},\ \bibinfo {pages} {772} (\bibinfo {year}
  {2009})}\BibitemShut {NoStop}%
\bibitem [{\citenamefont {Hafezi}\ \emph {et~al.}(2011)\citenamefont {Hafezi},
  \citenamefont {Demler}, \citenamefont {Lukin},\ and\ \citenamefont
  {Taylor}}]{Hafezi-11}%
  \BibitemOpen
  \bibfield  {author} {\bibinfo {author} {\bibfnamefont {M.}~\bibnamefont
  {Hafezi}}, \bibinfo {author} {\bibfnamefont {E.}~\bibnamefont {Demler}},
  \bibinfo {author} {\bibfnamefont {M.}~\bibnamefont {Lukin}}, \ and\ \bibinfo
  {author} {\bibfnamefont {J.~M.}\ \bibnamefont {Taylor}},\ }\bibfield  {title}
  {\enquote {\bibinfo {title} {{Robust Optical Delay Lines with Topological
  Protection}},}\ }\href@noop {} {\bibfield  {journal} {\bibinfo  {journal}
  {Nat. Phys.}\ }\textbf {\bibinfo {volume} {7}},\ \bibinfo {pages} {907}
  (\bibinfo {year} {2011})}\BibitemShut {NoStop}%
\bibitem [{\citenamefont {Fang}\ \emph {et~al.}(2012)\citenamefont {Fang},
  \citenamefont {Yu},\ and\ \citenamefont {Fan}}]{Fang-12}%
  \BibitemOpen
  \bibfield  {author} {\bibinfo {author} {\bibfnamefont {K.}~\bibnamefont
  {Fang}}, \bibinfo {author} {\bibfnamefont {Z.}~\bibnamefont {Yu}}, \ and\
  \bibinfo {author} {\bibfnamefont {S.}~\bibnamefont {Fan}},\ }\bibfield
  {title} {\enquote {\bibinfo {title} {{Realizing Effective Magnetic Field for
  Photons by Controlling the Phase of Dynamic Modulation}},}\ }\href@noop {}
  {\bibfield  {journal} {\bibinfo  {journal} {Nat. Photon.}\ }\textbf {\bibinfo
  {volume} {6}},\ \bibinfo {pages} {782} (\bibinfo {year} {2012})}\BibitemShut
  {NoStop}%
\bibitem [{\citenamefont {Khanikaev}\ \emph {et~al.}(2013)\citenamefont
  {Khanikaev}, \citenamefont {Mousavi}, \citenamefont {Tse}, \citenamefont
  {Kargarian}, \citenamefont {MacDonald},\ and\ \citenamefont
  {Shvets}}]{Khanikaev-13}%
  \BibitemOpen
  \bibfield  {author} {\bibinfo {author} {\bibfnamefont {A.~B.}\ \bibnamefont
  {Khanikaev}}, \bibinfo {author} {\bibfnamefont {S.~H.}\ \bibnamefont
  {Mousavi}}, \bibinfo {author} {\bibfnamefont {W.-K.}\ \bibnamefont {Tse}},
  \bibinfo {author} {\bibfnamefont {M.}~\bibnamefont {Kargarian}}, \bibinfo
  {author} {\bibfnamefont {A.~H.}\ \bibnamefont {MacDonald}}, \ and\ \bibinfo
  {author} {\bibfnamefont {G.}~\bibnamefont {Shvets}},\ }\bibfield  {title}
  {\enquote {\bibinfo {title} {{Photonic Topological Insulators}},}\
  }\href@noop {} {\bibfield  {journal} {\bibinfo  {journal} {Nat. Mater.}\
  }\textbf {\bibinfo {volume} {12}},\ \bibinfo {pages} {233} (\bibinfo {year}
  {2013})}\BibitemShut {NoStop}%
\bibitem [{\citenamefont {Rechtsman}\ \emph {et~al.}(2013)\citenamefont
  {Rechtsman}, \citenamefont {Zeuner}, \citenamefont {Plotnik}, \citenamefont
  {Lumer}, \citenamefont {Podolsky}, \citenamefont {Dreisow}, \citenamefont
  {Nolte}, \citenamefont {Segev},\ and\ \citenamefont
  {Szameit}}]{Rechtsman-13}%
  \BibitemOpen
  \bibfield  {author} {\bibinfo {author} {\bibfnamefont {M.}~\bibnamefont
  {Rechtsman}}, \bibinfo {author} {\bibfnamefont {J.~M.}\ \bibnamefont
  {Zeuner}}, \bibinfo {author} {\bibfnamefont {Y.}~\bibnamefont {Plotnik}},
  \bibinfo {author} {\bibfnamefont {Y.}~\bibnamefont {Lumer}}, \bibinfo
  {author} {\bibfnamefont {D.}~\bibnamefont {Podolsky}}, \bibinfo {author}
  {\bibfnamefont {F.}~\bibnamefont {Dreisow}}, \bibinfo {author} {\bibfnamefont
  {S.}~\bibnamefont {Nolte}}, \bibinfo {author} {\bibfnamefont
  {M.}~\bibnamefont {Segev}}, \ and\ \bibinfo {author} {\bibfnamefont
  {A.}~\bibnamefont {Szameit}},\ }\bibfield  {title} {\enquote {\bibinfo
  {title} {{Photonic Floquet Topological Insulators}},}\ }\href@noop {}
  {\bibfield  {journal} {\bibinfo  {journal} {Nature}\ }\textbf {\bibinfo
  {volume} {496}},\ \bibinfo {pages} {196} (\bibinfo {year}
  {2013})}\BibitemShut {NoStop}%
\bibitem [{\citenamefont {Hafezi}\ \emph {et~al.}(2013)\citenamefont {Hafezi},
  \citenamefont {Mittal}, \citenamefont {Fan}, \citenamefont {Migdall},\ and\
  \citenamefont {Taylor}}]{Hafezi-13}%
  \BibitemOpen
  \bibfield  {author} {\bibinfo {author} {\bibfnamefont {M.}~\bibnamefont
  {Hafezi}}, \bibinfo {author} {\bibfnamefont {S.}~\bibnamefont {Mittal}},
  \bibinfo {author} {\bibfnamefont {J.}~\bibnamefont {Fan}}, \bibinfo {author}
  {\bibfnamefont {A.}~\bibnamefont {Migdall}}, \ and\ \bibinfo {author}
  {\bibfnamefont {J.~M.}\ \bibnamefont {Taylor}},\ }\bibfield  {title}
  {\enquote {\bibinfo {title} {{Imaging Topological Edge States in Silicon
  Photonics}},}\ }\href@noop {} {\bibfield  {journal} {\bibinfo  {journal}
  {Nat. Photon.}\ }\textbf {\bibinfo {volume} {7}},\ \bibinfo {pages} {1001}
  (\bibinfo {year} {2013})}\BibitemShut {NoStop}%
\bibitem [{\citenamefont {Bliokh}\ \emph {et~al.}(2015)\citenamefont {Bliokh},
  \citenamefont {Smirnova},\ and\ \citenamefont {Nori}}]{Bliokh-15}%
  \BibitemOpen
  \bibfield  {author} {\bibinfo {author} {\bibfnamefont {K.~Y.}\ \bibnamefont
  {Bliokh}}, \bibinfo {author} {\bibfnamefont {D.}~\bibnamefont {Smirnova}}, \
  and\ \bibinfo {author} {\bibfnamefont {F.}~\bibnamefont {Nori}},\ }\bibfield
  {title} {\enquote {\bibinfo {title} {{Quantum Spin Hall Effect of Light}},}\
  }\href@noop {} {\bibfield  {journal} {\bibinfo  {journal} {Science}\ }\textbf
  {\bibinfo {volume} {348}},\ \bibinfo {pages} {1448} (\bibinfo {year}
  {2015})}\BibitemShut {NoStop}%
\bibitem [{\citenamefont {Lu}\ \emph {et~al.}(2014)\citenamefont {Lu},
  \citenamefont {Joannopoulos},\ and\ \citenamefont
  {Solja\u{c}i\'c}}]{Lu-review}%
  \BibitemOpen
  \bibfield  {author} {\bibinfo {author} {\bibfnamefont {L.}~\bibnamefont
  {Lu}}, \bibinfo {author} {\bibfnamefont {J.~D.}\ \bibnamefont
  {Joannopoulos}}, \ and\ \bibinfo {author} {\bibfnamefont {M.}~\bibnamefont
  {Solja\u{c}i\'c}},\ }\bibfield  {title} {\enquote {\bibinfo {title}
  {{Topological Photonics}},}\ }\href@noop {} {\bibfield  {journal} {\bibinfo
  {journal} {Nat. Photon.}\ }\textbf {\bibinfo {volume} {8}},\ \bibinfo {pages}
  {821} (\bibinfo {year} {2014})}\BibitemShut {NoStop}%
\bibitem [{\citenamefont {Ozawa}\ \emph {et~al.}(2019)\citenamefont {Ozawa},
  \citenamefont {Price}, \citenamefont {Amo}, \citenamefont {Goldman},
  \citenamefont {Hafezi}, \citenamefont {Lu}, \citenamefont {Rechtsman},
  \citenamefont {Schuster}, \citenamefont {Simon}, \citenamefont {Zilberberg},\
  and\ \citenamefont {Carusotto}}]{Ozawa-review}%
  \BibitemOpen
  \bibfield  {author} {\bibinfo {author} {\bibfnamefont {Tomoki}\ \bibnamefont
  {Ozawa}}, \bibinfo {author} {\bibfnamefont {Hannah~M.}\ \bibnamefont
  {Price}}, \bibinfo {author} {\bibfnamefont {Alberto}\ \bibnamefont {Amo}},
  \bibinfo {author} {\bibfnamefont {Nathan}\ \bibnamefont {Goldman}}, \bibinfo
  {author} {\bibfnamefont {Mohammad}\ \bibnamefont {Hafezi}}, \bibinfo {author}
  {\bibfnamefont {Ling}\ \bibnamefont {Lu}}, \bibinfo {author} {\bibfnamefont
  {Mikael~C.}\ \bibnamefont {Rechtsman}}, \bibinfo {author} {\bibfnamefont
  {David}\ \bibnamefont {Schuster}}, \bibinfo {author} {\bibfnamefont
  {Jonathan}\ \bibnamefont {Simon}}, \bibinfo {author} {\bibfnamefont {Oded}\
  \bibnamefont {Zilberberg}}, \ and\ \bibinfo {author} {\bibfnamefont {Iacopo}\
  \bibnamefont {Carusotto}},\ }\bibfield  {title} {\enquote {\bibinfo {title}
  {{Topological Photonics}},}\ }\href@noop {} {\bibfield  {journal} {\bibinfo
  {journal} {Rev. Mod. Phys.}\ }\textbf {\bibinfo {volume} {91}},\ \bibinfo
  {pages} {015006} (\bibinfo {year} {2019})}\BibitemShut {NoStop}%
\bibitem [{\citenamefont {Aidelsburger}\ \emph {et~al.}(2011)\citenamefont
  {Aidelsburger}, \citenamefont {Atala}, \citenamefont {Nascimb\`ene},
  \citenamefont {Trotzky}, \citenamefont {Chen},\ and\ \citenamefont
  {Bloch}}]{Aidelsburger-11}%
  \BibitemOpen
  \bibfield  {author} {\bibinfo {author} {\bibfnamefont {M.}~\bibnamefont
  {Aidelsburger}}, \bibinfo {author} {\bibfnamefont {M.}~\bibnamefont {Atala}},
  \bibinfo {author} {\bibfnamefont {S.}~\bibnamefont {Nascimb\`ene}}, \bibinfo
  {author} {\bibfnamefont {S.}~\bibnamefont {Trotzky}}, \bibinfo {author}
  {\bibfnamefont {Y.-A.}\ \bibnamefont {Chen}}, \ and\ \bibinfo {author}
  {\bibfnamefont {I.}~\bibnamefont {Bloch}},\ }\bibfield  {title} {\enquote
  {\bibinfo {title} {{Experimental Realization of Strong Effective Magnetic
  Fields in an Optical Lattice}},}\ }\href@noop {} {\bibfield  {journal}
  {\bibinfo  {journal} {Phys. Rev. Lett.}\ }\textbf {\bibinfo {volume} {107}},\
  \bibinfo {pages} {255301} (\bibinfo {year} {2011})}\BibitemShut {NoStop}%
\bibitem [{\citenamefont {Aidelsburger}\ \emph {et~al.}(2013)\citenamefont
  {Aidelsburger}, \citenamefont {Atala}, \citenamefont {Lohse}, \citenamefont
  {Barreiro}, \citenamefont {Paredes},\ and\ \citenamefont
  {Bloch}}]{Aidelsburger-13}%
  \BibitemOpen
  \bibfield  {author} {\bibinfo {author} {\bibfnamefont {M.}~\bibnamefont
  {Aidelsburger}}, \bibinfo {author} {\bibfnamefont {M.}~\bibnamefont {Atala}},
  \bibinfo {author} {\bibfnamefont {M.}~\bibnamefont {Lohse}}, \bibinfo
  {author} {\bibfnamefont {J.~T.}\ \bibnamefont {Barreiro}}, \bibinfo {author}
  {\bibfnamefont {B.}~\bibnamefont {Paredes}}, \ and\ \bibinfo {author}
  {\bibfnamefont {I.}~\bibnamefont {Bloch}},\ }\bibfield  {title} {\enquote
  {\bibinfo {title} {{Realization of the Hofstadter Hamiltonian with Ultracold
  Atoms in Optical Lattices}},}\ }\href@noop {} {\bibfield  {journal} {\bibinfo
   {journal} {Phys. Rev. Lett.}\ }\textbf {\bibinfo {volume} {111}},\ \bibinfo
  {pages} {185301} (\bibinfo {year} {2013})}\BibitemShut {NoStop}%
\bibitem [{\citenamefont {Atala}\ \emph {et~al.}(2013)\citenamefont {Atala},
  \citenamefont {Aidelsburger}, \citenamefont {Barreiro}, \citenamefont
  {Abanin}, \citenamefont {Kitagawa}, \citenamefont {Demler},\ and\
  \citenamefont {Bloch}}]{Atala-13}%
  \BibitemOpen
  \bibfield  {author} {\bibinfo {author} {\bibfnamefont {M.}~\bibnamefont
  {Atala}}, \bibinfo {author} {\bibfnamefont {M.}~\bibnamefont {Aidelsburger}},
  \bibinfo {author} {\bibfnamefont {J.~T.}\ \bibnamefont {Barreiro}}, \bibinfo
  {author} {\bibfnamefont {D.}~\bibnamefont {Abanin}}, \bibinfo {author}
  {\bibfnamefont {T.}~\bibnamefont {Kitagawa}}, \bibinfo {author}
  {\bibfnamefont {E.}~\bibnamefont {Demler}}, \ and\ \bibinfo {author}
  {\bibfnamefont {I.}~\bibnamefont {Bloch}},\ }\bibfield  {title} {\enquote
  {\bibinfo {title} {{Direct Measurement of the Zak Phase in Topological Bloch
  Bands}},}\ }\href@noop {} {\bibfield  {journal} {\bibinfo  {journal} {Nat.
  Phys.}\ }\textbf {\bibinfo {volume} {9}},\ \bibinfo {pages} {795} (\bibinfo
  {year} {2013})}\BibitemShut {NoStop}%
\bibitem [{\citenamefont {Jotzu}\ \emph {et~al.}(2014)\citenamefont {Jotzu},
  \citenamefont {Messer}, \citenamefont {Desbuquois}, \citenamefont {Lebrat},
  \citenamefont {Uehlinger}, \citenamefont {Greif},\ and\ \citenamefont
  {Esslinger}}]{Jotzu-14}%
  \BibitemOpen
  \bibfield  {author} {\bibinfo {author} {\bibfnamefont {G.}~\bibnamefont
  {Jotzu}}, \bibinfo {author} {\bibfnamefont {M.}~\bibnamefont {Messer}},
  \bibinfo {author} {\bibfnamefont {R.}~\bibnamefont {Desbuquois}}, \bibinfo
  {author} {\bibfnamefont {M.}~\bibnamefont {Lebrat}}, \bibinfo {author}
  {\bibfnamefont {T.}~\bibnamefont {Uehlinger}}, \bibinfo {author}
  {\bibfnamefont {D.}~\bibnamefont {Greif}}, \ and\ \bibinfo {author}
  {\bibfnamefont {T.}~\bibnamefont {Esslinger}},\ }\bibfield  {title} {\enquote
  {\bibinfo {title} {{Experimental Realization of the Topological Haldane Model
  with Ultracold Fermions}},}\ }\href@noop {} {\bibfield  {journal} {\bibinfo
  {journal} {Nature}\ }\textbf {\bibinfo {volume} {515}},\ \bibinfo {pages}
  {237} (\bibinfo {year} {2014})}\BibitemShut {NoStop}%
\bibitem [{\citenamefont {Mancini}\ \emph {et~al.}(2015)\citenamefont
  {Mancini}, \citenamefont {Pagano}, \citenamefont {Cappellini}, \citenamefont
  {Livi}, \citenamefont {Rider}, \citenamefont {Catani}, \citenamefont {Sias},
  \citenamefont {Zoller}, \citenamefont {Inguscio}, \citenamefont {Dalmonte},\
  and\ \citenamefont {Fallani}}]{Mancini-15}%
  \BibitemOpen
  \bibfield  {author} {\bibinfo {author} {\bibfnamefont {M.}~\bibnamefont
  {Mancini}}, \bibinfo {author} {\bibfnamefont {G.}~\bibnamefont {Pagano}},
  \bibinfo {author} {\bibfnamefont {G.}~\bibnamefont {Cappellini}}, \bibinfo
  {author} {\bibfnamefont {L.}~\bibnamefont {Livi}}, \bibinfo {author}
  {\bibfnamefont {M.}~\bibnamefont {Rider}}, \bibinfo {author} {\bibfnamefont
  {J.}~\bibnamefont {Catani}}, \bibinfo {author} {\bibfnamefont
  {C.}~\bibnamefont {Sias}}, \bibinfo {author} {\bibfnamefont {P.}~\bibnamefont
  {Zoller}}, \bibinfo {author} {\bibfnamefont {M.}~\bibnamefont {Inguscio}},
  \bibinfo {author} {\bibfnamefont {M.}~\bibnamefont {Dalmonte}}, \ and\
  \bibinfo {author} {\bibfnamefont {L.}~\bibnamefont {Fallani}},\ }\bibfield
  {title} {\enquote {\bibinfo {title} {{Observation of Chiral Edge States with
  Neutral Fermions in Synthetic Hall Ribbons}},}\ }\href@noop {} {\bibfield
  {journal} {\bibinfo  {journal} {Science}\ }\textbf {\bibinfo {volume}
  {349}},\ \bibinfo {pages} {1510} (\bibinfo {year} {2015})}\BibitemShut
  {NoStop}%
\bibitem [{\citenamefont {Stuhl}\ \emph {et~al.}(2015)\citenamefont {Stuhl},
  \citenamefont {Lu}, \citenamefont {Aycock}, \citenamefont {Genkina},\ and\
  \citenamefont {Spielman}}]{Stuhl-15}%
  \BibitemOpen
  \bibfield  {author} {\bibinfo {author} {\bibfnamefont {B.~K.}\ \bibnamefont
  {Stuhl}}, \bibinfo {author} {\bibfnamefont {H.-I.}\ \bibnamefont {Lu}},
  \bibinfo {author} {\bibfnamefont {L.~M.}\ \bibnamefont {Aycock}}, \bibinfo
  {author} {\bibfnamefont {D.}~\bibnamefont {Genkina}}, \ and\ \bibinfo
  {author} {\bibfnamefont {I.~B.}\ \bibnamefont {Spielman}},\ }\bibfield
  {title} {\enquote {\bibinfo {title} {{Visualizing Edge States with an Atomic
  Bose Gas in the Quantum Hall Regime}},}\ }\href@noop {} {\bibfield  {journal}
  {\bibinfo  {journal} {Science}\ }\textbf {\bibinfo {volume} {349}},\ \bibinfo
  {pages} {1514} (\bibinfo {year} {2015})}\BibitemShut {NoStop}%
\bibitem [{\citenamefont {Aidelsburger}\ \emph {et~al.}(2015)\citenamefont
  {Aidelsburger}, \citenamefont {Lohse}, \citenamefont {Schweizer},
  \citenamefont {Atala}, \citenamefont {Barreiro}, \citenamefont
  {Nascimb\`ene}, \citenamefont {Cooper}, \citenamefont {Bloch},\ and\
  \citenamefont {Goldman}}]{Aidelsburger-15}%
  \BibitemOpen
  \bibfield  {author} {\bibinfo {author} {\bibfnamefont {M.}~\bibnamefont
  {Aidelsburger}}, \bibinfo {author} {\bibfnamefont {M.}~\bibnamefont {Lohse}},
  \bibinfo {author} {\bibfnamefont {C.}~\bibnamefont {Schweizer}}, \bibinfo
  {author} {\bibfnamefont {M.}~\bibnamefont {Atala}}, \bibinfo {author}
  {\bibfnamefont {J.~T.}\ \bibnamefont {Barreiro}}, \bibinfo {author}
  {\bibfnamefont {S.}~\bibnamefont {Nascimb\`ene}}, \bibinfo {author}
  {\bibfnamefont {N.~R.}\ \bibnamefont {Cooper}}, \bibinfo {author}
  {\bibfnamefont {I.}~\bibnamefont {Bloch}}, \ and\ \bibinfo {author}
  {\bibfnamefont {N.}~\bibnamefont {Goldman}},\ }\bibfield  {title} {\enquote
  {\bibinfo {title} {{Measuring the Chern Number of Hofstadter Bands with
  Ultracold Bosonic Atoms}},}\ }\href@noop {} {\bibfield  {journal} {\bibinfo
  {journal} {Nat. Phys.}\ }\textbf {\bibinfo {volume} {11}},\ \bibinfo {pages}
  {162} (\bibinfo {year} {2015})}\BibitemShut {NoStop}%
\bibitem [{\citenamefont {Nakajima}\ \emph {et~al.}(2016)\citenamefont
  {Nakajima}, \citenamefont {Tomita}, \citenamefont {Taie}, \citenamefont
  {Ichinose}, \citenamefont {Ozawa}, \citenamefont {Wang}, \citenamefont
  {Troyer},\ and\ \citenamefont {Takahashi}}]{Nakajima-16}%
  \BibitemOpen
  \bibfield  {author} {\bibinfo {author} {\bibfnamefont {S.}~\bibnamefont
  {Nakajima}}, \bibinfo {author} {\bibfnamefont {T.}~\bibnamefont {Tomita}},
  \bibinfo {author} {\bibfnamefont {S.}~\bibnamefont {Taie}}, \bibinfo {author}
  {\bibfnamefont {T.}~\bibnamefont {Ichinose}}, \bibinfo {author}
  {\bibfnamefont {H.}~\bibnamefont {Ozawa}}, \bibinfo {author} {\bibfnamefont
  {L.}~\bibnamefont {Wang}}, \bibinfo {author} {\bibfnamefont {M.}~\bibnamefont
  {Troyer}}, \ and\ \bibinfo {author} {\bibfnamefont {Y.}~\bibnamefont
  {Takahashi}},\ }\bibfield  {title} {\enquote {\bibinfo {title} {{Topological
  Thouless Pumping of Ultracold Fermions}},}\ }\href@noop {} {\bibfield
  {journal} {\bibinfo  {journal} {Nat. Phys.}\ }\textbf {\bibinfo {volume}
  {12}},\ \bibinfo {pages} {296} (\bibinfo {year} {2016})}\BibitemShut
  {NoStop}%
\bibitem [{\citenamefont {Lohse}\ \emph {et~al.}(2016)\citenamefont {Lohse},
  \citenamefont {Schweizer}, \citenamefont {Zilberberg}, \citenamefont
  {Aidelsburger},\ and\ \citenamefont {Bloch}}]{Lohse-16}%
  \BibitemOpen
  \bibfield  {author} {\bibinfo {author} {\bibfnamefont {M.}~\bibnamefont
  {Lohse}}, \bibinfo {author} {\bibfnamefont {C.}~\bibnamefont {Schweizer}},
  \bibinfo {author} {\bibfnamefont {O.}~\bibnamefont {Zilberberg}}, \bibinfo
  {author} {\bibfnamefont {M.}~\bibnamefont {Aidelsburger}}, \ and\ \bibinfo
  {author} {\bibfnamefont {I.}~\bibnamefont {Bloch}},\ }\bibfield  {title}
  {\enquote {\bibinfo {title} {{A Thouless Quantum Pump with Ultracold Bosonic
  Atoms in an Optical Superlattice}},}\ }\href@noop {} {\bibfield  {journal}
  {\bibinfo  {journal} {Nat. Phys.}\ }\textbf {\bibinfo {volume} {12}},\
  \bibinfo {pages} {350} (\bibinfo {year} {2016})}\BibitemShut {NoStop}%
\bibitem [{\citenamefont {Goldman}\ \emph {et~al.}(2016)\citenamefont
  {Goldman}, \citenamefont {Budich},\ and\ \citenamefont
  {Zoller}}]{Goldman-review}%
  \BibitemOpen
  \bibfield  {author} {\bibinfo {author} {\bibfnamefont {N.}~\bibnamefont
  {Goldman}}, \bibinfo {author} {\bibfnamefont {J.~C.}\ \bibnamefont {Budich}},
  \ and\ \bibinfo {author} {\bibfnamefont {P.}~\bibnamefont {Zoller}},\
  }\bibfield  {title} {\enquote {\bibinfo {title} {{Topological Quantum Matter
  with Ultracold Gases in Optical Lattices}},}\ }\href@noop {} {\bibfield
  {journal} {\bibinfo  {journal} {Nat. Phys.}\ }\textbf {\bibinfo {volume}
  {12}},\ \bibinfo {pages} {639} (\bibinfo {year} {2016})}\BibitemShut
  {NoStop}%
\bibitem [{\citenamefont {Cooper}\ \emph {et~al.}(2019)\citenamefont {Cooper},
  \citenamefont {Dalibard},\ and\ \citenamefont {Spielman}}]{Cooper-review}%
  \BibitemOpen
  \bibfield  {author} {\bibinfo {author} {\bibfnamefont {N.~R.}\ \bibnamefont
  {Cooper}}, \bibinfo {author} {\bibfnamefont {J.}~\bibnamefont {Dalibard}}, \
  and\ \bibinfo {author} {\bibfnamefont {I.~B.}\ \bibnamefont {Spielman}},\
  }\bibfield  {title} {\enquote {\bibinfo {title} {{Topological Bands for
  Ultracold Atoms}},}\ }\href@noop {} {\bibfield  {journal} {\bibinfo
  {journal} {Rev. Mod. Phys.}\ }\textbf {\bibinfo {volume} {91}},\ \bibinfo
  {pages} {015005} (\bibinfo {year} {2019})}\BibitemShut {NoStop}%
\bibitem [{\citenamefont {Altland}\ and\ \citenamefont
  {Zirnbauer}(1997)}]{AZ-97}%
  \BibitemOpen
  \bibfield  {author} {\bibinfo {author} {\bibfnamefont {A.}~\bibnamefont
  {Altland}}\ and\ \bibinfo {author} {\bibfnamefont {M.~R.}\ \bibnamefont
  {Zirnbauer}},\ }\bibfield  {title} {\enquote {\bibinfo {title} {{N}onstandard
  {S}ymmetry {C}lasses in {M}esoscopic {N}ormal-{S}uperconducting {H}ybrid
  {S}tructures},}\ }\href@noop {} {\bibfield  {journal} {\bibinfo  {journal}
  {Phys. Rev. B}\ }\textbf {\bibinfo {volume} {55}},\ \bibinfo {pages} {1142}
  (\bibinfo {year} {1997})}\BibitemShut {NoStop}%
\bibitem [{\citenamefont {Schnyder}\ \emph {et~al.}(2008)\citenamefont
  {Schnyder}, \citenamefont {Ryu}, \citenamefont {Furusaki},\ and\
  \citenamefont {Ludwig}}]{Schnyder-08}%
  \BibitemOpen
  \bibfield  {author} {\bibinfo {author} {\bibfnamefont {A.~P.}\ \bibnamefont
  {Schnyder}}, \bibinfo {author} {\bibfnamefont {S.}~\bibnamefont {Ryu}},
  \bibinfo {author} {\bibfnamefont {A.}~\bibnamefont {Furusaki}}, \ and\
  \bibinfo {author} {\bibfnamefont {A.~W.~W.}\ \bibnamefont {Ludwig}},\
  }\bibfield  {title} {\enquote {\bibinfo {title} {{Classification of
  Topological Insulators and Superconductors in Three Spatial Dimensions}},}\
  }\href@noop {} {\bibfield  {journal} {\bibinfo  {journal} {Phys. Rev. B}\
  }\textbf {\bibinfo {volume} {78}},\ \bibinfo {pages} {195125(R)} (\bibinfo
  {year} {2008})}\BibitemShut {NoStop}%
\bibitem [{\citenamefont {Kitaev}(2009)}]{Kitaev-09}%
  \BibitemOpen
  \bibfield  {author} {\bibinfo {author} {\bibfnamefont {A.}~\bibnamefont
  {Kitaev}},\ }\bibfield  {title} {\enquote {\bibinfo {title} {{Periodic Table
  for Topological Insulators and Superconductors}},}\ }\href@noop {} {\bibfield
   {journal} {\bibinfo  {journal} {AIP Conf. Proc.}\ }\textbf {\bibinfo
  {volume} {1134}},\ \bibinfo {pages} {22} (\bibinfo {year}
  {2009})}\BibitemShut {NoStop}%
\bibitem [{\citenamefont {Ryu}\ \emph {et~al.}(2010)\citenamefont {Ryu},
  \citenamefont {Schnyder}, \citenamefont {Furusaki},\ and\ \citenamefont
  {Ludwig}}]{Ryu-10}%
  \BibitemOpen
  \bibfield  {author} {\bibinfo {author} {\bibfnamefont {S.}~\bibnamefont
  {Ryu}}, \bibinfo {author} {\bibfnamefont {A.~P.}\ \bibnamefont {Schnyder}},
  \bibinfo {author} {\bibfnamefont {A.}~\bibnamefont {Furusaki}}, \ and\
  \bibinfo {author} {\bibfnamefont {A.~W.~W.}\ \bibnamefont {Ludwig}},\
  }\bibfield  {title} {\enquote {\bibinfo {title} {{Topological Insulators and
  Superconductors: Tenfold Way and Dimensional Hierarchy}},}\ }\href@noop {}
  {\bibfield  {journal} {\bibinfo  {journal} {New J. Phys.}\ }\textbf {\bibinfo
  {volume} {12}},\ \bibinfo {pages} {065010} (\bibinfo {year}
  {2010})}\BibitemShut {NoStop}%
\bibitem [{\citenamefont {Teo}\ and\ \citenamefont {Kane}(2010)}]{Teo-10}%
  \BibitemOpen
  \bibfield  {author} {\bibinfo {author} {\bibfnamefont {J.~C.~Y.}\
  \bibnamefont {Teo}}\ and\ \bibinfo {author} {\bibfnamefont {C.~L.}\
  \bibnamefont {Kane}},\ }\bibfield  {title} {\enquote {\bibinfo {title}
  {{Topological Defects and Gapless Modes in Insulators and
  Superconductors}},}\ }\href@noop {} {\bibfield  {journal} {\bibinfo
  {journal} {Phys. Rev. B}\ }\textbf {\bibinfo {volume} {82}},\ \bibinfo
  {pages} {115120} (\bibinfo {year} {2010})}\BibitemShut {NoStop}%
\bibitem [{\citenamefont {Slager}\ \emph {et~al.}(2013)\citenamefont {Slager},
  \citenamefont {Mesaros}, \citenamefont {Juri\u{u}i\'c},\ and\ \citenamefont
  {Zaanen}}]{Slager-13}%
  \BibitemOpen
  \bibfield  {author} {\bibinfo {author} {\bibfnamefont {R.-J.}\ \bibnamefont
  {Slager}}, \bibinfo {author} {\bibfnamefont {A.}~\bibnamefont {Mesaros}},
  \bibinfo {author} {\bibfnamefont {V.}~\bibnamefont {Juri\u{u}i\'c}}, \ and\
  \bibinfo {author} {\bibfnamefont {J.}~\bibnamefont {Zaanen}},\ }\bibfield
  {title} {\enquote {\bibinfo {title} {{The Space Group Classification of
  Topological Band-Insulators}},}\ }\href@noop {} {\bibfield  {journal}
  {\bibinfo  {journal} {Nat. Phys.}\ }\textbf {\bibinfo {volume} {9}},\
  \bibinfo {pages} {98} (\bibinfo {year} {2013})}\BibitemShut {NoStop}%
\bibitem [{\citenamefont {Chiu}\ \emph {et~al.}(2013)\citenamefont {Chiu},
  \citenamefont {Yao},\ and\ \citenamefont {Ryu}}]{Chiu-13}%
  \BibitemOpen
  \bibfield  {author} {\bibinfo {author} {\bibfnamefont {C.-K.}\ \bibnamefont
  {Chiu}}, \bibinfo {author} {\bibfnamefont {H.}~\bibnamefont {Yao}}, \ and\
  \bibinfo {author} {\bibfnamefont {S.}~\bibnamefont {Ryu}},\ }\bibfield
  {title} {\enquote {\bibinfo {title} {{Classification of Topological
  Insulators and Superconductors in the Presence of Reflection Symmetry}},}\
  }\href@noop {} {\bibfield  {journal} {\bibinfo  {journal} {Phys. Rev. B}\
  }\textbf {\bibinfo {volume} {88}},\ \bibinfo {pages} {075142} (\bibinfo
  {year} {2013})}\BibitemShut {NoStop}%
\bibitem [{\citenamefont {Morimoto}\ and\ \citenamefont
  {Furusaki}(2013)}]{Morimoto-13}%
  \BibitemOpen
  \bibfield  {author} {\bibinfo {author} {\bibfnamefont {T.}~\bibnamefont
  {Morimoto}}\ and\ \bibinfo {author} {\bibfnamefont {A.}~\bibnamefont
  {Furusaki}},\ }\bibfield  {title} {\enquote {\bibinfo {title} {{Topological
  Classification with Additional Symmetries from Clifford Algebra}},}\
  }\href@noop {} {\bibfield  {journal} {\bibinfo  {journal} {Phys. Rev. B}\
  }\textbf {\bibinfo {volume} {88}},\ \bibinfo {pages} {125129} (\bibinfo
  {year} {2013})}\BibitemShut {NoStop}%
\bibitem [{\citenamefont {Shiozaki}\ and\ \citenamefont
  {Sato}(2014)}]{Shiozaki-Sato-14}%
  \BibitemOpen
  \bibfield  {author} {\bibinfo {author} {\bibfnamefont {K.}~\bibnamefont
  {Shiozaki}}\ and\ \bibinfo {author} {\bibfnamefont {M.}~\bibnamefont
  {Sato}},\ }\bibfield  {title} {\enquote {\bibinfo {title} {{Topology of
  Crystalline Insulators and Superconductors}},}\ }\href@noop {} {\bibfield
  {journal} {\bibinfo  {journal} {Phys. Rev. B}\ }\textbf {\bibinfo {volume}
  {90}},\ \bibinfo {pages} {165114} (\bibinfo {year} {2014})}\BibitemShut
  {NoStop}%
\bibitem [{\citenamefont {Shiozaki}\ \emph {et~al.}(2016)\citenamefont
  {Shiozaki}, \citenamefont {Sato},\ and\ \citenamefont {Gomi}}]{Shiozaki-16}%
  \BibitemOpen
  \bibfield  {author} {\bibinfo {author} {\bibfnamefont {K.}~\bibnamefont
  {Shiozaki}}, \bibinfo {author} {\bibfnamefont {M.}~\bibnamefont {Sato}}, \
  and\ \bibinfo {author} {\bibfnamefont {K.}~\bibnamefont {Gomi}},\ }\bibfield
  {title} {\enquote {\bibinfo {title} {{Topology of Nonsymmorphic Crystalline
  Insulators and Superconductors}},}\ }\href@noop {} {\bibfield  {journal}
  {\bibinfo  {journal} {Phys. Rev. B}\ }\textbf {\bibinfo {volume} {93}},\
  \bibinfo {pages} {195413} (\bibinfo {year} {2016})}\BibitemShut {NoStop}%
\bibitem [{\citenamefont {Shiozaki}\ \emph {et~al.}(2017)\citenamefont
  {Shiozaki}, \citenamefont {Sato},\ and\ \citenamefont {Gomi}}]{Shiozaki-17}%
  \BibitemOpen
  \bibfield  {author} {\bibinfo {author} {\bibfnamefont {K.}~\bibnamefont
  {Shiozaki}}, \bibinfo {author} {\bibfnamefont {M.}~\bibnamefont {Sato}}, \
  and\ \bibinfo {author} {\bibfnamefont {K.}~\bibnamefont {Gomi}},\ }\bibfield
  {title} {\enquote {\bibinfo {title} {{Topological Crystalline Materials:
  General Formulation, Module Structure, and Wallpaper Groups}},}\ }\href@noop
  {} {\bibfield  {journal} {\bibinfo  {journal} {{Phys. Rev. B}}\ }\textbf
  {\bibinfo {volume} {95}},\ \bibinfo {pages} {235425} (\bibinfo {year}
  {2017})}\BibitemShut {NoStop}%
\bibitem [{\citenamefont {Shiozaki}\ \emph {et~al.}()\citenamefont {Shiozaki},
  \citenamefont {Sato},\ and\ \citenamefont {Gomi}}]{Shiozaki-18}%
  \BibitemOpen
  \bibfield  {author} {\bibinfo {author} {\bibfnamefont {K.}~\bibnamefont
  {Shiozaki}}, \bibinfo {author} {\bibfnamefont {M.}~\bibnamefont {Sato}}, \
  and\ \bibinfo {author} {\bibfnamefont {K.}~\bibnamefont {Gomi}},\ }\href@noop
  {} {\enquote {\bibinfo {title} {{Atiyah-Hirzebruch Spectral Sequence in Band
  Topology: General Formalism and Topological Invariants for 230 Space
  Groups}},}\ }\bibinfo {note} {{arXiv: 1802.06694.}}\BibitemShut {Stop}%
\bibitem [{\citenamefont {Po}\ \emph {et~al.}(2017)\citenamefont {Po},
  \citenamefont {Vishwanath},\ and\ \citenamefont {Watanabe}}]{Po-17}%
  \BibitemOpen
  \bibfield  {author} {\bibinfo {author} {\bibfnamefont {H.~C.}\ \bibnamefont
  {Po}}, \bibinfo {author} {\bibfnamefont {A.}~\bibnamefont {Vishwanath}}, \
  and\ \bibinfo {author} {\bibfnamefont {H.}~\bibnamefont {Watanabe}},\
  }\bibfield  {title} {\enquote {\bibinfo {title} {{Symmetry-Based Indicators
  of Band Topology in the 230 Space Groups}},}\ }\href@noop {} {\bibfield
  {journal} {\bibinfo  {journal} {Nat. Commun.}\ }\textbf {\bibinfo {volume}
  {8}},\ \bibinfo {pages} {50} (\bibinfo {year} {2017})}\BibitemShut {NoStop}%
\bibitem [{\citenamefont {Watanabe}\ \emph {et~al.}(2018)\citenamefont
  {Watanabe}, \citenamefont {Po},\ and\ \citenamefont
  {Vishwanath}}]{Watanabe-18}%
  \BibitemOpen
  \bibfield  {author} {\bibinfo {author} {\bibfnamefont {H.}~\bibnamefont
  {Watanabe}}, \bibinfo {author} {\bibfnamefont {H.~C.}\ \bibnamefont {Po}}, \
  and\ \bibinfo {author} {\bibfnamefont {A.}~\bibnamefont {Vishwanath}},\
  }\bibfield  {title} {\enquote {\bibinfo {title} {{Structure and Topology of
  Band Structures in the 1651 Magnetic Space Groups}},}\ }\href@noop {}
  {\bibfield  {journal} {\bibinfo  {journal} {Sci. Adv.}\ }\textbf {\bibinfo
  {volume} {4}},\ \bibinfo {pages} {eaat8685} (\bibinfo {year}
  {2018})}\BibitemShut {NoStop}%
\bibitem [{\citenamefont {Ono}\ and\ \citenamefont {Watanabe}(2018)}]{Ono-18}%
  \BibitemOpen
  \bibfield  {author} {\bibinfo {author} {\bibfnamefont {S.}~\bibnamefont
  {Ono}}\ and\ \bibinfo {author} {\bibfnamefont {H.}~\bibnamefont {Watanabe}},\
  }\bibfield  {title} {\enquote {\bibinfo {title} {{Unified Understanding of
  Symmetry Indicators for All Internal Symmetry Classes}},}\ }\href@noop {}
  {\bibfield  {journal} {\bibinfo  {journal} {Phys. Rev. B}\ }\textbf {\bibinfo
  {volume} {98}},\ \bibinfo {pages} {115150} (\bibinfo {year}
  {2018})}\BibitemShut {NoStop}%
\bibitem [{\citenamefont {Tang}\ \emph {et~al.}(2019)\citenamefont {Tang},
  \citenamefont {Po}, \citenamefont {Vishwanath},\ and\ \citenamefont
  {Wan}}]{Tang-19}%
  \BibitemOpen
  \bibfield  {author} {\bibinfo {author} {\bibfnamefont {F.}~\bibnamefont
  {Tang}}, \bibinfo {author} {\bibfnamefont {H.~C.}\ \bibnamefont {Po}},
  \bibinfo {author} {\bibfnamefont {A.}~\bibnamefont {Vishwanath}}, \ and\
  \bibinfo {author} {\bibfnamefont {X.}~\bibnamefont {Wan}},\ }\bibfield
  {title} {\enquote {\bibinfo {title} {{Comprehensive Search for Topological
  Materials Using Symmetry Indicators}},}\ }\href@noop {} {\bibfield  {journal}
  {\bibinfo  {journal} {Nature}\ }\textbf {\bibinfo {volume} {566}},\ \bibinfo
  {pages} {486} (\bibinfo {year} {2019})}\BibitemShut {NoStop}%
\bibitem [{\citenamefont {Ono}\ \emph {et~al.}()\citenamefont {Ono},
  \citenamefont {Yanase},\ and\ \citenamefont {Watanabe}}]{OYW-18}%
  \BibitemOpen
  \bibfield  {author} {\bibinfo {author} {\bibfnamefont {S.}~\bibnamefont
  {Ono}}, \bibinfo {author} {\bibfnamefont {Y.}~\bibnamefont {Yanase}}, \ and\
  \bibinfo {author} {\bibfnamefont {H.}~\bibnamefont {Watanabe}},\ }\href@noop
  {} {\enquote {\bibinfo {title} {{Symmetry Indicators for Topological
  Superconductors}},}\ }\bibinfo {note} {{arXiv: 1811.08712.}}\BibitemShut
  {Stop}%
\bibitem [{\citenamefont {Bradlyn}\ \emph {et~al.}(2017)\citenamefont
  {Bradlyn}, \citenamefont {Elcoro}, \citenamefont {Cano}, \citenamefont
  {Vergniory}, \citenamefont {Wang}, \citenamefont {Felser}, \citenamefont
  {Aroyo},\ and\ \citenamefont {Bernevig}}]{Bradlyn-17}%
  \BibitemOpen
  \bibfield  {author} {\bibinfo {author} {\bibfnamefont {B.}~\bibnamefont
  {Bradlyn}}, \bibinfo {author} {\bibfnamefont {L.}~\bibnamefont {Elcoro}},
  \bibinfo {author} {\bibfnamefont {J.}~\bibnamefont {Cano}}, \bibinfo {author}
  {\bibfnamefont {M.~G.}\ \bibnamefont {Vergniory}}, \bibinfo {author}
  {\bibfnamefont {Z.}~\bibnamefont {Wang}}, \bibinfo {author} {\bibfnamefont
  {C.}~\bibnamefont {Felser}}, \bibinfo {author} {\bibfnamefont {M.~I.}\
  \bibnamefont {Aroyo}}, \ and\ \bibinfo {author} {\bibfnamefont {B.~A.}\
  \bibnamefont {Bernevig}},\ }\bibfield  {title} {\enquote {\bibinfo {title}
  {{Topological Quantum Chemistry}},}\ }\href@noop {} {\bibfield  {journal}
  {\bibinfo  {journal} {Nature}\ }\textbf {\bibinfo {volume} {547}},\ \bibinfo
  {pages} {298} (\bibinfo {year} {2017})}\BibitemShut {NoStop}%
\bibitem [{\citenamefont {Vergniory}\ \emph {et~al.}(2019)\citenamefont
  {Vergniory}, \citenamefont {Elcoro}, \citenamefont {Felser}, \citenamefont
  {Regnault}, \citenamefont {Bernevig},\ and\ \citenamefont
  {Wang}}]{Vergniory-19}%
  \BibitemOpen
  \bibfield  {author} {\bibinfo {author} {\bibfnamefont {M.~G.}\ \bibnamefont
  {Vergniory}}, \bibinfo {author} {\bibfnamefont {L.}~\bibnamefont {Elcoro}},
  \bibinfo {author} {\bibfnamefont {C.}~\bibnamefont {Felser}}, \bibinfo
  {author} {\bibfnamefont {N.}~\bibnamefont {Regnault}}, \bibinfo {author}
  {\bibfnamefont {B.~A.}\ \bibnamefont {Bernevig}}, \ and\ \bibinfo {author}
  {\bibfnamefont {Z.}~\bibnamefont {Wang}},\ }\bibfield  {title} {\enquote
  {\bibinfo {title} {{A Complete Catalogue of High-Quality Topological
  Materials}},}\ }\href@noop {} {\bibfield  {journal} {\bibinfo  {journal}
  {{Nature}}\ }\textbf {\bibinfo {volume} {566}},\ \bibinfo {pages} {480}
  (\bibinfo {year} {2019})}\BibitemShut {NoStop}%
\bibitem [{\citenamefont {Kruthoff}\ \emph {et~al.}(2017)\citenamefont
  {Kruthoff}, \citenamefont {de~Boer}, \citenamefont {van Wezel}, \citenamefont
  {Kane},\ and\ \citenamefont {Slager}}]{Kruthoff-17}%
  \BibitemOpen
  \bibfield  {author} {\bibinfo {author} {\bibfnamefont {J.}~\bibnamefont
  {Kruthoff}}, \bibinfo {author} {\bibfnamefont {J.}~\bibnamefont {de~Boer}},
  \bibinfo {author} {\bibfnamefont {J.}~\bibnamefont {van Wezel}}, \bibinfo
  {author} {\bibfnamefont {C.~L.}\ \bibnamefont {Kane}}, \ and\ \bibinfo
  {author} {\bibfnamefont {R.-J.}\ \bibnamefont {Slager}},\ }\bibfield  {title}
  {\enquote {\bibinfo {title} {{Topological Classification of Crystalline
  Insulators through Band Structure Combinatorics}},}\ }\href@noop {}
  {\bibfield  {journal} {\bibinfo  {journal} {Phys. Rev. X}\ }\textbf {\bibinfo
  {volume} {7}},\ \bibinfo {pages} {041069} (\bibinfo {year}
  {2017})}\BibitemShut {NoStop}%
\bibitem [{\citenamefont {Zhang}\ \emph {et~al.}(2019)\citenamefont {Zhang},
  \citenamefont {Jiang}, \citenamefont {Song}, \citenamefont {Huang},
  \citenamefont {He}, \citenamefont {Fang},\ and\ \citenamefont
  {H.~Weng}}]{Zhang-19}%
  \BibitemOpen
  \bibfield  {author} {\bibinfo {author} {\bibfnamefont {T.}~\bibnamefont
  {Zhang}}, \bibinfo {author} {\bibfnamefont {Y.}~\bibnamefont {Jiang}},
  \bibinfo {author} {\bibfnamefont {Z.}~\bibnamefont {Song}}, \bibinfo {author}
  {\bibfnamefont {H.}~\bibnamefont {Huang}}, \bibinfo {author} {\bibfnamefont
  {Y.}~\bibnamefont {He}}, \bibinfo {author} {\bibfnamefont {Z.}~\bibnamefont
  {Fang}}, \ and\ \bibinfo {author} {\bibfnamefont {C.~Fang}\ \bibnamefont
  {H.~Weng}},\ }\bibfield  {title} {\enquote {\bibinfo {title} {{Catalogue of
  Topological Electronic Materials}},}\ }\href@noop {} {\bibfield  {journal}
  {\bibinfo  {journal} {Nature}\ }\textbf {\bibinfo {volume} {566}},\ \bibinfo
  {pages} {475} (\bibinfo {year} {2019})}\BibitemShut {NoStop}%
\bibitem [{\citenamefont {Chiu}\ \emph {et~al.}(2016)\citenamefont {Chiu},
  \citenamefont {Teo}, \citenamefont {Schnyder},\ and\ \citenamefont
  {Ryu}}]{Schnyder-Ryu-review}%
  \BibitemOpen
  \bibfield  {author} {\bibinfo {author} {\bibfnamefont {C.-K.}\ \bibnamefont
  {Chiu}}, \bibinfo {author} {\bibfnamefont {J.~C.~Y.}\ \bibnamefont {Teo}},
  \bibinfo {author} {\bibfnamefont {A.~P.}\ \bibnamefont {Schnyder}}, \ and\
  \bibinfo {author} {\bibfnamefont {S.}~\bibnamefont {Ryu}},\ }\bibfield
  {title} {\enquote {\bibinfo {title} {{Classification of Topological Quantum
  Matter with Symmetries}},}\ }\href@noop {} {\bibfield  {journal} {\bibinfo
  {journal} {Rev. Mod. Phys.}\ }\textbf {\bibinfo {volume} {88}},\ \bibinfo
  {pages} {035005} (\bibinfo {year} {2016})}\BibitemShut {NoStop}%
\bibitem [{\citenamefont
  {Mostafazadeh}(2002{\natexlab{a}})}]{Mostafazadeh-02-1}%
  \BibitemOpen
  \bibfield  {author} {\bibinfo {author} {\bibfnamefont {A.}~\bibnamefont
  {Mostafazadeh}},\ }\bibfield  {title} {\enquote {\bibinfo {title}
  {{Pseudo-Hermiticity versus PT Symmetry: The Necessary Condition for the
  Reality of the Spectrum of a Non-Hermitian Hamiltonian}},}\ }\href@noop {}
  {\bibfield  {journal} {\bibinfo  {journal} {J. Math. Phys.}\ }\textbf
  {\bibinfo {volume} {43}},\ \bibinfo {pages} {205} (\bibinfo {year}
  {2002}{\natexlab{a}})}\BibitemShut {NoStop}%
\bibitem [{\citenamefont
  {Mostafazadeh}(2002{\natexlab{b}})}]{Mostafazadeh-02-2}%
  \BibitemOpen
  \bibfield  {author} {\bibinfo {author} {\bibfnamefont {A.}~\bibnamefont
  {Mostafazadeh}},\ }\bibfield  {title} {\enquote {\bibinfo {title}
  {{Pseudo-Hermiticity versus PT Symmetry II: A Complete Characterization of
  Non-Hermitian Hamiltonians with a Real Spectrum}},}\ }\href@noop {}
  {\bibfield  {journal} {\bibinfo  {journal} {{J. Math. Phys.}}\ }\textbf
  {\bibinfo {volume} {43}},\ \bibinfo {pages} {2814} (\bibinfo {year}
  {2002}{\natexlab{b}})}\BibitemShut {NoStop}%
\bibitem [{\citenamefont
  {Mostafazadeh}(2002{\natexlab{c}})}]{Mostafazadeh-02-3}%
  \BibitemOpen
  \bibfield  {author} {\bibinfo {author} {\bibfnamefont {A.}~\bibnamefont
  {Mostafazadeh}},\ }\bibfield  {title} {\enquote {\bibinfo {title}
  {{Pseudo-Hermiticity versus PT Symmetry III: Equivalence of
  Pseudo-Hermiticity and the Presence of Antilinear Symmetries}},}\ }\href@noop
  {} {\bibfield  {journal} {\bibinfo  {journal} {J. Math. Phys.}\ }\textbf
  {\bibinfo {volume} {43}},\ \bibinfo {pages} {3944} (\bibinfo {year}
  {2002}{\natexlab{c}})}\BibitemShut {NoStop}%
\bibitem [{\citenamefont {Mostafazadeh}(2003)}]{Mostafazadeh-03}%
  \BibitemOpen
  \bibfield  {author} {\bibinfo {author} {\bibfnamefont {A.}~\bibnamefont
  {Mostafazadeh}},\ }\bibfield  {title} {\enquote {\bibinfo {title} {{Exact
  \textit{PT}-Symmetry is Equivalent to Hermiticity}},}\ }\href@noop {}
  {\bibfield  {journal} {\bibinfo  {journal} {J. Phys. A}\ }\textbf {\bibinfo
  {volume} {36}},\ \bibinfo {pages} {7081} (\bibinfo {year}
  {2003})}\BibitemShut {NoStop}%
\bibitem [{\citenamefont {Brody}(2016)}]{Brody-16}%
  \BibitemOpen
  \bibfield  {author} {\bibinfo {author} {\bibfnamefont {D.~C.}\ \bibnamefont
  {Brody}},\ }\bibfield  {title} {\enquote {\bibinfo {title} {{Consistency of
  PT-Symmetric Quantum Mechanics}},}\ }\href@noop {} {\bibfield  {journal}
  {\bibinfo  {journal} {J. Phys. A}\ }\textbf {\bibinfo {volume} {49}},\
  \bibinfo {pages} {10LT03} (\bibinfo {year} {2016})}\BibitemShut {NoStop}%
\bibitem [{Ber()}]{Bernard-LeClair-02}%
  \BibitemOpen
  \href@noop {} {}\bibinfo {note} {{D. Bernard and A. LeClair, ``A
  Classification of Non-Hermitian Random Matrices," in {\it Statistical Field
  Theories} edited by A. Cappelli and G. Mussardo (Springer, Dordrecht,
  2002).}}\BibitemShut {Stop}%
\bibitem [{\citenamefont {Magnea}(2008)}]{Magnea-08}%
  \BibitemOpen
  \bibfield  {author} {\bibinfo {author} {\bibfnamefont {U.}~\bibnamefont
  {Magnea}},\ }\bibfield  {title} {\enquote {\bibinfo {title} {{Random Matrices
  beyond the Cartan Classification}},}\ }\href@noop {} {\bibfield  {journal}
  {\bibinfo  {journal} {J. Phys. A}\ }\textbf {\bibinfo {volume} {41}},\
  \bibinfo {pages} {045203} (\bibinfo {year} {2008})}\BibitemShut {NoStop}%
\bibitem [{\citenamefont {Karoubi}(2008)}]{Karoubi}%
  \BibitemOpen
  \bibfield  {author} {\bibinfo {author} {\bibfnamefont {M.}~\bibnamefont
  {Karoubi}},\ }\href@noop {} {\emph {\bibinfo {title} {{\textit{K}-theory: An
  Introduction}}}}\ (\bibinfo  {publisher} {Springer, Berlin},\ \bibinfo {year}
  {2008})\BibitemShut {NoStop}%
\bibitem [{HS-()}]{HS-pc}%
  \BibitemOpen
  \href@noop {} {}\bibinfo {note} {{H. Schomerus (private
  communication).}}\BibitemShut {Stop}%
\bibitem [{\citenamefont {Roy}\ and\ \citenamefont {Harper}(2017)}]{Roy-17}%
  \BibitemOpen
  \bibfield  {author} {\bibinfo {author} {\bibfnamefont {R.}~\bibnamefont
  {Roy}}\ and\ \bibinfo {author} {\bibfnamefont {F.}~\bibnamefont {Harper}},\
  }\bibfield  {title} {\enquote {\bibinfo {title} {{Periodic Table for Floquet
  Topological Insulators}},}\ }\href@noop {} {\bibfield  {journal} {\bibinfo
  {journal} {Phys. Rev. B}\ }\textbf {\bibinfo {volume} {96}},\ \bibinfo
  {pages} {155118} (\bibinfo {year} {2017})}\BibitemShut {NoStop}%
\bibitem [{\citenamefont {Ryu}\ and\ \citenamefont {Hatsugai}(2002)}]{Ryu-02}%
  \BibitemOpen
  \bibfield  {author} {\bibinfo {author} {\bibfnamefont {S.}~\bibnamefont
  {Ryu}}\ and\ \bibinfo {author} {\bibfnamefont {Y.}~\bibnamefont {Hatsugai}},\
  }\bibfield  {title} {\enquote {\bibinfo {title} {{Topological Origin of
  Zero-Energy Edge States in Particle-Hole Symmetric Systems}},}\ }\href@noop
  {} {\bibfield  {journal} {\bibinfo  {journal} {Phys. Rev. Lett.}\ }\textbf
  {\bibinfo {volume} {89}},\ \bibinfo {pages} {077002} (\bibinfo {year}
  {2002})}\BibitemShut {NoStop}%
\bibitem [{\citenamefont {Horn}\ and\ \citenamefont
  {Johnson}(1985)}]{MatrixAnalysis}%
  \BibitemOpen
  \bibfield  {author} {\bibinfo {author} {\bibfnamefont {R.~A.}\ \bibnamefont
  {Horn}}\ and\ \bibinfo {author} {\bibfnamefont {C.~R.}\ \bibnamefont
  {Johnson}},\ }\href@noop {} {\emph {\bibinfo {title} {{Matrix Analysis}}}}\
  (\bibinfo  {publisher} {Cambridge University Press, Cambridge},\ \bibinfo
  {year} {1985})\BibitemShut {NoStop}%
\bibitem [{\citenamefont {Nayak}\ \emph {et~al.}(2008)\citenamefont {Nayak},
  \citenamefont {Simon}, \citenamefont {Stern}, \citenamefont {Freedman},\ and\
  \citenamefont {Sarma}}]{Nayak-review}%
  \BibitemOpen
  \bibfield  {author} {\bibinfo {author} {\bibfnamefont {C.}~\bibnamefont
  {Nayak}}, \bibinfo {author} {\bibfnamefont {S.~H.}\ \bibnamefont {Simon}},
  \bibinfo {author} {\bibfnamefont {A.}~\bibnamefont {Stern}}, \bibinfo
  {author} {\bibfnamefont {M.}~\bibnamefont {Freedman}}, \ and\ \bibinfo
  {author} {\bibfnamefont {S.~Das}\ \bibnamefont {Sarma}},\ }\bibfield  {title}
  {\enquote {\bibinfo {title} {{Non-Abelian Anyons and Topological Quantum
  Computation}},}\ }\href@noop {} {\bibfield  {journal} {\bibinfo  {journal}
  {Rev. Mod. Phys.}\ }\textbf {\bibinfo {volume} {80}},\ \bibinfo {pages}
  {1083} (\bibinfo {year} {2008})}\BibitemShut {NoStop}%
\bibitem [{\citenamefont {Zhou}\ and\ \citenamefont {Lee}(2019)}]{ZL-18}%
  \BibitemOpen
  \bibfield  {author} {\bibinfo {author} {\bibfnamefont {H.}~\bibnamefont
  {Zhou}}\ and\ \bibinfo {author} {\bibfnamefont {J.~Y.}\ \bibnamefont {Lee}},\
  }\bibfield  {title} {\enquote {\bibinfo {title} {{Periodic Table for
  Topological Bands with Non-Hermitian Symmetries}},}\ }\href@noop {}
  {\bibfield  {journal} {\bibinfo  {journal} {Phys. Rev. B}\ }\textbf {\bibinfo
  {volume} {99}},\ \bibinfo {pages} {235112} (\bibinfo {year}
  {2019})}\BibitemShut {NoStop}%
\bibitem [{Lee()}]{Lee-pc}%
  \BibitemOpen
  \href@noop {} {}\bibinfo {note} {{J. Y. Lee (private
  communication).}}\BibitemShut {Stop}%
\bibitem [{\citenamefont {Freed}\ and\ \citenamefont {Moore}(2013)}]{Freed-13}%
  \BibitemOpen
  \bibfield  {author} {\bibinfo {author} {\bibfnamefont {D.~S.}\ \bibnamefont
  {Freed}}\ and\ \bibinfo {author} {\bibfnamefont {G.~W.}\ \bibnamefont
  {Moore}},\ }\bibfield  {title} {\enquote {\bibinfo {title} {{Twisted
  Equivariant Matter}},}\ }\href@noop {} {\bibfield  {journal} {\bibinfo
  {journal} {Ann. Henri Poincar\'e}\ }\textbf {\bibinfo {volume} {14}},\
  \bibinfo {pages} {1927} (\bibinfo {year} {2013})}\BibitemShut {NoStop}%
\bibitem [{\citenamefont {Gomi}()}]{Gomi-17}%
  \BibitemOpen
  \bibfield  {author} {\bibinfo {author} {\bibfnamefont {K.}~\bibnamefont
  {Gomi}},\ }\href@noop {} {\enquote {\bibinfo {title} {{Freed-Moore
  K-Theory}},}\ }\bibinfo {note} {{arXiv: 1705.09134}}\BibitemShut {NoStop}%
\bibitem [{\citenamefont {Ishikawa}\ and\ \citenamefont
  {Matsuyama}(1986)}]{Ishikawa-Matsuyama}%
  \BibitemOpen
  \bibfield  {author} {\bibinfo {author} {\bibfnamefont {K.}~\bibnamefont
  {Ishikawa}}\ and\ \bibinfo {author} {\bibfnamefont {T.}~\bibnamefont
  {Matsuyama}},\ }\bibfield  {title} {\enquote {\bibinfo {title} {{Magnetic
  Field Induced Multi-Component $\mathrm{QED}_{3}$ and Quantum Hall Effect}},}\
  }\href@noop {} {\bibfield  {journal} {\bibinfo  {journal} {Z. Phys. C}\
  }\textbf {\bibinfo {volume} {33}},\ \bibinfo {pages} {41} (\bibinfo {year}
  {1986})}\BibitemShut {NoStop}%
\bibitem [{\citenamefont {Volovik}(2003)}]{Volovik-textbook}%
  \BibitemOpen
  \bibfield  {author} {\bibinfo {author} {\bibfnamefont {G.~E.}\ \bibnamefont
  {Volovik}},\ }\href@noop {} {\emph {\bibinfo {title} {{The Universe in a
  Helium Droplet}}}}\ (\bibinfo  {publisher} {Oxford University Press,
  Oxford},\ \bibinfo {year} {2003})\BibitemShut {NoStop}%
\bibitem [{\citenamefont {Qi}\ \emph {et~al.}(2010)\citenamefont {Qi},
  \citenamefont {Hughes},\ and\ \citenamefont {Zhang}}]{QHZ-10}%
  \BibitemOpen
  \bibfield  {author} {\bibinfo {author} {\bibfnamefont {X.-L.}\ \bibnamefont
  {Qi}}, \bibinfo {author} {\bibfnamefont {T.~L.}\ \bibnamefont {Hughes}}, \
  and\ \bibinfo {author} {\bibfnamefont {S.-C.}\ \bibnamefont {Zhang}},\
  }\bibfield  {title} {\enquote {\bibinfo {title} {{Topological Invariants for
  the Fermi Surface of a Time-Reversal-Invariant Superconductor}},}\
  }\href@noop {} {\bibfield  {journal} {\bibinfo  {journal} {Phys. Rev. B}\
  }\textbf {\bibinfo {volume} {81}},\ \bibinfo {pages} {134508} (\bibinfo
  {year} {2010})}\BibitemShut {NoStop}%
\bibitem [{\citenamefont {Schnyder}\ and\ \citenamefont {Ryu}(2011)}]{SR-11}%
  \BibitemOpen
  \bibfield  {author} {\bibinfo {author} {\bibfnamefont {A.~P.}\ \bibnamefont
  {Schnyder}}\ and\ \bibinfo {author} {\bibfnamefont {S.}~\bibnamefont {Ryu}},\
  }\bibfield  {title} {\enquote {\bibinfo {title} {{Topological Phases and
  Surface Flat Bands in Superconductors without Inversion Symmetry}},}\
  }\href@noop {} {\bibfield  {journal} {\bibinfo  {journal} {Phys. Rev. B}\
  }\textbf {\bibinfo {volume} {84}},\ \bibinfo {pages} {060504} (\bibinfo
  {year} {2011})}\BibitemShut {NoStop}%
\bibitem [{\citenamefont {Budich}\ and\ \citenamefont
  {Ardonne}(2013)}]{Budich-13}%
  \BibitemOpen
  \bibfield  {author} {\bibinfo {author} {\bibfnamefont {J.~C.}\ \bibnamefont
  {Budich}}\ and\ \bibinfo {author} {\bibfnamefont {E.}~\bibnamefont
  {Ardonne}},\ }\bibfield  {title} {\enquote {\bibinfo {title} {{Topological
  Invariant for Generic One-Dimensional Time-Reversal-Symmetric Superconductors
  in Class DIII}},}\ }\href@noop {} {\bibfield  {journal} {\bibinfo  {journal}
  {Phys. Rev. B}\ }\textbf {\bibinfo {volume} {88}},\ \bibinfo {pages} {134523}
  (\bibinfo {year} {2013})}\BibitemShut {NoStop}%
\bibitem [{\citenamefont {Wang}\ \emph {et~al.}(2010)\citenamefont {Wang},
  \citenamefont {Qi},\ and\ \citenamefont {Zhang}}]{WQZ-10}%
  \BibitemOpen
  \bibfield  {author} {\bibinfo {author} {\bibfnamefont {Z.}~\bibnamefont
  {Wang}}, \bibinfo {author} {\bibfnamefont {X.-L.}\ \bibnamefont {Qi}}, \ and\
  \bibinfo {author} {\bibfnamefont {S.-C.}\ \bibnamefont {Zhang}},\ }\bibfield
  {title} {\enquote {\bibinfo {title} {{Equivalent Topological Invariants of
  Topological Insulators}},}\ }\href@noop {} {\bibfield  {journal} {\bibinfo
  {journal} {New J. Phys.}\ }\textbf {\bibinfo {volume} {12}},\ \bibinfo
  {pages} {065007} (\bibinfo {year} {2010})}\BibitemShut {NoStop}%
\bibitem [{\citenamefont {Katsura}\ \emph {et~al.}(2015)\citenamefont
  {Katsura}, \citenamefont {Schuricht},\ and\ \citenamefont
  {Takahashi}}]{Katsura-15}%
  \BibitemOpen
  \bibfield  {author} {\bibinfo {author} {\bibfnamefont {H.}~\bibnamefont
  {Katsura}}, \bibinfo {author} {\bibfnamefont {D.}~\bibnamefont {Schuricht}},
  \ and\ \bibinfo {author} {\bibfnamefont {M.}~\bibnamefont {Takahashi}},\
  }\bibfield  {title} {\enquote {\bibinfo {title} {{Exact Ground States and
  Topological Order in Interacting Kitaev/Majorana Chains}},}\ }\href@noop {}
  {\bibfield  {journal} {\bibinfo  {journal} {Phys. Rev. B}\ }\textbf {\bibinfo
  {volume} {92}},\ \bibinfo {pages} {115137} (\bibinfo {year}
  {2015})}\BibitemShut {NoStop}%
\bibitem [{\citenamefont {Kawabata}\ \emph
  {et~al.}(2017{\natexlab{b}})\citenamefont {Kawabata}, \citenamefont
  {Kobayashi}, \citenamefont {Wu},\ and\ \citenamefont {Katsura}}]{KKWK-17}%
  \BibitemOpen
  \bibfield  {author} {\bibinfo {author} {\bibfnamefont {K.}~\bibnamefont
  {Kawabata}}, \bibinfo {author} {\bibfnamefont {R.}~\bibnamefont {Kobayashi}},
  \bibinfo {author} {\bibfnamefont {N.}~\bibnamefont {Wu}}, \ and\ \bibinfo
  {author} {\bibfnamefont {H.}~\bibnamefont {Katsura}},\ }\bibfield  {title}
  {\enquote {\bibinfo {title} {{Exact Zero Modes in Twisted Kitaev Chains}},}\
  }\href@noop {} {\bibfield  {journal} {\bibinfo  {journal} {{Phys. Rev. B}}\
  }\textbf {\bibinfo {volume} {95}},\ \bibinfo {pages} {195140} (\bibinfo
  {year} {2017}{\natexlab{b}})}\BibitemShut {NoStop}%
\bibitem [{\citenamefont {Alase}\ \emph {et~al.}(2016)\citenamefont {Alase},
  \citenamefont {Cobanera}, \citenamefont {Ortiz},\ and\ \citenamefont
  {Viola}}]{Alase-16}%
  \BibitemOpen
  \bibfield  {author} {\bibinfo {author} {\bibfnamefont {A.}~\bibnamefont
  {Alase}}, \bibinfo {author} {\bibfnamefont {E.}~\bibnamefont {Cobanera}},
  \bibinfo {author} {\bibfnamefont {G.}~\bibnamefont {Ortiz}}, \ and\ \bibinfo
  {author} {\bibfnamefont {L.}~\bibnamefont {Viola}},\ }\bibfield  {title}
  {\enquote {\bibinfo {title} {{Exact Solution of Quadratic Fermionic
  Hamiltonians for Arbitrary Boundary Conditions}},}\ }\href@noop {} {\bibfield
   {journal} {\bibinfo  {journal} {Phys. Rev. Lett.}\ }\textbf {\bibinfo
  {volume} {117}},\ \bibinfo {pages} {076804} (\bibinfo {year}
  {2016})}\BibitemShut {NoStop}%
\bibitem [{\citenamefont {Alase}\ \emph {et~al.}(2017)\citenamefont {Alase},
  \citenamefont {Cobanera}, \citenamefont {Ortiz},\ and\ \citenamefont
  {Viola}}]{Alase-17}%
  \BibitemOpen
  \bibfield  {author} {\bibinfo {author} {\bibfnamefont {A.}~\bibnamefont
  {Alase}}, \bibinfo {author} {\bibfnamefont {E.}~\bibnamefont {Cobanera}},
  \bibinfo {author} {\bibfnamefont {G.}~\bibnamefont {Ortiz}}, \ and\ \bibinfo
  {author} {\bibfnamefont {L.}~\bibnamefont {Viola}},\ }\bibfield  {title}
  {\enquote {\bibinfo {title} {{Generalization of Bloch's Theorem for Arbitrary
  Boundary Conditions: Theory}},}\ }\href@noop {} {\bibfield  {journal}
  {\bibinfo  {journal} {Phys. Rev. B}\ }\textbf {\bibinfo {volume} {96}},\
  \bibinfo {pages} {195133} (\bibinfo {year} {2017})}\BibitemShut {NoStop}%
\bibitem [{\citenamefont {Cobanera}\ \emph {et~al.}(2018)\citenamefont
  {Cobanera}, \citenamefont {Alase}, \citenamefont {Ortiz},\ and\ \citenamefont
  {Viola}}]{Cobanera-18}%
  \BibitemOpen
  \bibfield  {author} {\bibinfo {author} {\bibfnamefont {E.}~\bibnamefont
  {Cobanera}}, \bibinfo {author} {\bibfnamefont {A.}~\bibnamefont {Alase}},
  \bibinfo {author} {\bibfnamefont {G.}~\bibnamefont {Ortiz}}, \ and\ \bibinfo
  {author} {\bibfnamefont {L.}~\bibnamefont {Viola}},\ }\bibfield  {title}
  {\enquote {\bibinfo {title} {{Generalization of Bloch's Theorem for Arbitrary
  Boundary Conditions: Interfaces and Topological Surface Band Structure}},}\
  }\href@noop {} {\bibfield  {journal} {\bibinfo  {journal} {{Phys. Rev. B}}\
  }\textbf {\bibinfo {volume} {98}},\ \bibinfo {pages} {245423} (\bibinfo
  {year} {2018})}\BibitemShut {NoStop}%
\end{thebibliography}%

\end{document}